\documentclass[lettersize,journal]{IEEEtran}
\usepackage{amsmath,amsfonts}
\usepackage{algorithmic}
\usepackage{array}
\usepackage{textcomp}
\usepackage{stfloats}
\usepackage{times}
\usepackage{epsfig}
\usepackage{graphicx}
\usepackage{amsmath}
\usepackage{amssymb}
\usepackage{subcaption}
\usepackage{xcolor}
\usepackage{tikz}
\usetikzlibrary{decorations.pathreplacing}
\usetikzlibrary{arrows, fit, backgrounds, spy, calc}
\usepackage{booktabs}
\usepackage{tabularx}
\usepackage{multirow}
\usepackage{arydshln}
\definecolor{darkgreen}{rgb}{0,0.6,0}
\usepackage{url}
\usepackage{verbatim}
\usepackage{graphicx}
\usepackage{makecell}
\def\BibTeX{{\rm B\kern-.05em{\sc i\kern-.025em b}\kern-.08em
		T\kern-.1667em\lower.7ex\hbox{E}\kern-.125emX}}
\usepackage{balance}

\usepackage{fancyhdr}
\usepackage{textcomp}

\begin{document}
	\title{Disparity Estimation of Planar Reflective Surfaces Using Specular Reflections From a Single Light Source}
	\author{Katja Kossira \IEEEmembership{Graduate Student Member, IEEE}, Frank Sippel \IEEEmembership{Graduate Student Member, IEEE}, J\"urgen Seiler \IEEEmembership{Senior Member, IEEE} and Andr\'e Kaup \IEEEmembership{Fellow, IEEE}
	\thanks{Manuscript created November, 2025; manuscript revised April and August 2026. The authors are with the Chair of Multimedia Communications and Signal Processing, Department of Electrical Engineering, Electronics and Information Technology, Friedrich-Alexander University of Erlangen-Nürnberg (FAU), 91058 Erlangen, Germany.}}
	
 \markboth{IEEE Transactions on Instrumentation and Measurement,~Vol.~75, ~2026}{}
\maketitle
\thispagestyle{fancy}

\begin{abstract}
Multi-camera imaging and camera arrays have become ubiquitous in many applications, such as autonomous driving, robot control, or virtual reality, and accurate disparity maps of objects and their environment are essential for reliable operation. Despite recent advances in neural networks, correctly estimating the disparity of flat and textureless objects remains challenging. In particular, we consider a scenario defined by flat, textureless surfaces illuminated by a single fixed light source, resulting in one dominant specular reflection visible on the object surface. Under these conditions, reliable geometric and photometric cues are missing, and the specular reflection often causes mispredictions, especially when using conventional methods that rely on texture information to match corresponding pixels. To address this issue, the novel Specular Reflection Disparity Estimation SRDE algorithm is introduced, which is specifically designed for the constrained scenario of planar, textureless objects and single-source illumination. Unlike conventional stereo matching methods, SRDE ignores texture and instead leverages the geometric properties of specular reflections by incorporating the position information of the reflective region, the light source, and the camera setup. We show that SRDE outperforms existing methods by a notable margin, achieving more than a 52\% improvement in End Point Error on synthetic images. Further tests demonstrate superior performance on real-world data. Furthermore, we integrate SRDE into existing neural disparity estimation pipelines by selectively replacing predictions in specular regions without modifying the backbone model. This hybrid strategy enables additional performance gains without requiring network retraining.
\end{abstract} 

\vspace{-0.6cm}
\section{Introduction}
\label{sec:Introduction}
\IEEEPARstart{I}{n} modern imaging, multi-camera systems are employed for a wide range of applications. One of the main reasons for using more than one camera is the ability to compute disparity, and consequently the depth of the scene, by analyzing pixel shifts, allowing the extraction of 3D information from multiple 2D images. 

Camera array setups are used in the field of autonomous driving to provide real-time depth information for obstacle detection \cite{AutonomousDriving}, distance estimation \cite{Foggy}, and navigation \cite{AutonomousDriving2}, which is crucial to ensure reliable scene understanding, enhancing the safety and efficiency of self-driving vehicles. In robotics, disparity estimation is employed to enable robots to perceive and understand their environment in three dimensions \cite{Robotics2}. Other applications include multispectral \cite{CAMSI} and hyperspectral \cite{HAHSI} imaging, where disparity estimation is a crucial step to detect, register and reconstruct occluded pixel information. These approaches have proven valuable in diverse domains such as bruise detection in fruits and vegetables \cite{BruiseDetection}, cancer cell diagnosis in human tissue \cite{CancerEstimation}, or separation of plastic types for recycling purposes \cite{PlasticSeparation}.
	
Disparity and depth estimation have been the focus of extensive research, leading to the development of numerous methods based on both hardware and software. 
	\begin{figure}[t]
	\centering
	\begin{tikzpicture}
	\node(Oriim)[]{\includegraphics[width=0.145\textwidth]{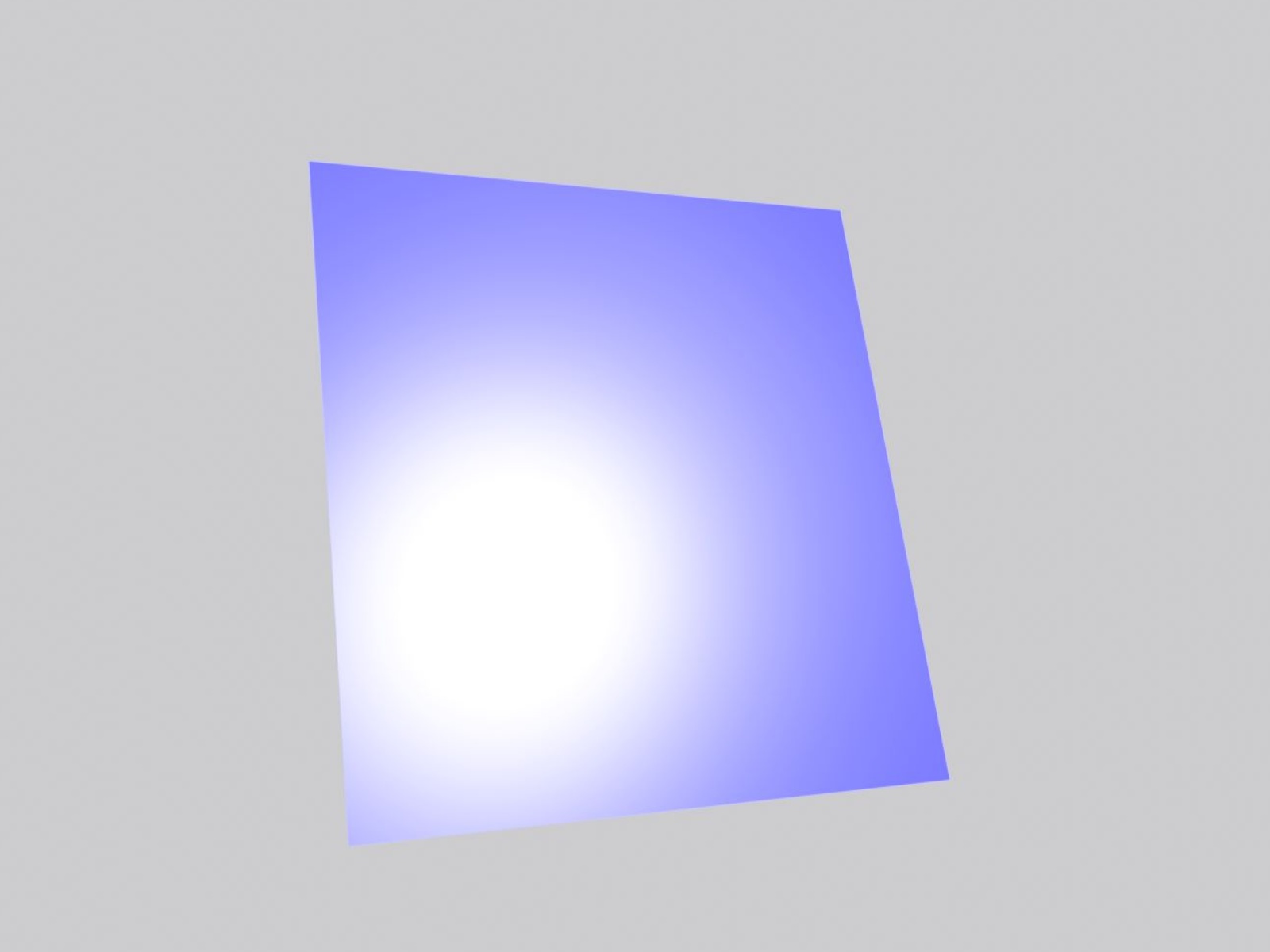}};
	\node(Oriname)[align=center, above of=Oriim, yshift=0.15cm]{Original};
	\node(GTdepth)[right of=Oriim, xshift=1.7cm]{\includegraphics[width=0.145\textwidth]{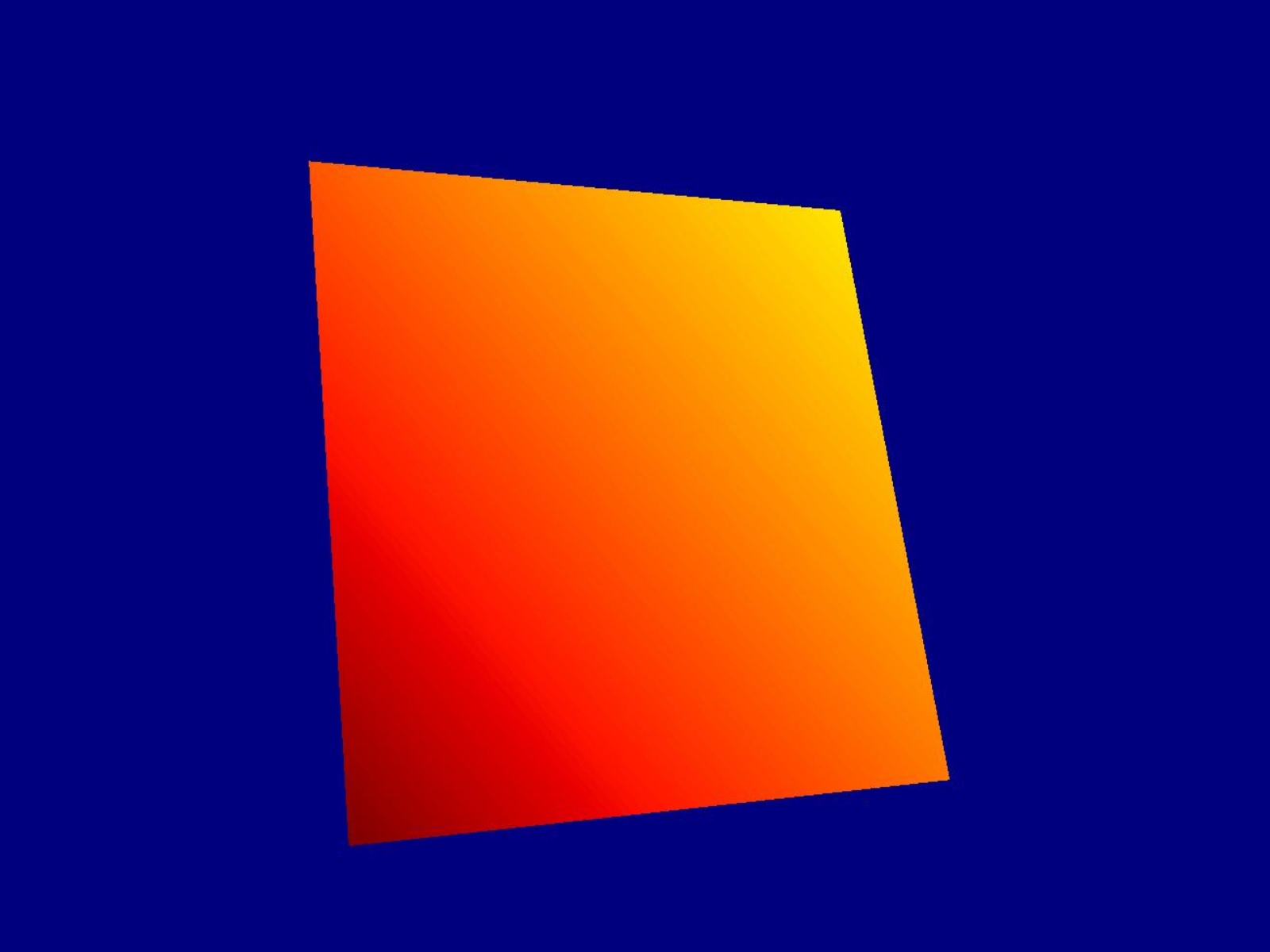}};
	\node(GTname)[align=center, above of=GTdepth, yshift=0.15cm]{Ground truth};
	\node(Oursdepth)[right of=GTdepth, xshift=1.7cm]{\includegraphics[width=0.145\textwidth]{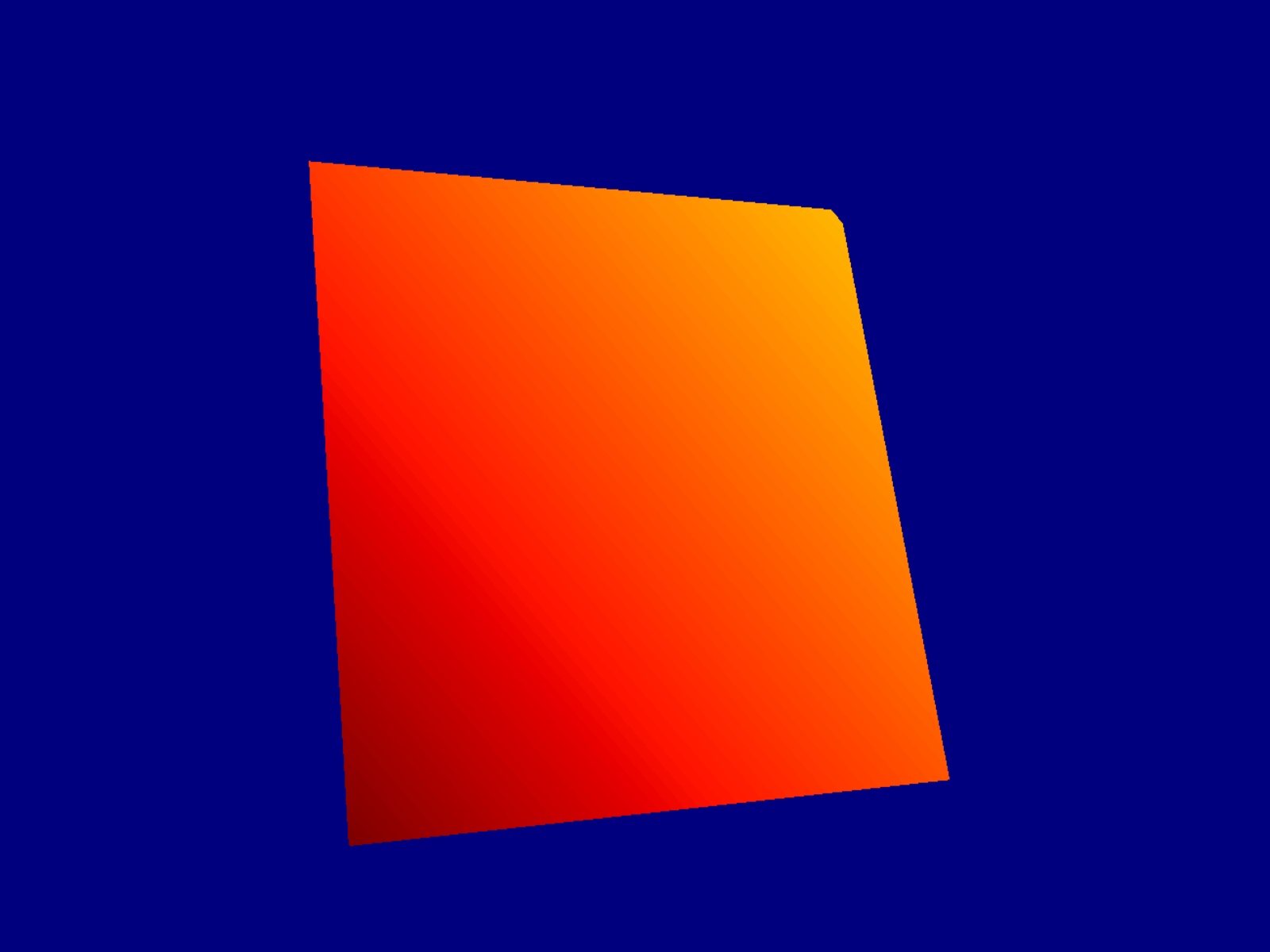}};
	\node(oursName)[align=center,above of=Oursdepth, yshift=0.15cm]{SRDE (Ours)};
	
	\node(NMRFdepth)[below of=Oriim, yshift=-1.03cm]{\includegraphics[width=0.145\textwidth]{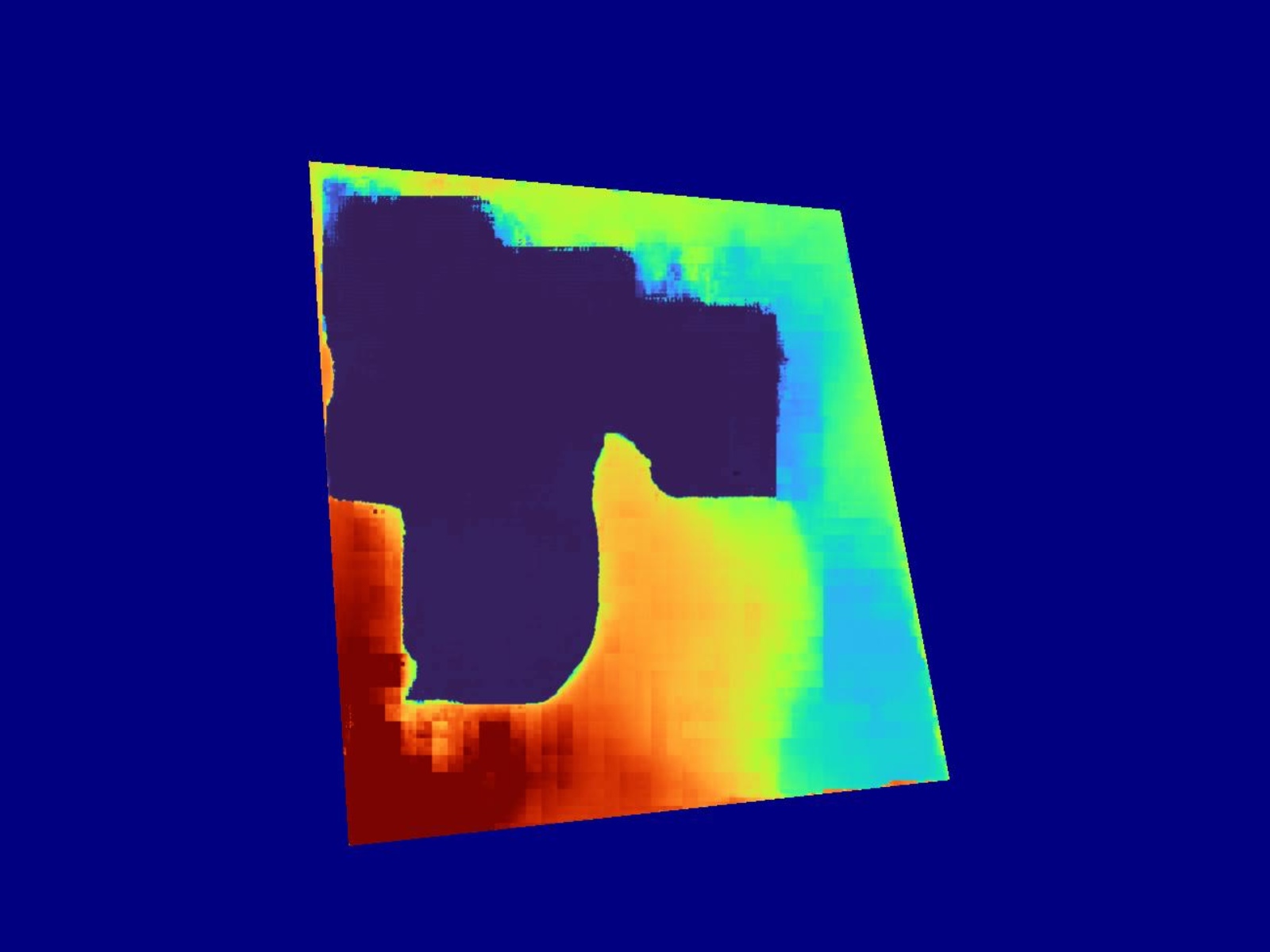}};
	\node(NMRFname)[align=center, below of=NMRFdepth, yshift=-0.2cm]{NMRF \cite{NMRF}};
	\node(CREdepth)[below of=GTdepth, yshift=-1.03cm]{\includegraphics[width=0.145\textwidth]{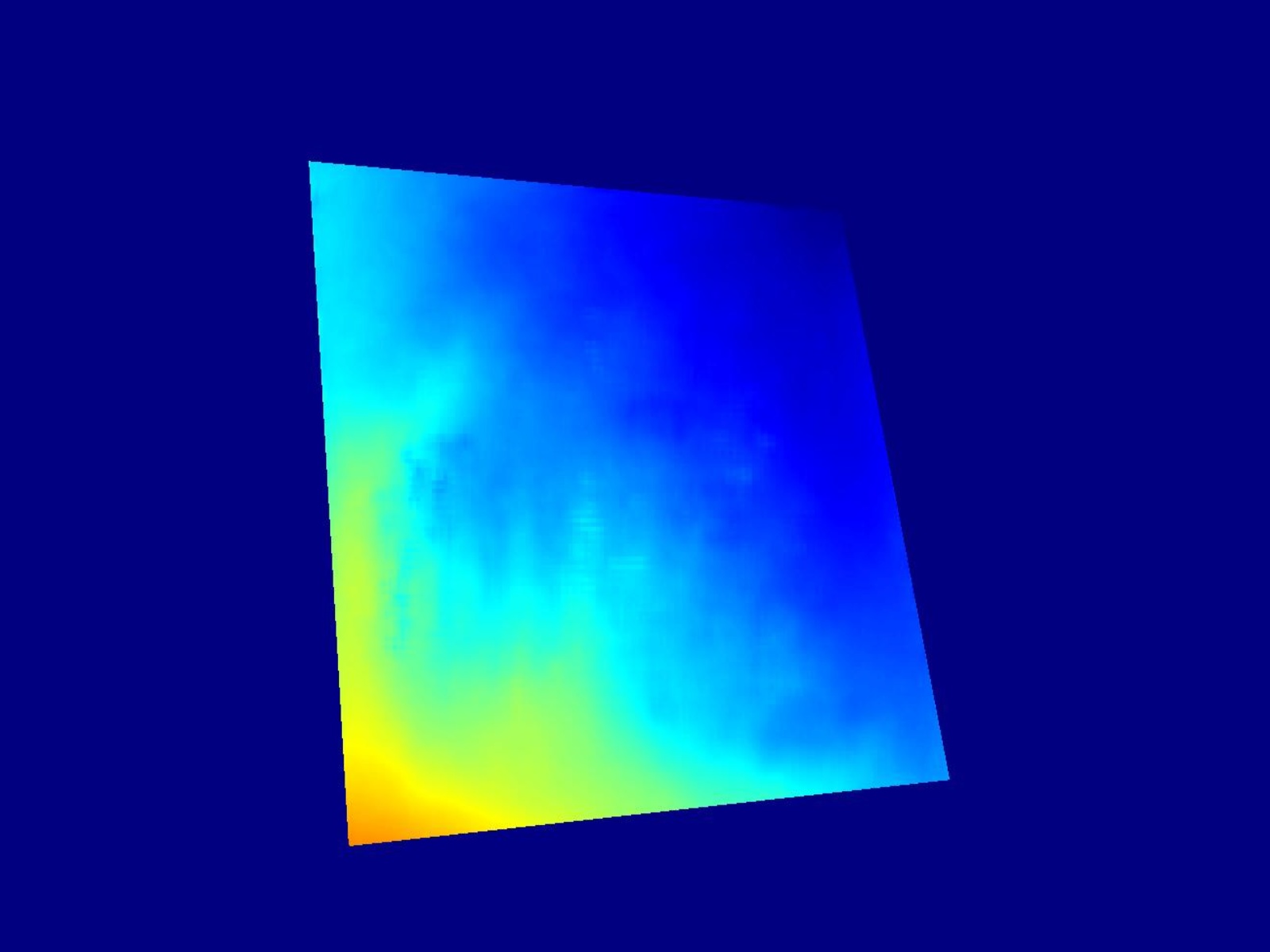}};
	\node(CREname)[align=center, below of=CREdepth, yshift=-0.2cm]{CREStereo \cite{CREStereo}};
	\node(IGEVdepth)[below of=Oursdepth, yshift=-1.03cm]{\includegraphics[width=0.145\textwidth]{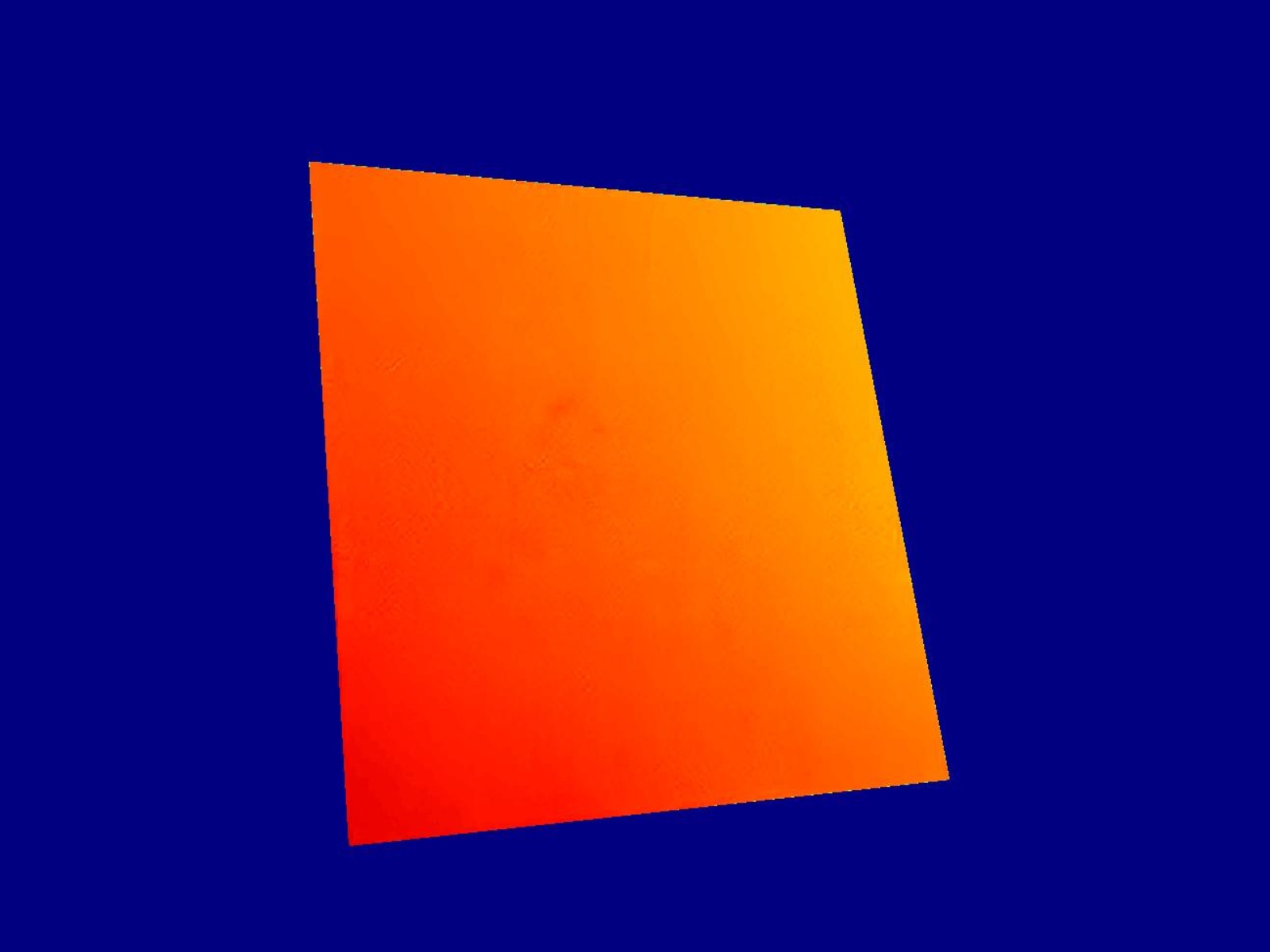}};
	\node(IGEVname)[align=center,below of=IGEVdepth, yshift=-0.2cm]{IGEV \cite{IgevDE}};
\end{tikzpicture}
\caption{Disparity estimation maps of a flat, textureless object with a specular reflection. The original image along with the ground truth disparity (GT) and our proposed solution SRDE is shown in the top row. The bottom row displays the disparity maps obtained with NMRF \cite{NMRF}, CREStereo \cite{CREStereo} and IGEV \cite{IgevDE}.}
\label{fig:IntroductionExample}
\vspace{-0.5cm}
\end{figure}
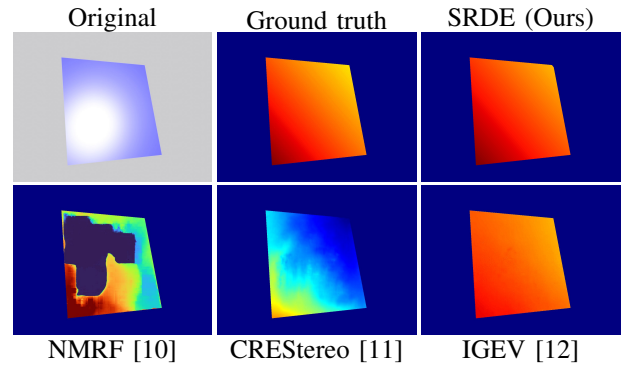
While many of these approaches have demonstrated excellent results in structured scenes with significant texture, they encounter $\text{substantial}$ limitations in scenarios where specular reflections occur on objects. This becomes particularly evident in applications like plastic separation, where the target objects are often smooth and textureless. Such surfaces, which are challenging for traditional methods, are further prone to specular reflections appearing as bright, overexposed regions in the images due to amplitude clipping. Existing neural networks typically fail in these regions as such objects lack the structural cues required for the underlying matching. These reflections, however, provide additional cues that are ignored in existing approaches. Fig. \ref{fig:IntroductionExample} shows a sample image of a flat, textureless object and the \-spe\-cular reflection of the light source, captured with a virtual camera array rendered in Blender \cite{Blender}. In addition, the ground truth (GT) disparity map and the results of the state-of-the-art disparity estimation networks Neural Markov Random Field (NMRF) \cite{NMRF}, Cascaded Recurrent Stereo Matching Network (CREStereo) \cite{CREStereo}, and Iterative Geometry Encoding Volume (IGEV) \cite{IgevDE} as well as our proposed Specular Reflection Disparity Estimation (SRDE) method are presented. The key innovation of the proposed SRDE method lies in analyzing the patterns of the specular reflections. SRDE incorporates the geometric properties and mathematical relations of the camera array and the scene into the disparity estimation process. Instead of treating these reflections as image regions without usable information, SRDE leverages the shift of the specular reflection across different views to recover the object's orientation and disparity map even in the complete absence of surface textures.
\vspace{-0.3cm}

\subsection{Scope and Applicability}
\label{sec:Scope and Applicability}
\noindent SRDE is specifically designed to handle flat, textureless objects, which is a scenario where neural algorithms typically fail. SRDE is intended for the analysis of individual objects rather than the entire, complex scenes. It operates under three fundamental constraints: (1) The target surface must be flat, where flat refers to surfaces that can be well approximated by a plane, (2) the scene must be illuminated by a single dominant light source, resulting in a specular reflection on the object surface, and (3) the light source position must be known or can be estimated.
	
As SRDE can not handle complex backgrounds of scenes, it can be integrated into neural networks. In such a hybrid configuration, the network handles complex and textured scene regions, while SRDE is selectively applied to generate the disparity map for the flat specular objects where the network lacks reliability.

\subsection{Key Contributions}
\label{sec:Key Contributions}
\noindent Our novel contributions can be summarized as follows:
\begin{itemize}
	\item We exploit the geometric properties of specular reflections by modeling their inter-view displacement. Combined with the known camera configuration and light source position, this allows us to extract accurate disparity maps of flat, textureless objects.
	\item We propose a robust method to estimate the light source position for real-world applications such as industrial sorting and visual inspection systems. By formulating a least-squares optimization over a system of reflection rays, our approach actively mitigates minor measurement inaccuracies and ensures the necessary robustness.
	\item We demonstrate the seamless integration of SRDE into existing neural disparity estimation frameworks. This hybrid approach enables reliable disparity estimation in specular regions without the need for retraining the backbone model.
\end{itemize}

\section{Related Work}
\label{sec:RelatedWork}
\subsection{Specular Reflection Detection}
\noindent Specular reflections manifest as intense highlights on surfaces, resulting from the coherent reflection of incident light \cite{MonovsStereo}. They occur predominantly on smooth surfaces like glass, plastics, or polished materials, where light rays are reflected in a directional manner. The angle of incidence relative to the surface normal significantly influences the direction and intensity of these reflections. Further, they can be divided into a diffuse and a specular component.
The first contains the intrinsic properties of the surface, while the latter is considerably affected by the illumination of the scene. Since reflections disturb the image and may hide important spectral information, many different approaches have been introduced to locate the reflective regions based on varying camera setups. Several approaches are rooted in intensity thresholding, as specular reflections typically have much higher intensities than the surrounding pixels. In the work of Stehle \cite{HistogramDetection}, the image is converted to the YUV color space, where Y represents the brightness and U, V depict the chrominance channels, and a Y-channel histogram is generated. Due to the high brightness of the object's reflections, the rightmost peak of the histogram represents the specular reflection, thus a global threshold can be determined. This method was further improved by Arnold et al. \cite{AdaptiveThreshold}, who suggested using an adaptive threshold depending on the image scenery instead of a global one. Another solution by Oh et al. \cite{HSVDetection} exploits the properties of the HSV color space. Reflections from a shiny surface result in very bright regions, where the RGB values are high and nearly equal, resulting in white or near-white highlights. The image is therefore transferred into the HSV color space, where specularities are characterized by a low saturation (black spots) and a high value (white spots), while the hue does not yield any information in this regard \cite{AdaptiveReflDetandInpainting}. Since the specularities often cause distinct changes in intensity, Morgand and Tamaazousti \cite{GradientSpec} use gradients for their detection. In image processing, the gradient represents the rate of change of pixel intensity in an image and is computed using derivative filters, such as the Sobel operator. Thus, the gradient magnitude at each pixel indicates the strength or steepness of the intensity change at that location. Additional early color-dependent work from Lee et al. \cite{MonovsStereo} is based on assuming Lambertian consistency, which means that the reflection does not change its brightness and spectral content depending on viewing directions, but the mixture of Lambertian and specular reflections can change.  
Approaches setting different illumination for the same scene were proposed by Sato and Ikeuchi \cite{TempRefl} and Chen et al. \cite{Mesostructure}. Different from these color-based solutions, polarization-based methods take advantage of polarimetric information. In general, diffuse reflection is unpolarized, while the specular reflection is polarized. In a common method proposed by Wen et al. \cite{PolarizationGuidedModel}, the use of polarization filters in front of the recording cameras is exploited and has shown good performance in their setup. Other recent approaches are based on neural networks for specularity detection and removal \cite{AdRefDet, RefDetUltra}.

\subsection{Disparity Estimation}
\noindent Disparity estimation is defined as the process of determining the pixels in different camera views that correspond to the same 3D point in the scene. It is inversely related to depth: the greater the disparity, i.e., the larger the shift between the corresponding points in two images, the closer the object is to the cameras. Conversely, a smaller disparity indicates that the object is farther away. 

Disparity estimation is a fundamental task in computer vision and robotics, since it is crucial for applications in \-autonomous navigation \cite{DEStructuredLight}, augmented reality \cite{DEToF} or 3D reconstruction \cite{LightField, Ego}. As stereo camera setups or camera $\text{arrays}$ such as the Camera Array for Multispectral Imaging (CAMSI) \cite{CAMSI} deliver images from multiple positions, those images have to be mapped to the center view. Therefore, Genser et al. introduced a fast disparity estimation \cite{CAMSI}. Further, Convolutional Neural Networks (CNNs) have led to significant advancements in disparity estimation. Early approaches such as \cite{DECNN} concentrated on computing matching distances, while the subsequent disparity estimation relied on classical steps from the literature, including cost aggregation, disparity computation, and refinement \cite{SCA}. More recent CNNs such as CREStereo \cite{CREStereo} and IGEV \cite{IgevDE} adopt end-to-end learning approaches, which typically involve three main stages: first, Spatial Pyramid Pooling layers are used to construct a cost volume; second, convolutional layers aggregate the cost volume in a manner similar to classical semi-global matching; and third, a regression step generates the final disparity map \cite{DSR}. Neural approaches such as NMRF \cite{NMRF} further extend this line of work by incorporating structured modeling into the learning framework.

\section{Methods}
\label{sec:Methods}
\noindent In this section, we present the key components of our introduced SRDE method, consisting of the detection of specular reflections and the subsequent disparity estimation process. 
The proposed algorithm relies on the following physical assumptions: First, we assume a single light source which generates a single specular reflection on the object surface. Second, the object surface can be modeled as an Euclidean plane, which allows a single surface normal $\mathbf{n}$ to represent the entire object plane. Third, we assume the intrinsic camera parameters to be known.

\subsection{Specular Reflection Detection}
\label{sec:Specular Reflection Detection}
\begin{figure}
	\centering
	\begin{tikzpicture}
		\node(0)[inner sep=0]{\includegraphics[width=0.15\textwidth]{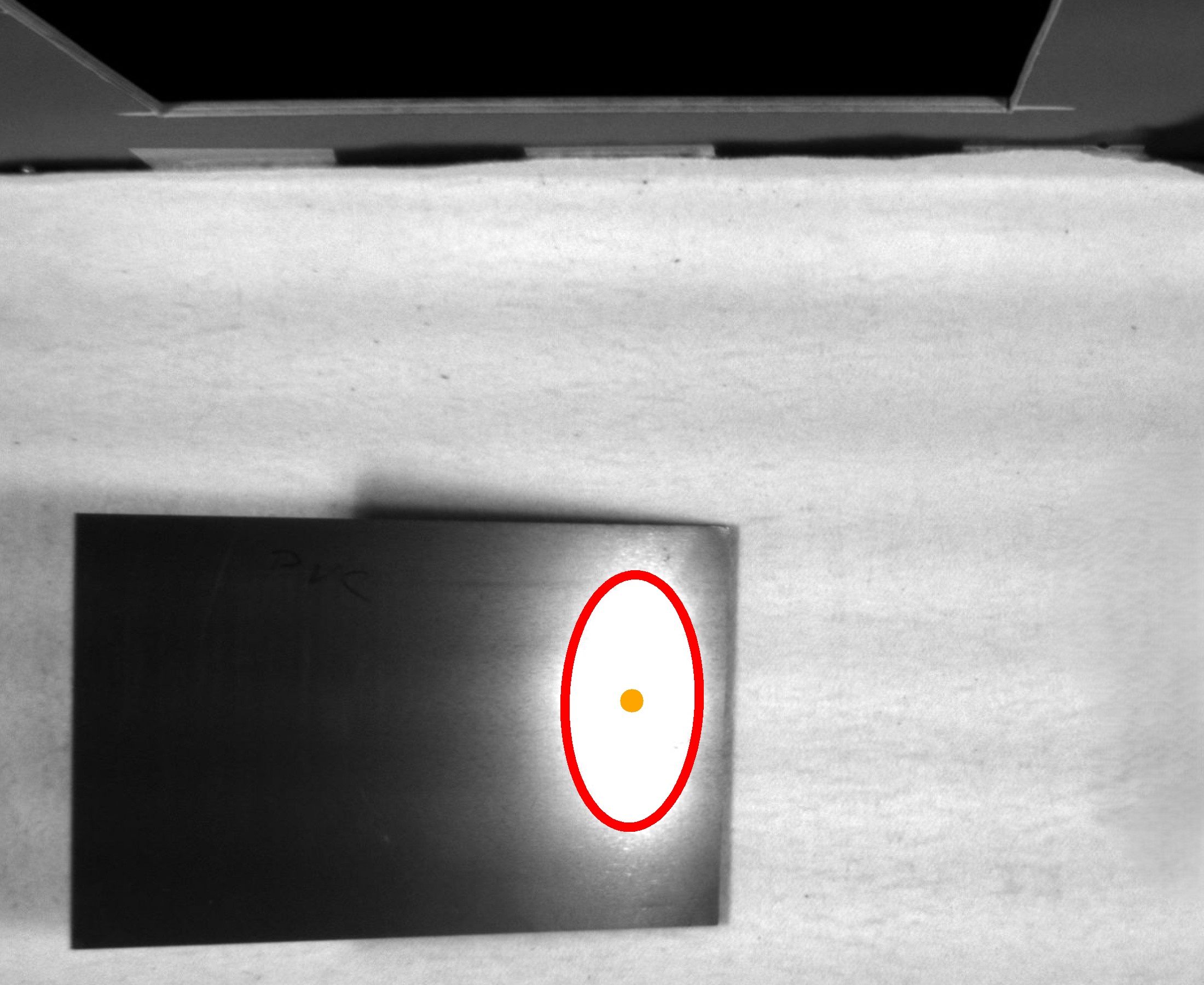}};
		\node(1)[right of=0,xshift=1.8cm,inner sep=0]{\includegraphics[width=0.15\textwidth]{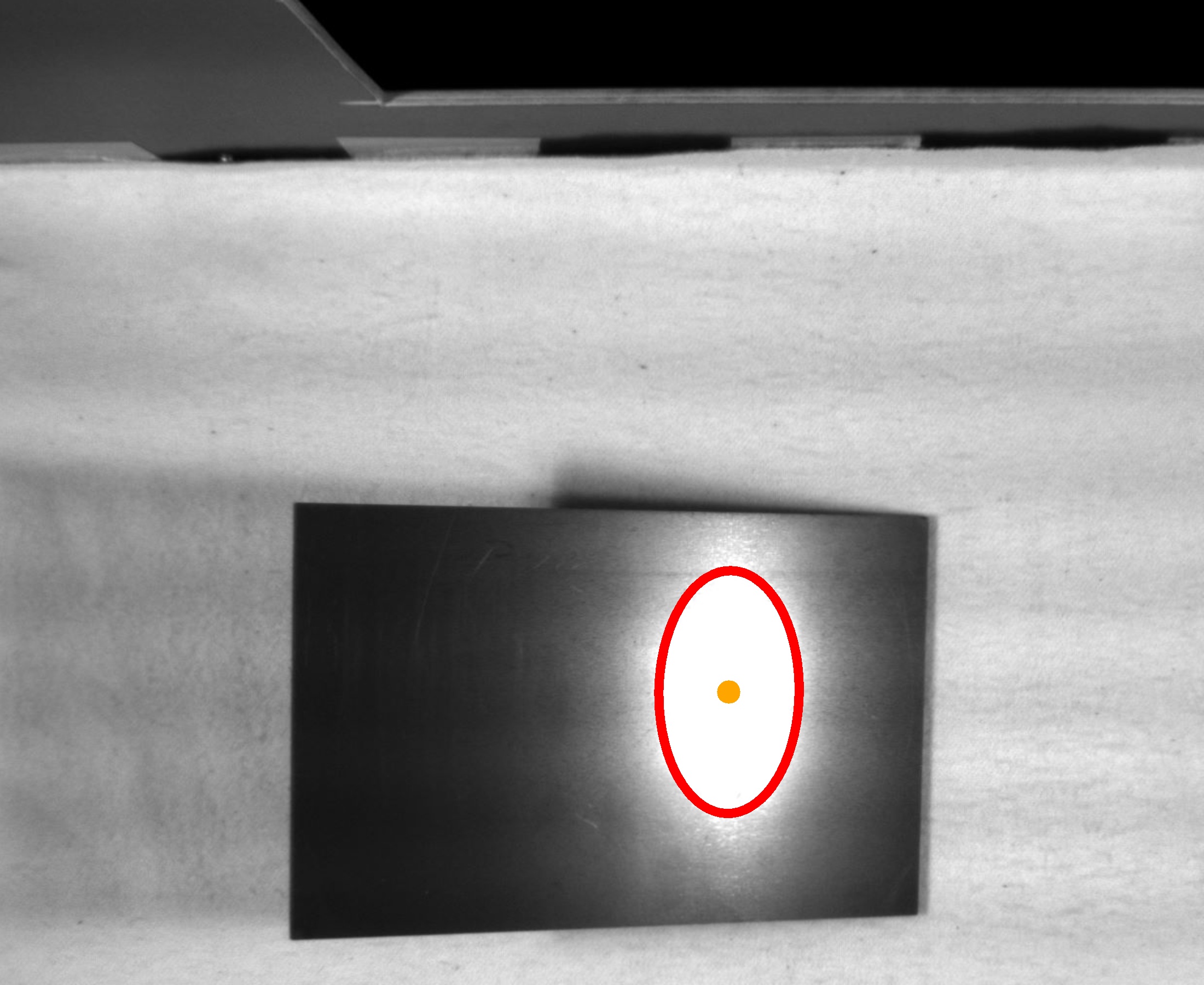}};
		\node(2)[right of=1,xshift=1.8cm,inner sep=0]{\includegraphics[width=0.15\textwidth]{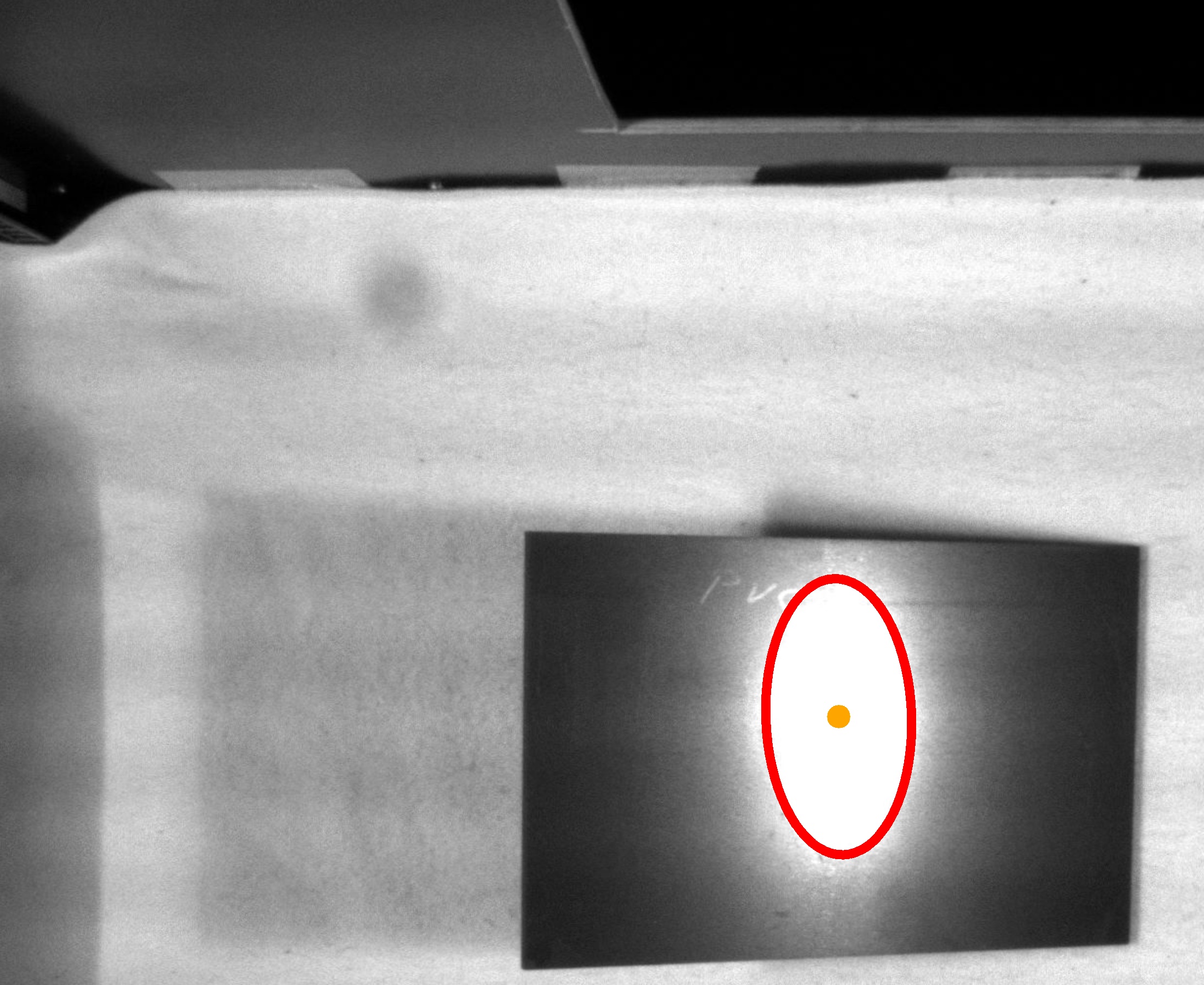}};
		\node(3)[below of=0,yshift=-1.3cm,inner sep=0]{\includegraphics[width=0.15\textwidth]{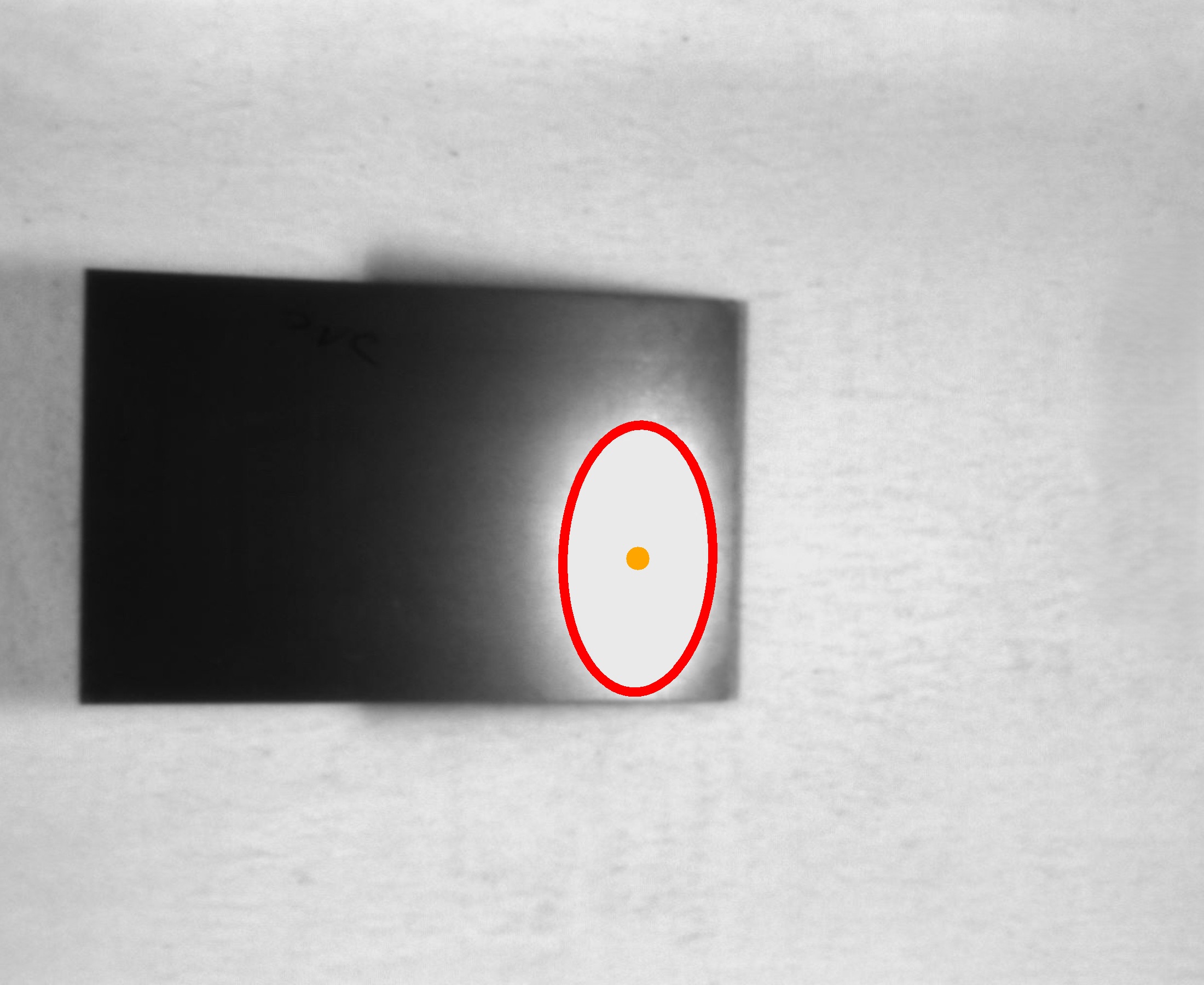}};
		\node(4)[right of=3,xshift=1.8cm,inner sep=0]{\includegraphics[width=0.15\textwidth]{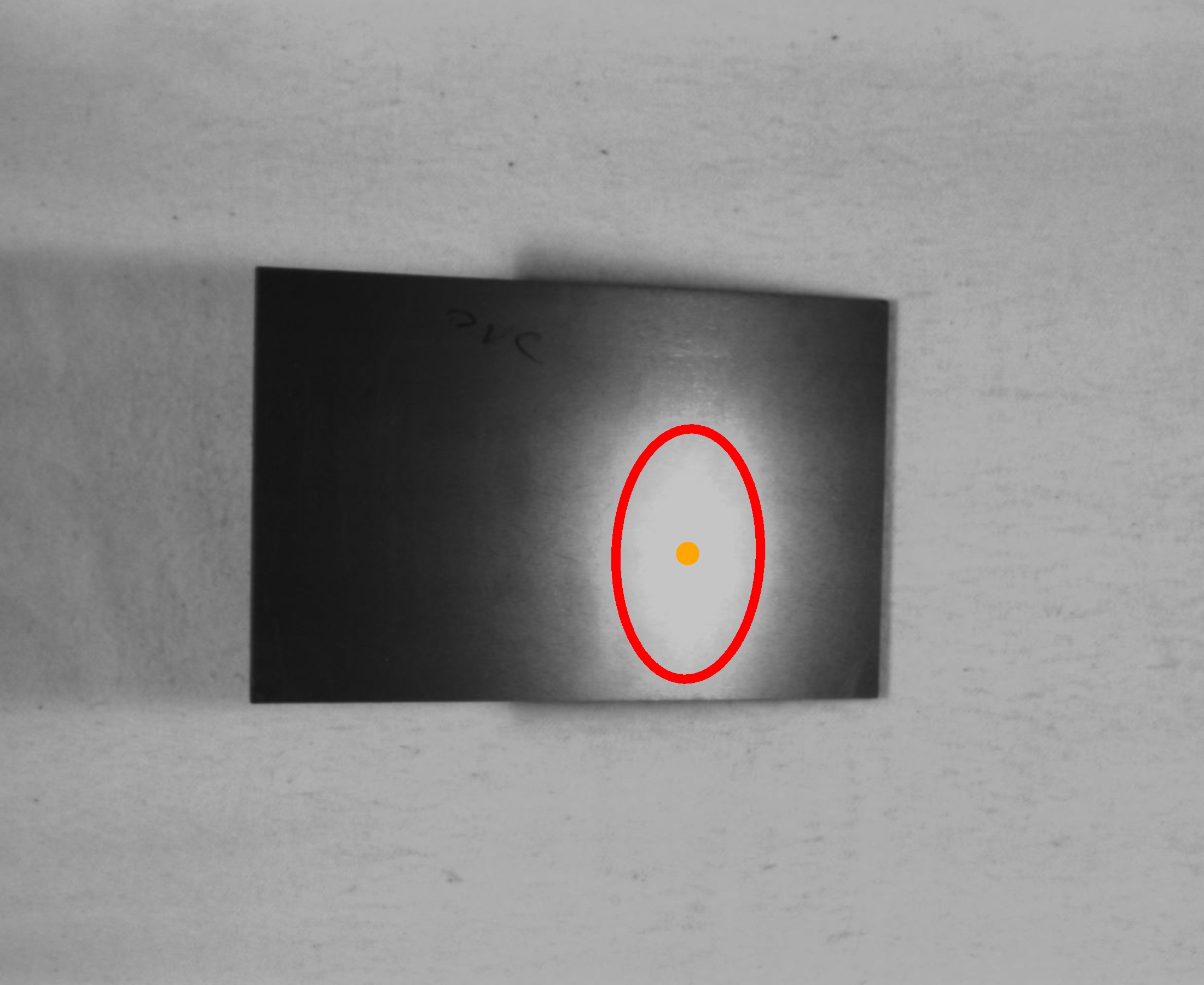}};
		\node(5)[right of=4,xshift=1.8cm,inner sep=0]{\includegraphics[width=0.15\textwidth]{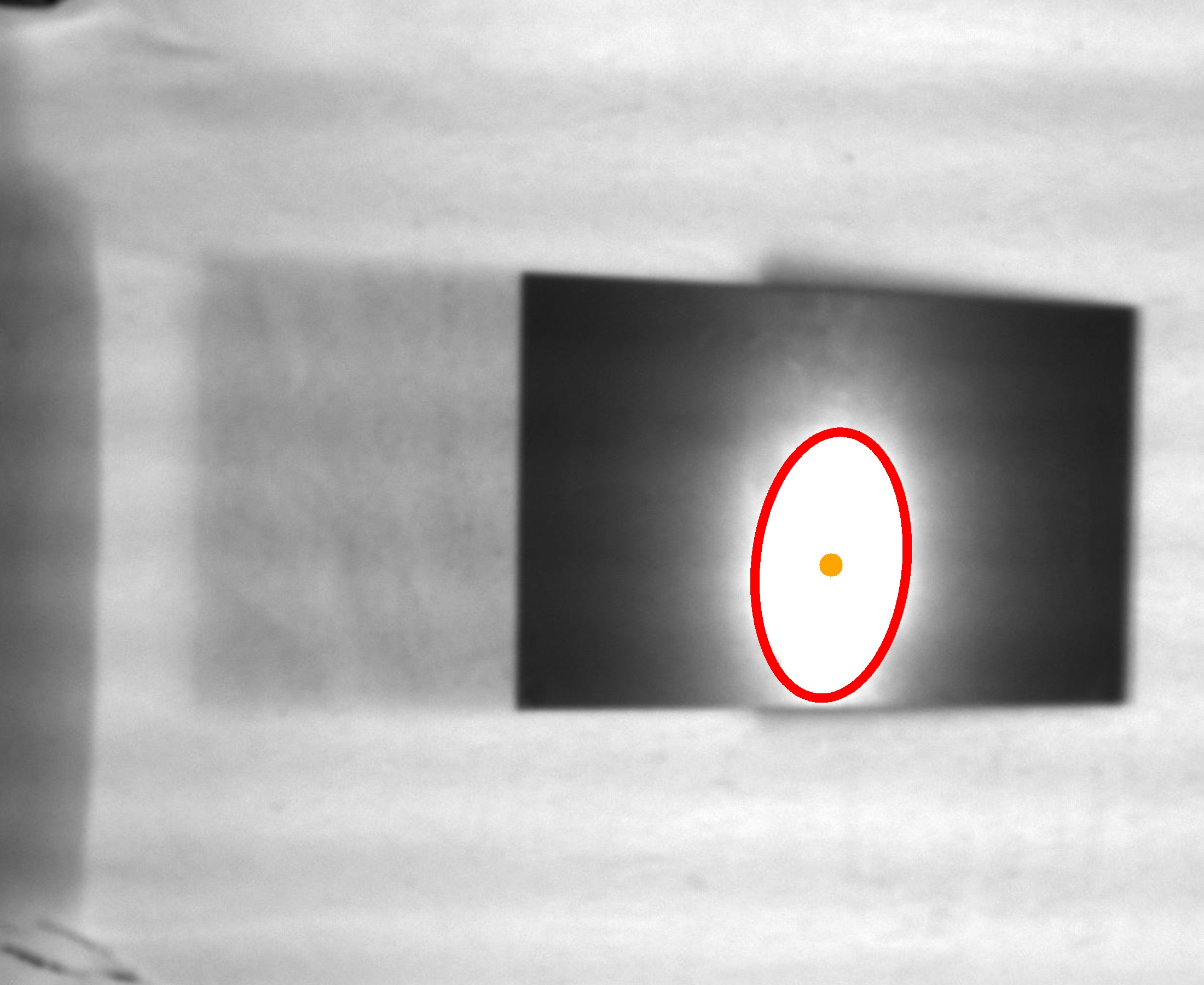}};
		\node(6)[below of=3,yshift=-1.3cm,inner sep=0]{\includegraphics[width=0.15\textwidth]{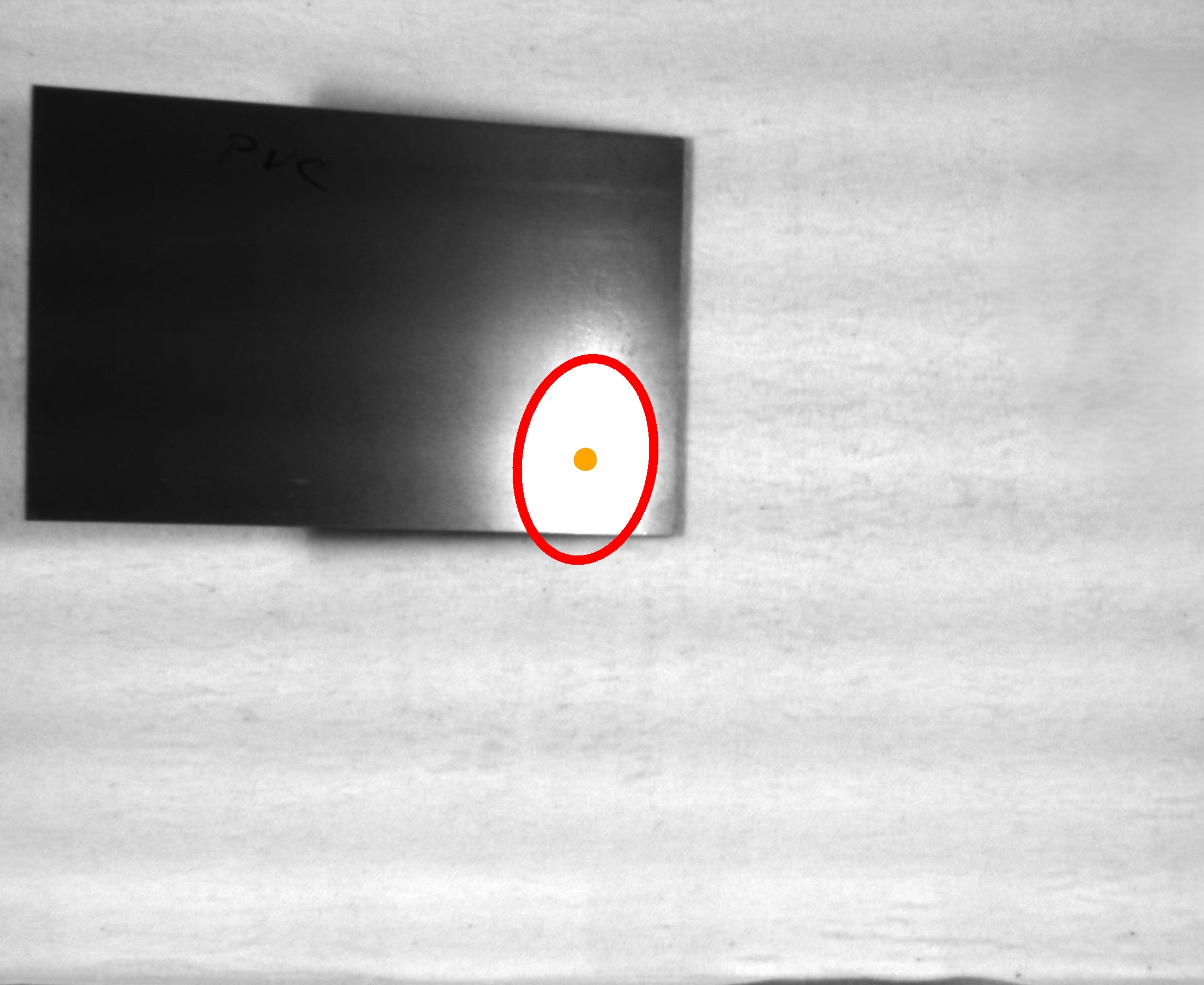}};
		\node(7)[right of=6,xshift=1.8cm,inner sep=0]{\includegraphics[width=0.15\textwidth]{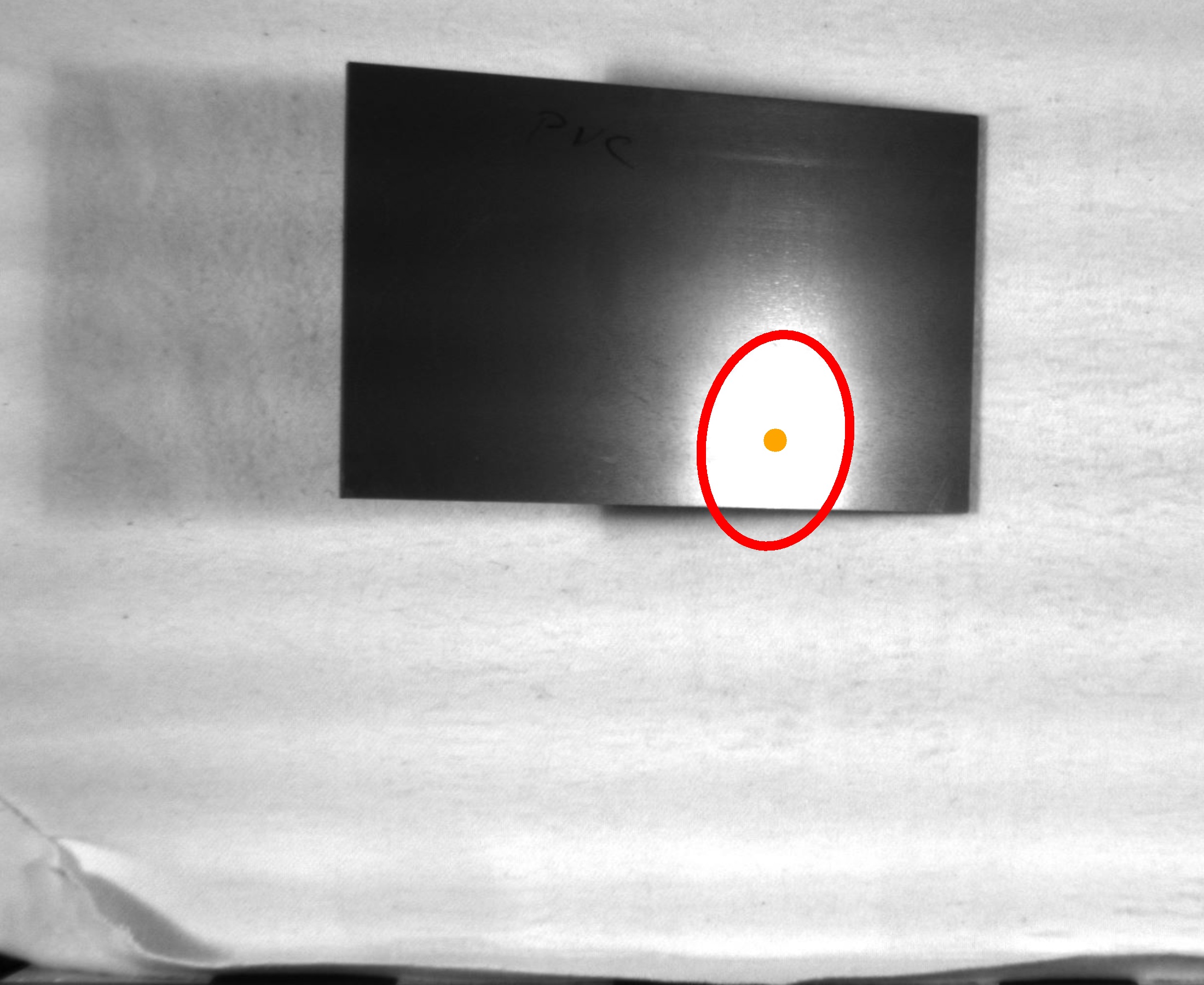}};
		\node(8)[right of=7,xshift=1.8cm,inner sep=0]{\includegraphics[width=0.15\textwidth]{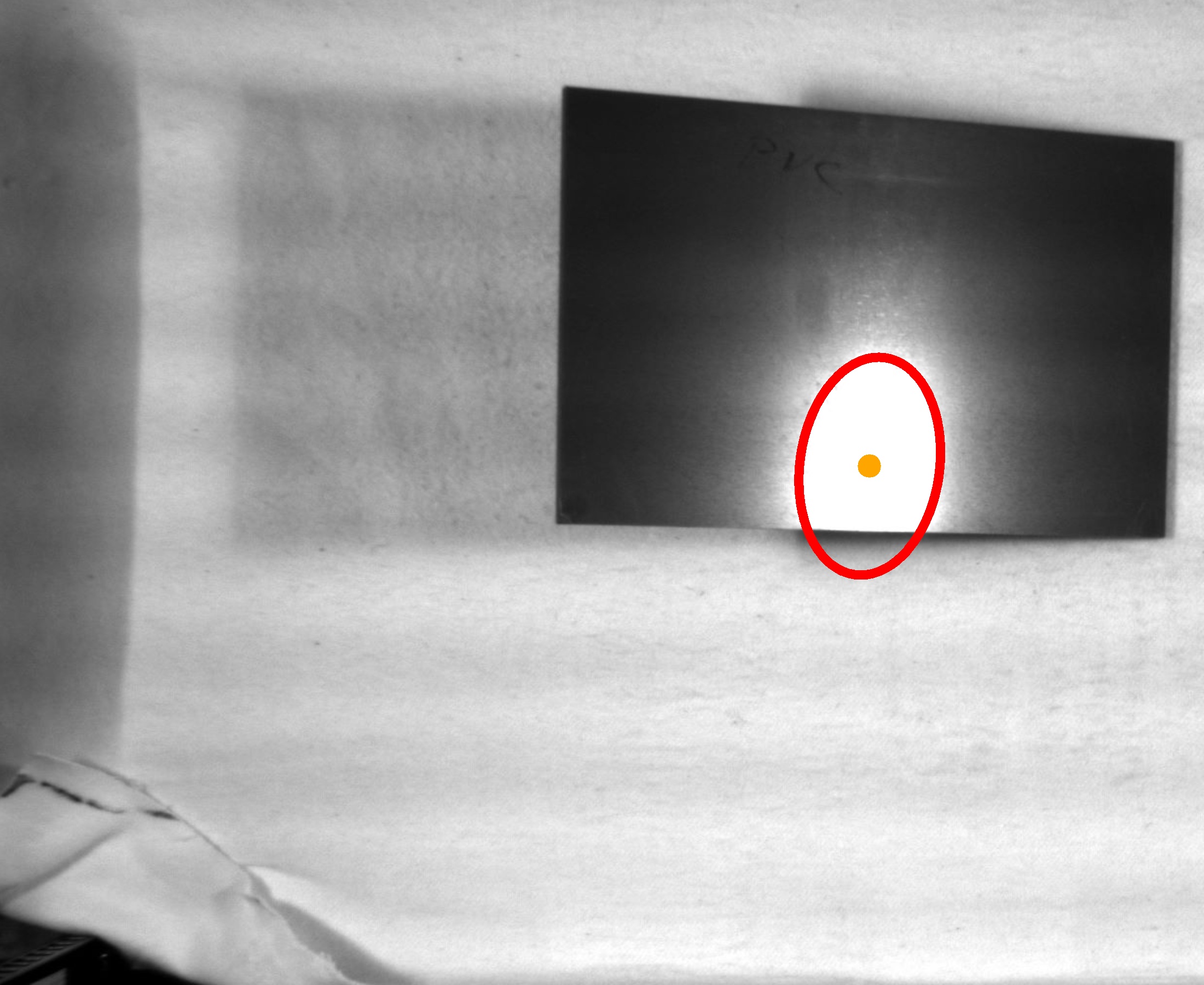}};
	
		\foreach \i in {0,...,8} {
			
			\tikzset{mystyle/.style={ultra thin, black, opacity=0.7}}
			
			\foreach \v in {1,...,5} {
				\draw[mystyle] ($(\i.south west)!\v/6!(\i.south east)$) -- ($(\i.north west)!\v/6!(\i.north east)$);
			}
			
			\foreach \h in {1,2,3} {
				\draw[mystyle] ($(\i.south west)!\h/4!(\i.north west)$) -- ($(\i.south east)!\h/4!(\i.north east)$);
			}
		}
		
		\begin{scope}[x={(0.south east)},y={(0.north west)}]
			\node[anchor=north west, fill=white, fill opacity=0.8, text opacity=1, inner sep=2pt] at (0.00,0.99) {\scriptsize Cam 0};
		\end{scope}
		\begin{scope}[x={(1.south east)},y={(1.north west)}]
			\node[anchor=north west, fill=white, fill opacity=0.8, text opacity=1, inner sep=2pt] at (0.00,0.99) {\scriptsize Cam 1};
		\end{scope}
		\begin{scope}[x={(2.south east)},y={(2.north west)}]
			\node[anchor=north west, fill=white, fill opacity=0.8, text opacity=1, inner sep=2pt] at (0.00,0.99) {\scriptsize Cam 2};
		\end{scope}
		\begin{scope}[x={(3.south east)},y={(3.north west)}]
			\node[anchor=north west, fill=white, fill opacity=0.8, text opacity=1, inner sep=2pt] at (0.00,0.99) {\scriptsize Cam 3};
		\end{scope}
		\begin{scope}[x={(4.south east)},y={(4.north west)}]
			\node[anchor=north west, fill=white, fill opacity=0.8, text opacity=1, inner sep=2pt] at (0.00,0.99) {\scriptsize Cam 4};
		\end{scope}
		\begin{scope}[x={(5.south east)},y={(5.north west)}]
			\node[anchor=north west, fill=white, fill opacity=0.8, text opacity=1, inner sep=2pt] at (0.00,0.99) {\scriptsize Cam 5};
		\end{scope}
		\begin{scope}[x={(6.south east)},y={(6.north west)}]
			\node[anchor=north west, fill=white, fill opacity=0.8, text opacity=1, inner sep=2pt] at (0.00,0.99) {\scriptsize Cam 6};
		\end{scope}
		\begin{scope}[x={(7.south east)},y={(7.north west)}]
			\node[anchor=north west, fill=white, fill opacity=0.8, text opacity=1, inner sep=2pt] at (0.00,0.99) {\scriptsize Cam 7};
		\end{scope}
		\begin{scope}[x={(8.south east)},y={(8.north west)}]
			\node[anchor=north west, fill=white, fill opacity=0.8, text opacity=1, inner sep=2pt] at (0.00,0.99) {\scriptsize Cam 8};
		\end{scope}
	\end{tikzpicture}
	\vspace{-0.1cm}
	\caption{Images captured with a $3 \times 3$ camera array showing a smooth plastic surface with a specular reflection. For each camera recording, the reflection marked with a red circle is located at a different position. The center of the reflection is marked with an orange circle.}
	\label{fig:ReflectionDifferences}
	\vspace{-0.4cm}
\end{figure}
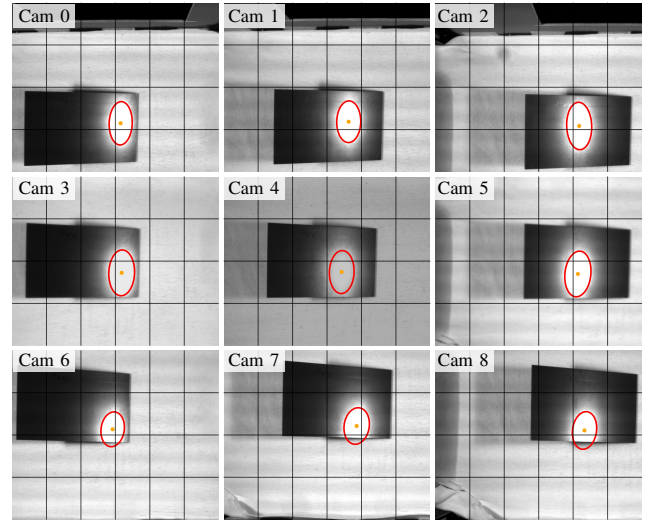

\noindent In order to use the geometric properties of specular reflections, it is crucial to know their positions in each of the available images, as they differ depending on the positions of the cameras. An example is shown in Fig. \ref{fig:ReflectionDifferences}, which was captured with a camera array with nine cameras arranged in a $3 \times 3$ grid layout. This specific grid topology is chosen to provide a diverse set of horizontal, vertical, and diagonal baselines. 
The practical necessity of the $3 \times 3$ camera setup for light source position estimation will be discussed in Section \ref{sec:LightSourcePositionEstimation}.

In a first processing step, the images are normalized using calibrated reflection standards (Zenith-Polymer) to ensure consistent intensity scaling across acquisitions. Afterwards, the specular reflections are identified by applying adaptive gradient thresholding, which combines adaptive thresholding \cite{AdaptiveThreshold} with gradient-based thresholding \cite{GradientSpec}. To quantify local intensity variations which serve as primary indicator for specular reflections, the gradient magnitude $\textit{G}[x,y]$ is computed for  each image as follows:
\begin{equation}
	\textit{G}[x,y] = |\nabla I[x,y]| = \sqrt{\left(\frac{\partial I}{\partial x}\right)^2 +
	\left(\frac{\partial I}{\partial y}\right)^2},
\end{equation}
where $I[x,y]$ denotes the image intensity at position $[x,y]$. The gradient operator $\nabla$ is defined by the first-order partial derivatives $\frac{\partial I}{\partial x}$ and $\frac{\partial I}{\partial y}$ in the horizontal and vertical directions, respectively, which are numerically approximated using a \mbox{$3 \times 3$} Sobel operator. This formulation assumes locally smooth intensity variations and is applicable to both grayscale and color images. For color images, the gradient is computed on the luminance channel to ensure robustness against chromatic noise. 

Unlike a fixed threshold, an adaptive threshold $\vartheta[x,y]$ is computed dynamically based on local gradient statistics. This local thresholding scheme accounts for varying illumination levels across the image, identifying significant edges that characterize specular reflections: 
\begin{equation}
\vartheta[x,y] = \mu_G(W[x,y]) + k\cdot \sigma_G(W[x,y])~~,
\end{equation}
where $\mu_G(W[x,y])$ and $\sigma_G(W[x,y])$ denote the local mean and standard deviation of the gradient magnitudes within a square window $W[x,y]$ of size $15 \times 15$ pixels. The window size is chosen as a trade-off between spatial localization and robustness to noise \cite{GradientSpec}, assuming that the neighborhood contains sufficient statistical information to distinguish reflections from background texture. The parameter $k$ serves as a scaling factor to control the sensitivity of the detector relative to the local contrast. For adaptive gradient thresholding schemes of the form $T = \mu + k\sigma$, values in the range $k \in [0.2,\dots,0.5]$ are commonly reported for intensity-based binarization \cite{BinarizationThresh}. In contrast, gradient-based detection requires larger values of $k$ in order to suppress texture-induced responses and emphasizes gradient peaks caused by specular reflections. In our setup, $k$ was determined by evaluating the detection performance against a subset of manually annotated samples. Sensitivity tests with varying $k$ across the range $[1.0, 2.0]$ in increments of 0.1 indicated stable results within the interval $k \in [1.2,\dots,1.8]$. We identified $k = 1.5$ as providing the optimal balance between suppressing texture noise and preserving reflection boundaries. Consequently, this value was fixed for all experiments. For different imaging environments, $k$ should be re-adjusted accordingly.

Using this adaptive threshold $\vartheta$, a binary mask $M[x,y]$ is generated to isolate the reflection from the background:
\begin{equation}
	M[x,y] = \left\{\begin{array}{ll} 1, & \text{if} ~~ \textit{G}[x,y] > \vartheta[x,y] \\
	0, & \text{else} \end{array}\right. .
\end{equation}
This decision rule assigns each pixel either to the reflection region ($M = 1$) or the background ($M = 0$). Additionally, connected component analysis \cite{CCA} is applied using 8-neighborhood connectivity to identify spatially contiguous regions corresponding to potential specular reflections. To ensure robustness against noise, only components with a minimum spatial extent of $10 \times 10$ pixels are retained. Although the considered setup assumes a single dominant specular reflection per image, additional small reflections may arise due to minor surface irregularities. These secondary reflections are typically limited in spatial extent and do not represent the specular reflection. Therefore, only the largest connected component is retained for further processing.

Let $\Omega$ denote the set of all pixels $[x,y]$ for which ${M[x,y] = 1}$. 
To determine the precise location of the specular reflection, the center $(c_x, c_y)$ of the detected specular reflection region $\Omega$ is computed as
\begin{equation}
	c_{x} = \frac{1}{|\Omega|}\sum_{(x,y \in \Omega)} x,~~~ c_{y} = \frac{1}{|\Omega|}\sum_{(x,y \in \Omega)}y ~~~,
\end{equation}
where $|\Omega|$ denotes the number of pixels in the segmented region.

Finally, the positional shift $\Delta_{x}, \Delta_{y}$ of the specular reflection between spatially adjacent cameras $i$ and $j$ is computed to quantify the specular reflection shift across the camera array as
\begin{equation}
\begin{aligned}	
	\Delta_{x,i,j} = c_{x,i} - c_{x,j} \\
	\Delta_{y,i,j} = c_{y,i} - c_{y,j}~~~~~.
	\label{eq:PositionDifferences}
\end{aligned}
\end{equation}
These differences serve as the geometric basis for the subsequent geometric reconstruction described in the following section.

\subsection{Disparity Estimation}
\label{sec:Disparity and Depth Estimation}
\begin{figure}[t!]
	\centering
	\begin{tikzpicture}[y=-1cm, >=triangle 60]
		\begin{scope}[shift={(-0.5, -3)}]
			\fill (-1,0) circle (0.09);
			\draw[line width=1.05pt, gray] (0,-1) -- (0,1);
			\node[align=center] at (0, 1.15) {\color{gray}{Camera 1}};
		\end{scope}
		\begin{scope}[shift={(-0.5, -0)}]
			\fill (-1,0) circle (0.09);
			\draw[line width=1.05pt, gray] (0,-1) -- (0,1);
			\node[align=center] at (0, 1.15) {\color{gray}{Camera 2}};
		\end{scope}
		
		\draw[darkgreen, line width=1.1pt] (1, -4) -- (4, 2);
		\node[align=center] at (4, 2.2) {\color{darkgreen}Object};
		\fill[black] (2.88, -0.24) circle (0.05);
		\node[align=center] at (3.2, -0.2) {\color{black}\textbf{H}$_1$};
		
		\fill[black] (2.15, -1.7) circle (0.05);
		\node[align=center] at (2.4, -1.9) {\color{black}\textbf{H}$_2$};
		
		\begin{scope}[shift={(1, 2)}]
			\draw[orange] (0, 0) circle (0.25);
			\draw[orange] (0.3, 0) -- (0.4, 0);
			\draw[orange] (0.21, 0.21) -- (0.28, 0.28);
			\draw[orange] (0, 0.3) -- (0, 0.4);
			\draw[orange] (-0.21, 0.21) -- (-0.28, 0.28);
			\draw[orange] (-0.3, 0) -- (-0.4, 0);
			\draw[orange] (-0.21, -0.21) -- (-0.28, -0.28);
			\draw[orange] (0, -0.3) -- (0, -0.4);
			\draw[orange] (0.21, -0.21) -- (0.28, -0.28);
			\node[align=center] at (0, 0.8) {Real light\\source \textbf{L}};
		\end{scope}
		\draw[dashed] (1, 2) -- (5.8, -0.4);
		\begin{scope}[shift={(5.8, -0.4)}]
			\draw[orange] (0, 0) circle (0.25);
			\draw[orange] (0.3, 0) -- (0.4, 0);
			\draw[orange] (0.21, 0.21) -- (0.28, 0.28);
			\draw[orange] (0, 0.3) -- (0, 0.4);
			\draw[orange] (-0.21, 0.21) -- (-0.28, 0.28);
			\draw[orange] (-0.3, 0) -- (-0.4, 0);
			\draw[orange] (-0.21, -0.21) -- (-0.28, -0.28);
			\draw[orange] (0, -0.3) -- (0, -0.4);
			\draw[orange] (0.21, -0.21) -- (0.28, -0.28);
			\node[align=center] at (0, 1) {Virtual reflected \\ light source \textbf{V}};
		\end{scope}
		
		\draw[->, dotted] (5.8, -0.4) -- (-1.5, -3);
		\draw[->, dotted] (5.8, -0.4) -- (-1.5, 0);
		
		\draw[solid] (2.88, -0.24) -- (-1.5, 0);
		\draw[solid] (2.15, -1.7) -- (-1.5, -3);
		
		\draw[->,solid] (1, 2) -- (2.88, -0.24);
		\draw[->,solid] (1, 2) -- (2.15, -1.7);
		
		\draw[blue, line width = 1.2pt] (-0.5, 0) -- (-0.5, -0.05);
		\draw[blue, decorate, decoration={brace, amplitude=3pt, mirror}, thick] (-0.5,0) -- (-0.5,-0.05) node[below, yshift=-0.15cm, right] {\color{blue}o};
		
		\draw[blue, line width = 1.2pt] (-0.5, -3) -- (-0.5, -2.65);
		\draw[blue, decorate, decoration={brace, amplitude=3pt}, thick] (-0.5,-3) -- (-0.5,-2.65) node[midway, right] {\color{blue}o};
		
		\draw[dashed] (1, 2) -- (5.8, -0.4);
		\fill[red] (3.4, 0.8) circle (0.09);
		\node[align=center] at (3.7,0.9) {\color{red}\textbf{s}};
		\draw[->, red] (3.4, 0.8) -- (2.8, 1.1);
		\node[align=center] at (2.8, 1.4) {\color{red}\textbf{n}};
		
		\draw[dotted] (-0.5, -3) -- (-1.5, -3);
		\draw[dotted] (-0.5, 0) -- (-1.5, 0);
		
		\draw[gray, <->, >=latex] (-2.2, -3) -- (-2.2, 0) node[midway, left] {\color{gray}$B$};
		\draw[blue, dotted] (-1.6, -3) -- (-2.4, -3);
		\draw[blue, dotted] (-1.6, 0) -- (-2.4, 0);
		
		\draw[gray, <->, >=latex] (-1.5, -4.4) -- (-0.5, -4.4) node[midway, above] {\color{gray}$f$};
		\draw[blue, dotted] (-1.5, -3.15) -- (-1.5, -4.6);
		\draw[blue, dotted] (-0.5, -4.05) -- (-0.5, -4.6);
		
	\end{tikzpicture}
	\caption{Geometric fundamentals and relations of our disparity estimation approach. The baseline $B$ and focal length $f$ define the basic camera geometry, while the pixel size $p$ is omitted for visual clarity. As shown, the method is based on estimating the scene geometry from the interaction of cameras, light source, and object.}
	\label{fig:CameraObjectRelation}
	\vspace{-0.4cm}
\end{figure}
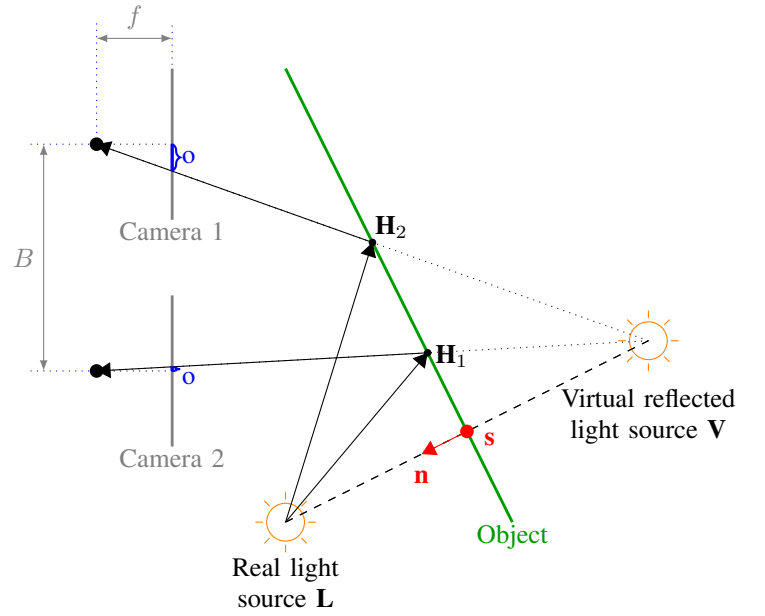

\noindent The disparity map estimation in SRDE differs from conventional methods because it explicitly considers the relative viewpoint differences of the cameras and assumes a flat object surface. The principle we propose is illustrated in Fig. 
\ref{fig:CameraObjectRelation}.
As shown, observing the object from different viewpoints, such as Camera 1 and Camera 2, the specular reflection appears on distinct physical points on the object surface, denoted as \textbf{H}$_1$ and \textbf{H}$_2$. Thus, specular reflections are extremely sensitive to small changes in the viewing direction, and their apparent position on the object surface may shift horizontally, vertically, or diagonally. 

To utilize these shifts for 3D reconstruction, we model the specular reflection as an image of a virtual reflected light source $\mathbf{V}$. According to the optical principles, a flat reflective surface acts like a mirror, creating a virtual image of the real light source $\mathbf{L}$ that appears to be located behind the object's surface. The fundamental idea of our approach is to first triangulate the 3D position of $\mathbf{V}$ using the camera array, and subsequently use the geometric relationship between $\mathbf{V}$ and $\mathbf{L}$ to determine the object surface.

The following calculations are performed independently for each adjacent camera pair.
First, the position of the virtual light source $\mathbf{V}$ $= (x_{\text{rel}}, y_{\text{rel}}, z_{\text{rel}})$ is determined. The virtual relative depth $z_{\text{rel}}$ is found by applying standard stereo triangulation to the observed specular reflections:
\begin{equation}
	z_{\text{rel}} = \frac{B \cdot f}{d \cdot p} ~~~,
\end{equation}
where the baseline $B$ represents the camera distances, and $f$ and $p$ denote the focal length and sensor pixel size, respectively. The disparity $d$ is defined as the pixel displacement between the reflection centers of adjacent camera pairs along the baseline connecting them. Hence, multiple disparity values are computed for each target camera with respect to its neighboring cameras. To determine the horizontal and vertical coordinates $x_{\text{rel}}$ and $y_{\text{rel}}$ of $\mathbf{V}$, the displacement of the specular reflection to the image center for each image pair is transferred to a physical displacement on the camera's sensor by  
\begin{equation}
	o_{x} = (c_u - c_{x}) \cdot p ~~,~~ o_{y} = (c_v - c_{y}) \cdot p.
\end{equation} 
$c_{u}, c_v$ denote the pixel coordinates of the image center of the camera, which serve as the origin of the camera's intrinsic coordinate system in the image plane, and $c_{x,y}$ represents the reflection center in the image, as determined from the detected specular reflection. This conversion is a necessary prerequisite for the subsequent 3D reconstructions, as it translates discrete pixel indices into metric units. Subsequently, the real-world horizontal coordinate $x_{\text{rel}}$ relative to the optical center of the camera for the virtual light source $\mathbf{V}$ is derived using the perspective projection of the pinhole camera model \cite{Pinhole}:
\begin{equation}
	x_{\text{rel}}  = \frac{o_{x} \cdot z_{\text{rel}}}{f}~~~.
\end{equation}
This projection scales the image coordinate by the triangulated distance $z_{\text{rel}}$. This step is applied analogously to the vertical direction $y_{\text{rel}}$ to determine the full 3D position of the virtual light source $\mathbf{V}$.

These observations from different cameras need to be combined into a unified global coordinate framework. Therefore, the pitch and roll angles required to align the virtual relative light $\textbf{V}$ with the real light source $\textbf{L}$ have to be determined. The angular deviations, $\theta_{\text{pitch}}$ and $\theta_{\text{roll}}$, are derived from the components of the normal vector $\textbf{n} = \textbf{V} - \textbf{L} = [n_{x}, n_{y}, n_{z}]^\top$, which represents the orientation of the reflective surface relative to the camera array:
\begin{equation}
	\theta_{\text{pitch}} = -\arctan\Bigg(\ \frac{n_{x}}{n_{z}}\Bigg) ,~~~
	\theta_{\text{roll}} = -\arctan\Bigg(\ \frac{n_{y}}{n_{z}}\Bigg)    .
\end{equation}
Because the virtual light source and the real light source are physically equidistant from a planar mirror, the object surface lies exactly halfway between the real virtual and real light sources. Thus, the midpoint $\textbf{s} = \frac{\textbf{V} + \textbf{L}}{2}$ is computed, which lies directly on the object plane. With the surface normal $\mathbf{n}$ defining the orientation and the midpoint $\mathbf{s}$ fixing its spatial position, the object plane is fully defined in 3D space.

The depth map is generated by finding the intersection of rays $\mathbf{k}$ emanating from the camera with the specified object plane in 3D space. Each pixel $[x, y]$ on the image sensor corresponds to a unique ray originating from the camera's optical center. The direction of this ray is determined by the pixel position, adjusted by the sensor pixel size $p$, the focal length $f$, and the camera's origin $(c_u, c_v)$, which represents the intersection of the optical axis with the image plane:
\begin{equation}
	\mathbf{k}[x,y] = \left( \frac{x - c_u}{f \cdot p}, \frac{y - c_v}{f \cdot p}, 1 \right).
	\label{eq:RayDirection}
\end{equation}
For each pixel ray, the absolute depth $z[x,y]$ is calculated using the ray-plane intersection formula \cite{IntersectionFormula}
\begin{equation}
	z[x,y] = \frac{\langle \textbf{n}, \textbf{s} \rangle }{\textbf{k}[x,y] \cdot \textbf{n}},
\end{equation}
where $z[x,y]$ denotes the absolute geometric depth of the intersection point for the corresponding pixel, $\mathbf{n}$ is the surface normal of the object, $\mathbf{s}$ is a point known to lie on the object plane (such as the midpoint between the real and virtual light source), and $\langle \cdot, \cdot \rangle$ denotes the scalar product. The corresponding 3D intersection point $\textbf{H}$ on the object surface is then defined as:
\begin{equation}
	\textbf{H}[x,c] = z[x,y] \times \textbf{k}[x,y]~~.
\end{equation} 

Finally, the the absolute depth is converted back into a pixel displacement to generate the disparity map $d[x,y]$ using the standard stereo vision relationship:
\begin{equation}
	d[x,y] = \frac{B \cdot f}{z[x,y]}~~.
\end{equation}
To generate the global disparity map, the independent disparity estimates from all camera pairs are aggregated by computing their pixel-wise median.

\section{Experimental Results}
\label{sec:Experiments}
\subsection{Synthetic and Real-World Data Generation}
\label{sec:Synthetic Data Generation}
\begin{figure}
	\centering
	\begin{tikzpicture}
		\node(img1)[]{\includegraphics[width=0.079\textwidth]{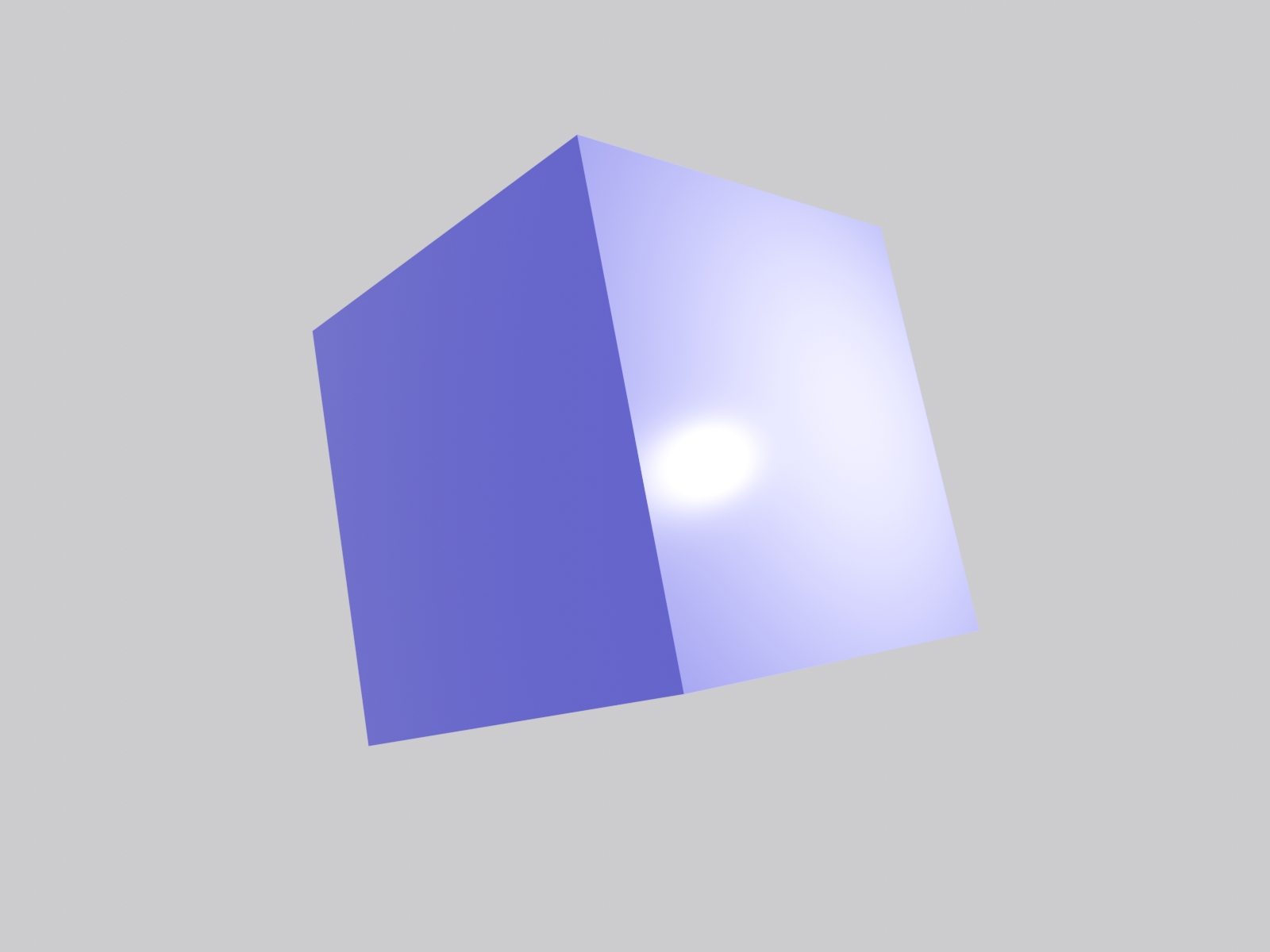}};
		\node(disp1)[below of=img1, yshift=-0.15cm]{\includegraphics[width=0.079\textwidth]{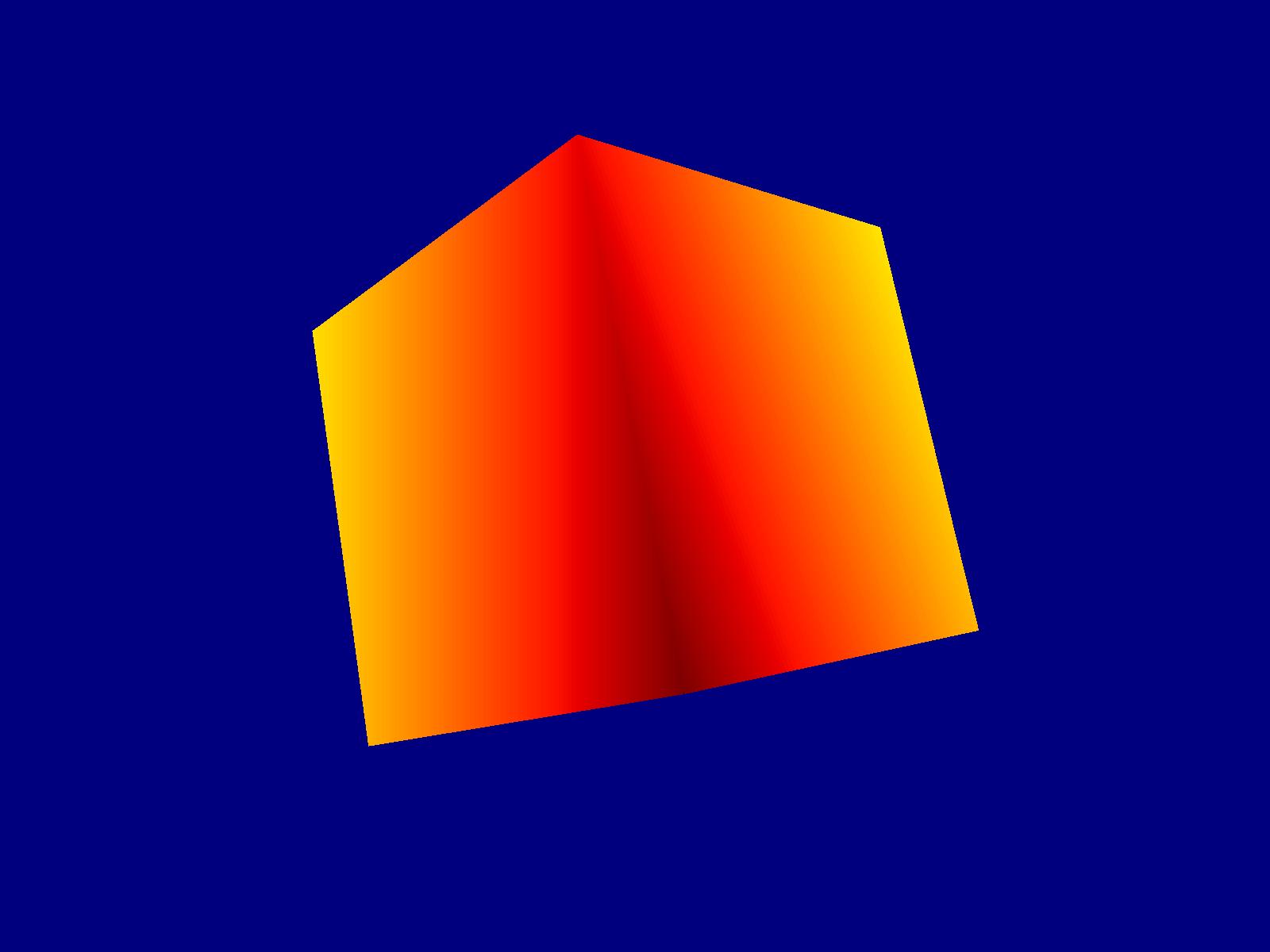}};
		\node(img2)[right of=img1, xshift=0.5cm]{\includegraphics[width=0.079\textwidth]{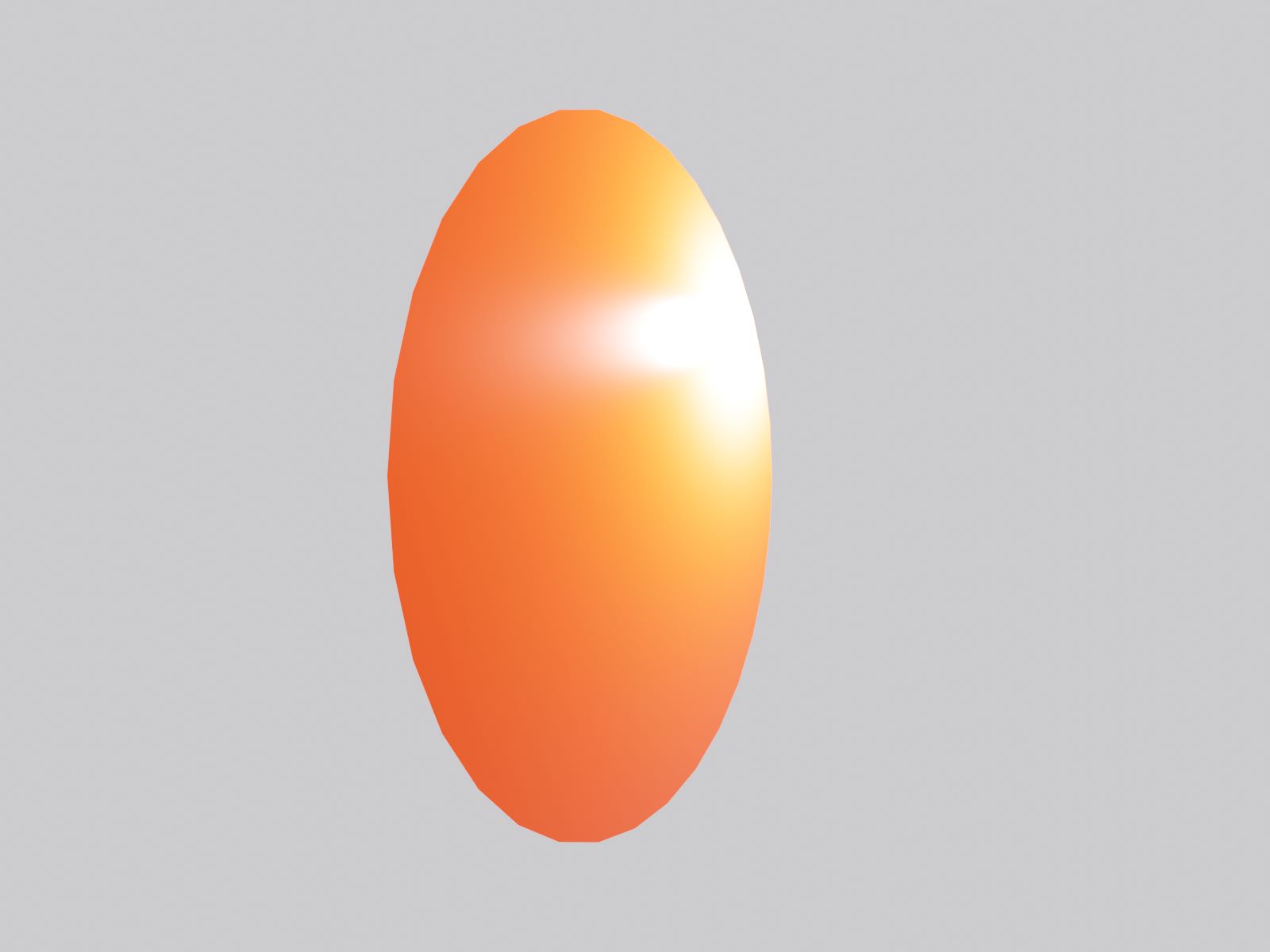}};
		\node(disp2)[below of=img2, yshift=-0.15cm]{\includegraphics[width=0.079\textwidth]{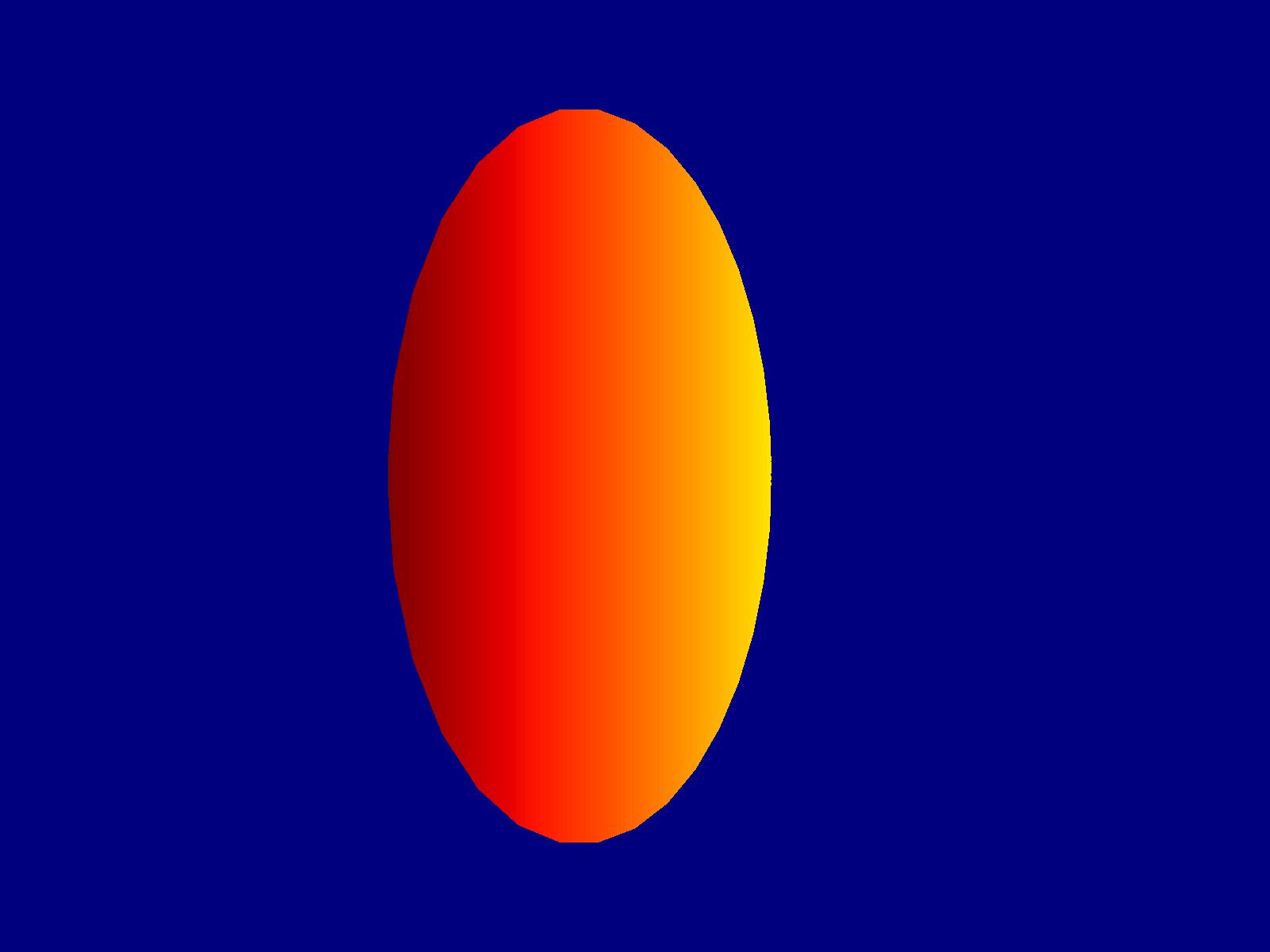}};
		\node(img3)[right of=img2, xshift=0.5cm]{\includegraphics[width=0.079\textwidth]{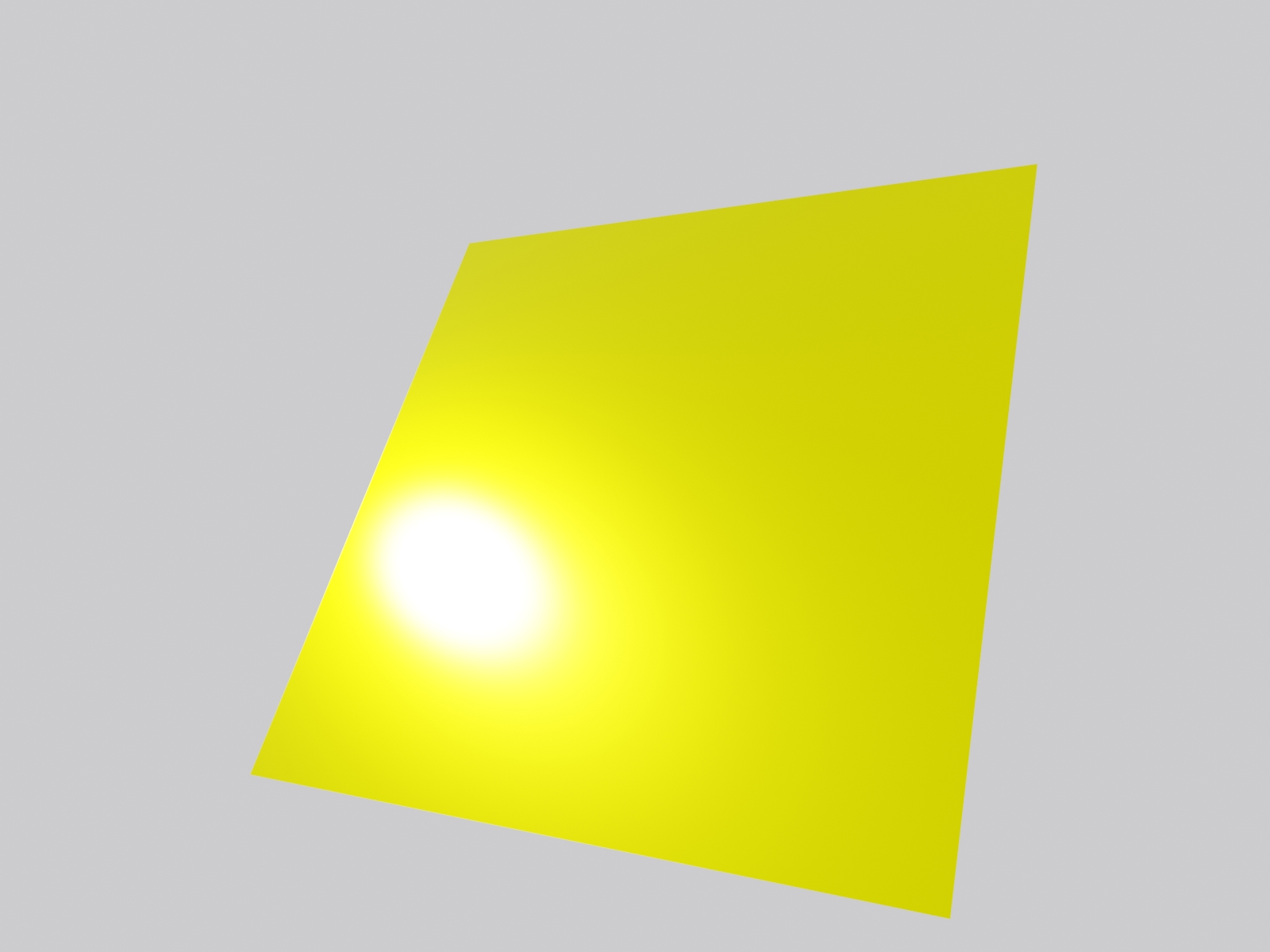}};
		\node(disp3)[below of=img3, yshift=-0.15cm]{\includegraphics[width=0.079\textwidth]{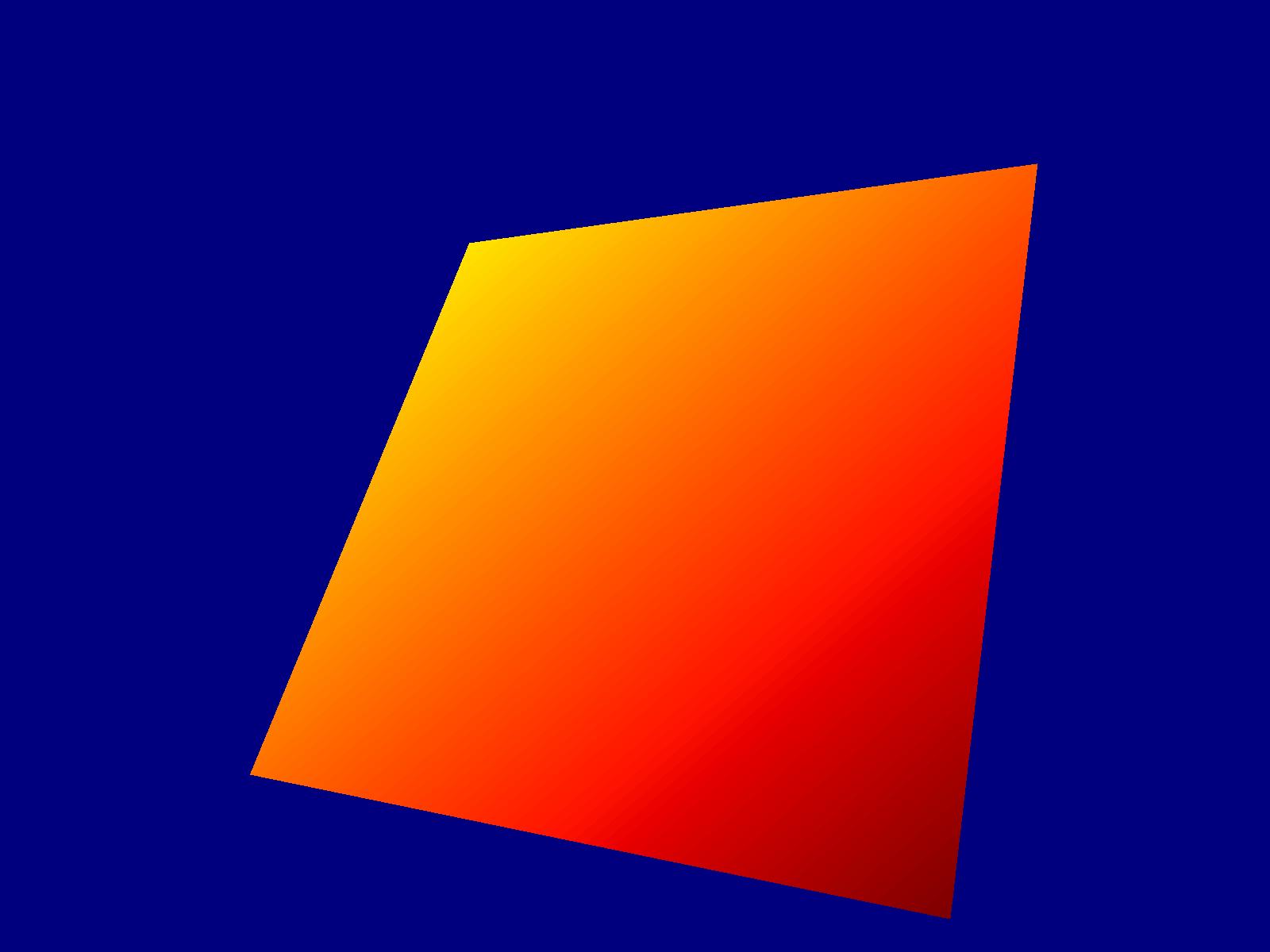}};
		\node(img4)[right of=img3, xshift=0.5cm]{\includegraphics[width=0.079\textwidth]{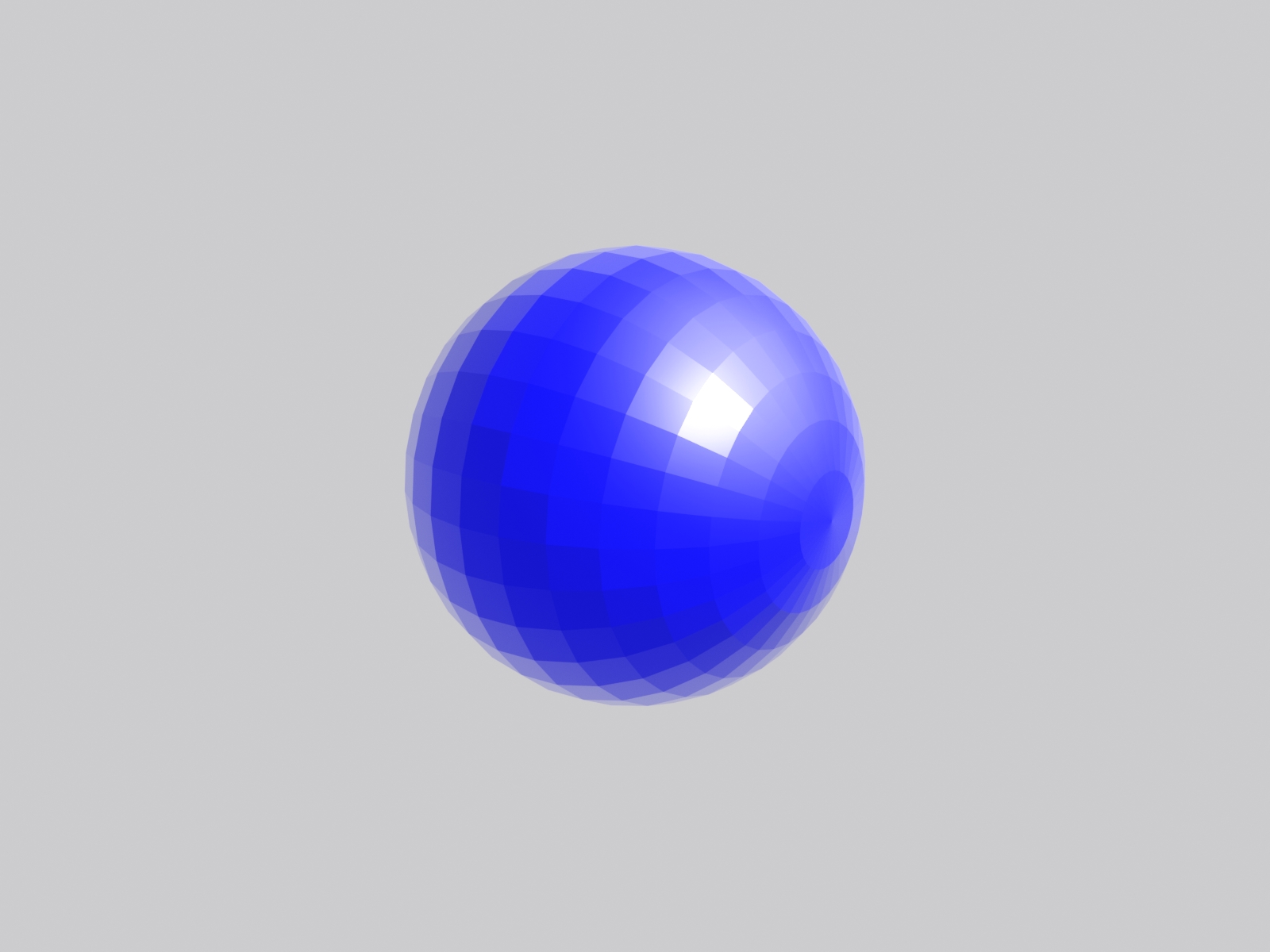}};
		\node(disp4)[below of=img4, yshift=-0.15cm]{\includegraphics[width=0.079\textwidth]{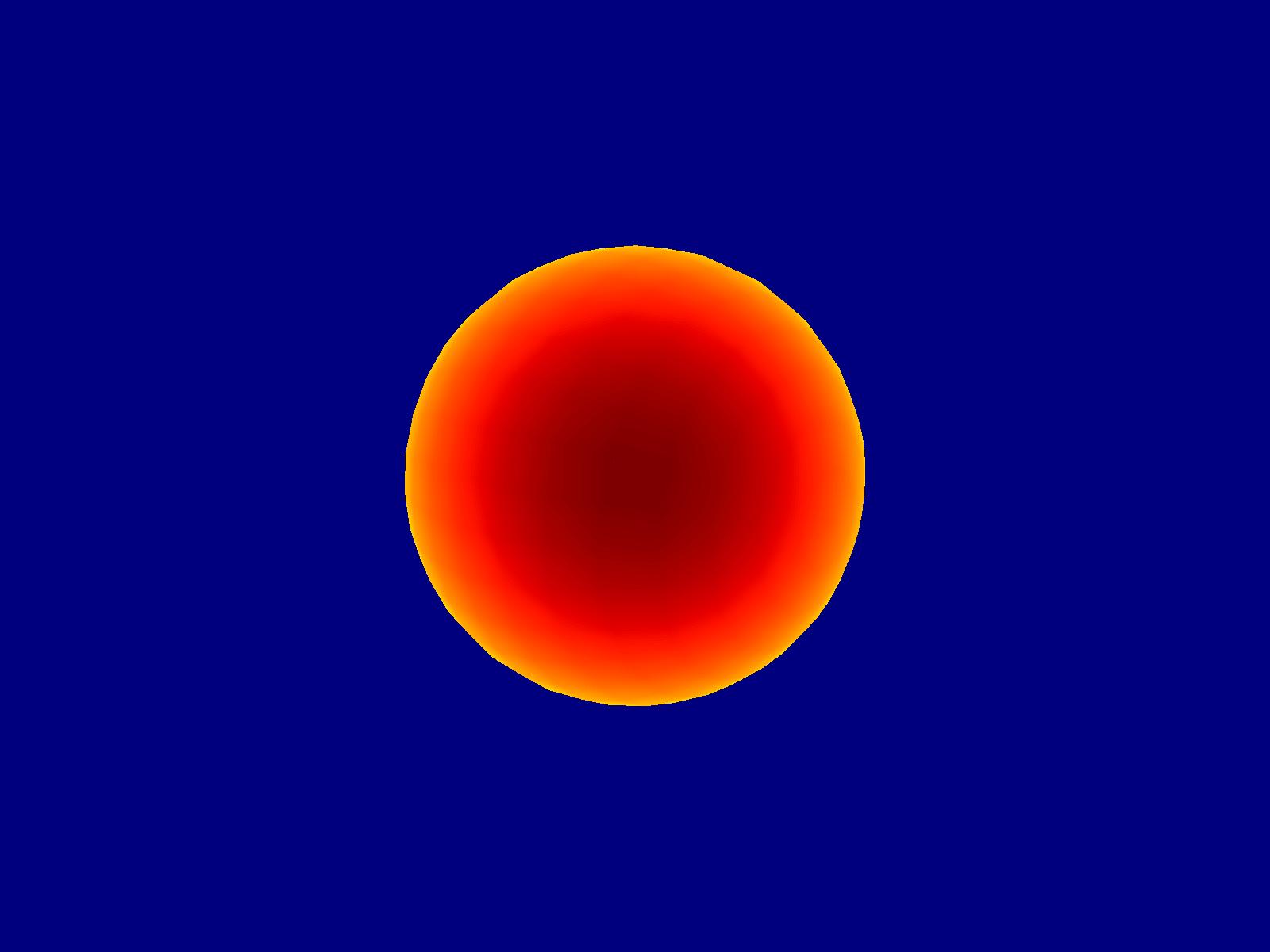}};
		\node(img5)[right of=img4, xshift=0.5cm]{\includegraphics[width=0.079\textwidth]{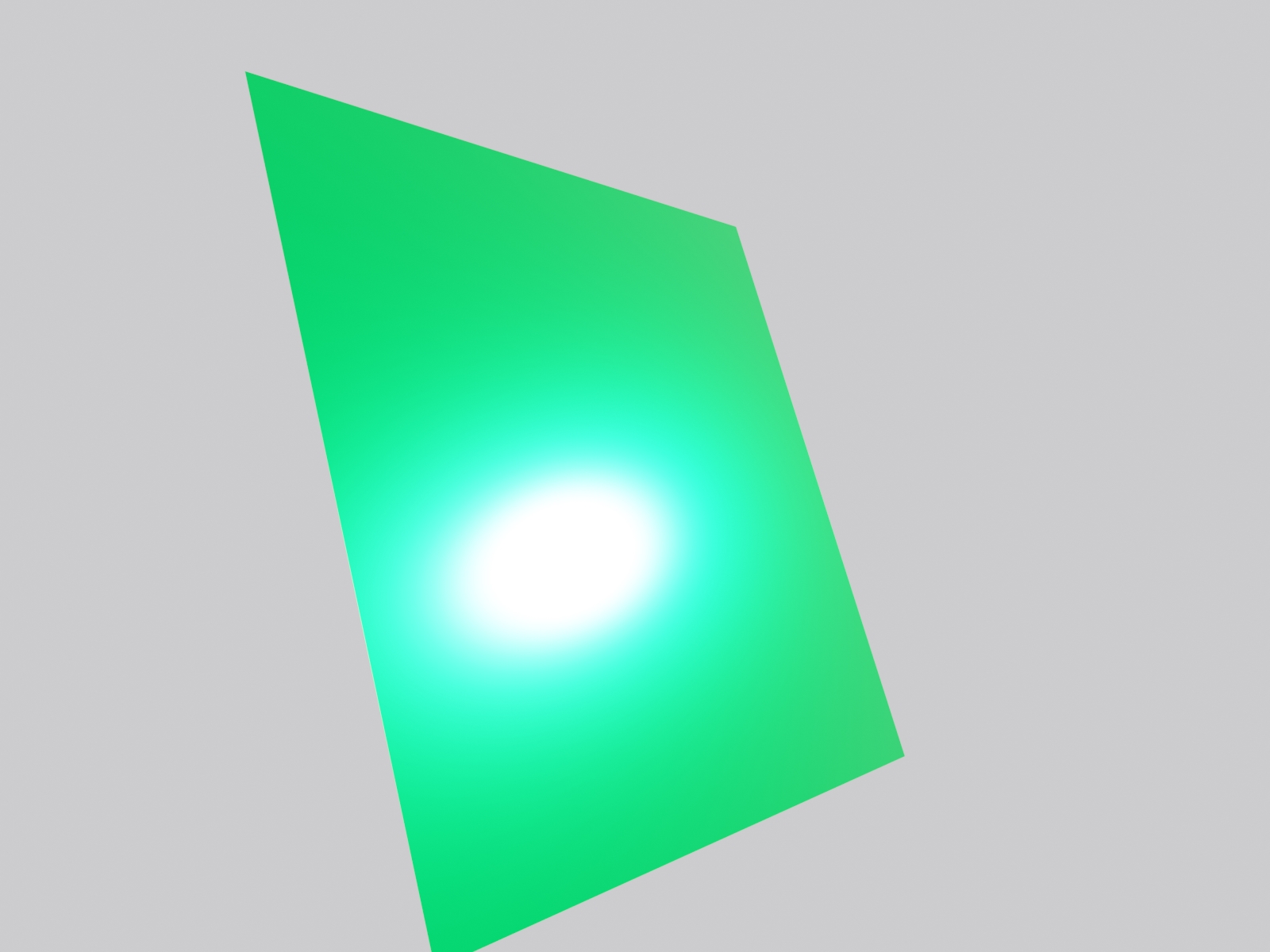}};
		\node(disp5)[below of=img5, yshift=-0.15cm]{\includegraphics[width=0.079\textwidth]{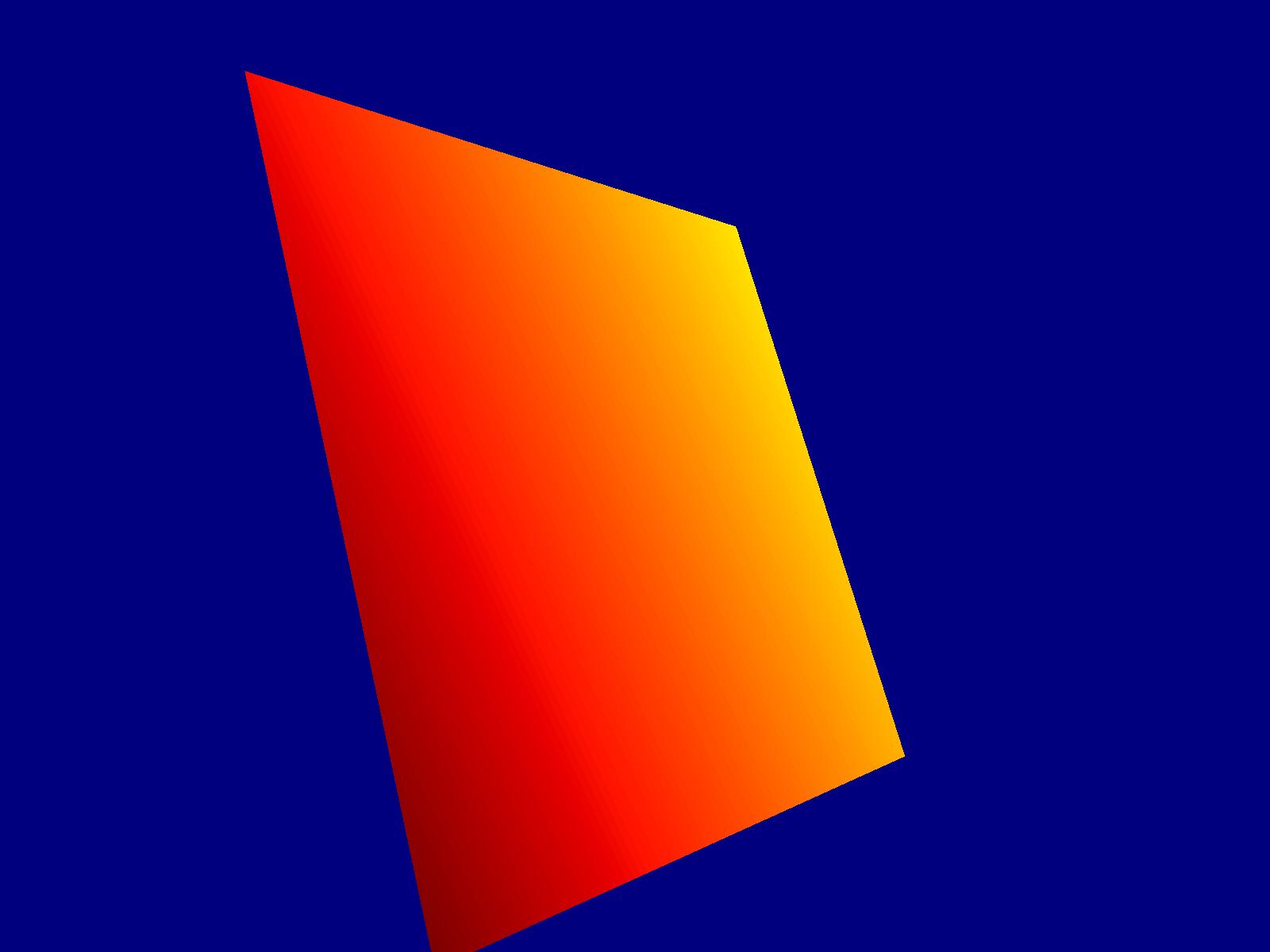}};
		\node(img6)[right of=img5, xshift=0.5cm]{\includegraphics[width=0.079\textwidth]{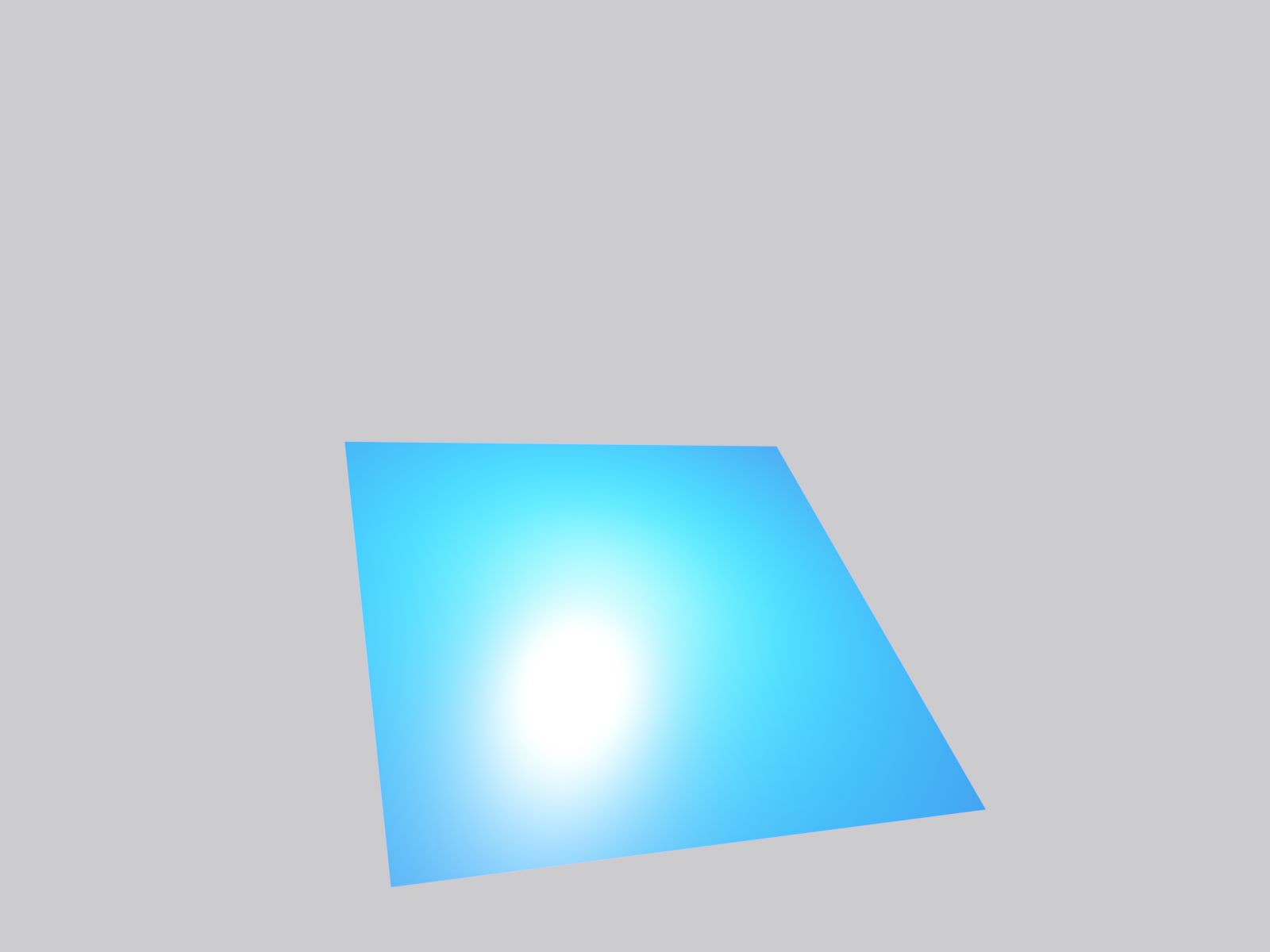}};
		\node(disp6)[below of=img6, yshift=-0.15cm]{\includegraphics[width=0.079\textwidth]{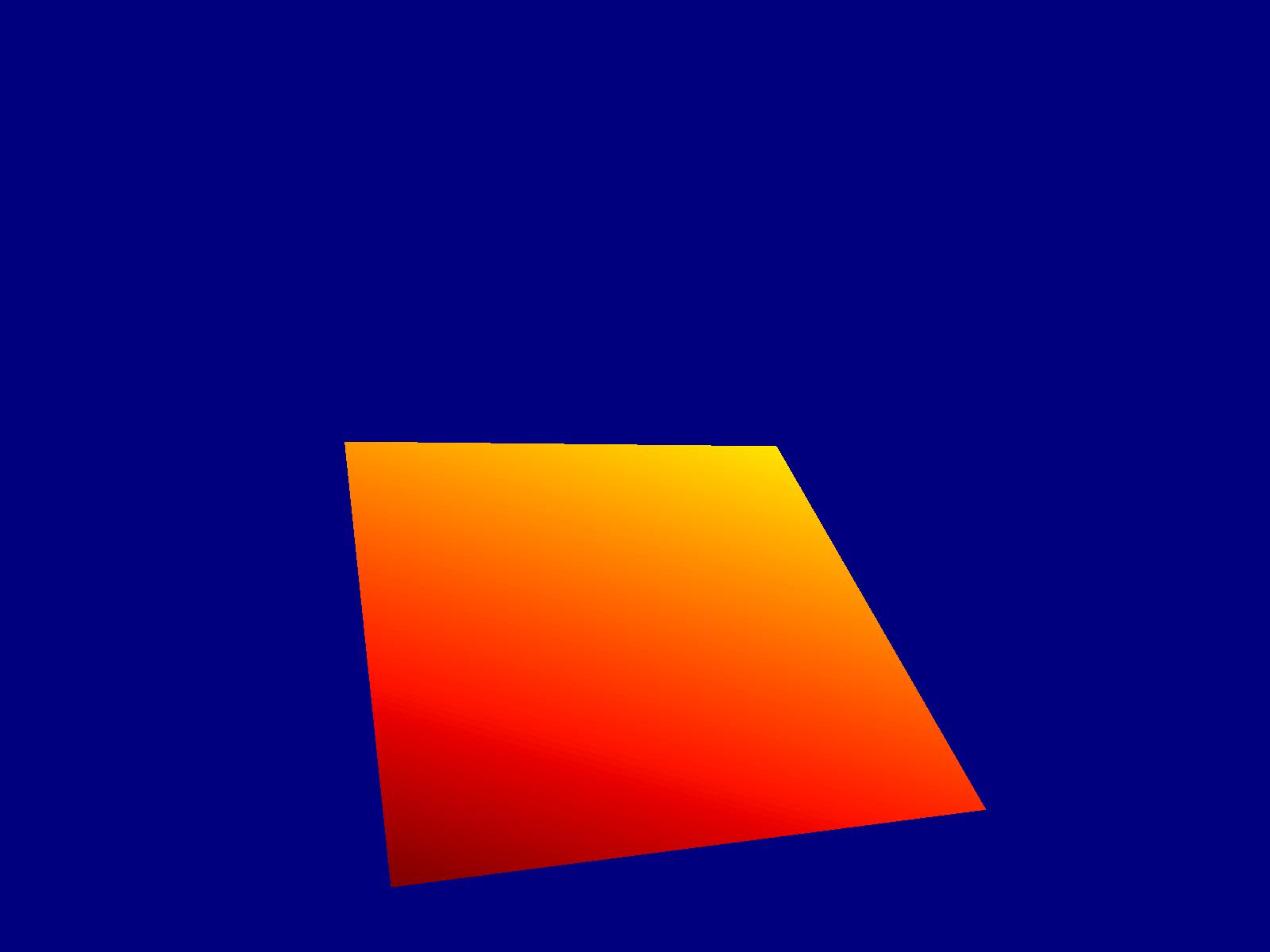}};
	\end{tikzpicture}
	\vspace{-0.5cm}
	\caption{Example image (top) and disparity (bottom) pairs of our synthetic data generated using Blender \cite{Blender}.}
	\label{fig:BlenderExamples}
	\vspace{-0.5cm}
\end{figure}

\noindent The introduced method is designed for disparity estimation of flat, textureless objects which are thus prone to specular reflections. Existing datasets might contain few such images but do not fully cover this specific scenario. Therefore, we generated a new synthetic dataset using Blender \cite{Blender}. The camera array can be precisely modeled and the ground truth disparity maps can be obtained, which are not directly available for real-world images. Since camera arrays consist of multiple cameras, our scenes are captured with nine virtual cameras arranged in a $3 \times 3$ array. The considered camera array exhibits a local displacement with a baseline of 85 mm in horizontal and vertical, and 85$\sqrt{2}$ mm in diagonal direction. Furthermore, one light source of random intensity, size, and color is placed at various positions, but is never directly captured by the cameras. Instead, only its reflection on the object surface is visible. The objects are placed at different positions in front of the camera array. While they are all textureless and reflect the light source, they vary in their color, shape, scale, and orientation. Some example images along with their ground truth disparity maps are shown in Fig. \ref{fig:BlenderExamples}. 

The real-world images were captured with a $3 \times 3$ camera array, matching the geometric configuration of the virtual setup. Since this contribution focuses on areas with specular reflections, as other regions can already be estimated well using conventional methods, the background is modeled as being at infinity and no notable disparity is present in this area.

\subsection{Experimental Setting}
\subsubsection{Implementation Details}
We test our proposed \-dis\-pa\-rity estimation method using synthetic as well as real-world data. For detecting the reflections, we apply the described adaptive gradient thresholding. The algorithms were executed on a NVIDIA RTX A4000 GPU and the runtime is measured as best out of three.
	
The quantitative and qualitative results reported below are obtained under the conditions for which SRDE is designed. Specifically, the evaluation focuses on flat, textureless objects illuminated by a single light source, resulting in a specular reflection that is visible on the object surface in the image.

\subsubsection{Evaluation Metrics}
For the evaluation of the accuracy of our SRDE solution and the benchmark networks we follow the popular metrics End Point Error (EPE), Root Mean Squared Error (RMSE), and Bad Matching Pixels, with 0.1, 0.01, and 0.001 pixel threshold (bmp0.1, bmp0.01, and bmp0.001).
For the evaluated synthetic images, the ground truth disparity maps are considered as a reference, while a subjective visual comparison is performed for the real-world data.
\begin{table}[t!]
	\centering
	\caption{List of available pretrained models. \linebreak \underline{Underlined}: Chosen models due to best performance.}
	\vspace{-0.1cm}
	\begin{tabular}{c|l}
		\toprule[1.5pt]
		Model & Dataset \\ \midrule
		\multirow{2}{*}{NMRF \cite{NMRF}} & \textit{KITTI} \cite{Kitti} \\
		& \textit{\underline{Sceneflow}} \cite{Sceneflow}\\[1.5ex]
		CREStereo \cite{CREStereo} & \textit{\underline{ETH3D}} \cite{eth3d} \\[1.5ex]
		\multirow{4}{*}{IGEV \cite{IgevDE}} & \textit{ETH3D} \cite{eth3d}\\
		& \textit{KITTI} \cite{Kitti} \\
		& \textit{Middlebury} \cite{Middlebury}\\
		& \textit{\underline{Sceneflow}} \cite{Sceneflow} \\
		\bottomrule[1.5pt]
	\end{tabular}
	\label{tab:PretrainedModelTests}
	\vspace{-0.5cm}
\end{table}
	
\subsubsection{Benchmark Comparison}
We evaluate our method in comparison to the three \-po\-pular benchmark models NMRF \cite{NMRF}, CREStereo \cite{CREStereo} and IGEV \cite{IgevDE}. The tests were performed considering the available pretrained models, which are listed in Table \ref{tab:PretrainedModelTests}. It is important to note that training or fine-tuning standard stereo networks on datasets consisting entirely of flat, textureless objects is practically ineffective, as these architectures fundamentally rely on texture and local spatial features to establish correspondences. Consequently, we evaluated the networks using their pretrained weights, and only the best performing models are used for evaluation and comparison. For NMRF and IGEV, Sceneflow \cite{Sceneflow} yields the best results, while it is ETH3D \cite{eth3d} for CREStereo. These datasets contain at least some instances of naturally occurring reflections such as glare on wet streets or direct sunlight captured by the camera, ensuring the networks had some prior exposure to specular reflections.

For the evaluation, object masks were generated using Reflection-aware Postprocessed Segmentation (RePoSeg) \cite{RePoSeg}, which reliably identifies objects exhibiting specular reflections. As SRDE computes the disparity by modeling the object surface as a plane, its geometric validity is limited to the object boundaries. Applying this planar model to a non-planar background would result in physically meaningless disparities and would not accurately reflect the performance of SRDE. While state-of-the-art neural networks perform well on textured backgrounds, they typically fail on these featureless, specular object regions. Evaluating the entire image would therefore obscure the specific problem SRDE is designed to solve and distort the comparison by including regions where SRDE is not intended to operate. To ensure a fair comparison of the disparity estimation on reflective surfaces, the following quantitative and qualitative results are computed exclusively within these masked regions, while the background is disregarded. For a comprehensive evaluation on whole scenes we perform experiments in Section \ref{sec:Integration}, showing how SRDE complements state-of-the-art models in complex environments.
	
	\begin{figure*}[ht!]
		\centering
	 \begin{tikzpicture}
			\node(rgb1)[]{\includegraphics[width=0.15\textwidth]{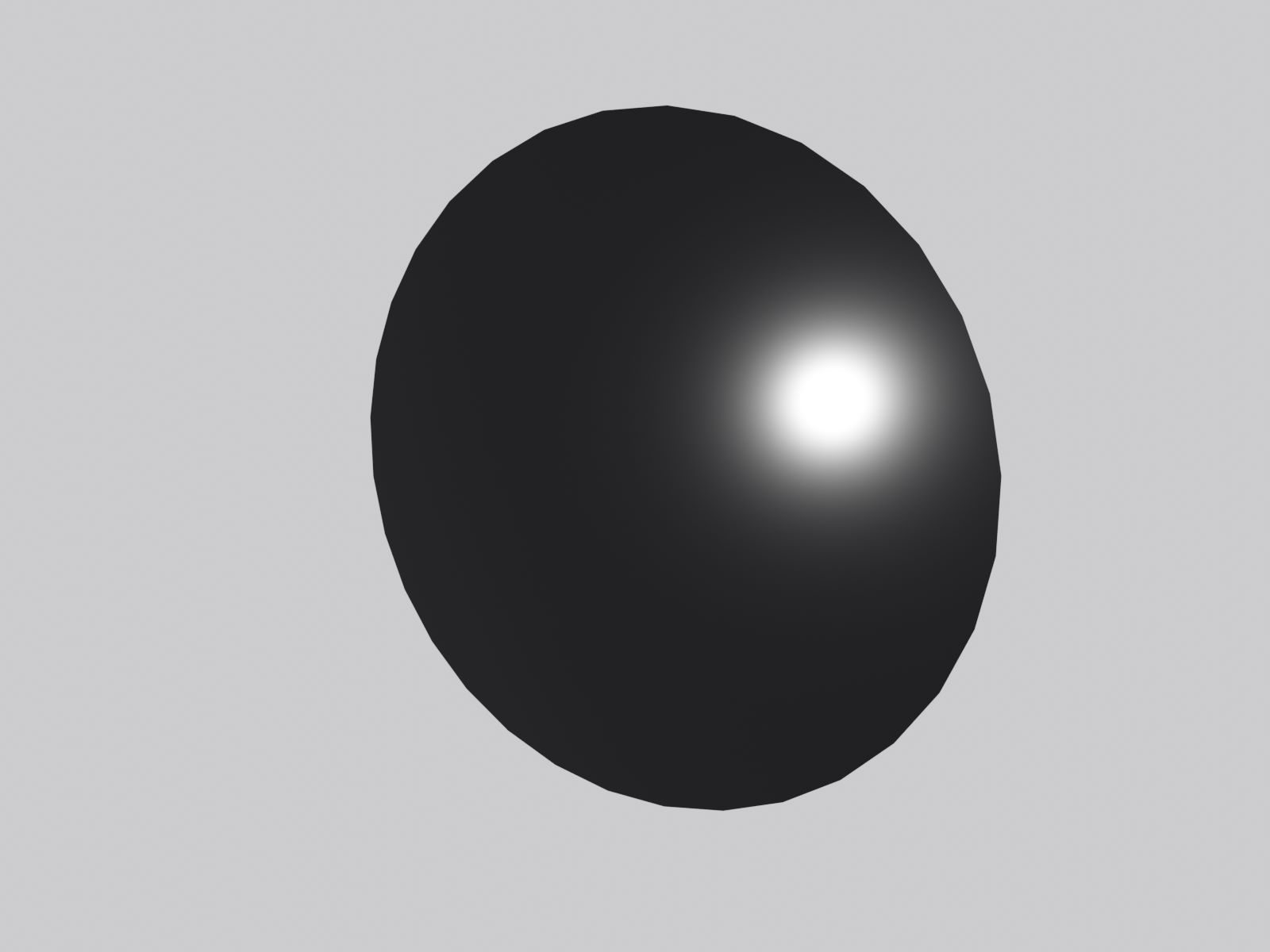}};
			\node(rgb2)[below of=rgb1,yshift=-1.12cm]{\includegraphics[width=0.15\textwidth]{Images/Circle4.jpg}};
			\node(rgb3)[below of=rgb2, yshift=-1.12cm]{\includegraphics[width=0.15\textwidth]{Images/Plane3.jpg}};
			\node(rgb4)[below of=rgb3, yshift=-1.12cm]{\includegraphics[width=0.15\textwidth]{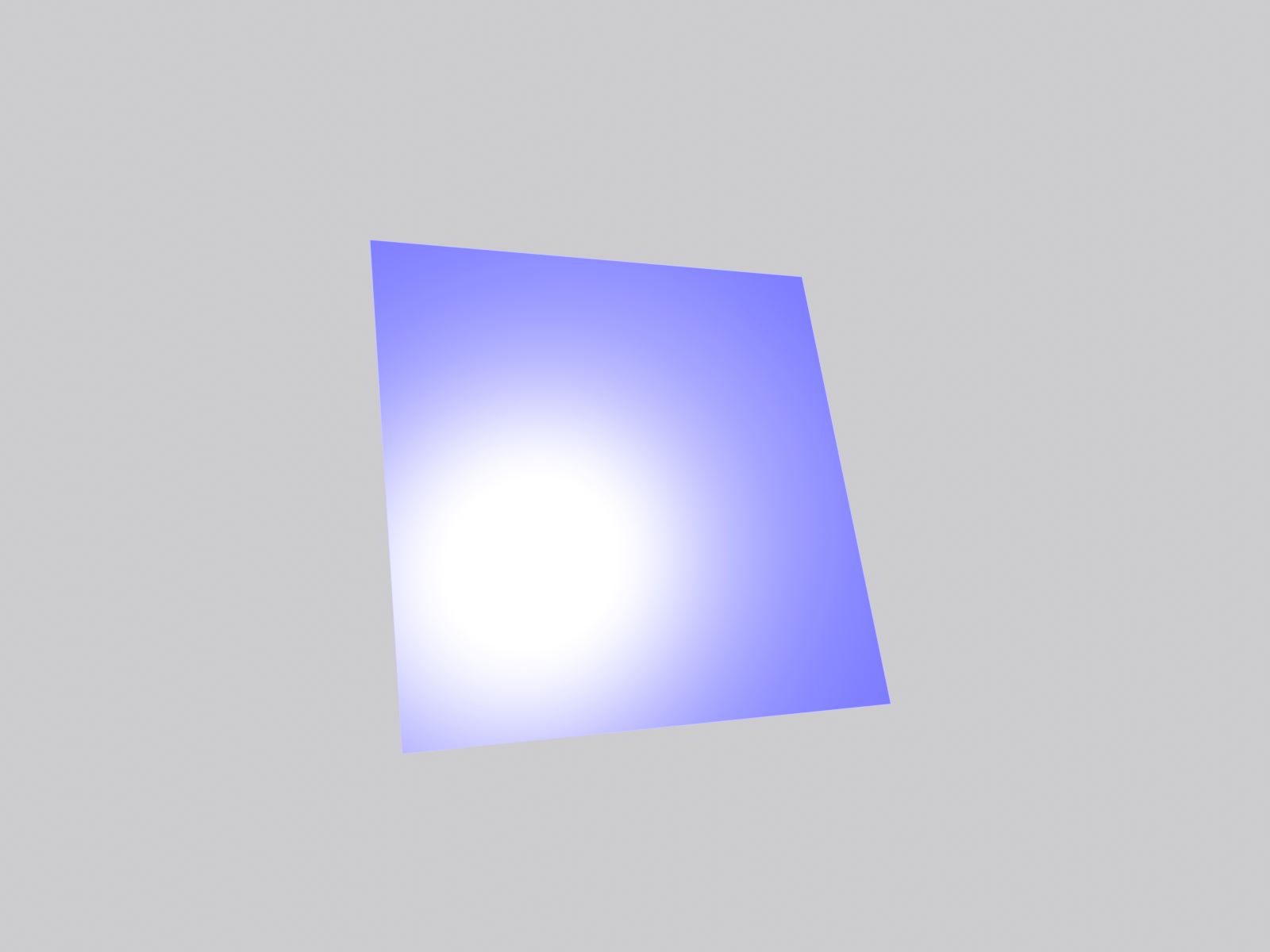}};
			\node(rgb5)[below of=rgb4, yshift=-1.12cm]{\includegraphics[width=0.15\textwidth]{Images/Rect8.jpg}};
			
			\node(gt1)[right of=rgb1, xshift=1.8cm]{\includegraphics[width=0.15\textwidth]{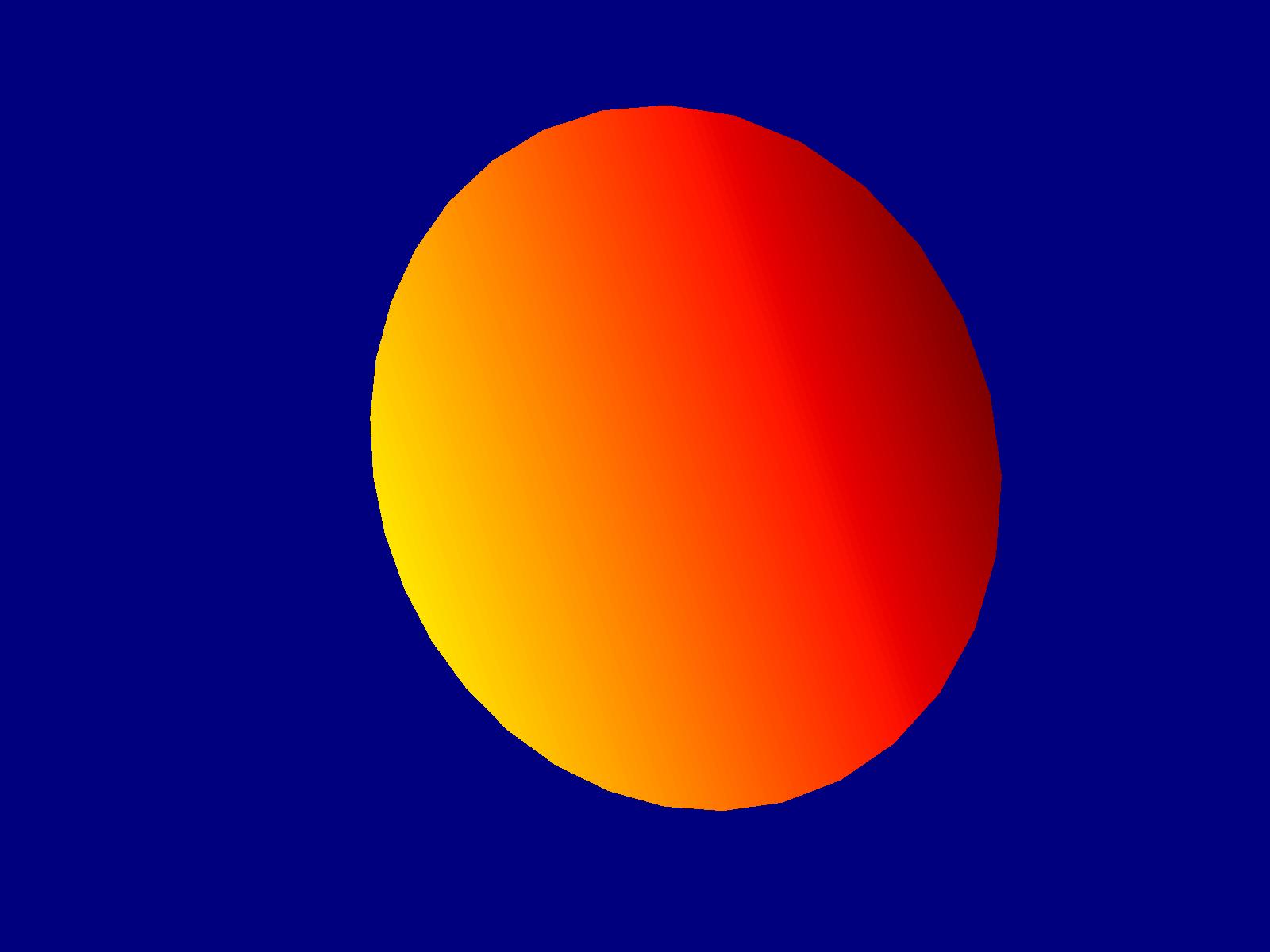}};
			\node(gt2)[right of=rgb2,xshift=1.8cm]{\includegraphics[width=0.15\textwidth]{Images/GT_Circle4.jpg}};
			\node(gt3)[right of=rgb3, xshift=1.8cm]{\includegraphics[width=0.15\textwidth]{Images/GT_Plane3.jpg}};
			\node(gt4)[right of=rgb4, xshift=1.8cm]{\includegraphics[width=0.15\textwidth]{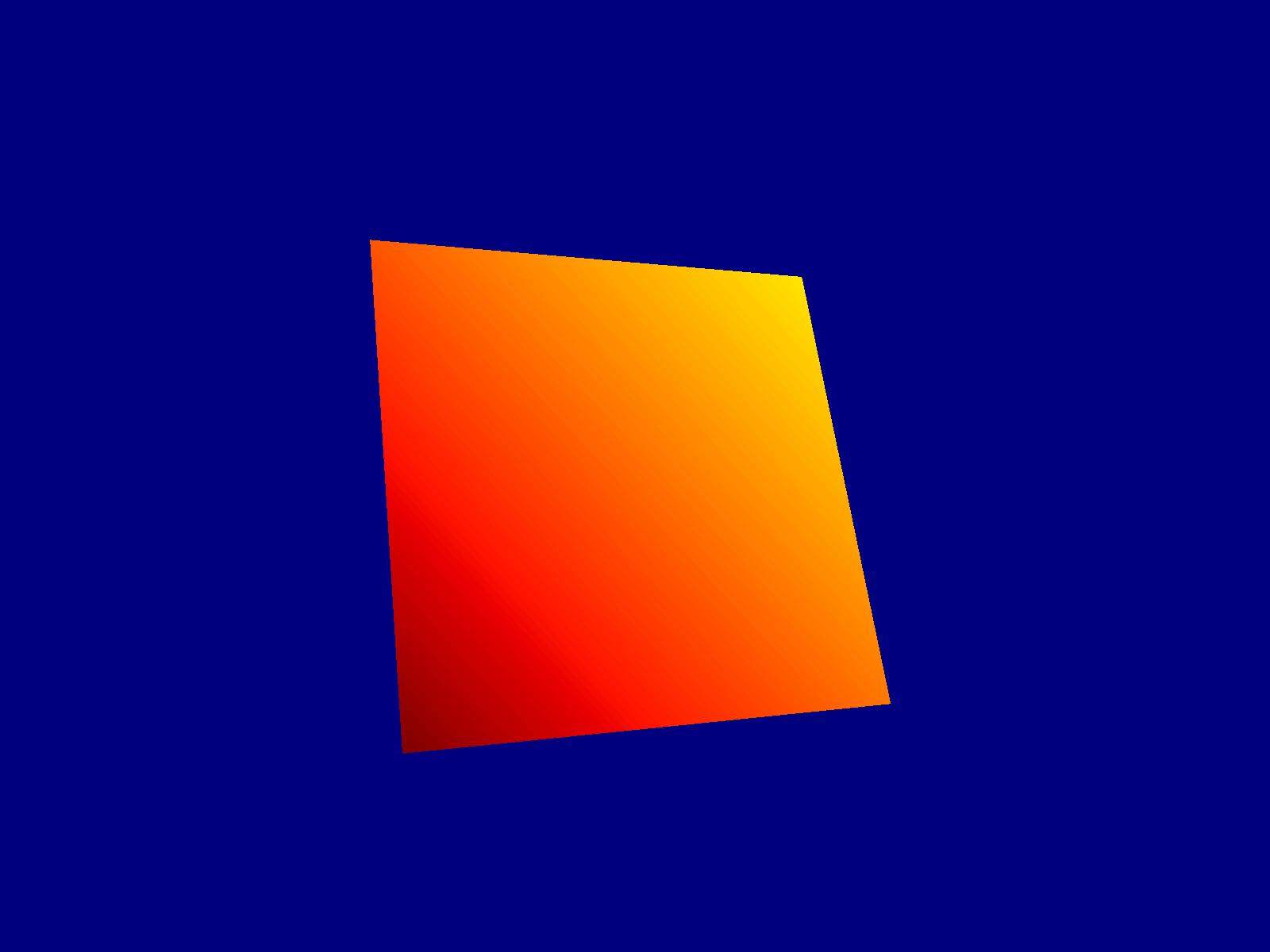}};
			\node(gt5)[right of=rgb5,xshift=1.8cm]{\includegraphics[width=0.15\textwidth]{Images/GT_Rect8.jpg}};
			
			\node(NMRF1)[right of=gt1,xshift=1.8cm]{\includegraphics[width=0.15\textwidth]{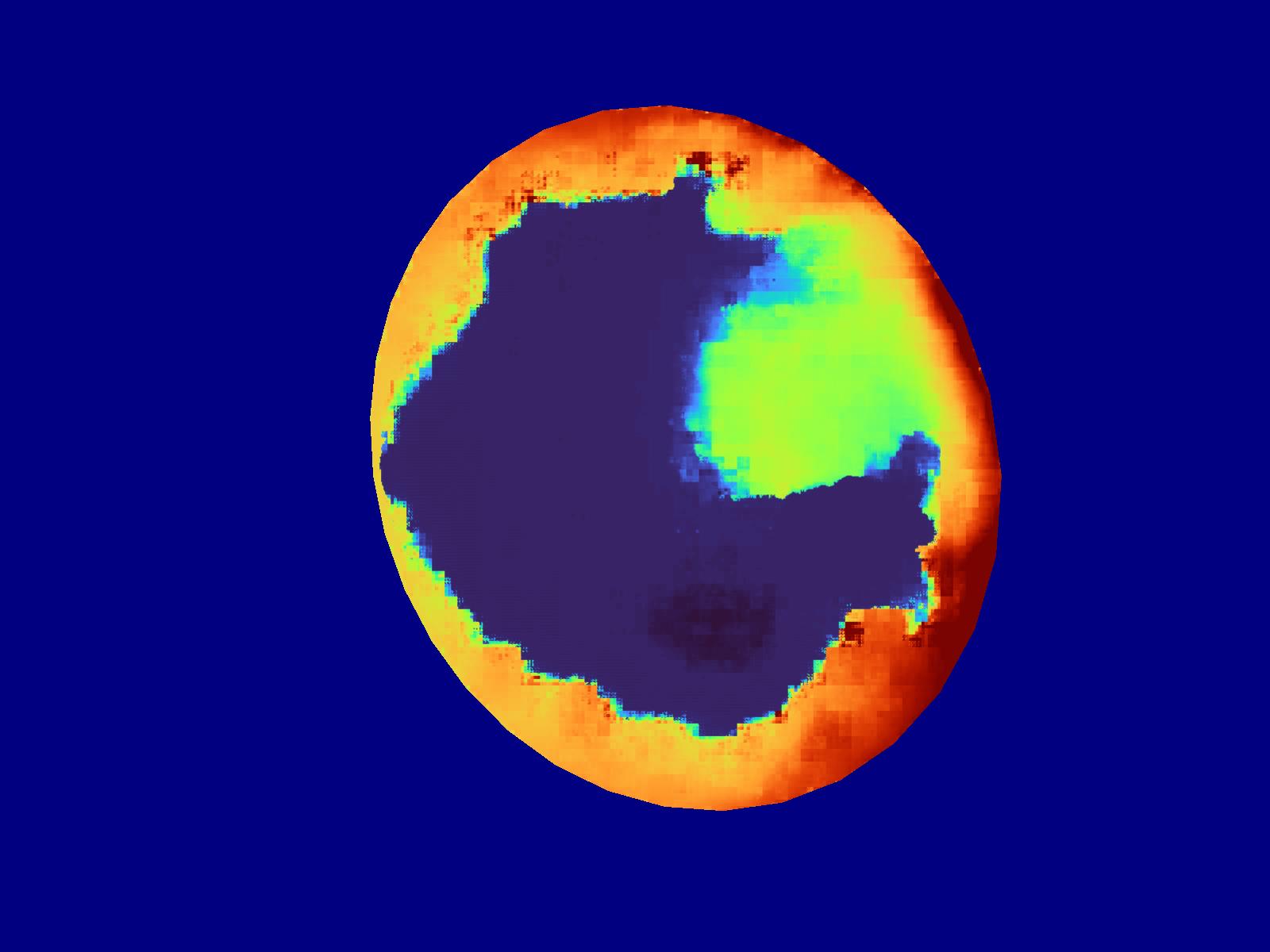}};
			\node(NMRF2)[right of=gt2,xshift=1.8cm]{\includegraphics[width=0.15\textwidth]{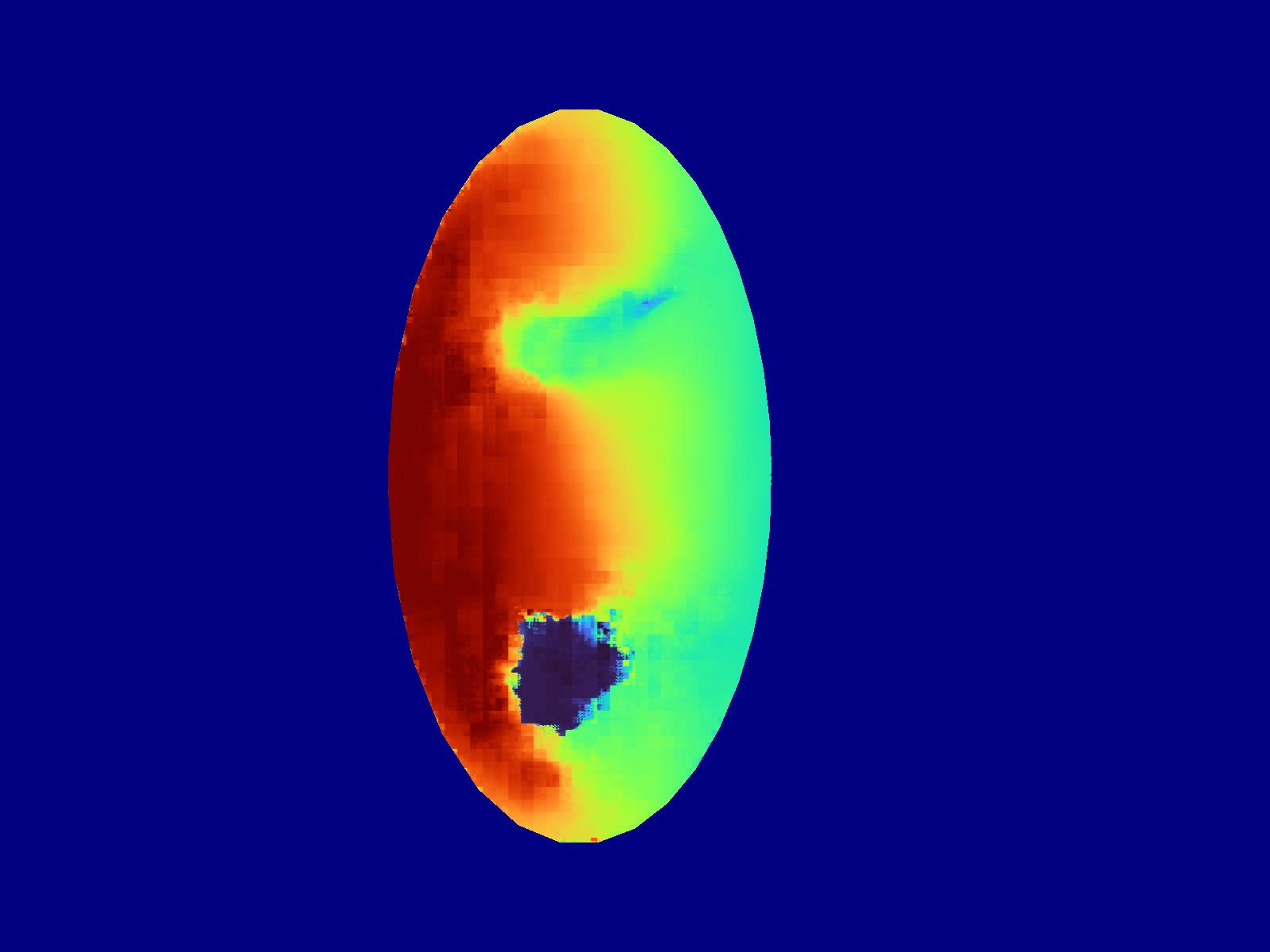}};
			\node(NMRF3)[right of=gt3,xshift=1.8cm]{\includegraphics[width=0.15\textwidth]{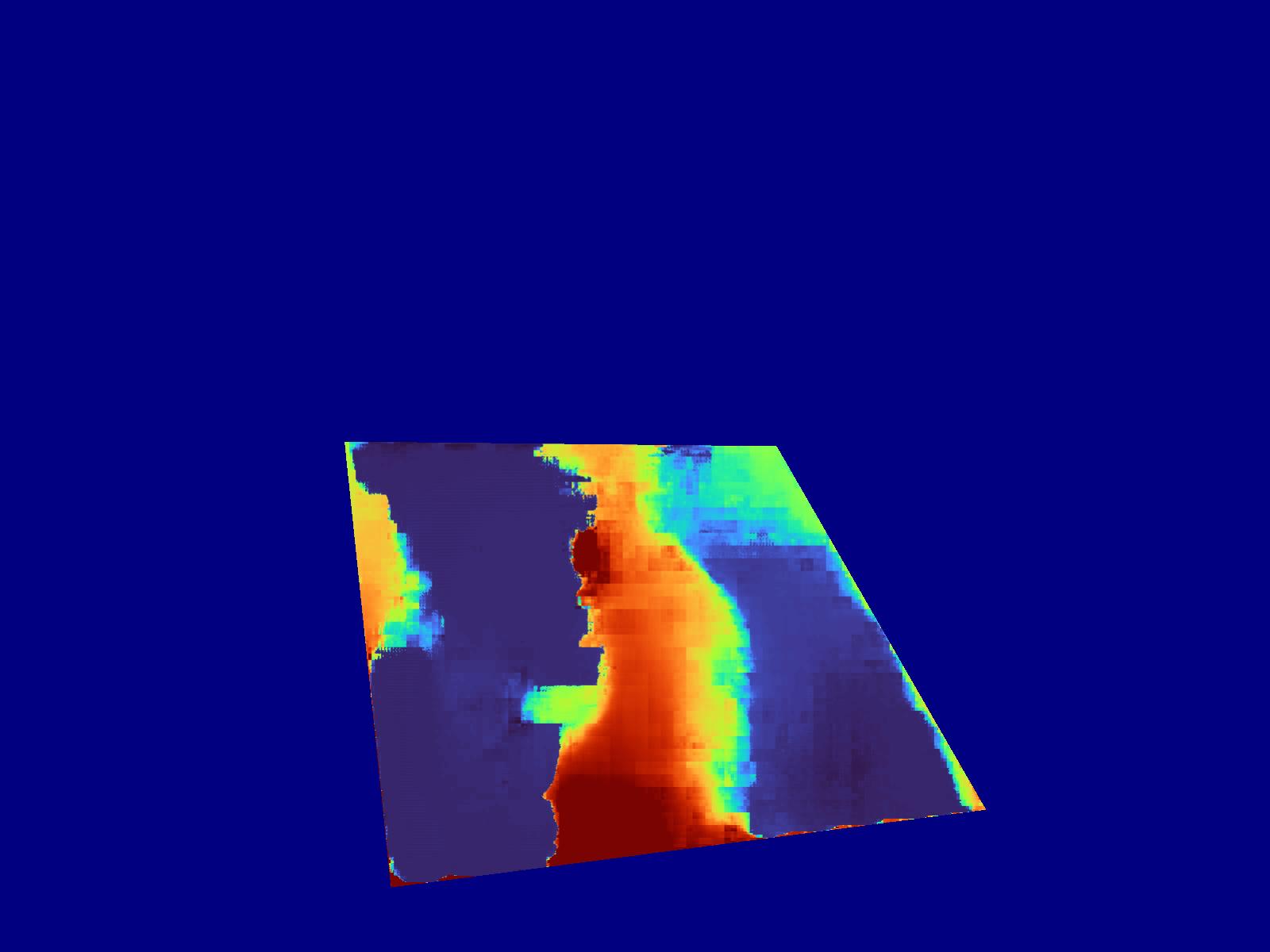}};
			\node(NMRF4)[right of=gt4,xshift=1.8cm]{\includegraphics[width=0.15\textwidth]{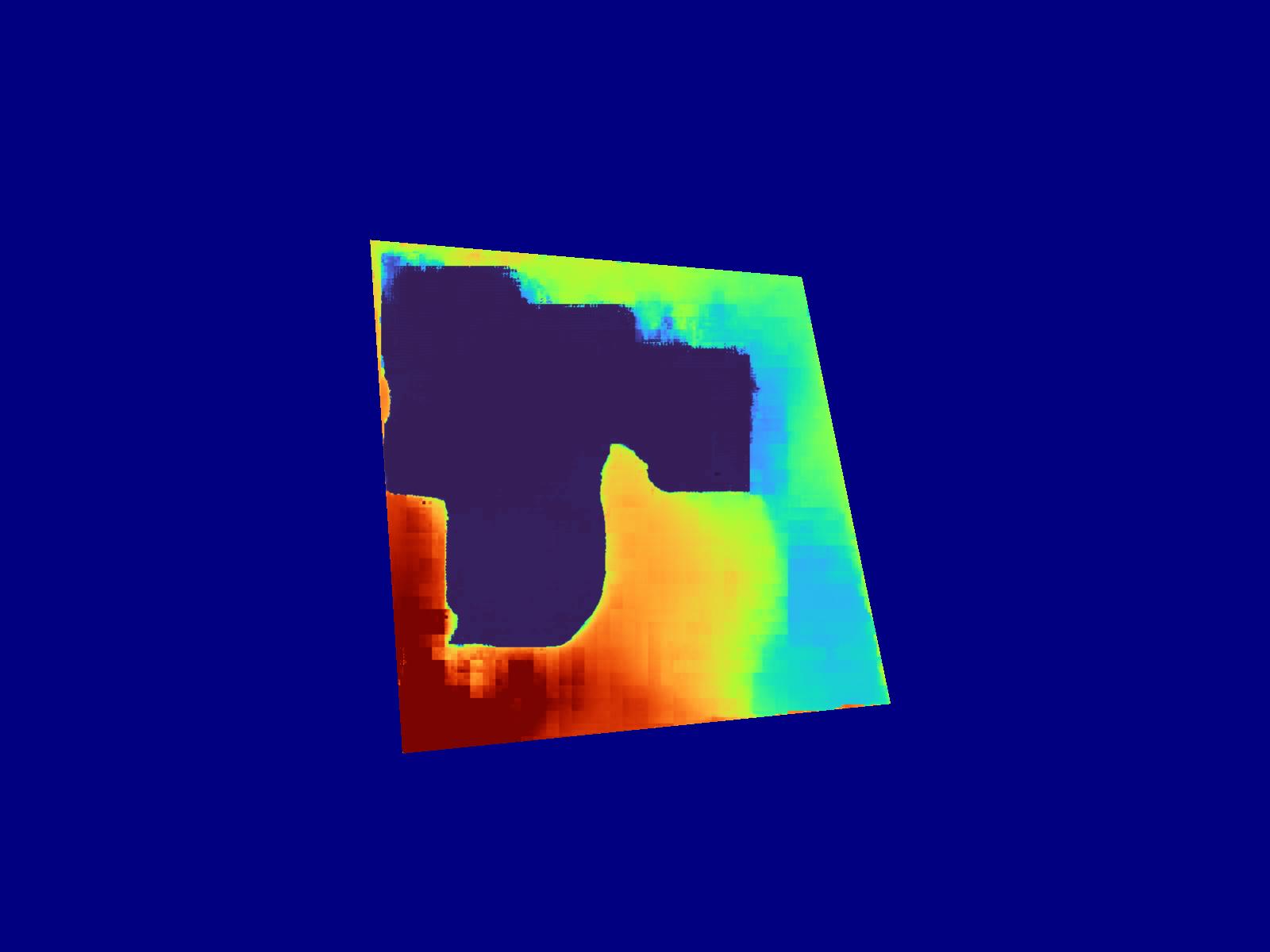}};
			\node(NMRF5)[right of=gt5,xshift=1.8cm]{\includegraphics[width=0.15\textwidth]{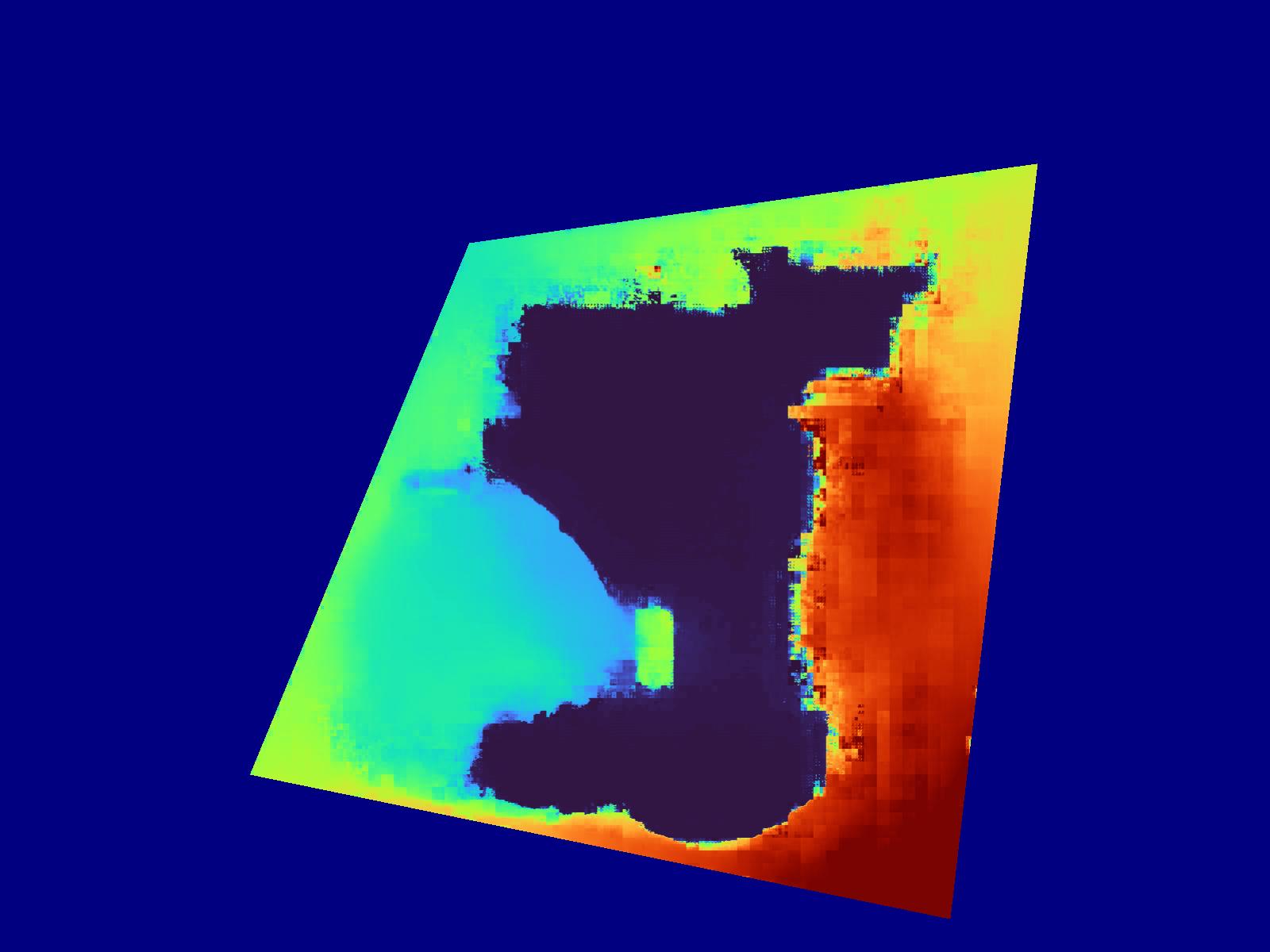}};
			
			\node(CRE1)[right of=NMRF1,xshift=1.8cm]{\includegraphics[width=0.15\textwidth]{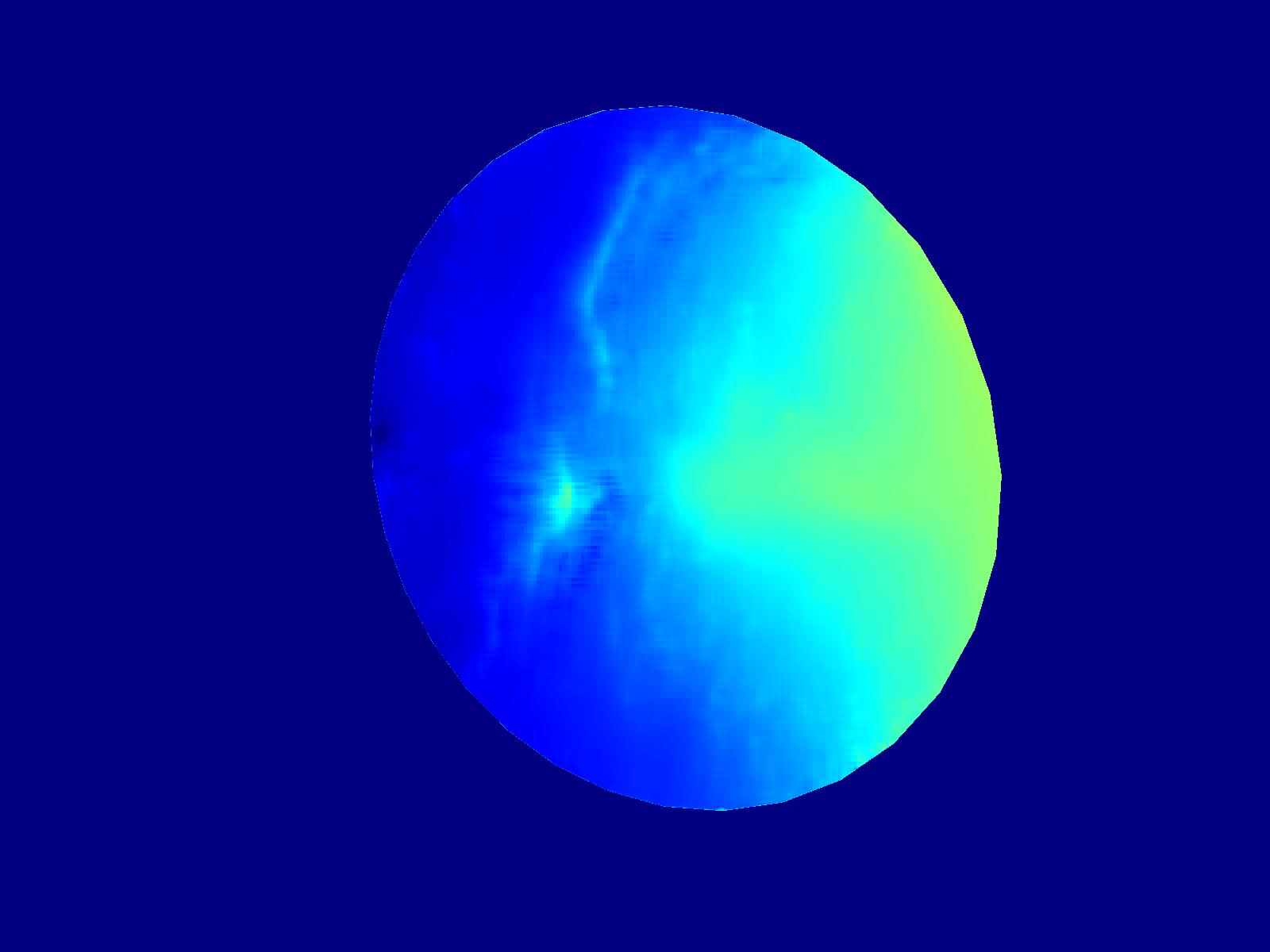}};
			\node(CRE2)[right of=NMRF2,xshift=1.8cm]{\includegraphics[width=0.15\textwidth]{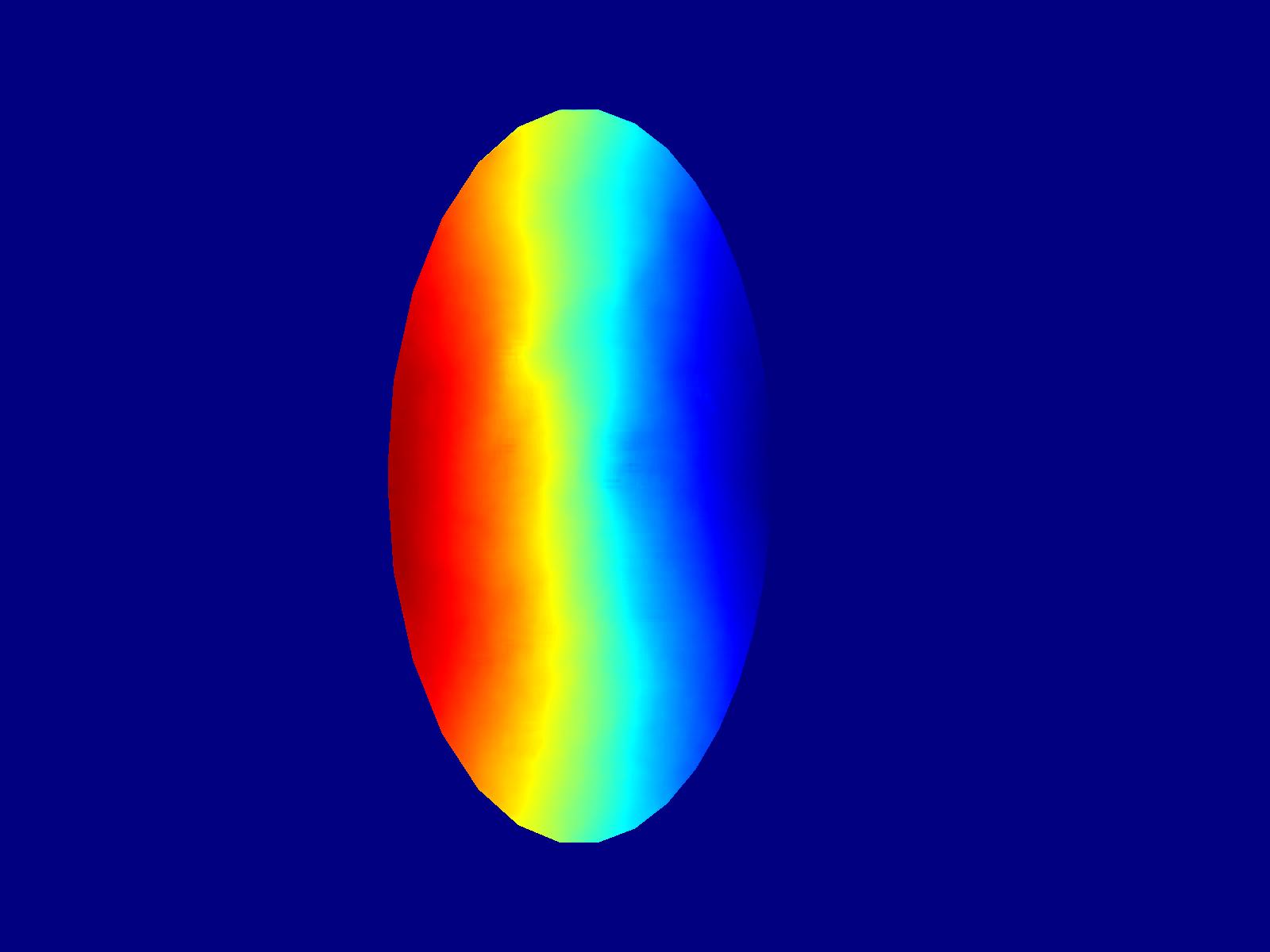}};
			\node(CRE3)[right of=NMRF3,xshift=1.8cm]{\includegraphics[width=0.15\textwidth]{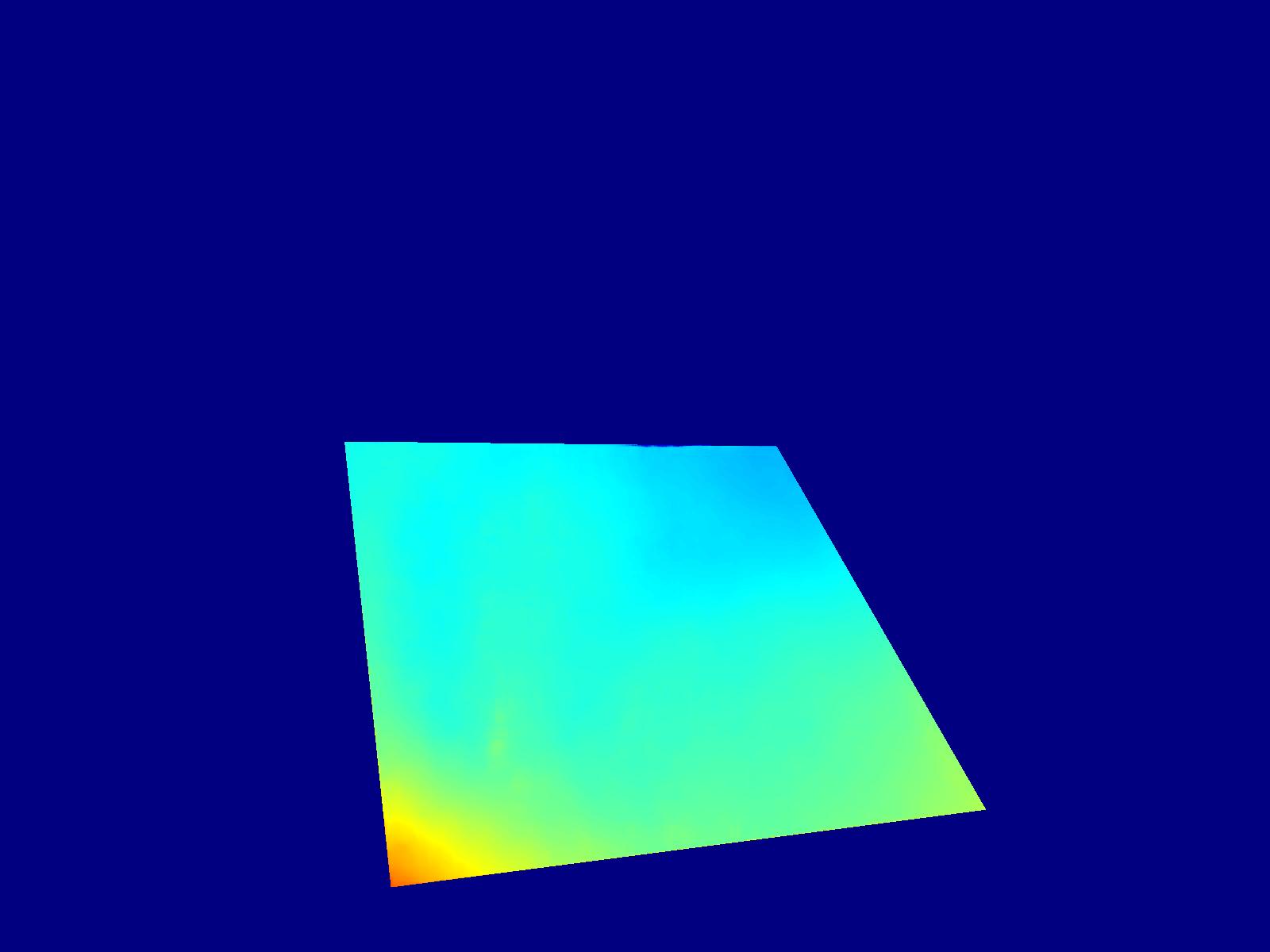}};
			\node(CRE4)[right of=NMRF4,xshift=1.8cm]{\includegraphics[width=0.15\textwidth]{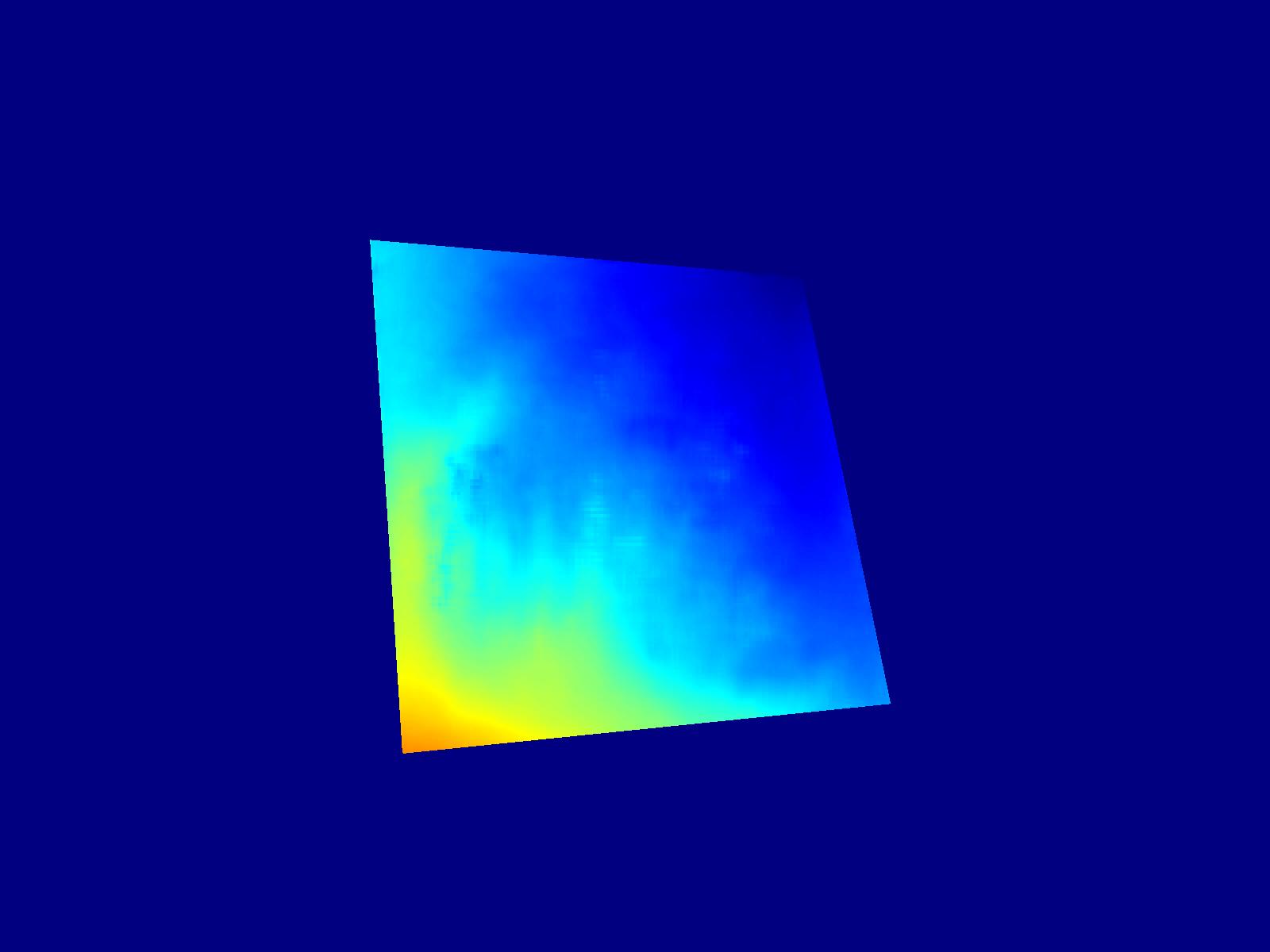}};
			\node(CRE5)[right of=NMRF5,xshift=1.8cm]{\includegraphics[width=0.15\textwidth]{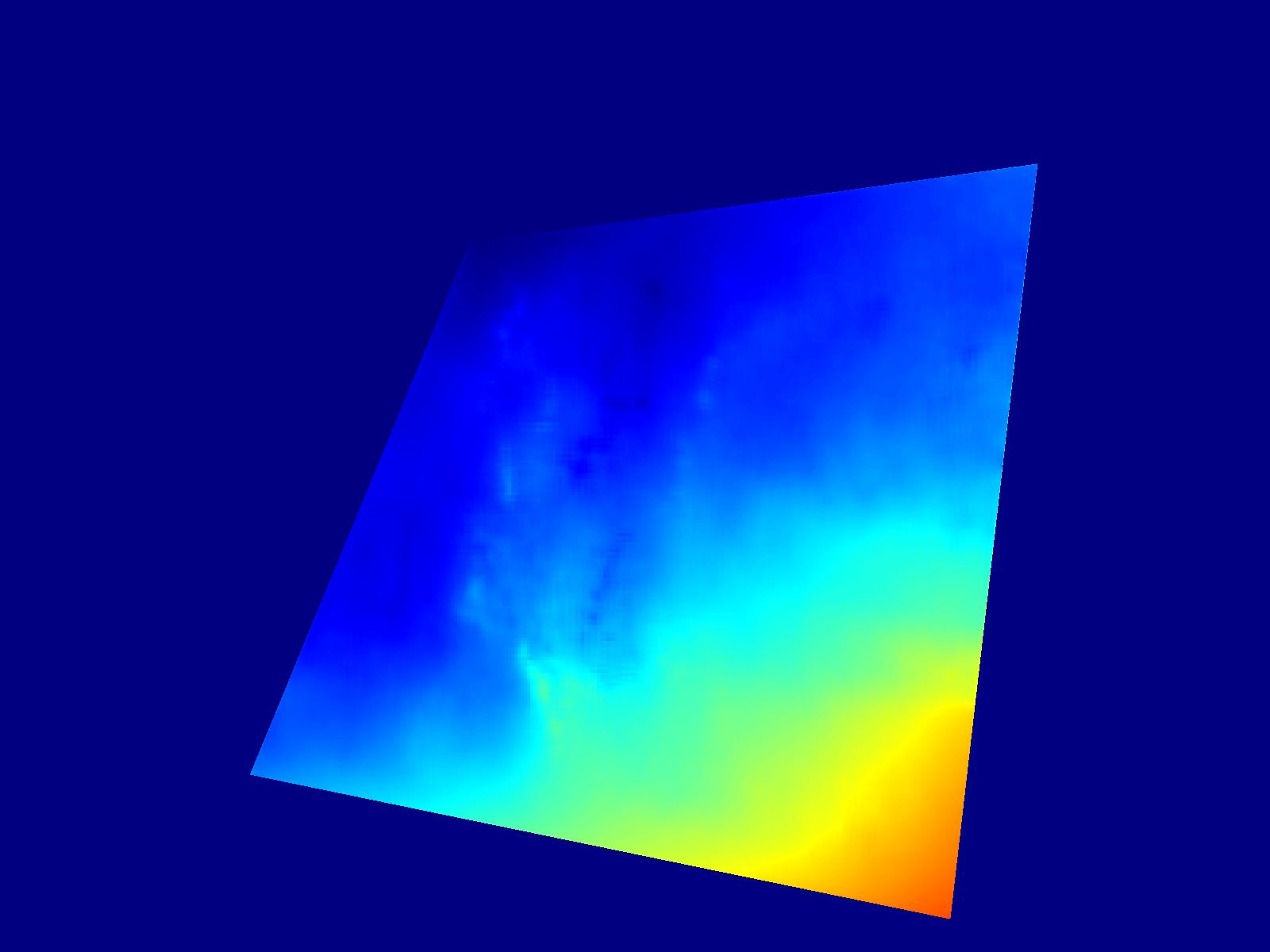}};
			
			\node(IGEV1)[right of=CRE1,xshift=1.8cm]{\includegraphics[width=0.15\textwidth]{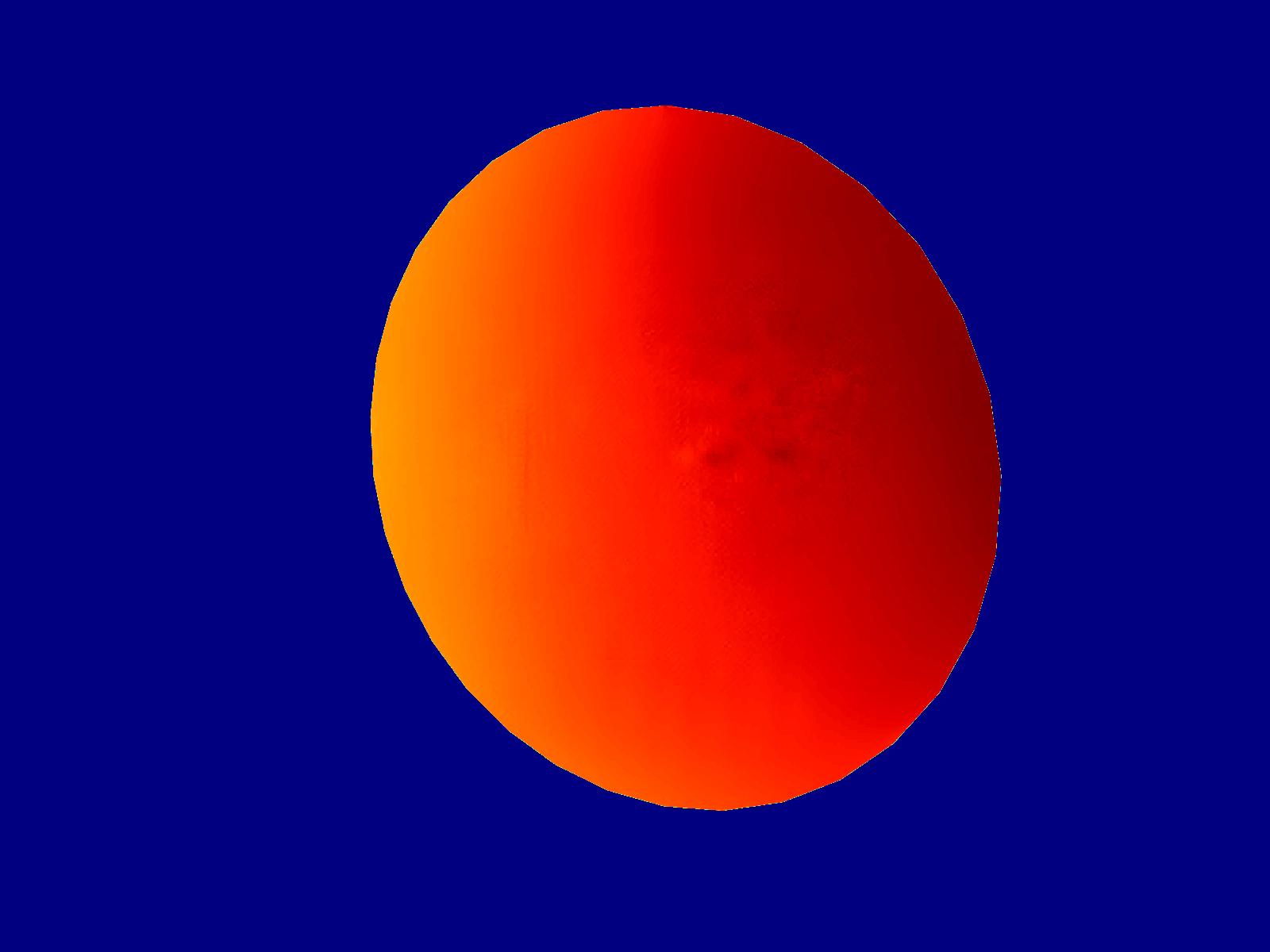}};
			\node(IGEV2)[right of=CRE2,xshift=1.8cm]{\includegraphics[width=0.15\textwidth]{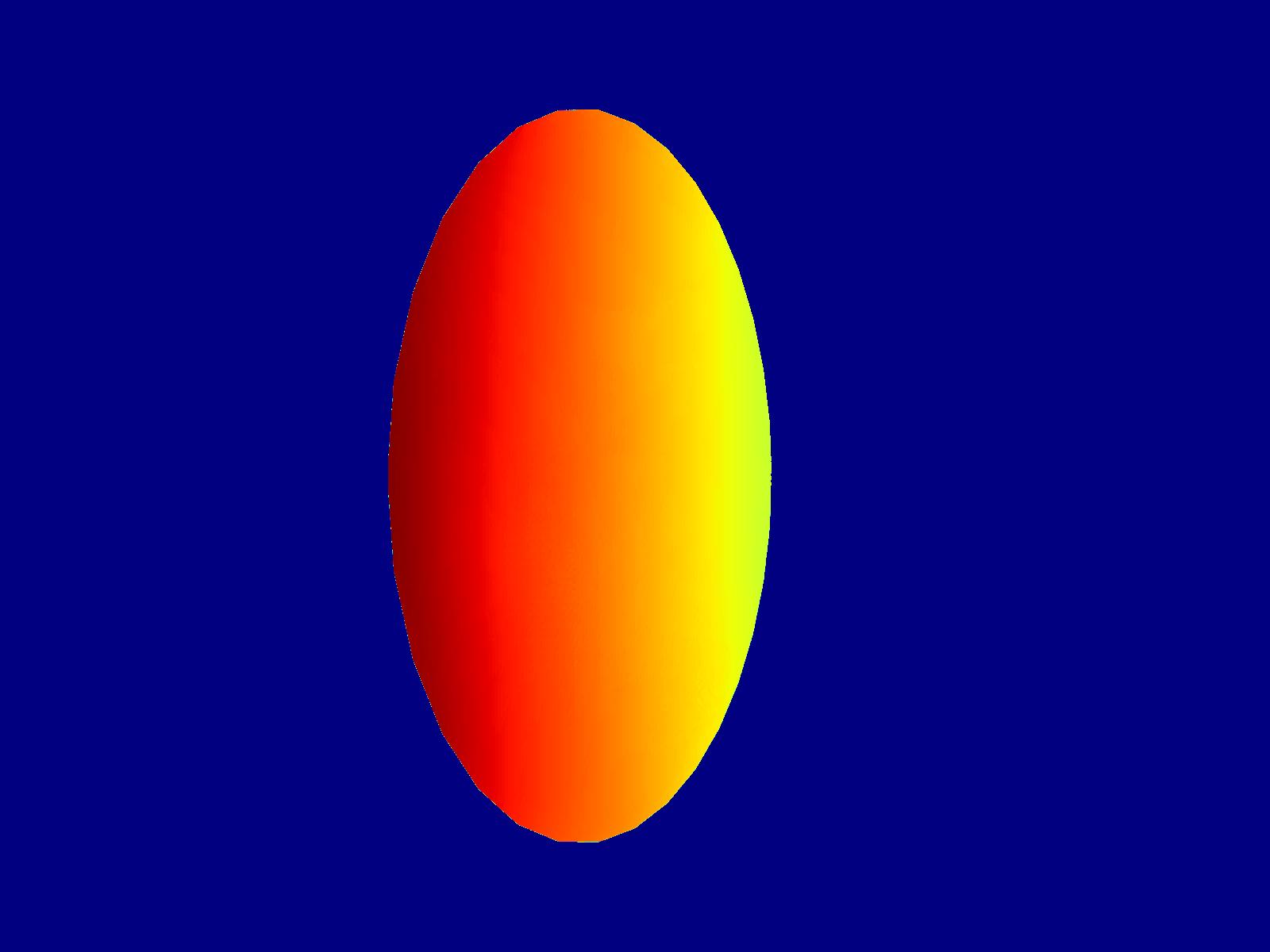}};
			\node(IGEV3)[right of=CRE3,xshift=1.8cm]{\includegraphics[width=0.15\textwidth]{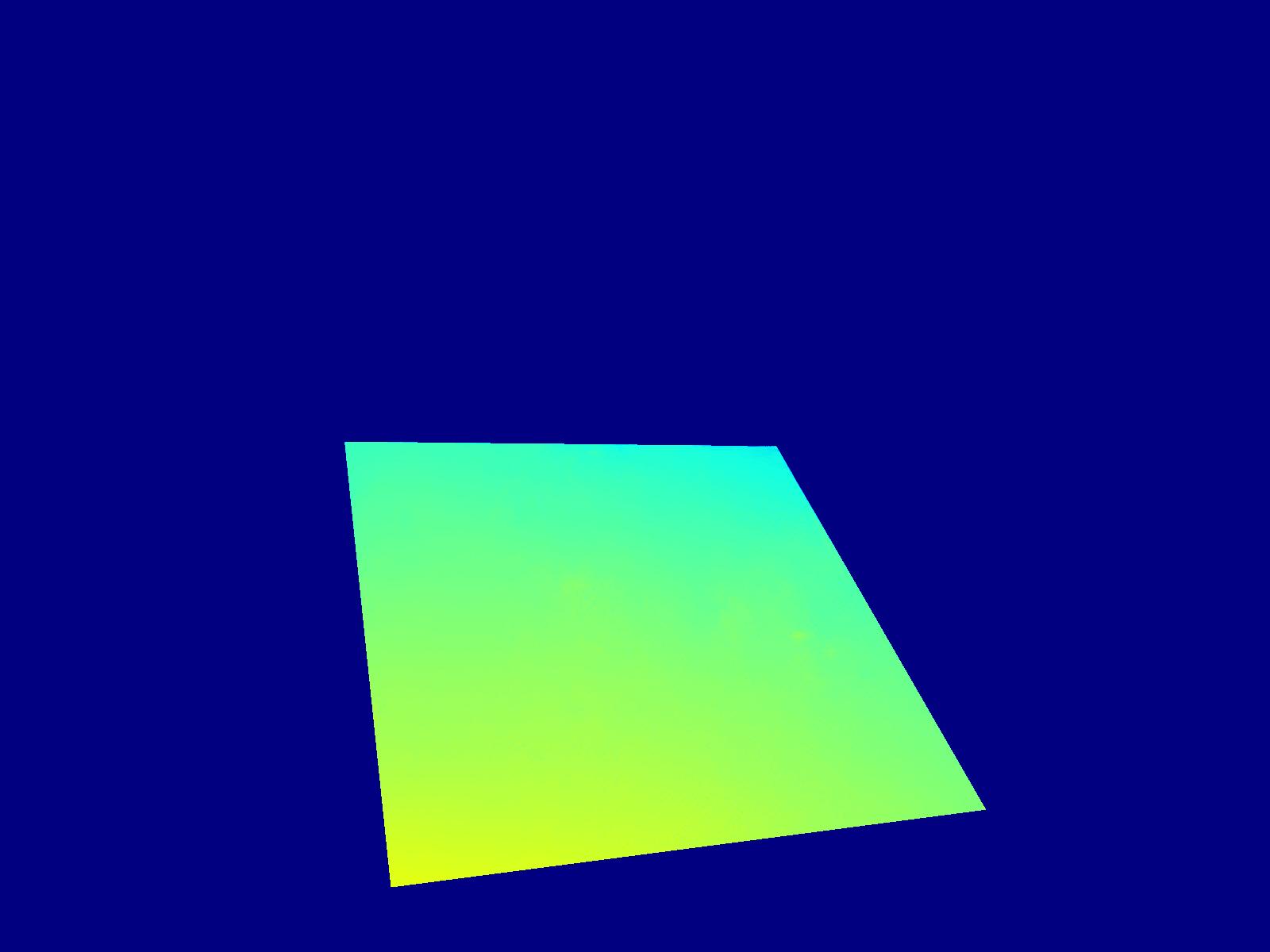}};
			\node(IGEV4)[right of=CRE4, xshift=1.8cm]{\includegraphics[width=0.15\textwidth]{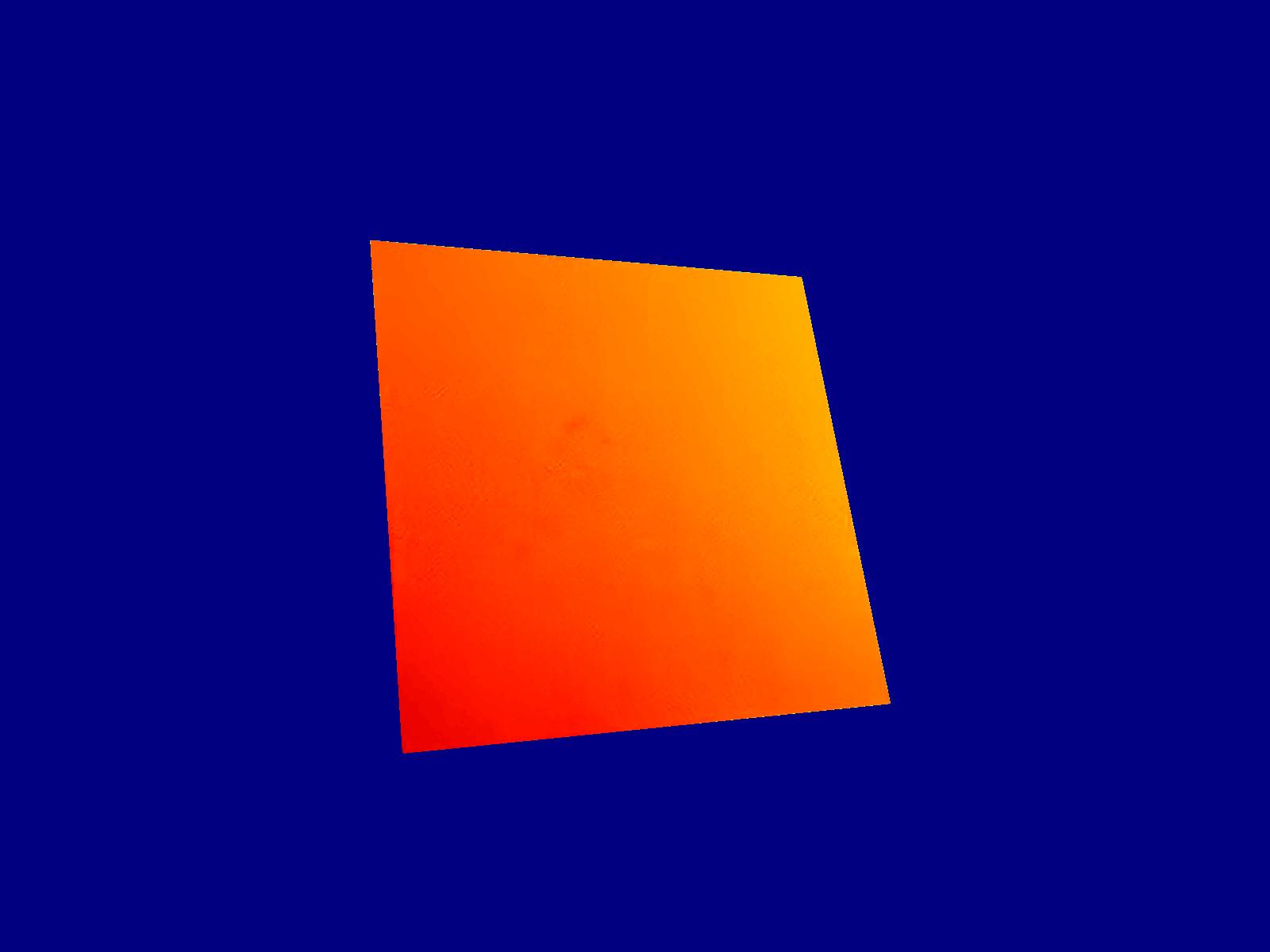}};
			\node(IGEV5)[right of=CRE5, xshift=1.8cm]{\includegraphics[width=0.15\textwidth]{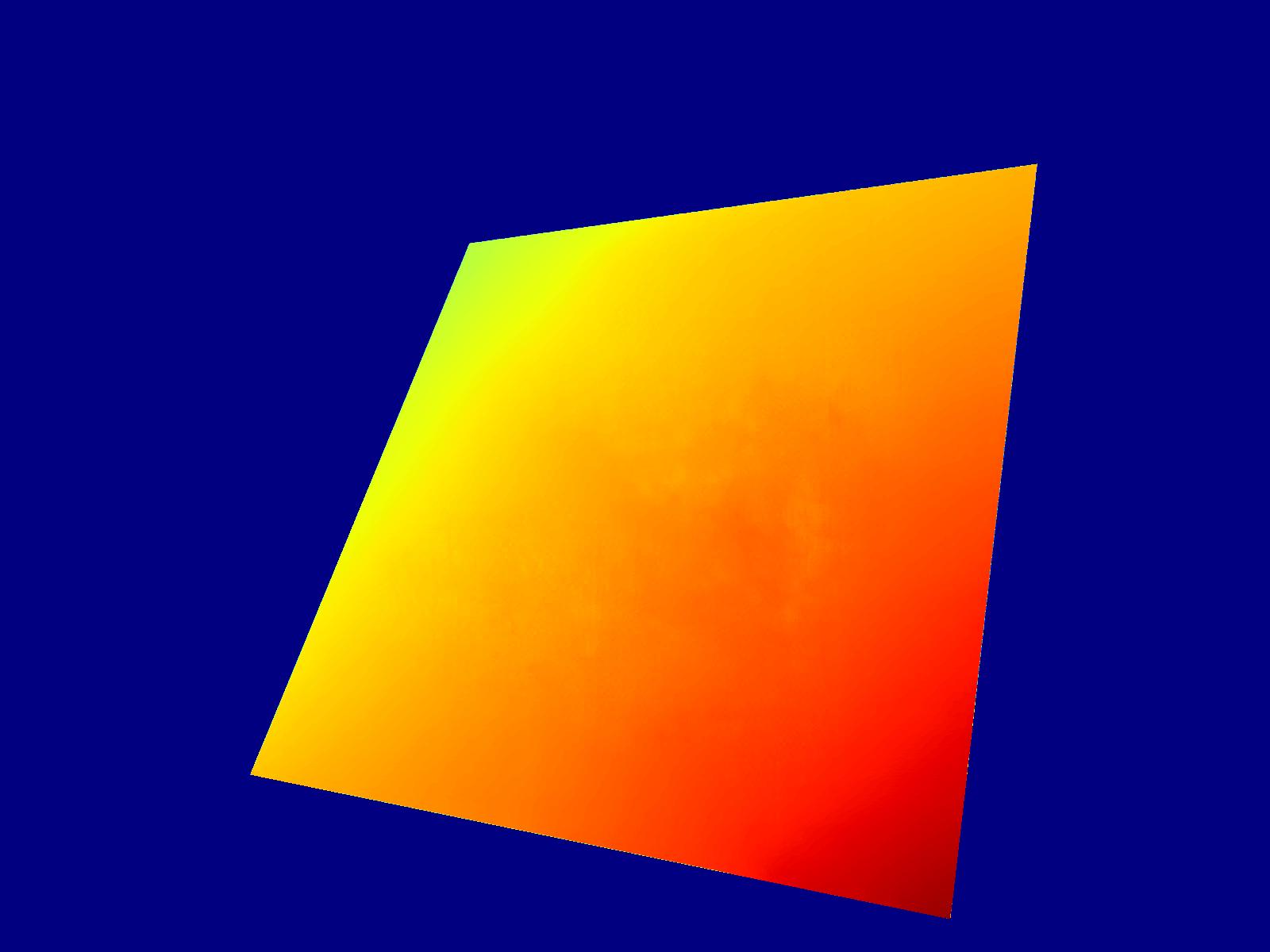}};
			
			\node(Ours1)[right of=IGEV1,xshift=1.8cm]{\includegraphics[width=0.15\textwidth]{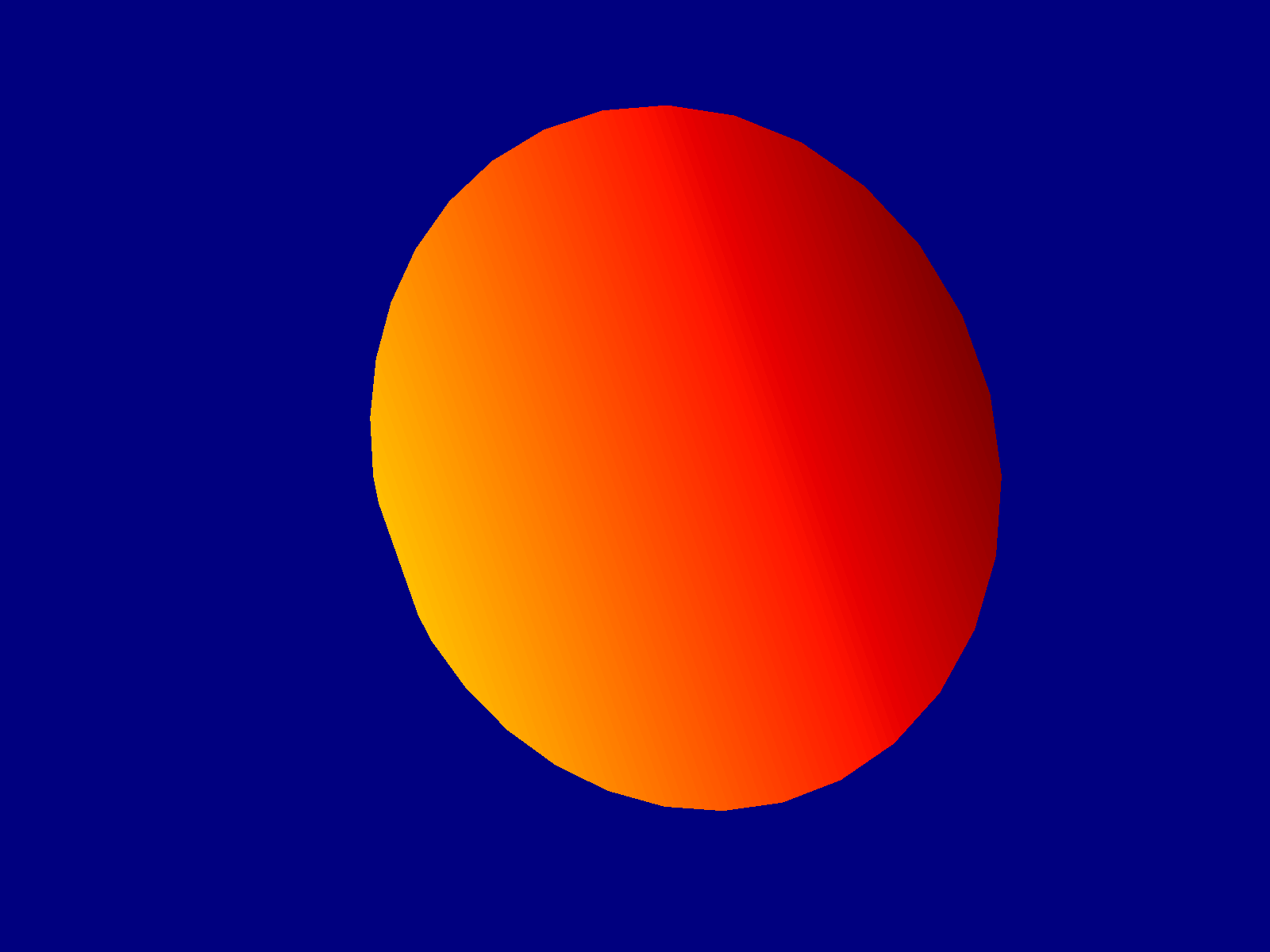}};
			\node(Ours2)[right of=IGEV2,xshift=1.8cm]{\includegraphics[width=0.15\textwidth]{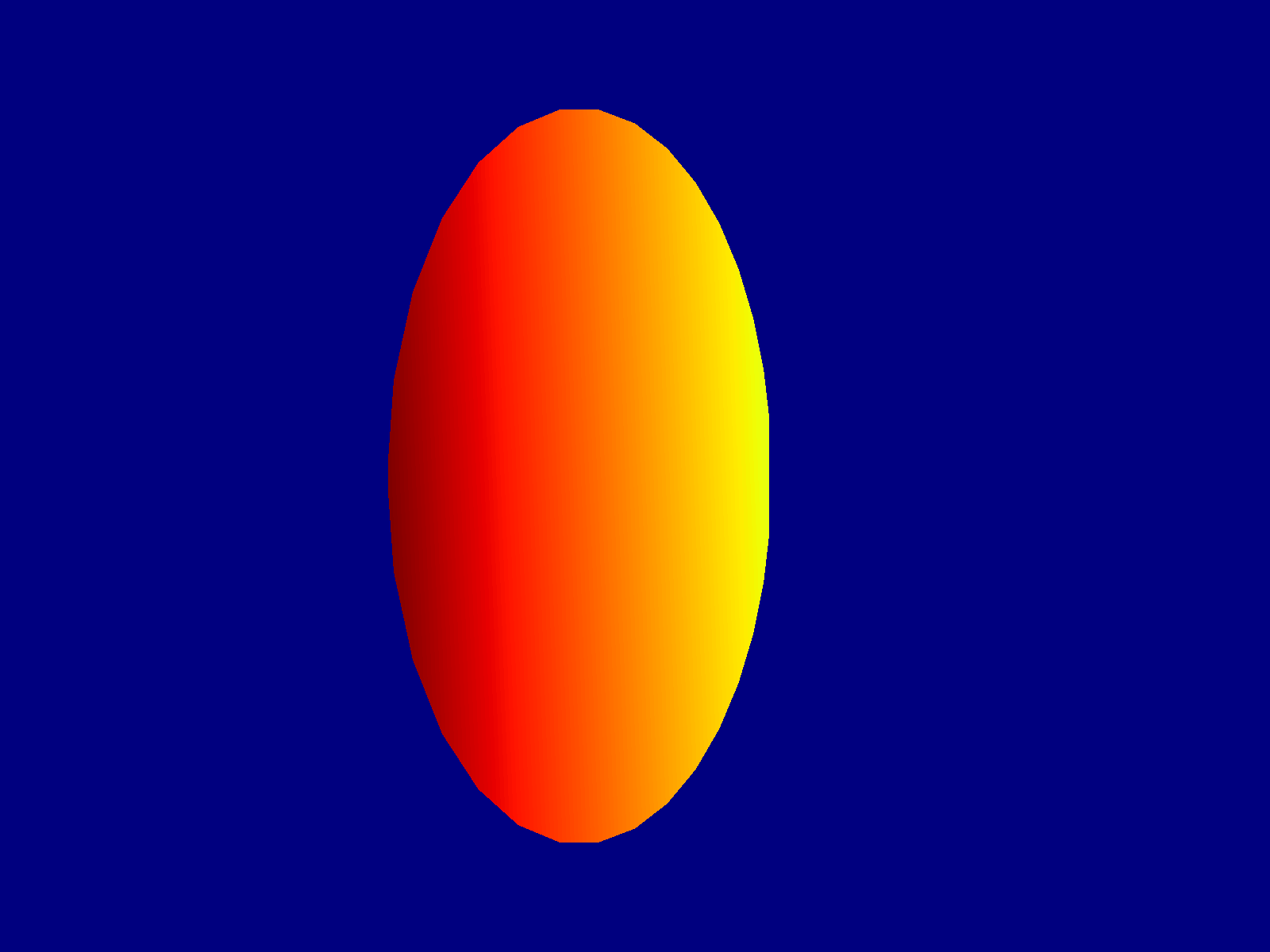}};
			\node(Ours3)[right of=IGEV3,xshift=1.8cm]{\includegraphics[width=0.15\textwidth]{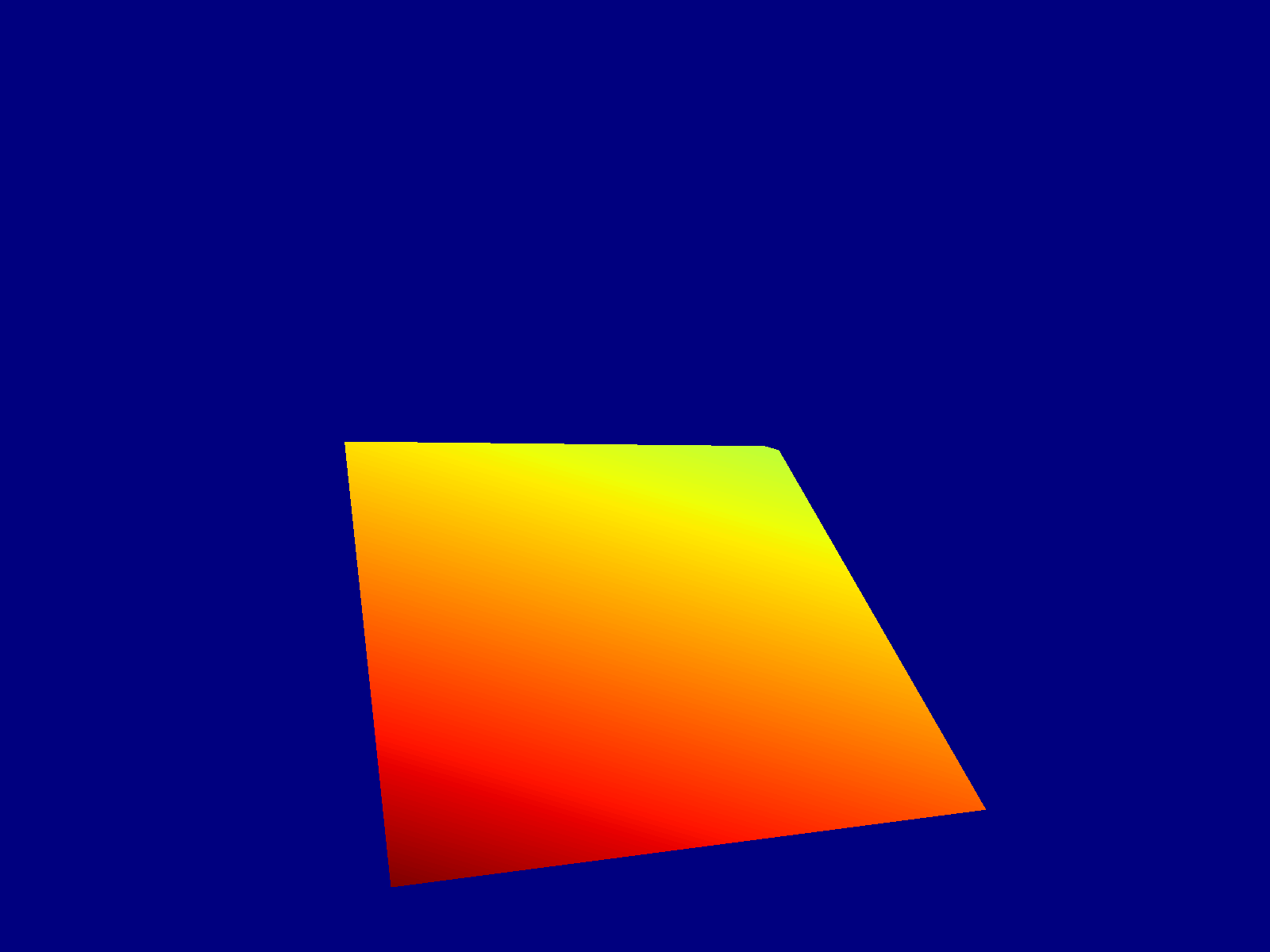}};
			\node(Ours4)[right of=IGEV4, xshift=1.8cm]{\includegraphics[width=0.15\textwidth]{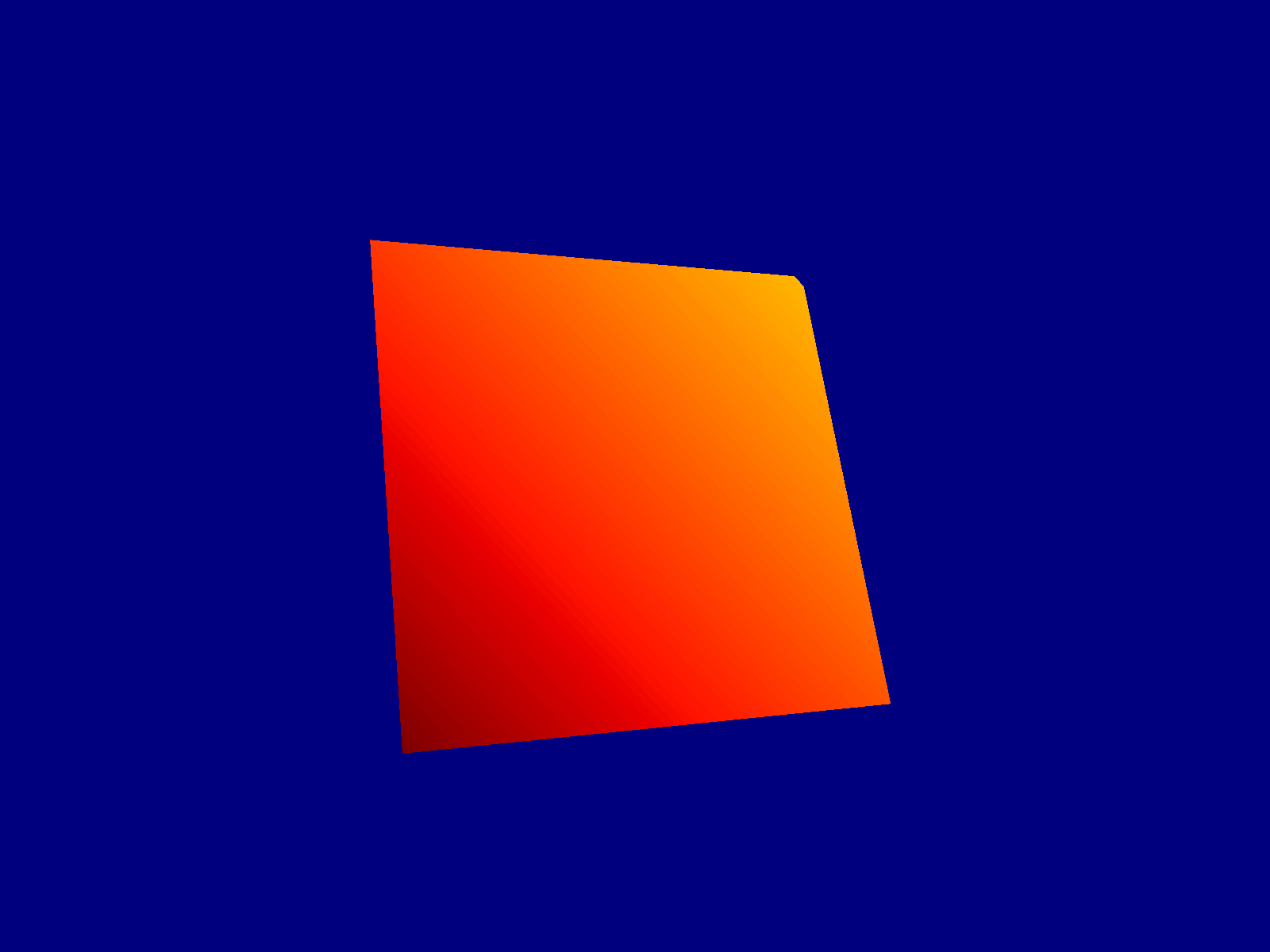}};
			\node(Ours5)[right of=IGEV5, xshift=1.8cm]{\includegraphics[width=0.15\textwidth]{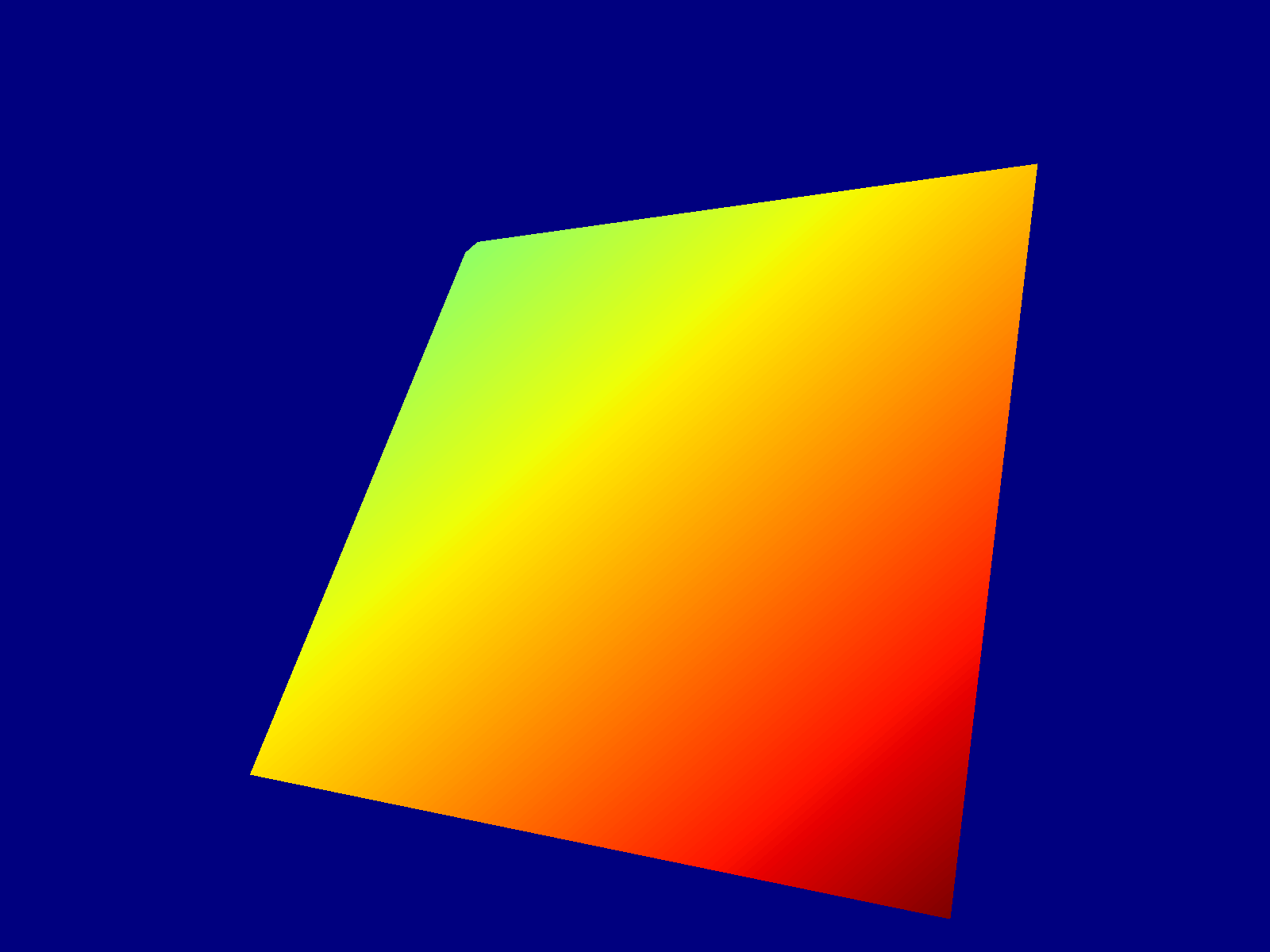}};	
			
			\node(title1)[above of=rgb1,yshift=0.2cm]{Input image};
			\node(title2)[above of=gt1,yshift=0.2cm]{Ground truth};
			\node(title3)[above of=NMRF1,yshift=0.2cm]{NMRF \cite{NMRF}};
			\node(title4)[above of=CRE1,yshift=0.2cm]{CREStereo \cite{CREStereo}};
			\node(title5)[above of=IGEV1,yshift=0.2cm]{IGEV \cite{IgevDE}};
			\node(title6)[above of=Ours1,yshift=0.2cm]{SRDE};
		\end{tikzpicture}
		\vspace{-0.1cm}
		\caption{Qualitative disparity map results on flat and textureless synthetic data. Shown are the input image, ground truth, and disparity maps produced by NMRF \cite{NMRF}, CREStereo \cite{CREStereo}, IGEV \cite{IgevDE} and SRDE (ours). Colors are normalized to depth.}
		\label{fig:ResultImagesDisp}
		\vspace{-0.5cm}
		\end{figure*}	
	
\vspace{-0.4cm}
\subsection{Performance on Synthetic Data}
\subsubsection{Quantitative Comparison}	
As shown in Table \ref{tab:ResultsBenchmarkModels}, with a quantitative performance improvement of more than 52\% for EPE, 25\% for bmp0.1, 11\% for bmp0.01, 76\% for bmp0.001,  and 43\% for RMSE, SRDE achieves the best results for all metrics compared to the state of the art. 
CREStereo \cite{CREStereo} and NMRF \cite{NMRF} yield lower results in quality.
IGEV \cite{IgevDE} demonstrates an improved performance and shows the best neural network results. 
Further, Table \ref{tab:ResultsBenchmarkModels} includes the standard deviation to address the variability of the results across different samples. For our proposed SRDE method, the variance in performance is primarily based on the specific geometric configuration of the scene, such as the spatial extent of the specular reflection and the distance to the virtual light source. Extreme angles or distorted reflections can slightly decrease the accuracy of the reflection center localization, leading to minor deviations in the final disparity estimate. However, as the standard deviations indicate, SRDE maintains a smaller error margin compared to the baseline models. They exhibit much higher volatility because their feature-matching mechanisms produce unpredictable artifacts when encountering specular regions of varying sizes and shapes.
\begin{table}[t!]
	\centering
	\caption{Quantitative evaluation of pretrained benchmark models on flat, textureless synthetic objects. Results are reported as mean ($\pm$ standard deviation) to illustrate performance variability across samples. The proposed SRDE method outperforms state-of-the-art models in all metrics. \textbf{Bold:} Best.}
	\vspace{-0.1cm}
	\begin{tabular}{lcccc}
		\toprule[1.5pt]
		\textbf{Method} 
		&\multicolumn{1}{c}{\parbox{1.1cm}{\centering \textbf{NMRF}\\\cite{NMRF}}} &  \multicolumn{1}{c}{\parbox{1.1cm}{\centering \textbf{CREStereo}\\\cite{CREStereo}}} & \multicolumn{1}{c}{\parbox{1.1cm}{\centering \textbf{IGEV}\\\cite{IgevDE}}} & \multicolumn{1}{c}{\parbox{1.1cm}{\centering \textbf{SRDE}\\(Ours)}}\\
		\midrule[1.2pt]
		\textbf{EPE} 
		& \makecell{0.0605 \\ ($\pm$ 0.034)}
		& \makecell{0.1023 \\ ($\pm$ 0.049)}
		& \makecell{0.0628 \\ ($\pm$ 0.053)} 
		& \makecell{\textbf{0.0290} \\ \textbf{($\pm$ 0.020)}} \\
		\addlinespace 
		\textbf{bmp0.1} 
		& \makecell{0.1205 \\ ($\pm$ 0.056)} 
		& \makecell{0.1440 \\ ($\pm$ 0.064)} 
		& \makecell{0.1143 \\ ($\pm$ 0.066)}
		& \makecell{\textbf{0.0852} \\ \textbf{($\pm$ 0.049)}} \\
		\addlinespace
		\textbf{bmp0.01} 
		& \makecell{0.1520 \\ ($\pm$ 0.064)}
		& \makecell{0.1545 \\ ($\pm$ 0.067)} 
		& \makecell{0.1327 \\ ($\pm$ 0.061)} 
		& \makecell{\textbf{0.1168} \\ \textbf{($\pm$ 0.047)}} \\
		\addlinespace
		\textbf{bmp0.001} 
		& \makecell{0.5321 \\ \textbf{($\pm$ 0.016)}} 
		& \makecell{0.5306 \\ ($\pm$ 0.017)} 
		& \makecell{0.5092 \\ ($\pm$ 0.029)} 
		& \makecell{\textbf{0.1190} \\ ($\pm$ 0.046)} \\
		\addlinespace
		\textbf{RMSE} 
		& \makecell{0.1764 \\ ($\pm$ 0.065)} 
		& \makecell{0.2767 \\ ($\pm$ 0.080)} 
		& \makecell{0.1682 \\ ($\pm$ 0.124)} 
		& \makecell{\textbf{0.0954} \\ \textbf{($\pm$ 0.051)}} \\ 
		\midrule
		\textbf{Runtime} (ms) & 2.33 & 263.87 & 2.92 & \textbf{0.39} \\
		\bottomrule[1.5pt]        
	\end{tabular}
	\label{tab:ResultsBenchmarkModels}
	\vspace{-0.3cm}
\end{table}
	
\subsubsection{Qualitative Comparison}
Fig. \ref{fig:ResultImagesDisp} shows some results for the synthetic dataset. The ground truth disparity map is displayed along with the outputs of NMRF \cite{NMRF}, CREStereo \cite{CREStereo}, IGEV \cite{IgevDE} and the proposed SRDE approach. The results show that SRDE clearly outperforms NMRF, CREStereo and IGEV for $\text{single}$, textureless, flat objects of any shape.

	\begin{figure}
		\centering
		\begin{tikzpicture}
			\node(synth)[]{\includegraphics[width=0.2\textwidth]{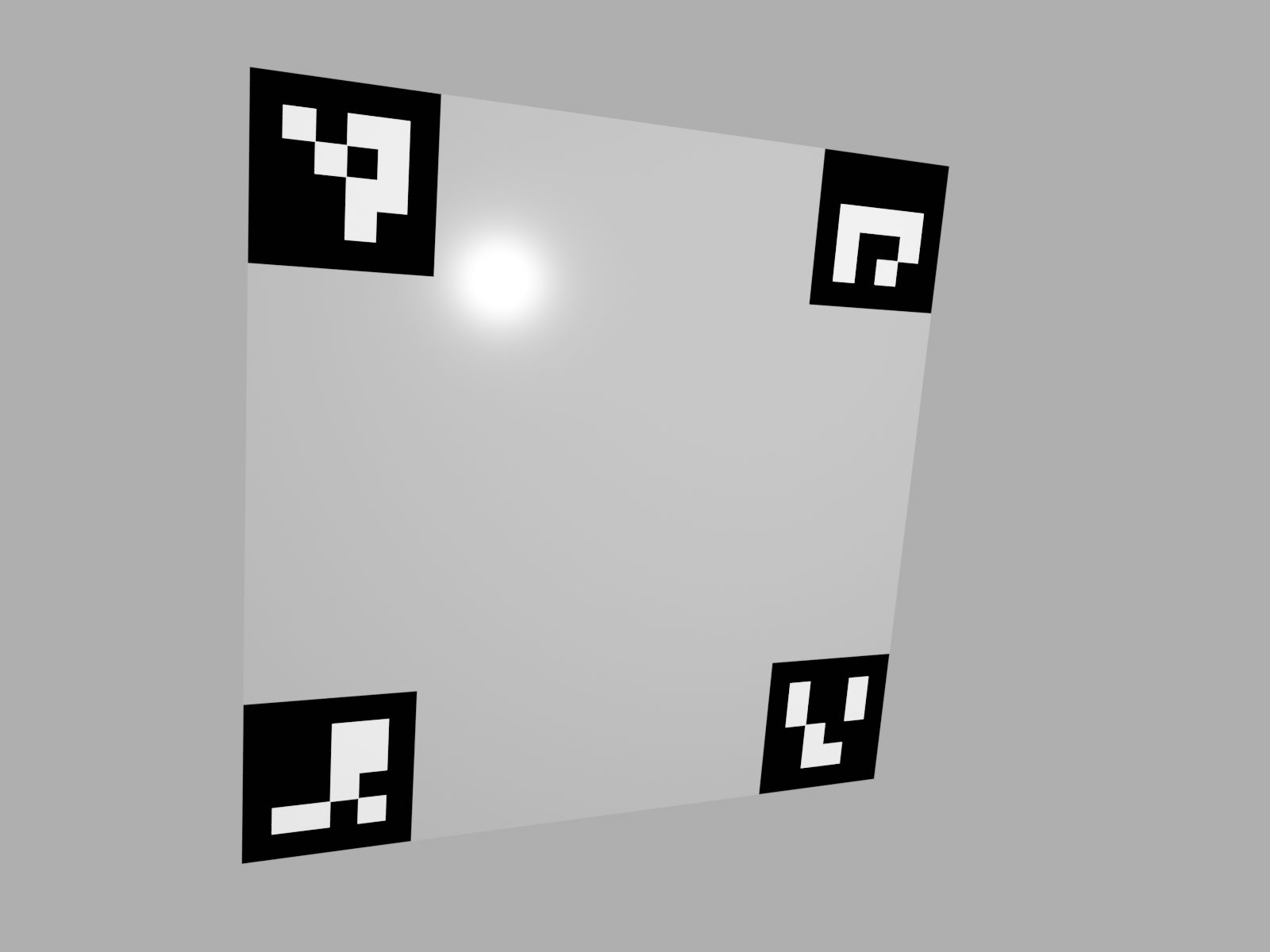}};
			\node(real)[right of = synth, xshift=2.6cm]{\includegraphics[width=0.183\textwidth]{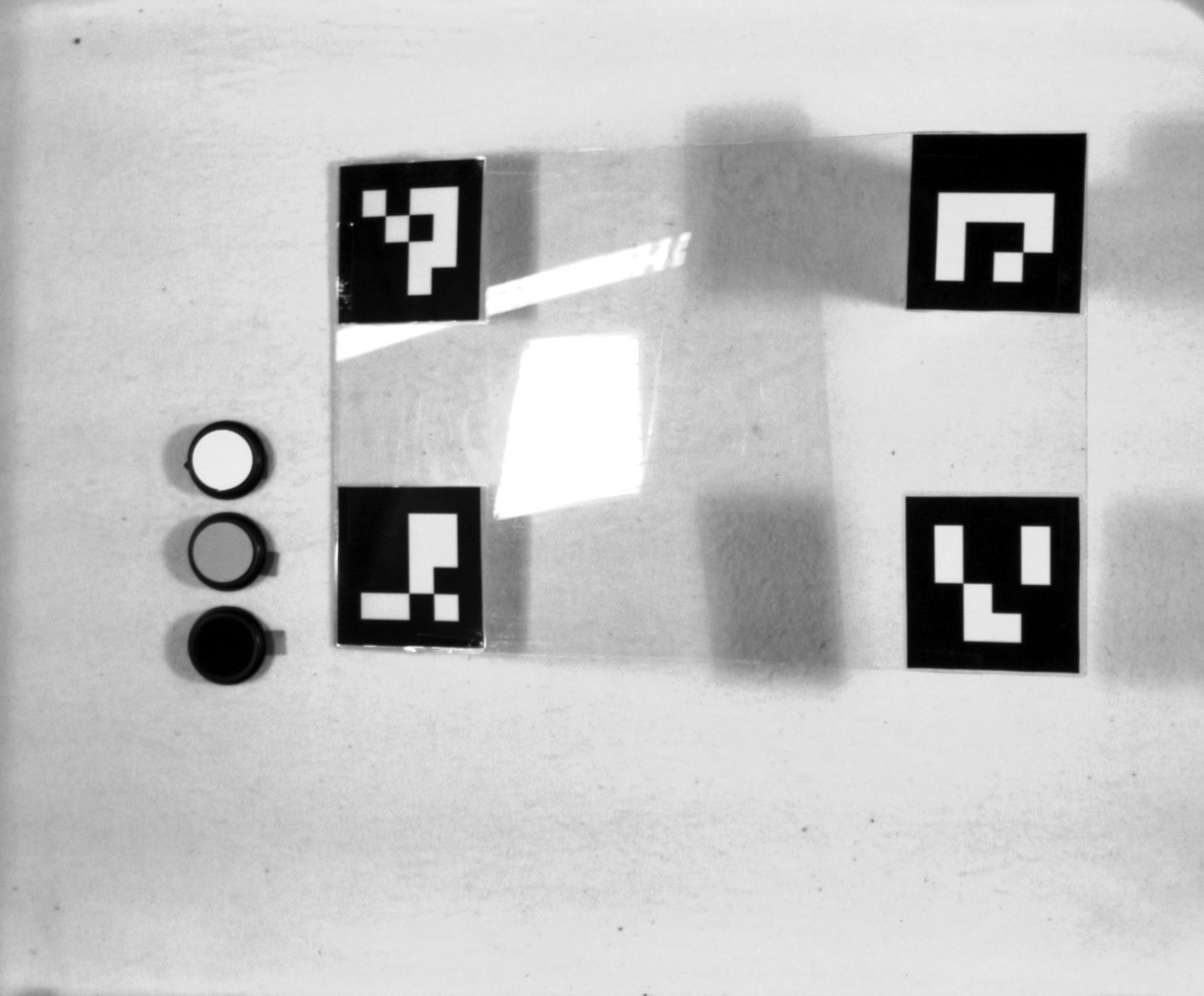}};
		\end{tikzpicture}
		\vspace{-0.2cm}
		\caption{Synthetic Blender ArUco calibration plate (left) and real ArUco PVC calibration plate (right) used for estimating the real light source position.}
		\label{fig:MarkersLightSourcePositionEstimation}
		\vspace{-0.4cm}
	\end{figure}
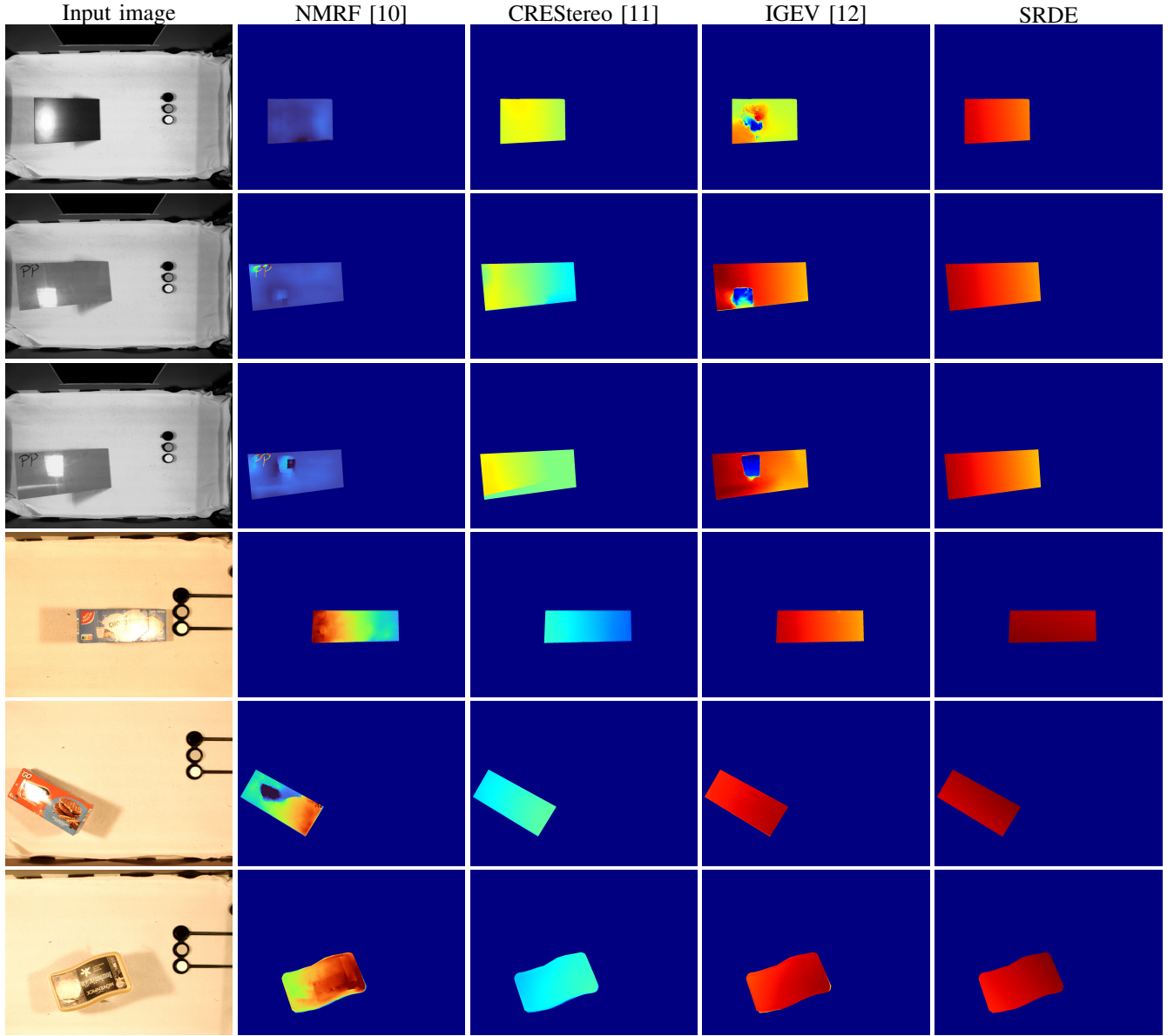
\begin{figure*}[h!]
	\centering
	\begin{tikzpicture}
		\node(rgb1)[]{\includegraphics[width=0.19\textwidth, height=2.5cm]{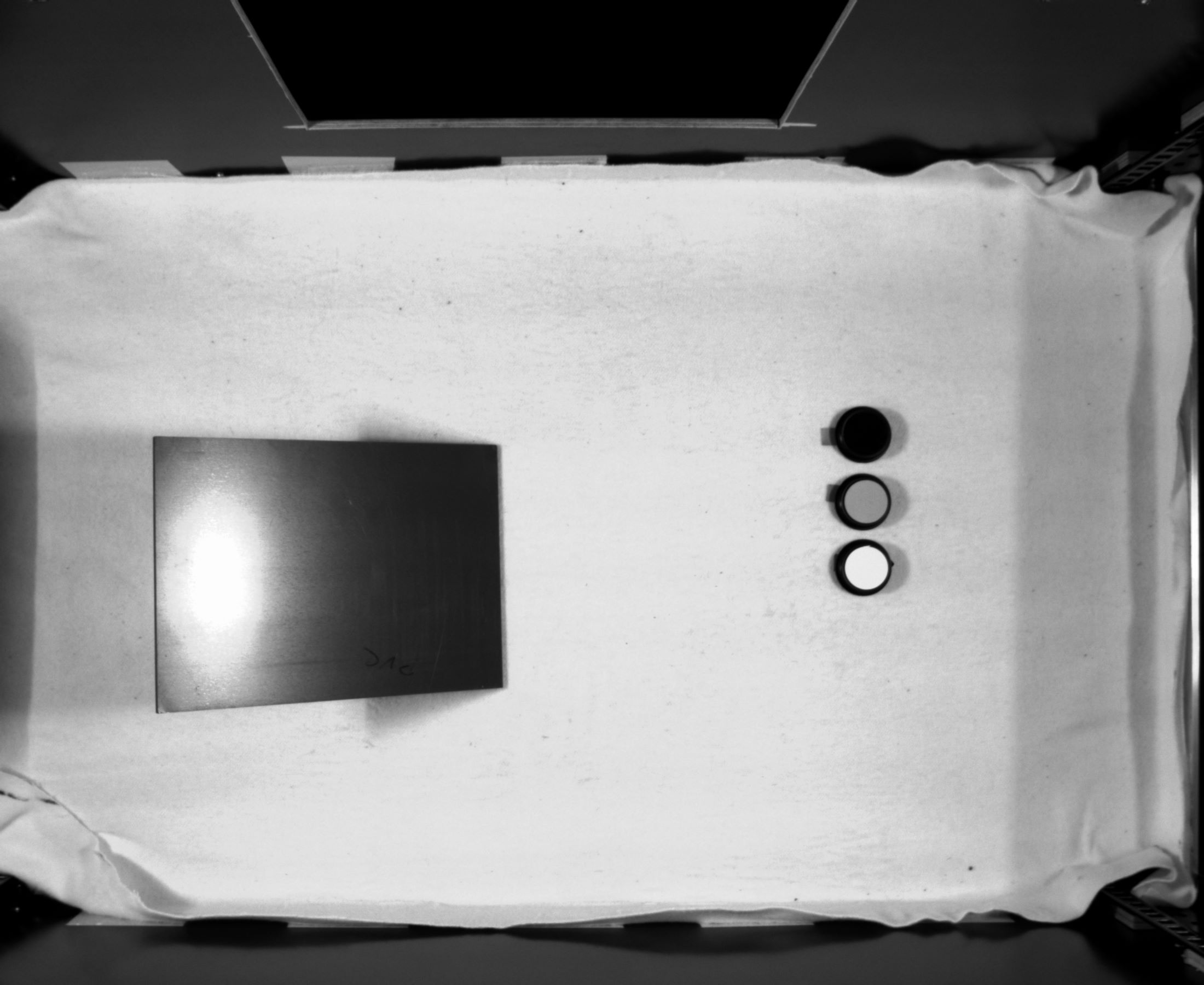}};
		\node(rgb2)[below of=rgb1,yshift=-1.57cm]{\includegraphics[width=0.19\textwidth,height=2.5cm]{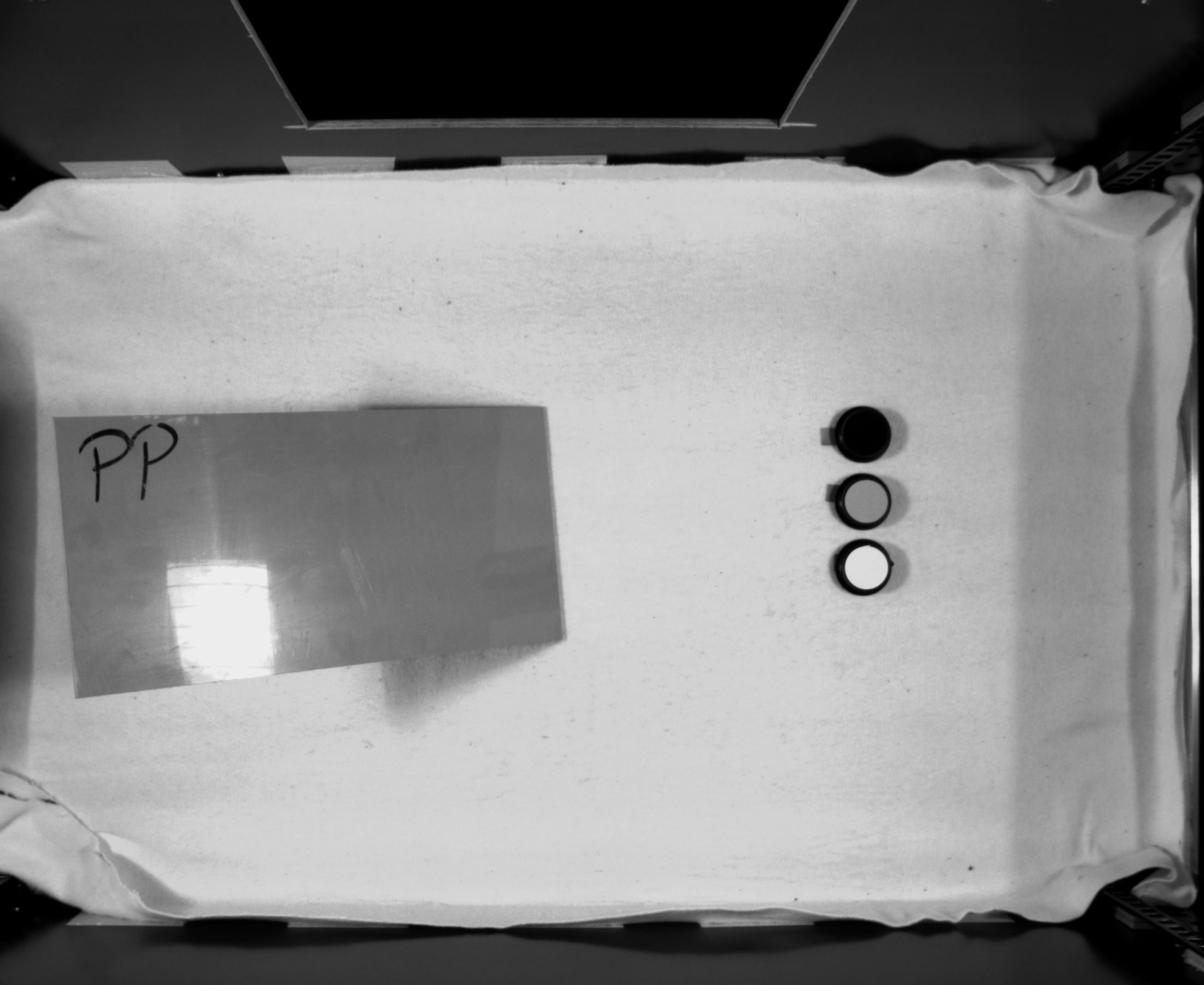}};
		\node(rgb3)[below of=rgb2,yshift=-1.57cm]{\includegraphics[width=0.19\textwidth,height=2.5cm]{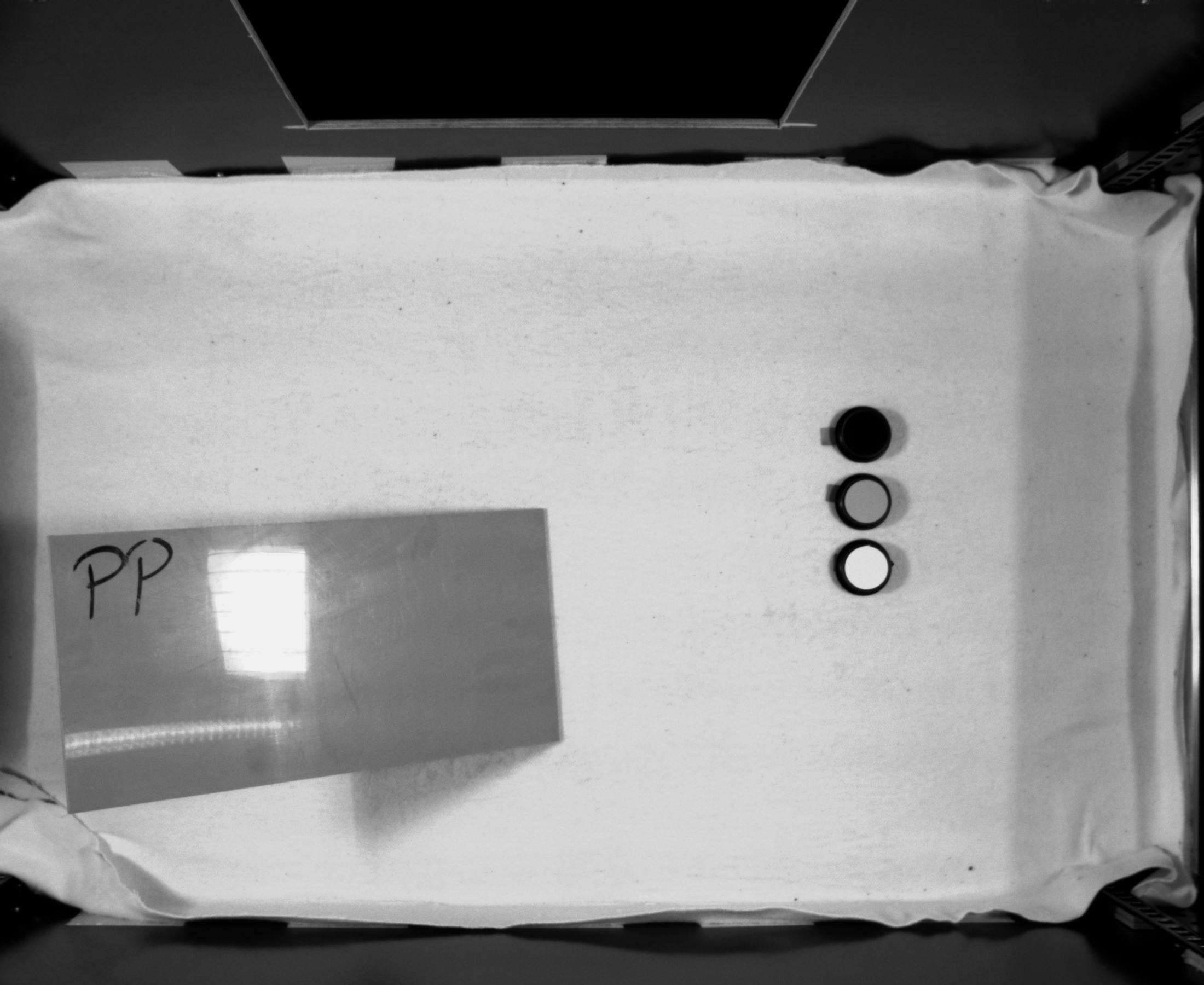}};
	
		\node(rgb4)[below of=rgb3,yshift=-1.57cm]{\includegraphics[width=0.19\textwidth,height=2.5cm]{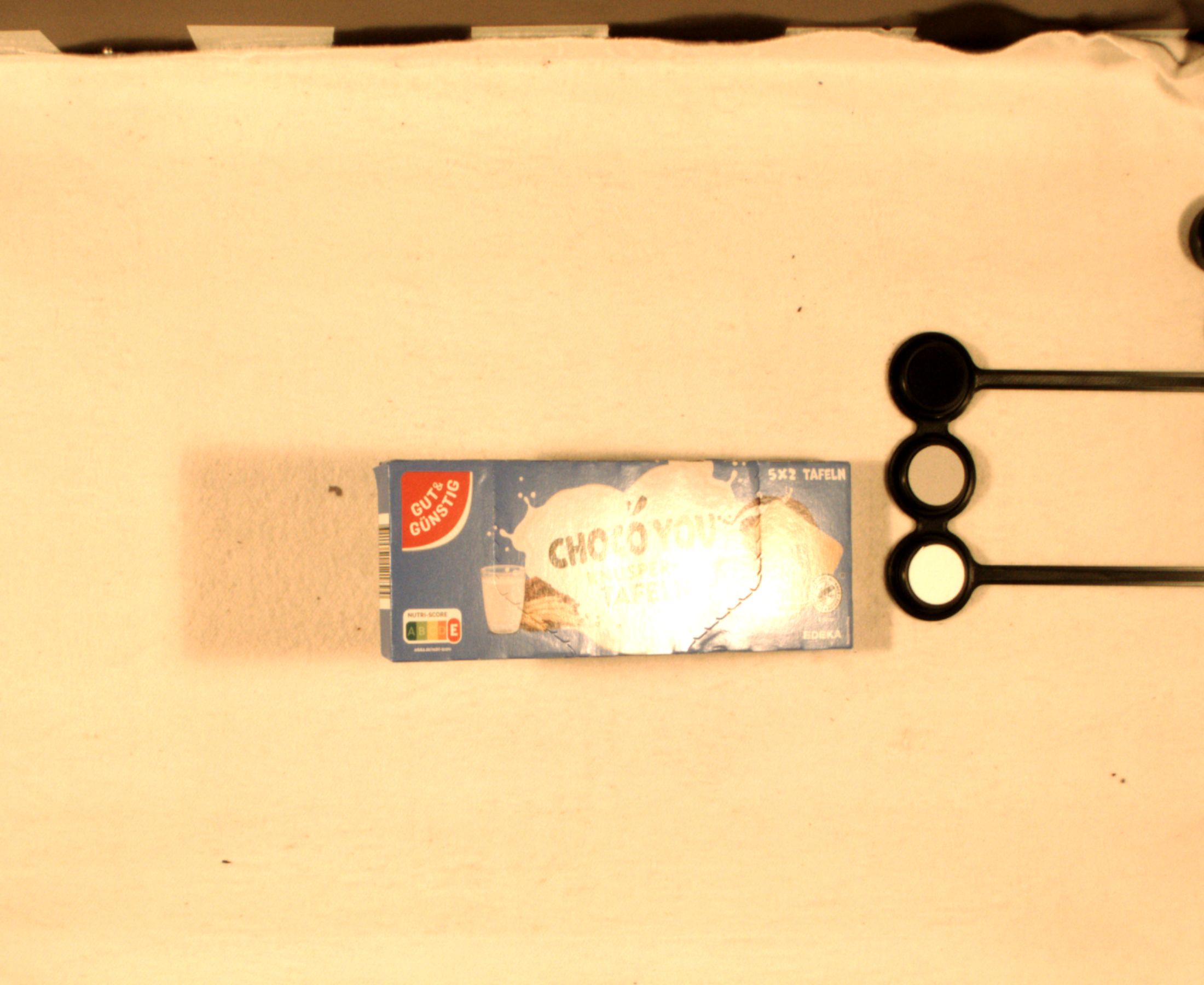}};
		\node(rgb5)[below of=rgb4,yshift=-1.57cm]{\includegraphics[width=0.19\textwidth,height=2.5cm]{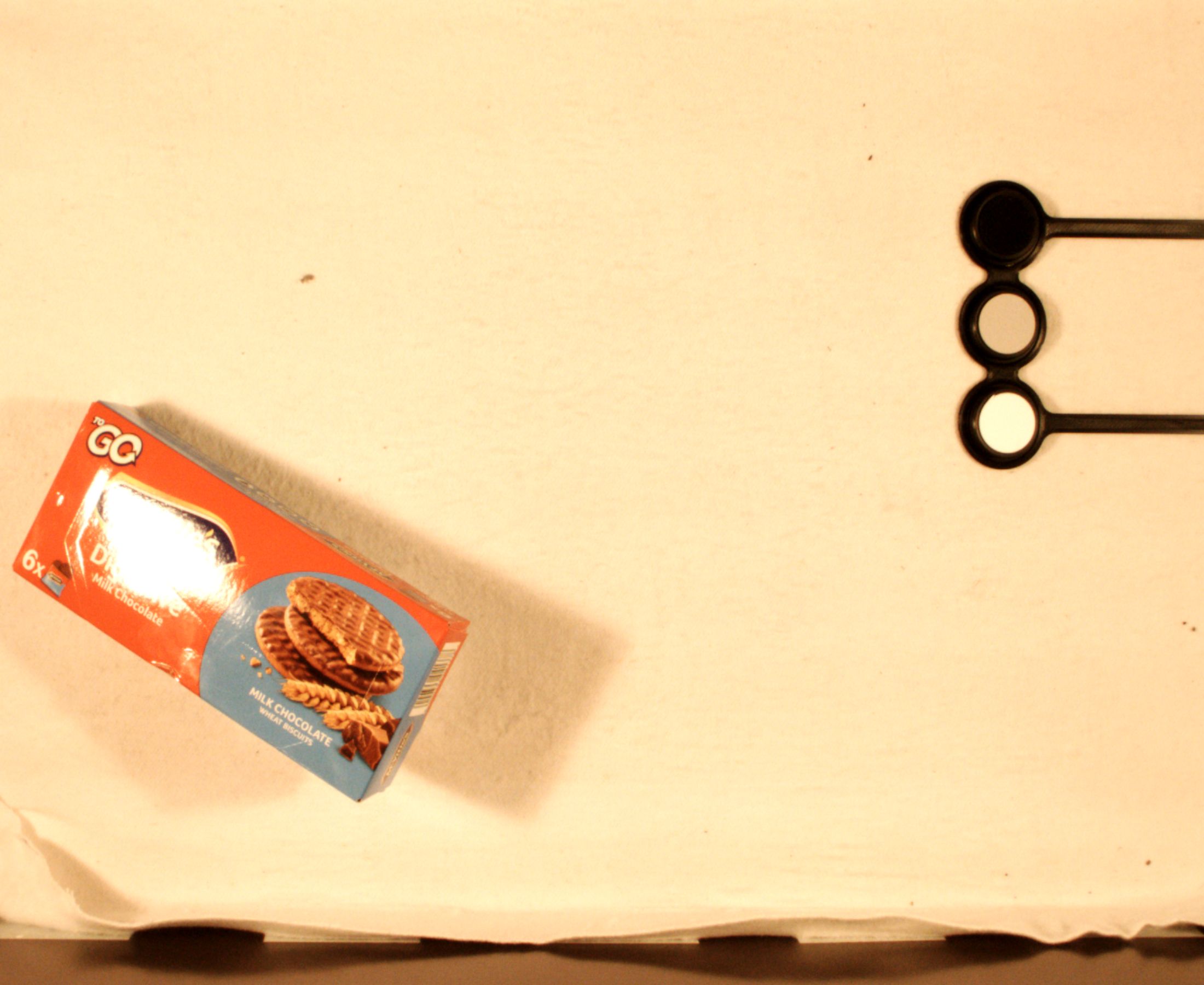}};
		\node(rgb6)[below of=rgb5,yshift=-1.57cm]{\includegraphics[width=0.19\textwidth,height=2.5cm]{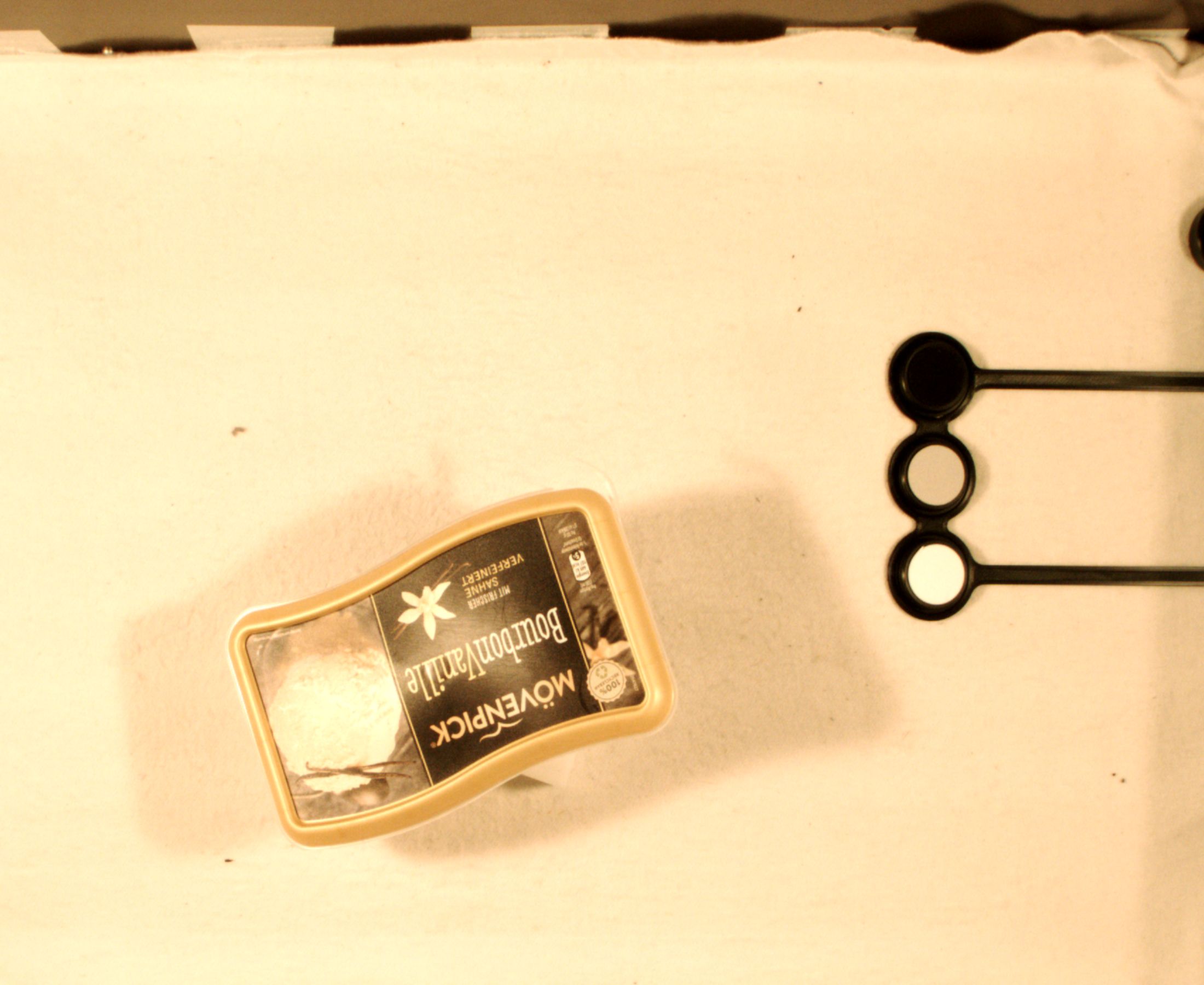}};

		\node(NMRF1)[right of=rgb1,xshift=2.53cm]{\includegraphics[width=0.19\textwidth,height=2.5cm]{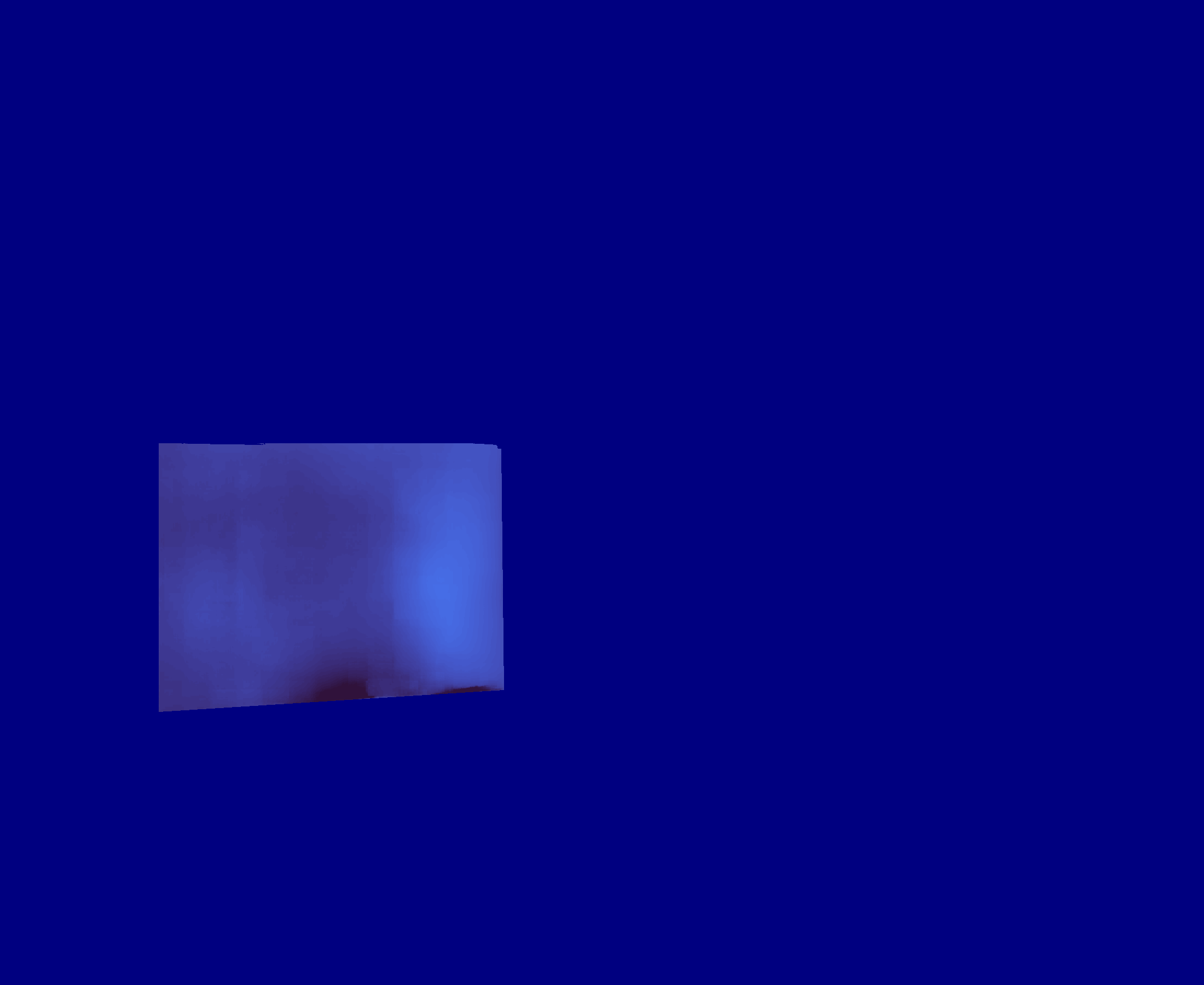}};
		\node(NMRF2)[right of=rgb2,xshift=2.53cm]{\includegraphics[width=0.19\textwidth,height=2.5cm]{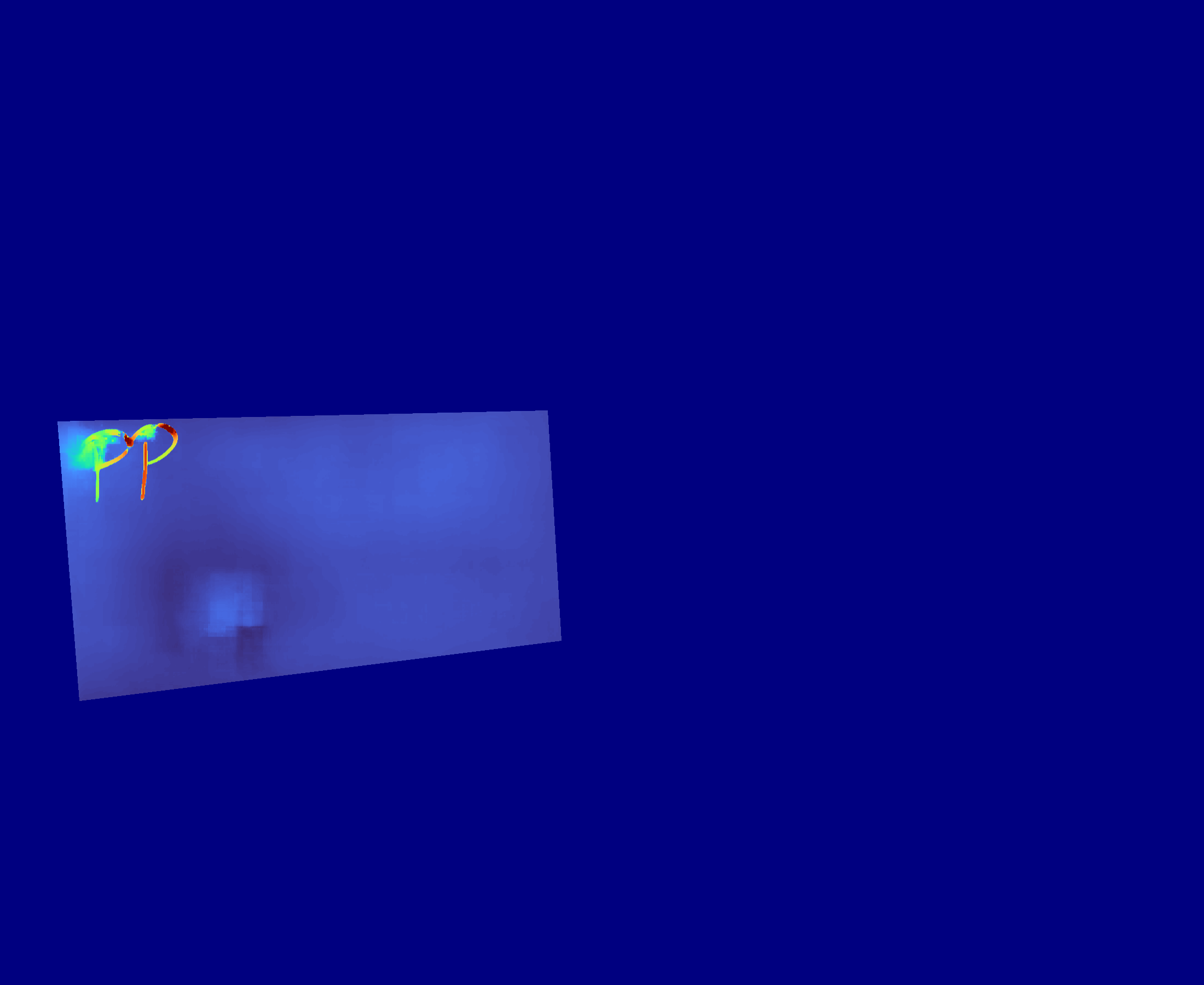}};
		\node(NMRF3)[right of=rgb3,xshift=2.53cm]{\includegraphics[width=0.19\textwidth,height=2.5cm]{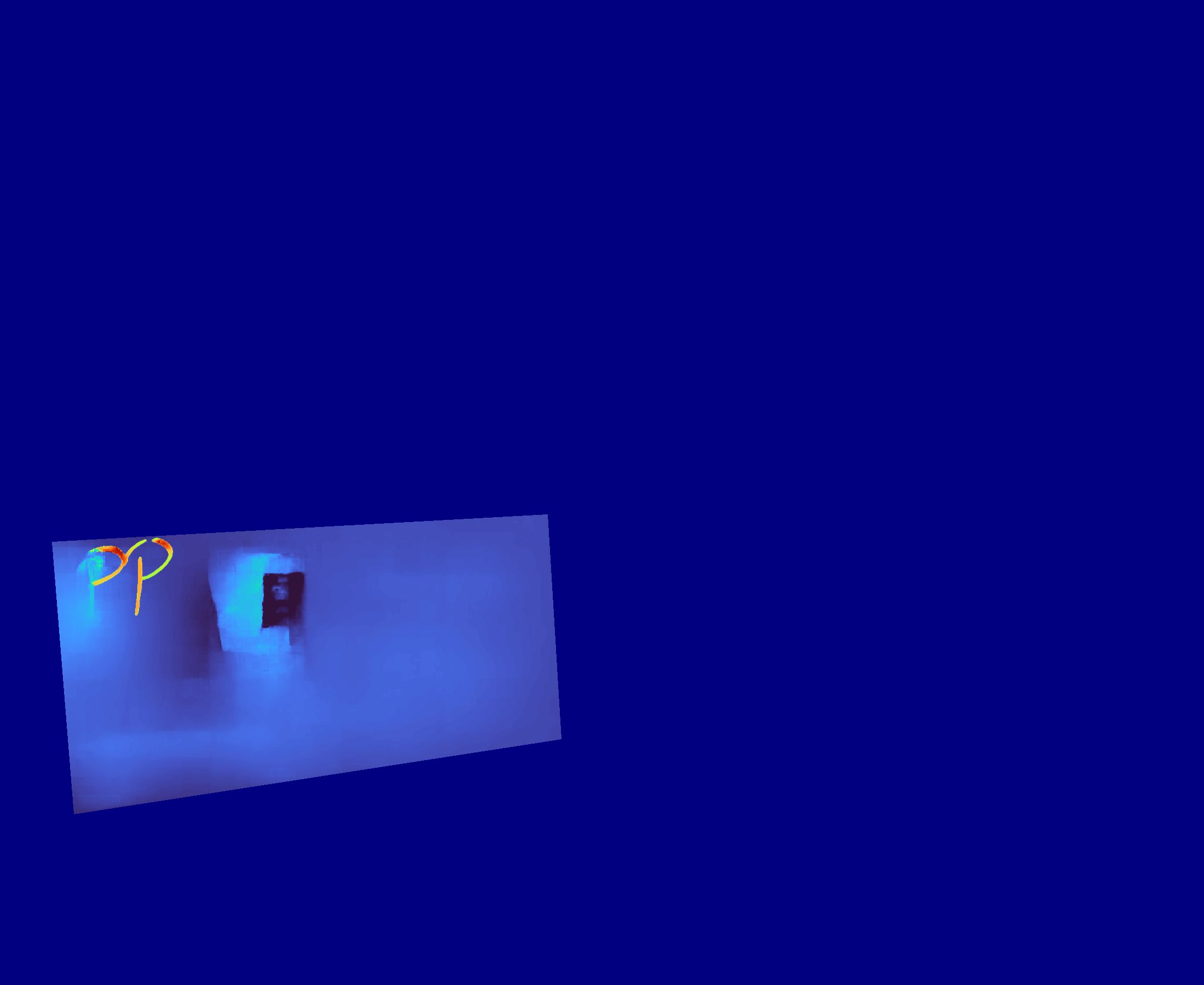}};
	
		\node(NMRF4)[right of=rgb4,xshift=2.53cm]{\includegraphics[width=0.19\textwidth,height=2.5cm]{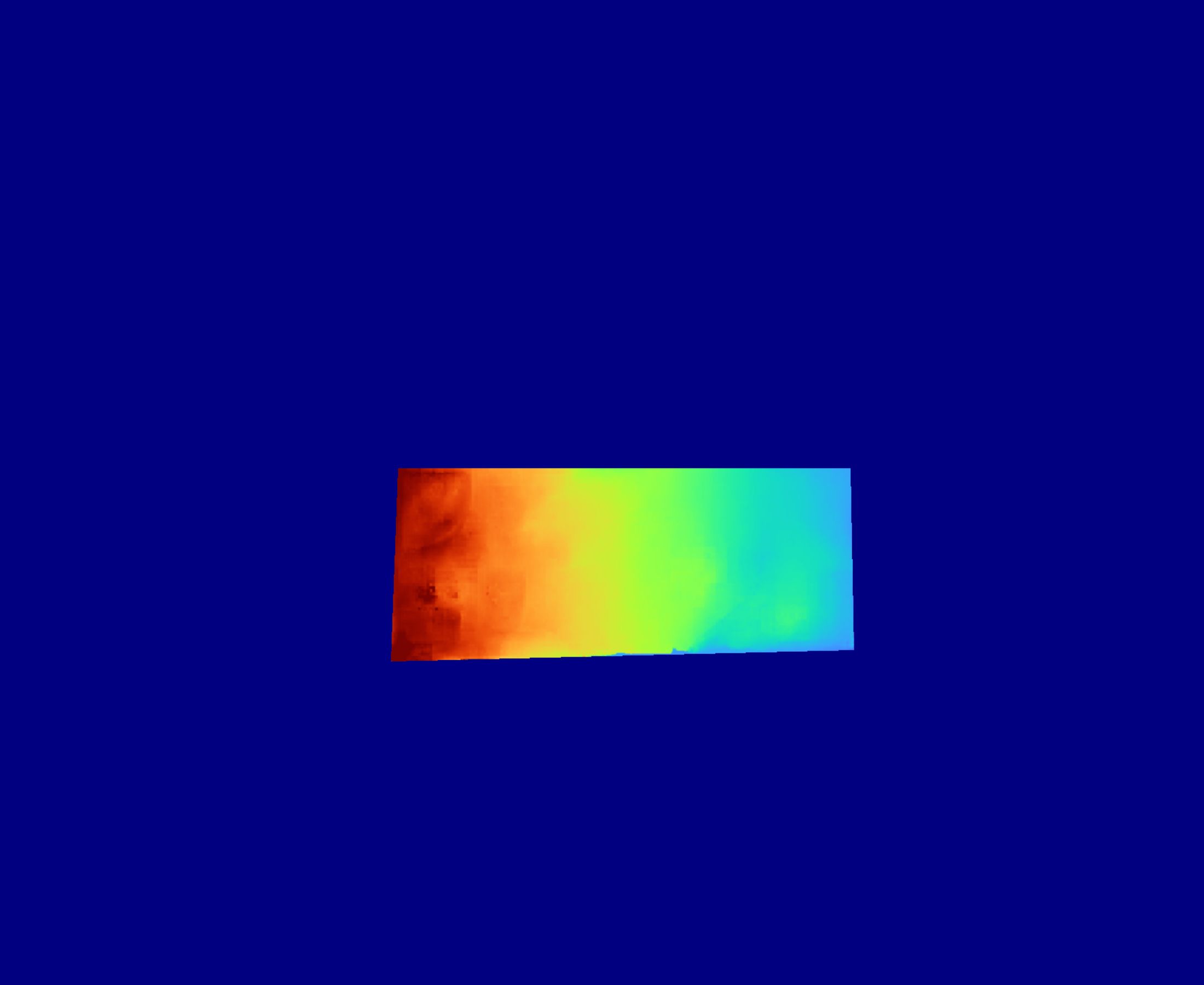}};
		\node(NMRF5)[right of=rgb5,xshift=2.53cm]{\includegraphics[width=0.19\textwidth,height=2.5cm]{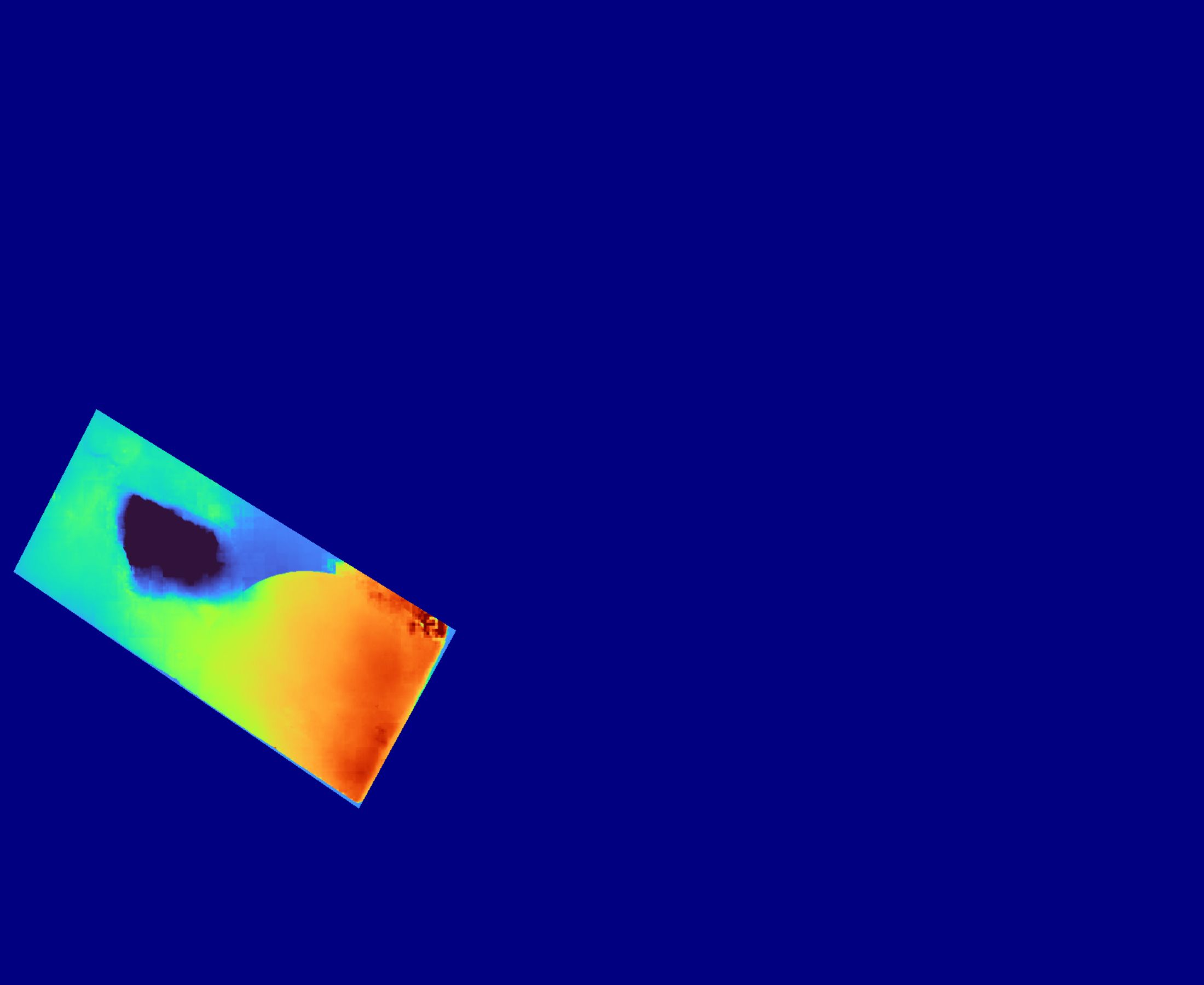}};
		\node(NMRF6)[right of=rgb6,xshift=2.53cm]{\includegraphics[width=0.19\textwidth,height=2.5cm]{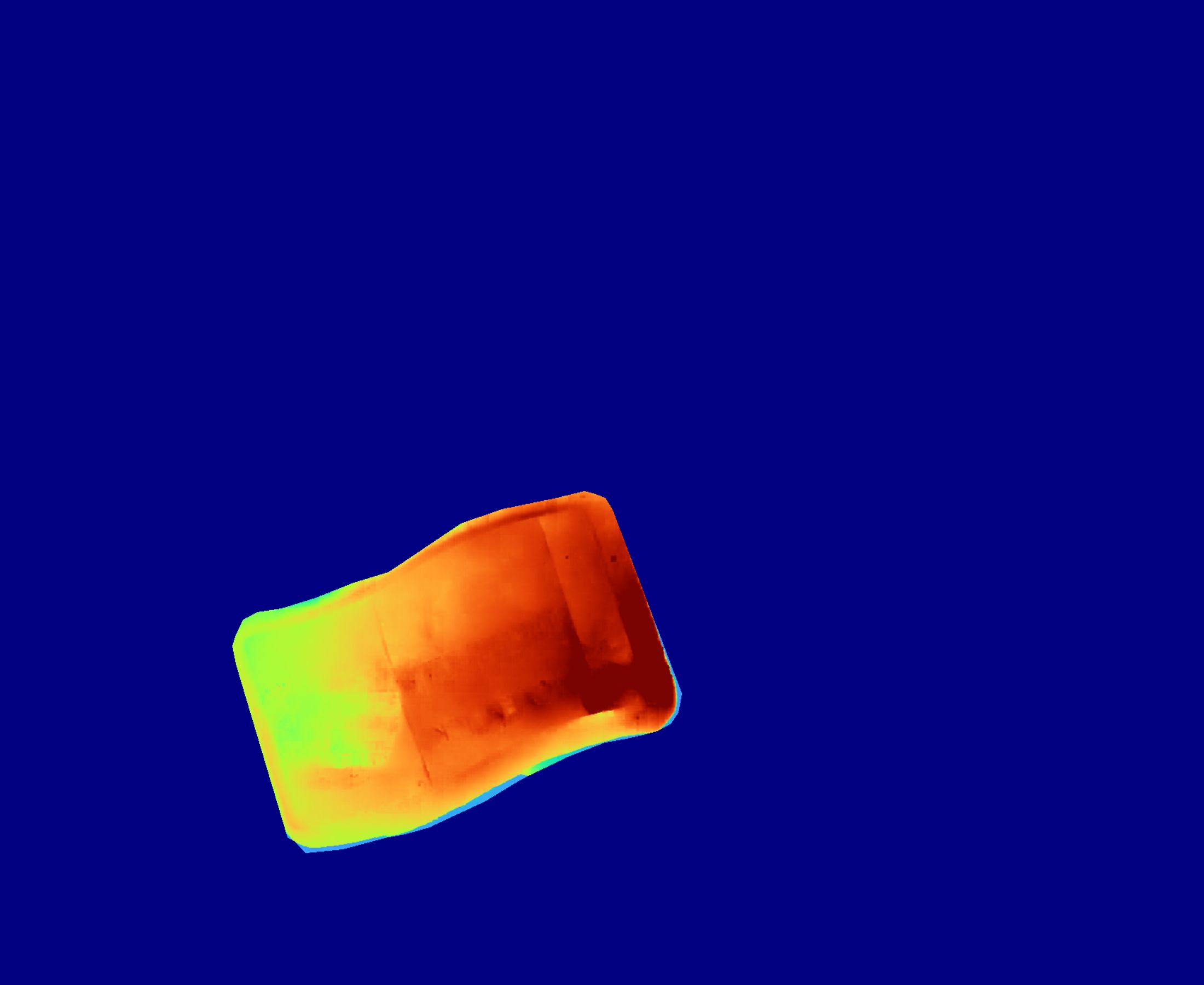}};

		\node(CRE1)[right of=NMRF1,xshift=2.53cm]{\includegraphics[width=0.19\textwidth,height=2.5cm]{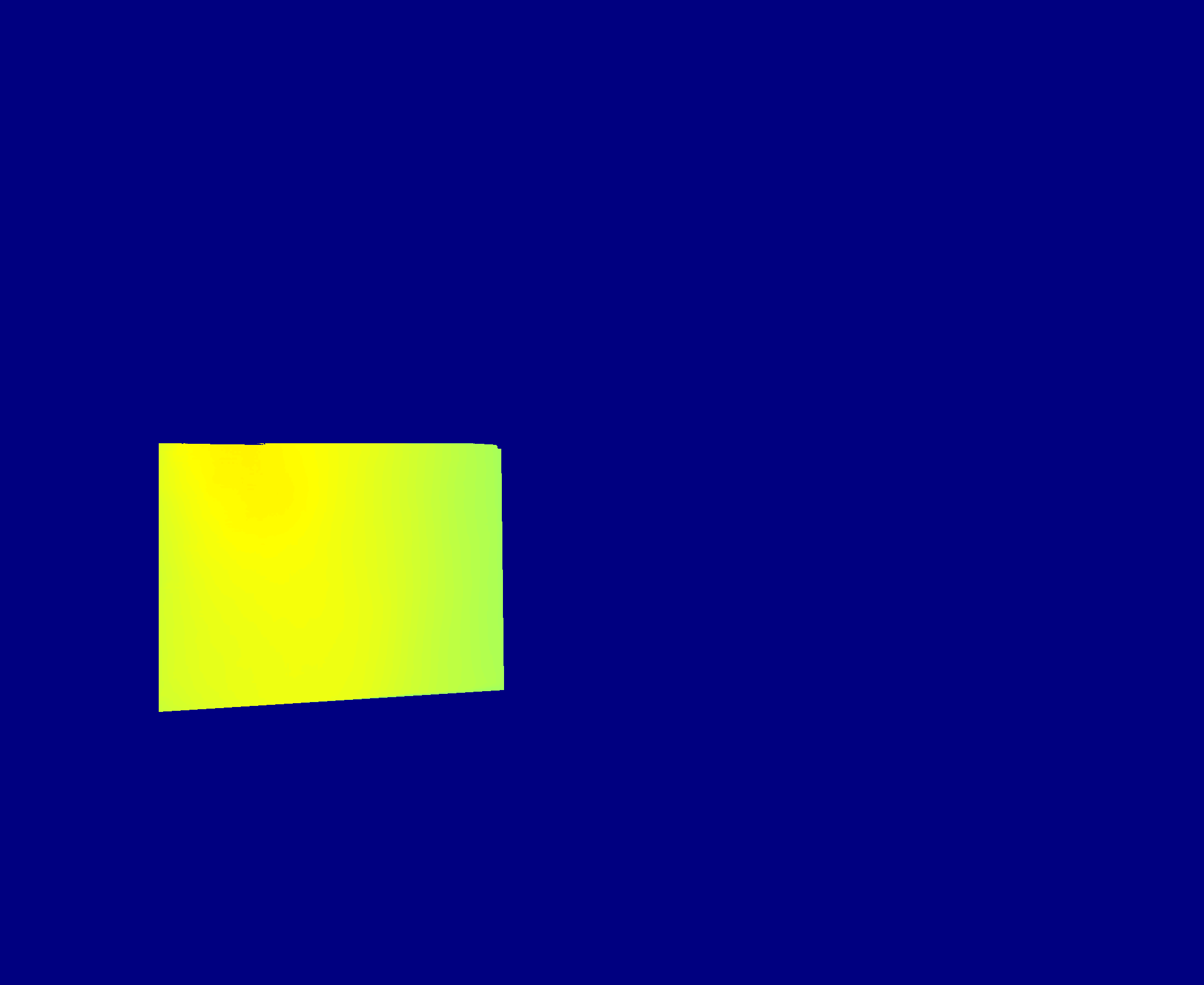}};
		\node(CRE2)[right of=NMRF2,xshift=2.53cm]{\includegraphics[width=0.19\textwidth,height=2.5cm]{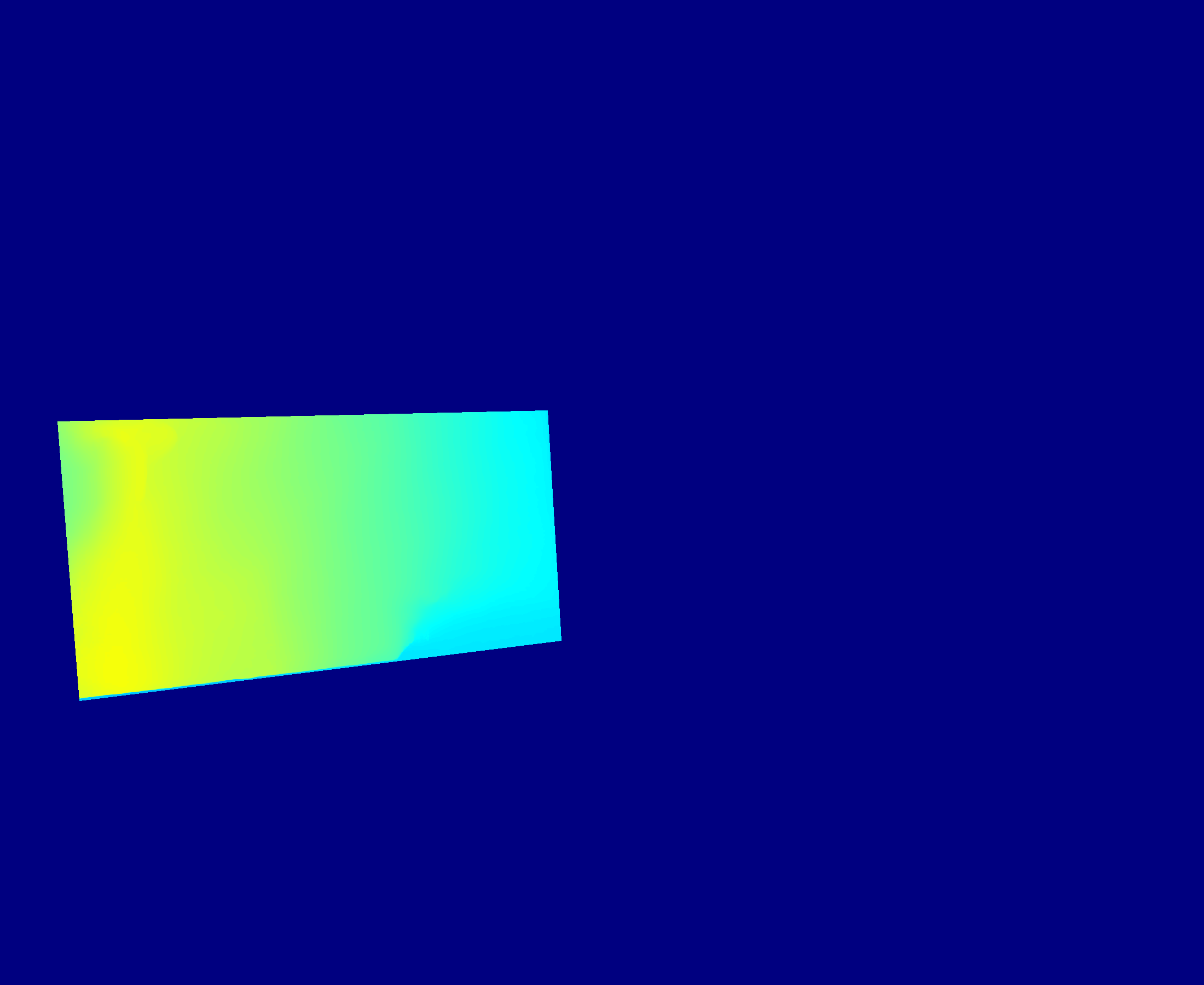}};
		\node(CRE3)[right of=NMRF3,xshift=2.53cm]{\includegraphics[width=0.19\textwidth,height=2.5cm]{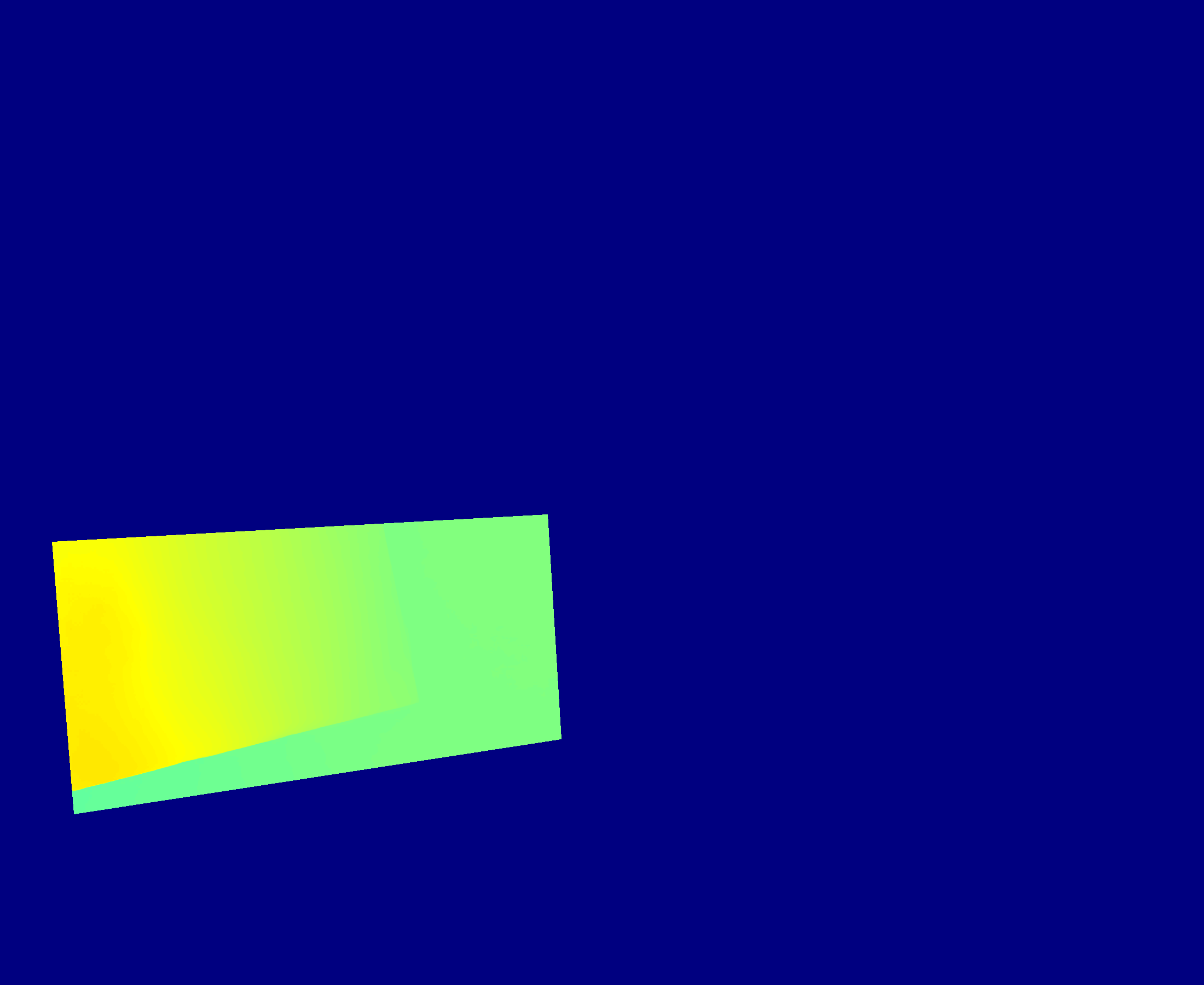}};
	
		\node(CRE4)[right of=NMRF4,xshift=2.53cm]{\includegraphics[width=0.19\textwidth,height=2.5cm]{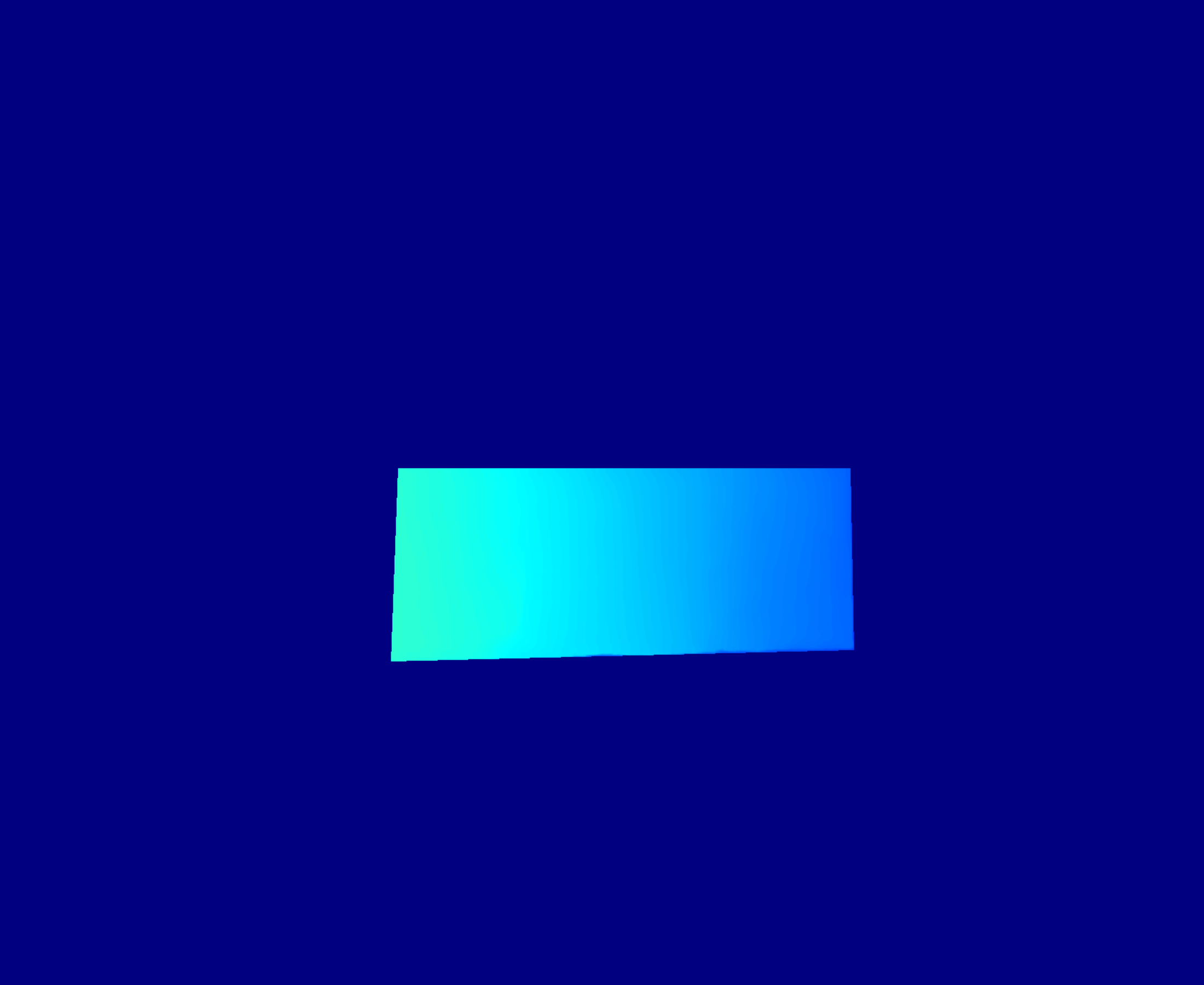}};
		\node(CRE5)[right of=NMRF5,xshift=2.53cm]{\includegraphics[width=0.19\textwidth,height=2.5cm]{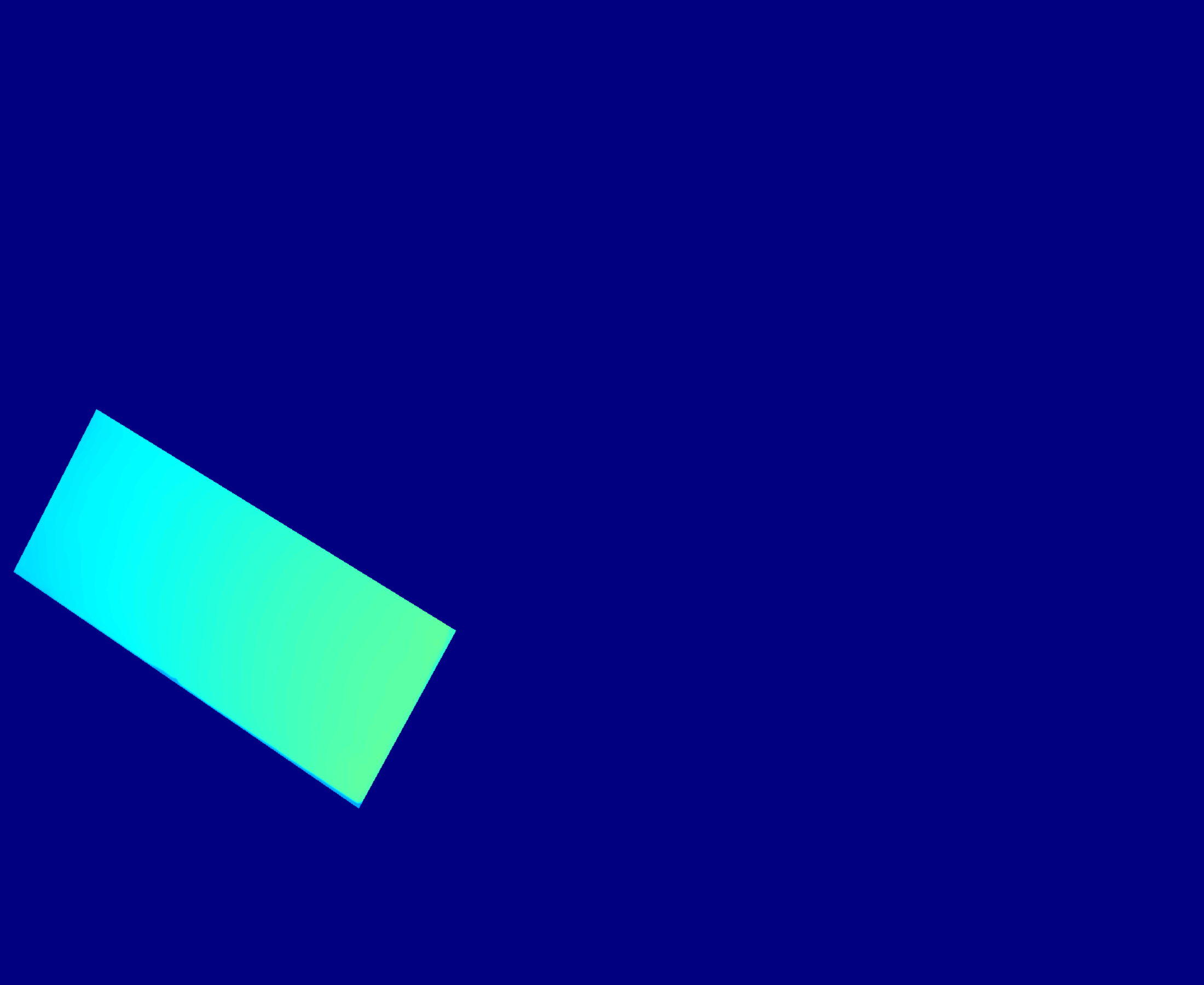}};
		\node(CRE6)[right of=NMRF6,xshift=2.53cm]{\includegraphics[width=0.19\textwidth,height=2.5cm]{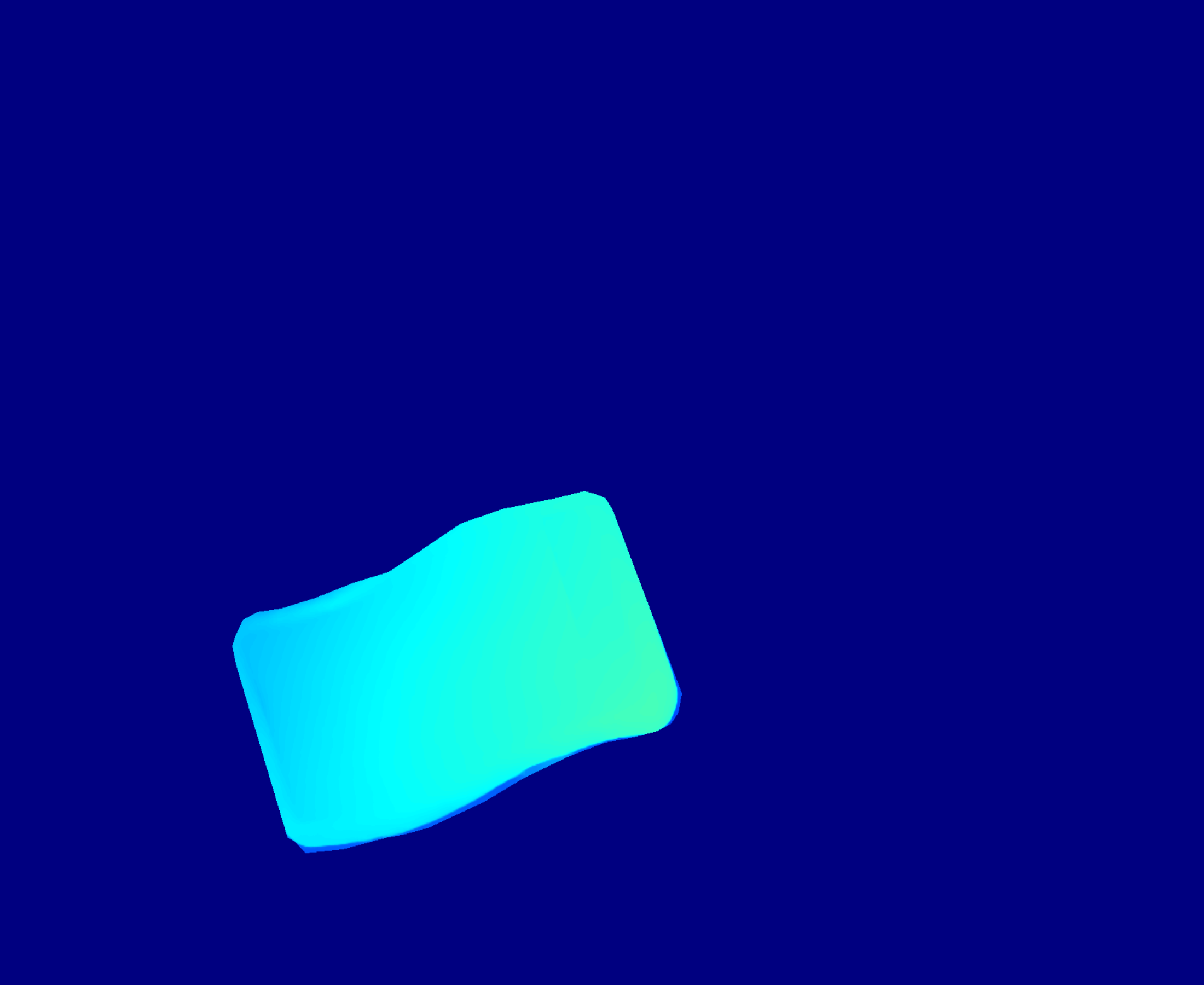}};

		\node(IGEV1)[right of=CRE1,xshift=2.53cm]{\includegraphics[width=0.19\textwidth,height=2.5cm]{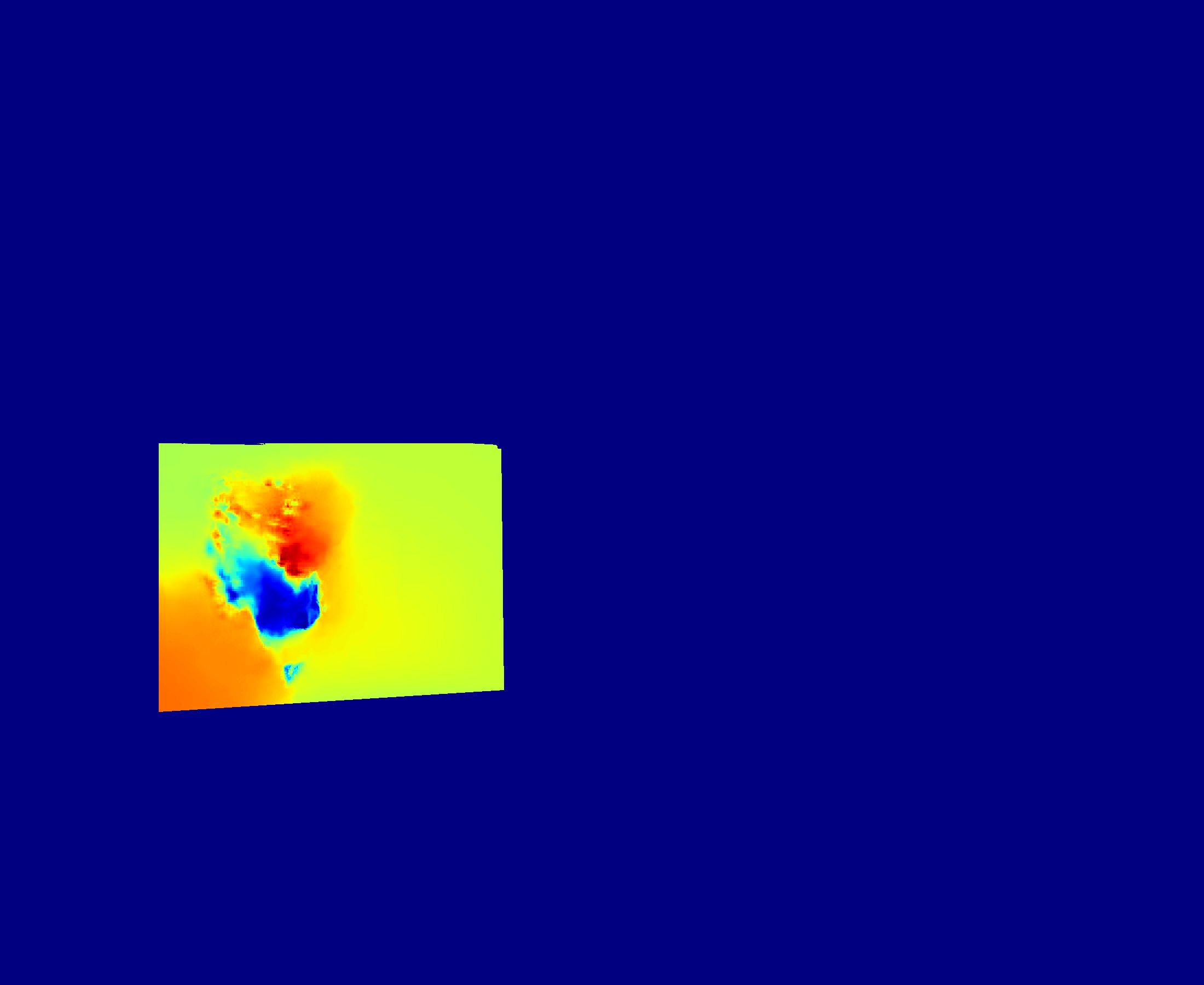}};
		\node(IGEV2)[right of=CRE2,xshift=2.53cm]{\includegraphics[width=0.19\textwidth,height=2.5cm]{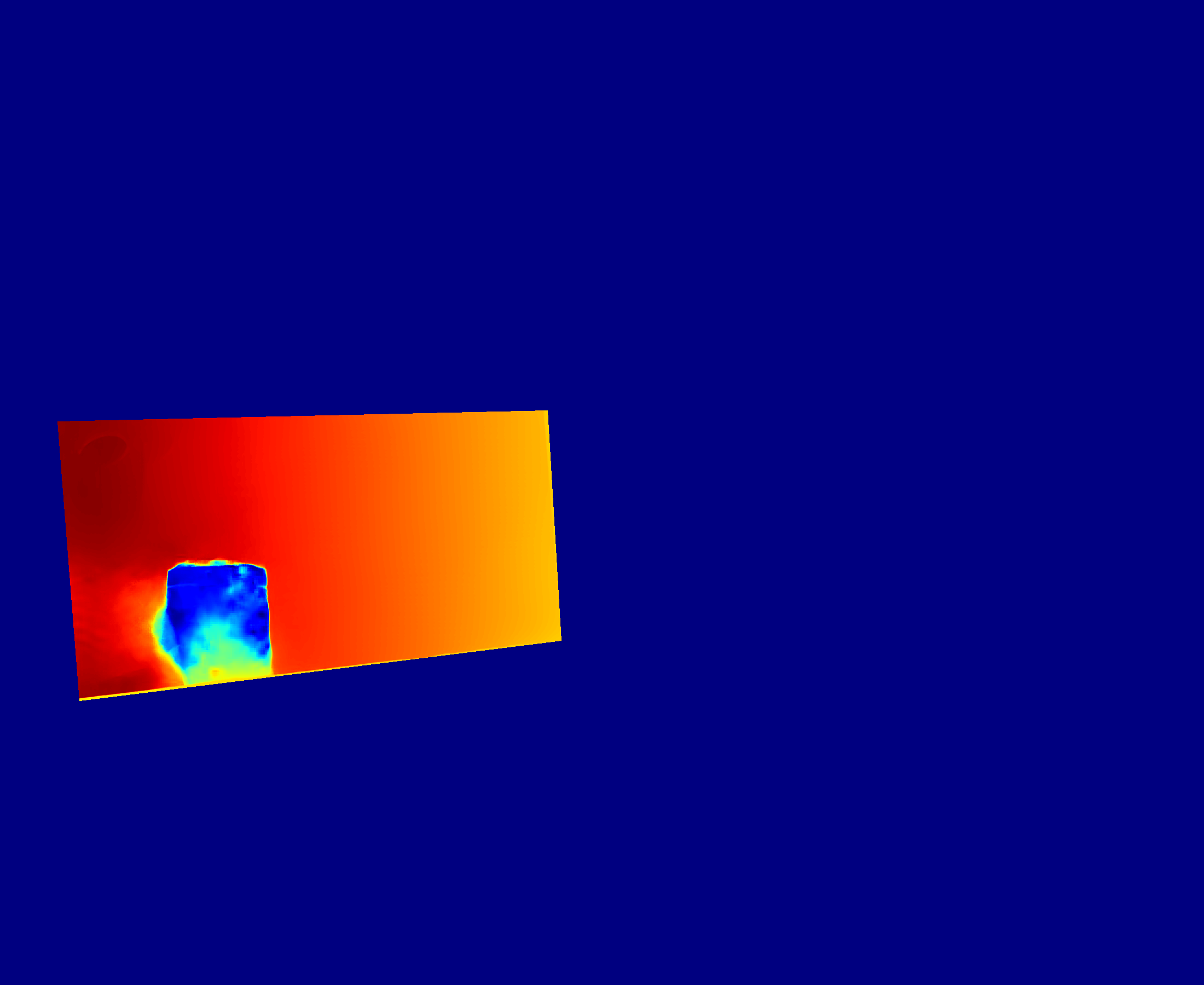}};
		\node(IGEV3)[right of=CRE3,xshift=2.53cm]{\includegraphics[width=0.19\textwidth,height=2.5cm]{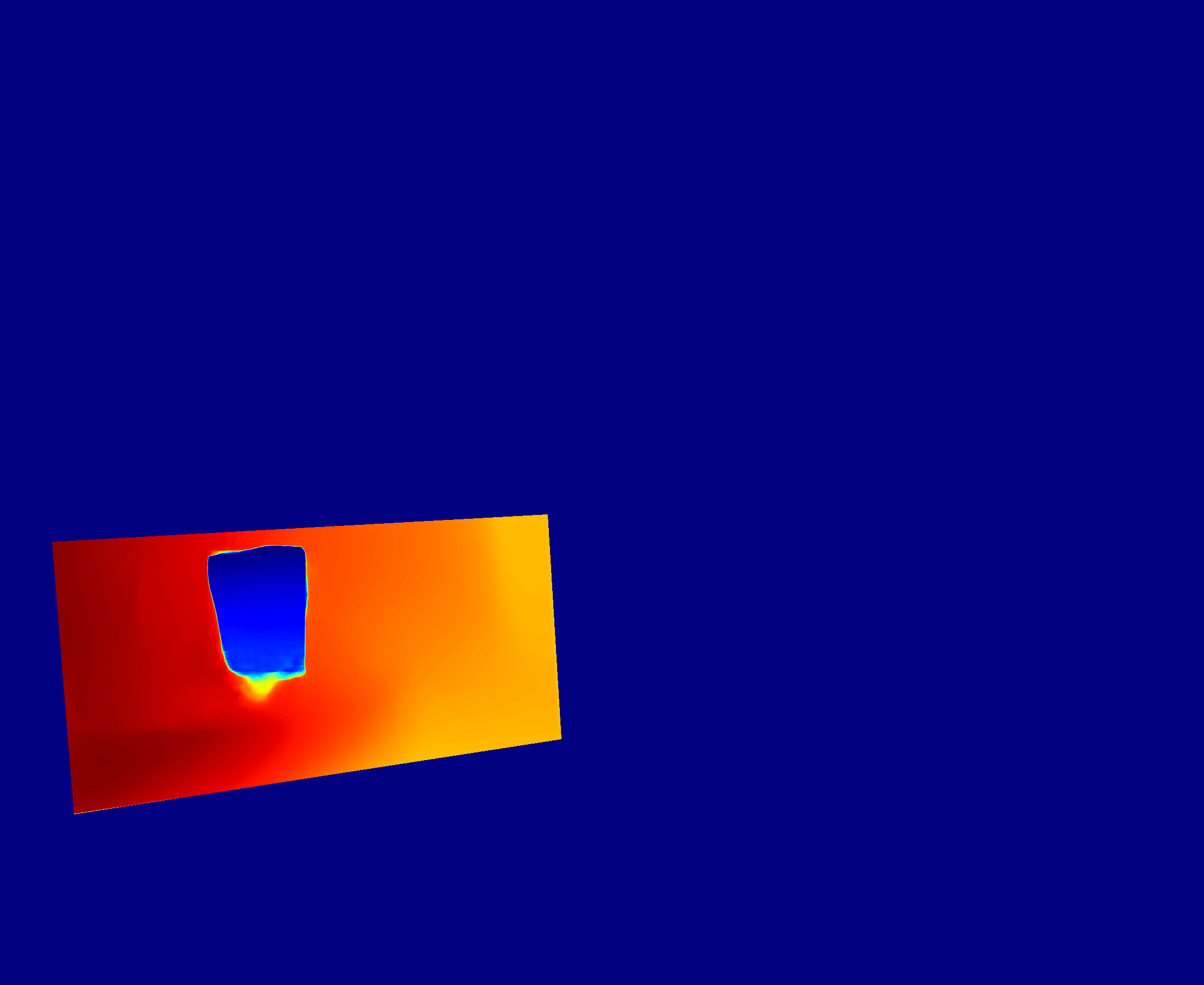}};
	
		\node(IGEV4)[right of=CRE4,xshift=2.53cm]{\includegraphics[width=0.19\textwidth,height=2.5cm]{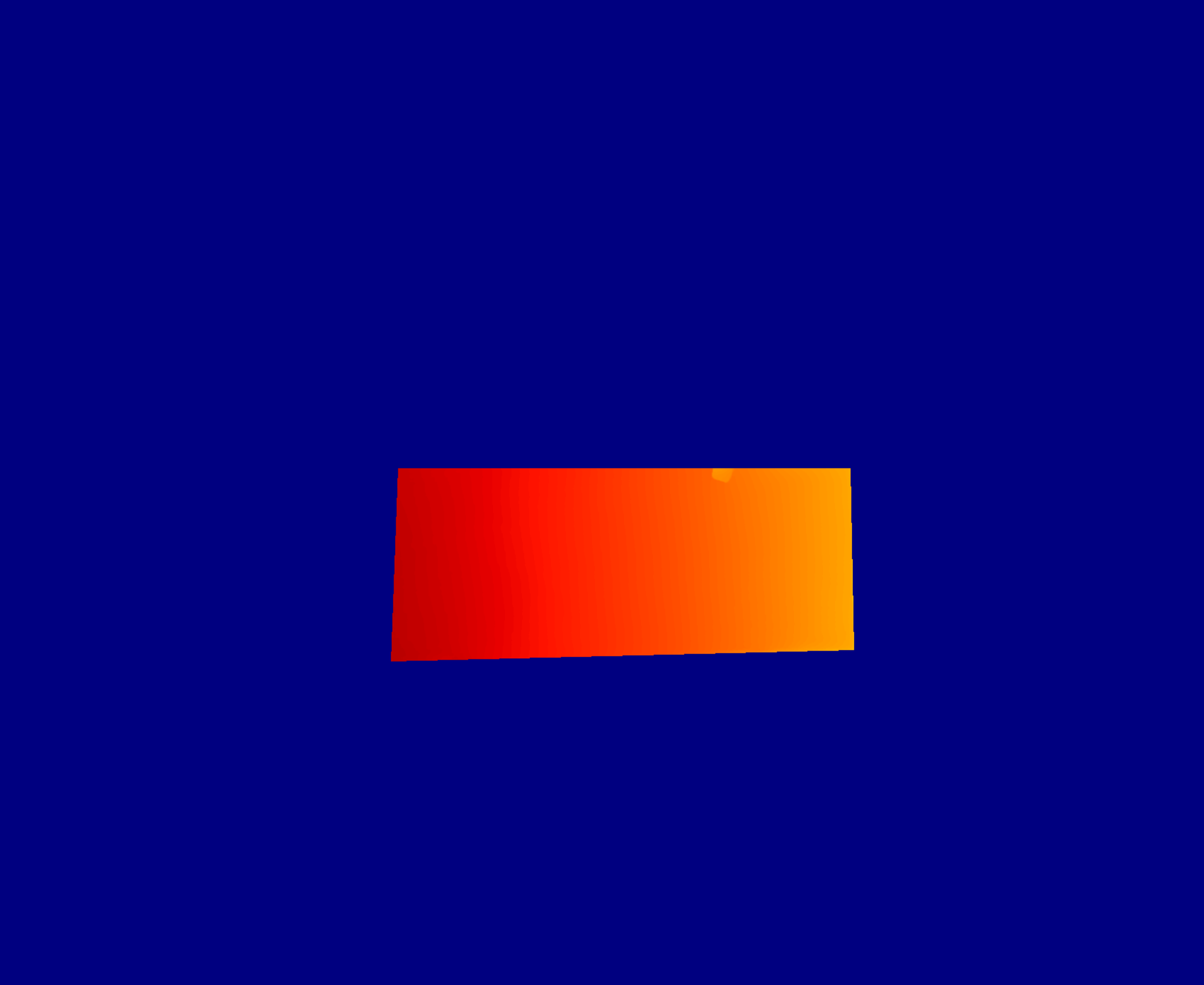}};
		\node(IGEV5)[right of=CRE5,xshift=2.53cm]{\includegraphics[width=0.19\textwidth,height=2.5cm]{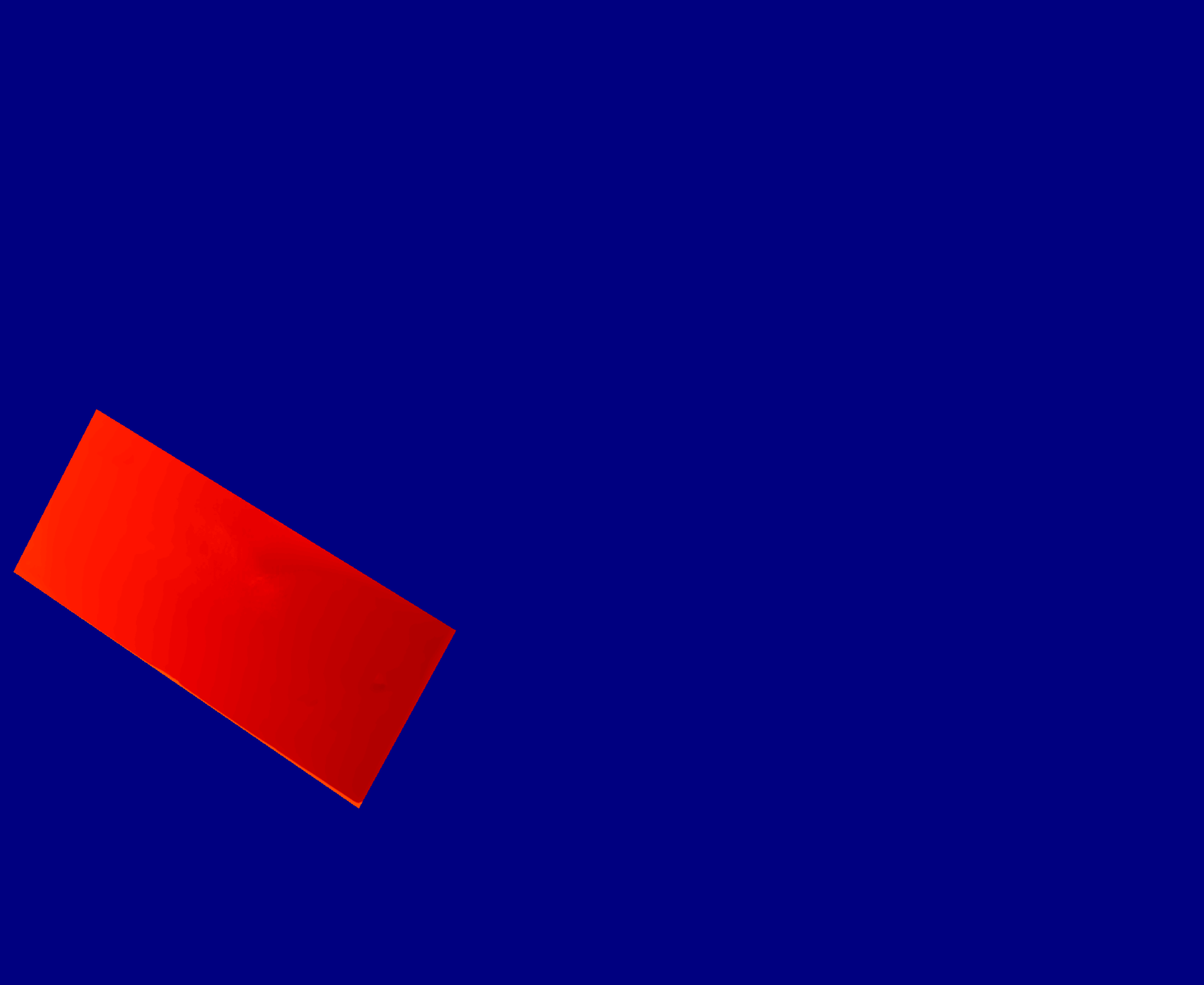}};
		\node(IGEV6)[right of=CRE6,xshift=2.53cm]{\includegraphics[width=0.19\textwidth,height=2.5cm]{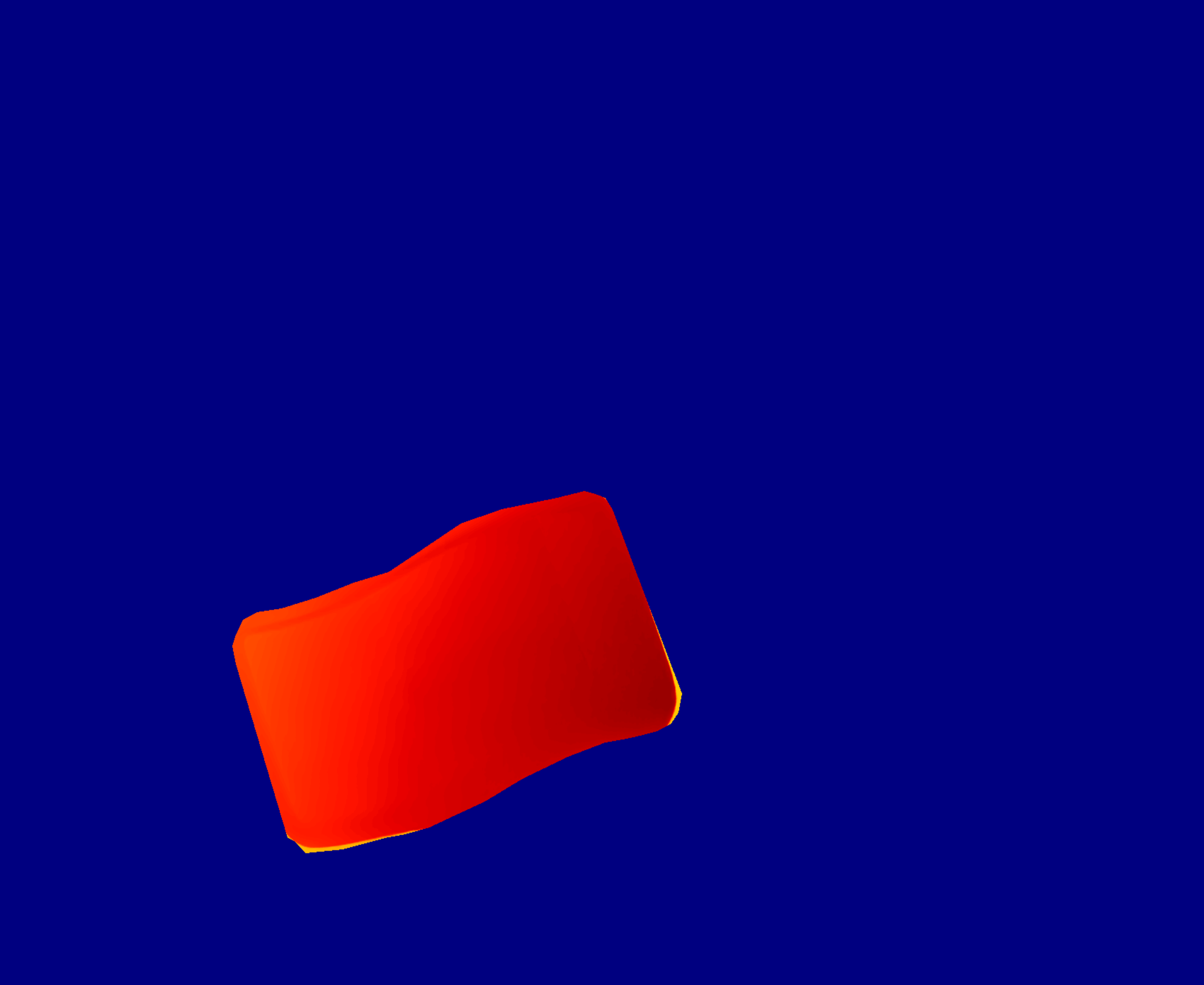}};
	
		\node(Ours1)[right of=IGEV1,xshift=2.53cm]{\includegraphics[width=0.19\textwidth,height=2.5cm]{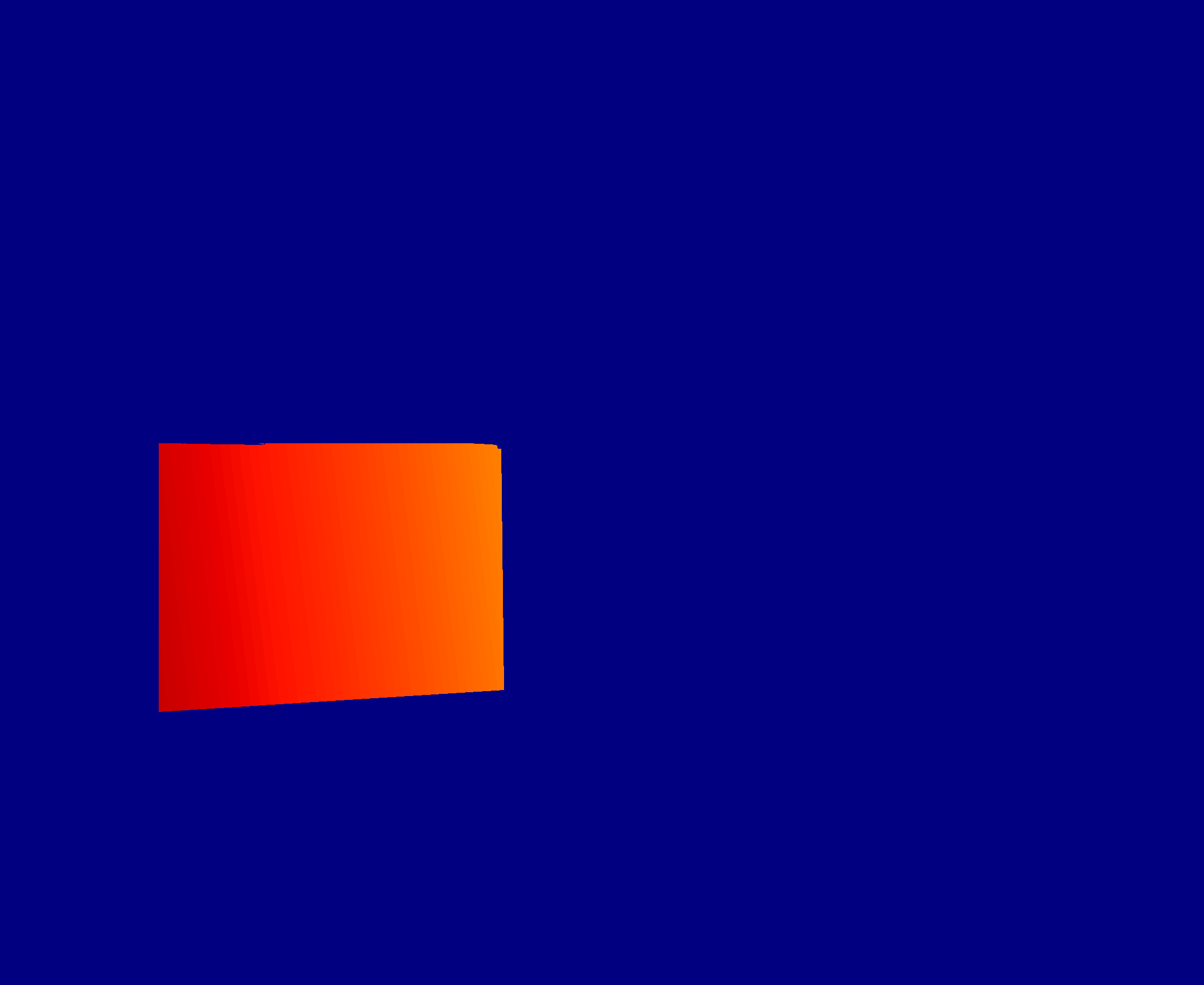}};
		\node(Ours2)[right of=IGEV2,xshift=2.53cm]{\includegraphics[width=0.19\textwidth,height=2.5cm]{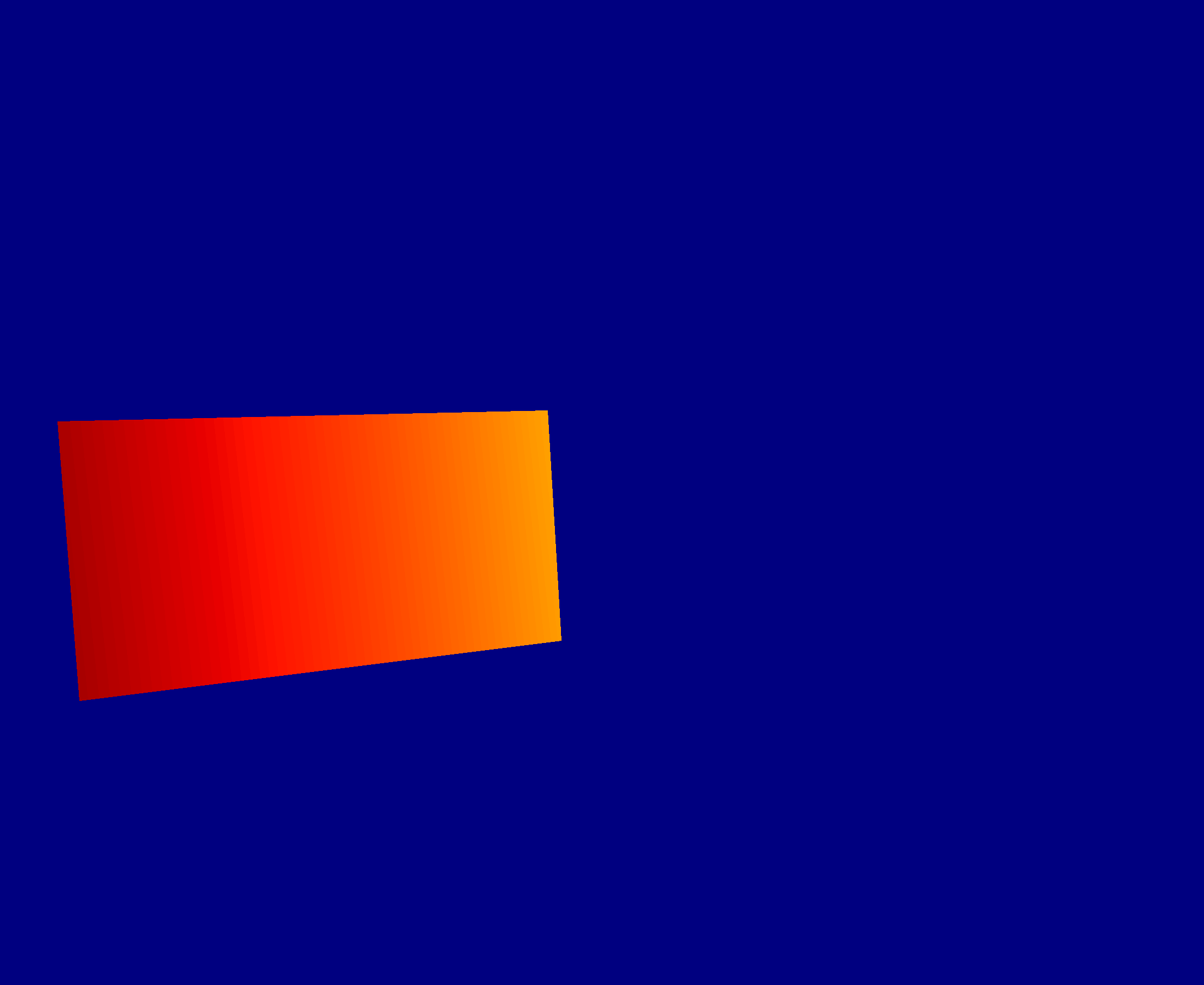}};
		\node(Ours3)[right of=IGEV3,xshift=2.53cm]{\includegraphics[width=0.19\textwidth,height=2.5cm]{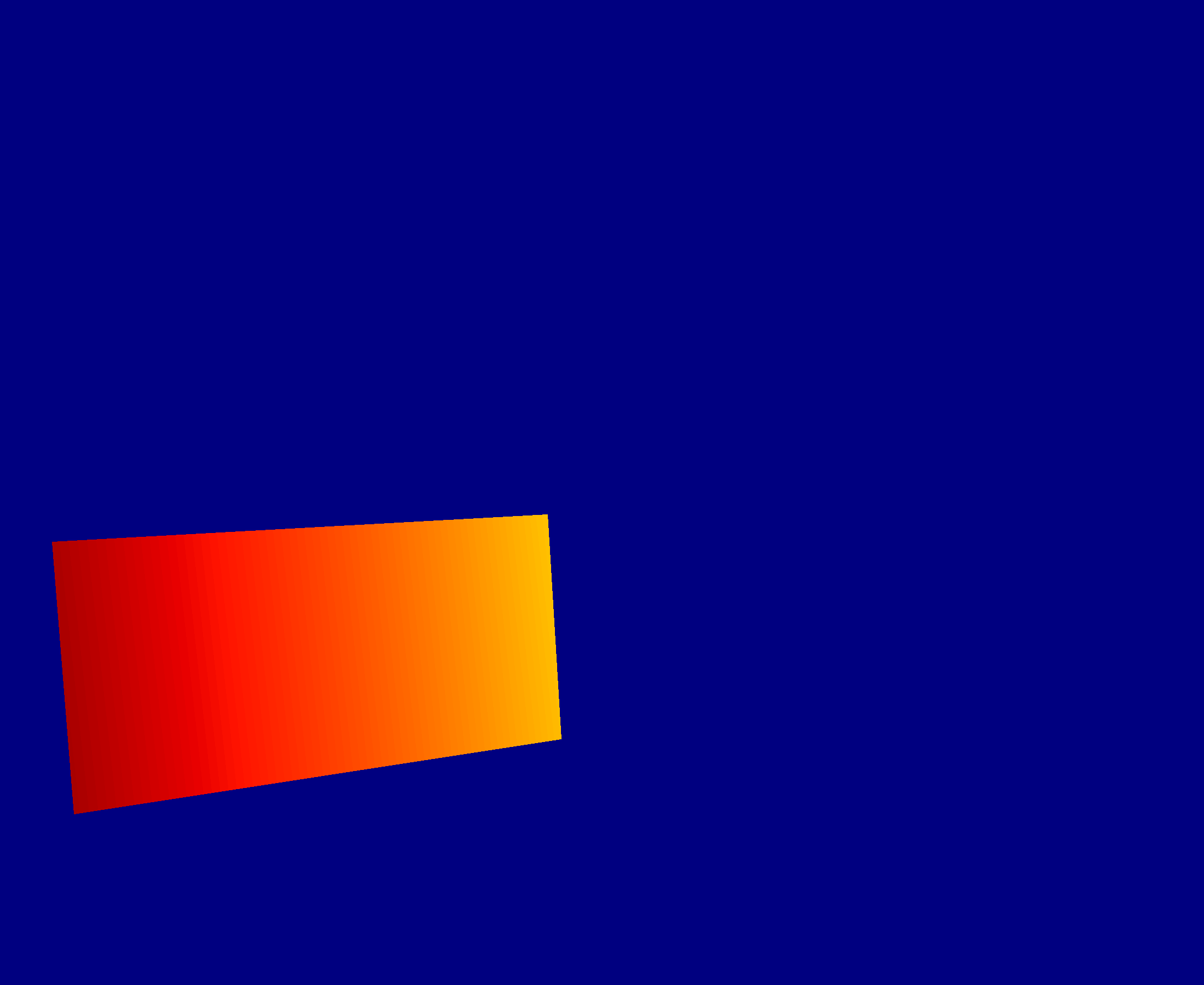}};
	
		\node(Ours4)[right of=IGEV4,xshift=2.53cm]{\includegraphics[width=0.19\textwidth,height=2.5cm]{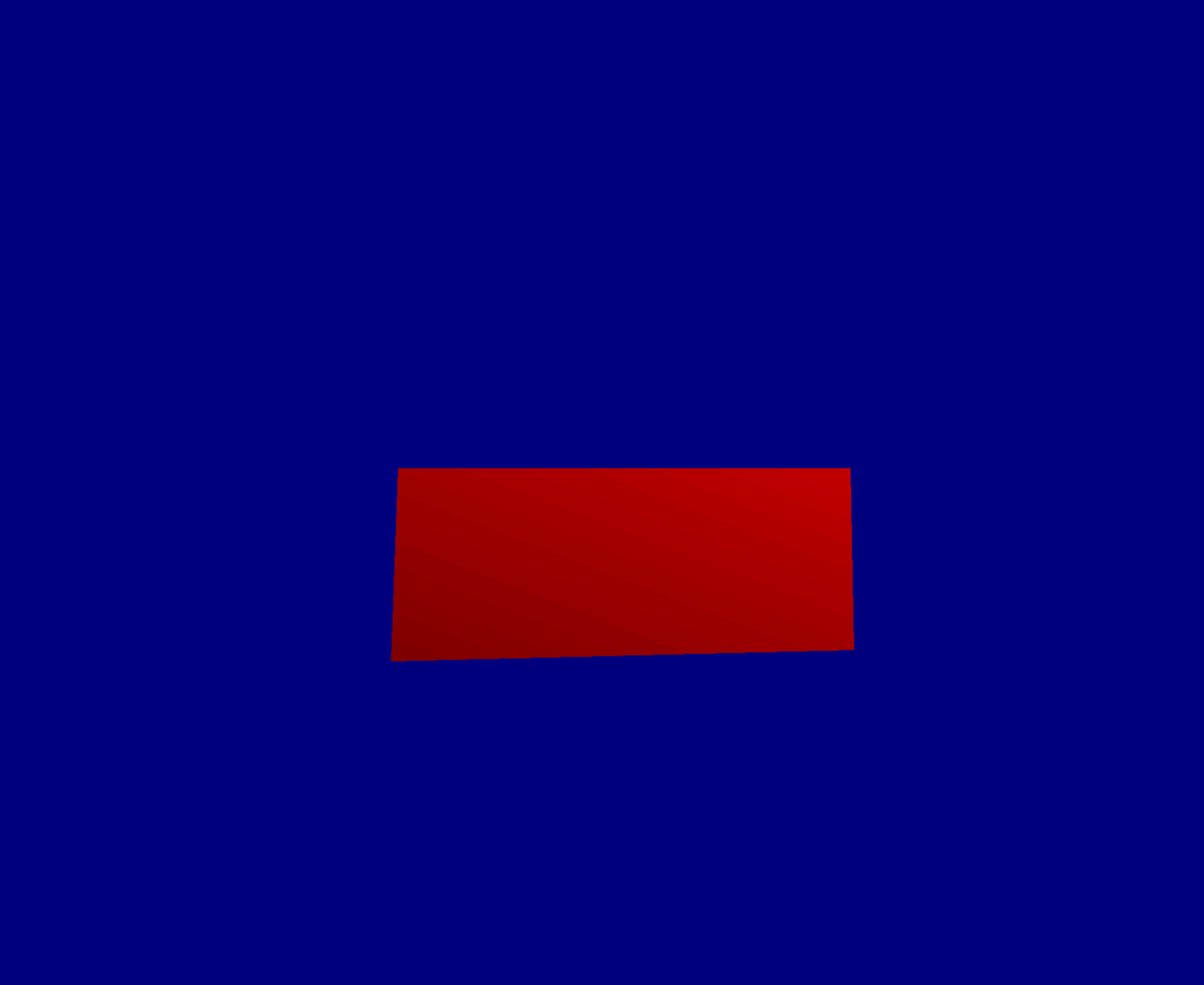}};
		\node(Ours5)[right of=IGEV5,xshift=2.53cm]{\includegraphics[width=0.19\textwidth,height=2.5cm]{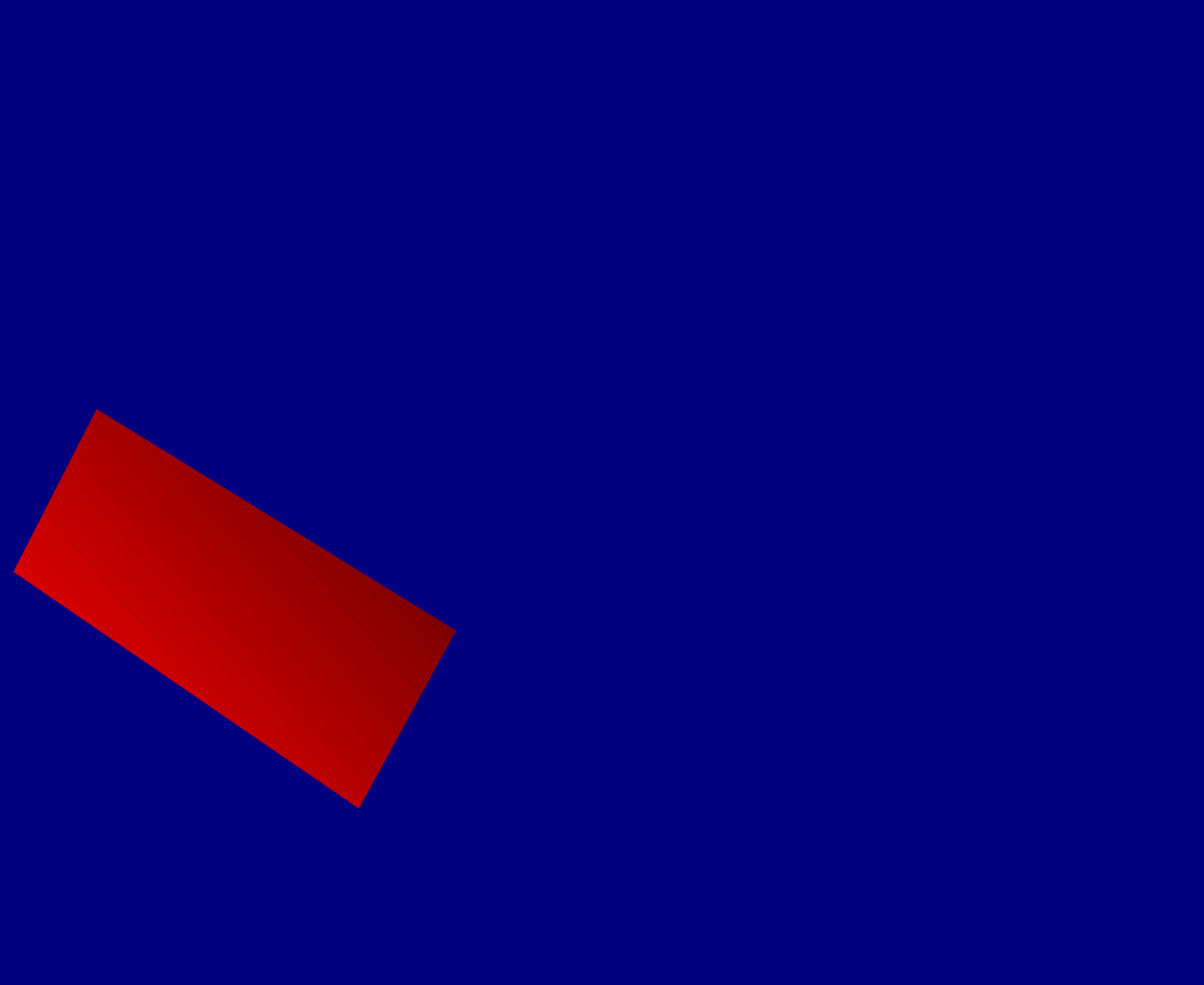}};
		\node(Ours6)[right of=IGEV6,xshift=2.53cm]{\includegraphics[width=0.19\textwidth,height=2.5cm]{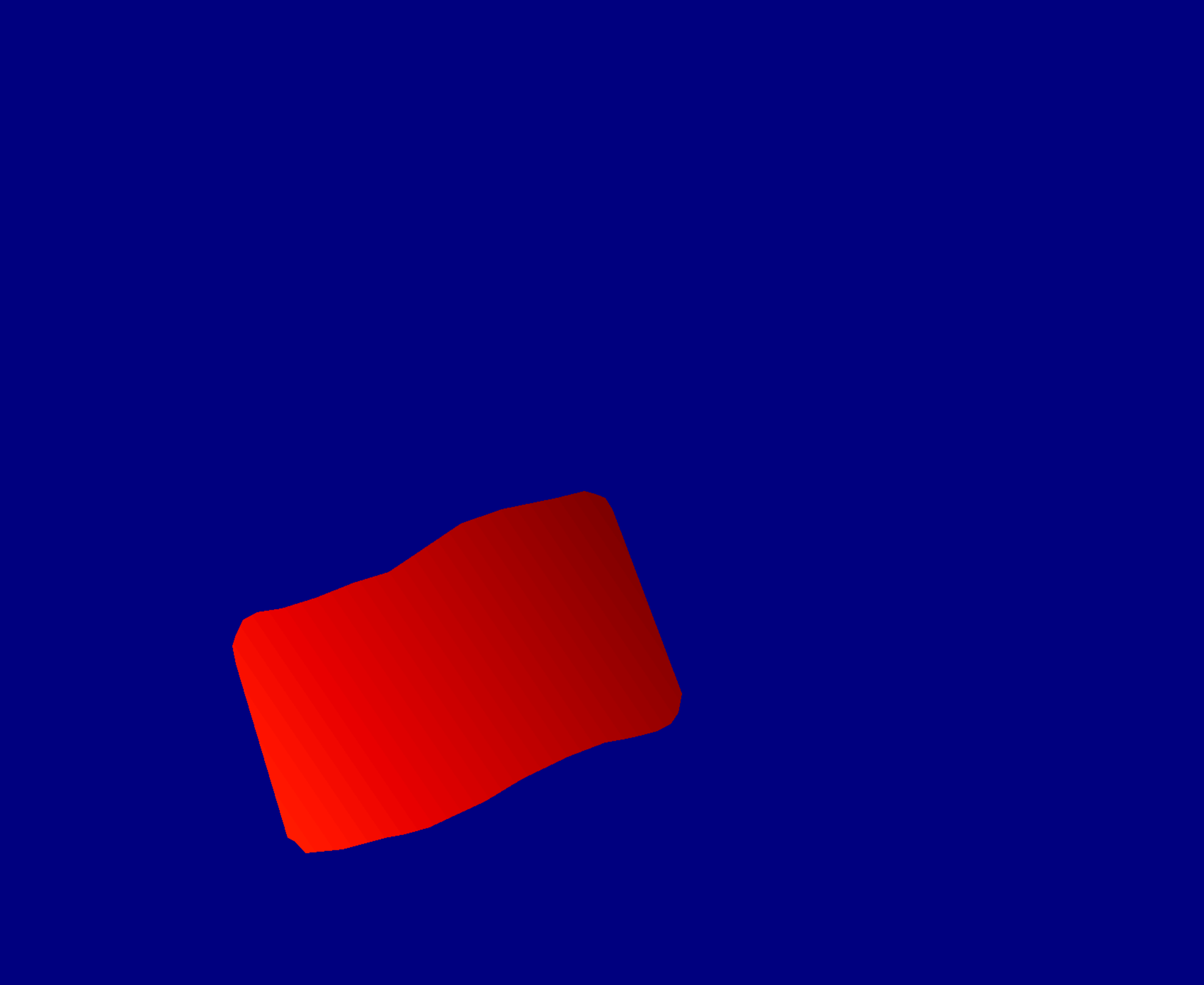}};

		\node(title1)[above of=rgb1,yshift=0.4cm]{Input image};
		\node(title3)[above of=NMRF1,yshift=0.4cm]{NMRF \cite{NMRF}};
		\node(title4)[above of=CRE1,yshift=0.4cm]{CREStereo \cite{CREStereo}};
		\node(title5)[above of=IGEV1,yshift=0.4cm]{IGEV \cite{IgevDE}};
		\node(title6)[above of=Ours1,yshift=0.4cm]{SRDE};
	\end{tikzpicture} 
		\vspace{-0.1cm}
		\caption{Qualitative disparity map results on real-world images captured with a camera array. Shown are input image and disparity maps produced by NMRF \cite{NMRF}, CREStereo \cite{CREStereo}, IGEV \cite{IgevDE} and SRDE (ours) after the light source position estimation. Colors are normalized to depth.}
		\label{fig:RealWorldResults}
		\vspace{-0.5cm}
	\end{figure*}

\subsection{Performance on Real Data}
\subsubsection{Light Source Position Estimation}
\label{sec:LightSourcePositionEstimation}
The proposed SRDE method relies on the knowledge of the real light source position $\textbf{L}$ relative to the camera array. In controlled or synthetic environments, these coordinates are available. In real-world applications, however, the light source position is unknown and must be estimated in an initial calibration step.\\
\\

For that, we first determine the orientation of a flat reflective calibration surface. We utilize a polyvinyl chloride (PVC) calibration plate where we placed ArUco markers \cite{ArUcoMarker} of known physical dimensions in each corner, as shown in Fig. \ref{fig:MarkersLightSourcePositionEstimation}. By detecting these markers in all nine available images of the camera array, the orientation and position of the calibration plate \cite{opencv} can be determined. Specifically, the detection algorithm extracts the sub-pixel corners of the ArUco markers across all views. Because the physical dimensions of the markers and their relative placement are precisely known, this allows for a robust estimation of the plate's 3D surface normal relative to the cameras. In industrial environments, the light source position remains stable, so a single calibration at the beginning is sufficient and the calibration process does not affect practical applications. Afterwards, the disparity estimation can be performed as described in the previous section, and the calibration plate is no longer required.

The plate's surface is modeled as a plane in Hessian normal form \cite{HNF}:
\begin{equation}
	\textbf{n}_p \cdot \textbf{Q} + D_p = 0~~,
\end{equation}
where $\textbf{n}_p = [n_x, n_y, n_z]^\top$ is the unit normal vector of the plane, $\textbf{Q}$ is a generic point on the plane, and $D_p$ is the distance from the camera origin to the plane.

For each camera $i \in \{1,...,9\}$, we identify the center of the specular reflection $(c_{x,i}, c_{y,i})$ on the plate as described in Section \ref{sec:Specular Reflection Detection}. Using the ray formulation from equation (10), we define the directional ray $\textbf{k}_i$ for each reflection center. The corresponding 3D intersection point $\textbf{H}_i$ on the calibration plate is then calculated using (11) and (12).

According to the Law of Reflection \cite{LawReflection}, the angle of incidence equals the angle of reflection. This means that the light ray traveling from the real light source $\textbf{L}$ to the surface point $\textbf{H}_i$ is reflected symmetrically along the surface normal $\textbf{n}_p$ towards the camera. The direction of the reflected ray $\textbf{r}_i$, which points back toward the light source, is calculated as
\begin{equation}
	\mathbf{r}_i = \mathbf{k}_i - 2(\mathbf{k}_i \cdot \mathbf{n}_p)\mathbf{n}_p.
\end{equation}

Ideally, all nine reflected rays would intersect at the real light source position $\mathbf{L}$. In practice, minor sensor noise and localization inaccuracies cause these rays to deviate slightly. Therefore, we estimate the light source position $\hat{\mathbf{L}}$ by finding the point that minimizes the distance to all nine rays using a least-squares optimization:
\begin{equation}
	\hat{\mathbf{L}} = \arg \min_{\mathbf{L}} \sum_{i=1}^{9} \| (\mathbf{L} - \mathbf{H}_i) \times \mathbf{r}_i\|^2~.
\end{equation}
Here, the symbol '$\times$' denotes the vector cross product. This objective function minimizes the sum of squared distances between the estimated point $\hat{\mathbf{L}}$ and the bundle of nine reflected rays, effectively finding the optimal 3D intersection point that represents the true physical location of the light source.

The robustness of SRDE is fundamentally linked to the number and spatial arrangement of viewpoints. We utilize a $3 \times 3$ camera array because it provides a highly overdetermined system of equations. While three non-collinear cameras could theoretically enable specular reflection-based disparity estimation if the light source position was exactly known, this condition is rarely satisfied in real-world scenarios. The $3 \times 3$ configuration yields nine rays that converge toward the light source position from multiple angles. This spatial diversity ensures that the least-squares optimization is well-conditioned, significantly reducing the sensitivity to small localization errors in the reflection center detection.

We first evaluated this principle for functionality using synthetic data. Subsequently, we applied it to real-world data. In real-world scenarios, several potential error sources can affect this calibration process. Inaccuracies in the ArUco marker detection, caused e.g. by imperfect lens distortion correction, can lead to slight misalignments in the estimated calibration plane normal. Additionally, errors in localizing the exact specular reflection center on the calibration plate introduce angular deviations in the reflected rays. If unmitigated, these geometric errors would propagate and result in an offset of the estimated light source position, which would subsequently lead to tilted disparity maps for the target objects. Consequently, the overdetermined system of the $3 \times 3$ camera array is essential. The least-squares optimization across all nine views effectively averages out these errors. By finding the optimal intersection point that minimizes the distance of the bundle of rays, the optimization significantly stabilizes the estimation against local inaccuracies.

To formally quantify the impact of the remaining positional uncertainties, the intercept theorem \cite{InterceptTheorem} and error propagation in angular measurements \cite{ErrorPropagation} can be applied. The intercept theorem is used to estimate the effect of positional inaccuracies in related geometries, while error propagation in angular measurements returns how positional uncertainties influence the resulting angle and disparity estimations. In our configuration, we found that a spatial accuracy of $\Delta p = \pm$5 cm (measured manually) for the light source position was achievable. Given a distance of $D = 115$ cm between the object an the light source, this offset results in an angle difference of $\Delta \theta = 2.49^\circ$, which falls within an acceptable range for accurate disparity estimation.
			
\subsubsection{Qualitative Comparison}
For real-world data, it is almost impossible to obtain valid ground truth disparity maps, hence this evaluation is limited to a subjective inspection of the results. Fig. \ref{fig:RealWorldResults} shows real images captured with a $3 \times 3$ multi-camera system, and the disparity estimation results of the tested benchmark models and our approach. The recorded scenes consist of typical industrial sorting line objects, such as plastic plates and various packaging items processed on conveyor belts in automated production systems. As the focus of the evaluation is solely on the object themselves, background regions are excluded from the analysis. The object masks were generated using RePoSeg \cite{RePoSeg}.
			
The evaluation reveals significant differences in the handling of reflections and overall disparity accuracy. IGEV \cite{IgevDE} prominently highlights reflections, resulting in inaccurate disparity maps at those positions. A possible reason for this outcome could be the sharp edges of the reflections, which might be treated as features or texture. NMRF \cite{NMRF} shows a slight improvement over IGEV, but still struggles with reflections. This is especially noticeable in the lower examples of Fig. \ref{fig:RealWorldResults}, where reflections are incorrectly interpreted as depth-relevant features. CREStereo \cite{CREStereo} produces more reliable results among the evaluated networks for real-world data. However, in the first row, there is almost no notable disparity difference. 
			
The proposed SRDE shows good results and coherent disparity maps, with plausible gradients reflecting the object orientations. Unlike conventional methods that treat specular reflections as regions without usable information, SRDE exploits the geometric properties to determine the orientation and position of planar surfaces in 3D space. Consequently, SRDE produces more accurate disparity maps, particularly for flat, textureless objects. 
		
\vspace{-0.3cm}	
\subsection{Integration into Existing Disparity Estimation Pipelines}
\label{sec:Integration}
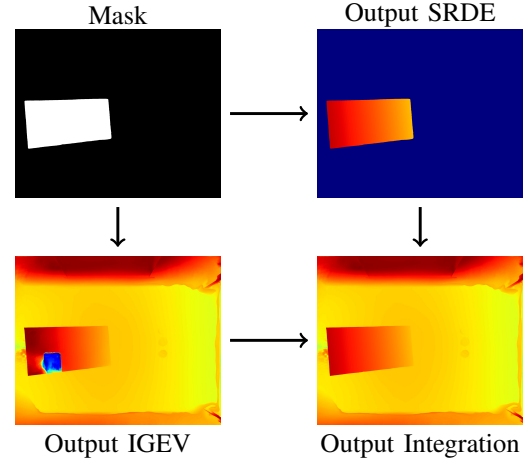
\begin{figure}
	\centering
	\begin{tikzpicture}
	\node(Mask)[]{\includegraphics[width=0.15\textwidth]{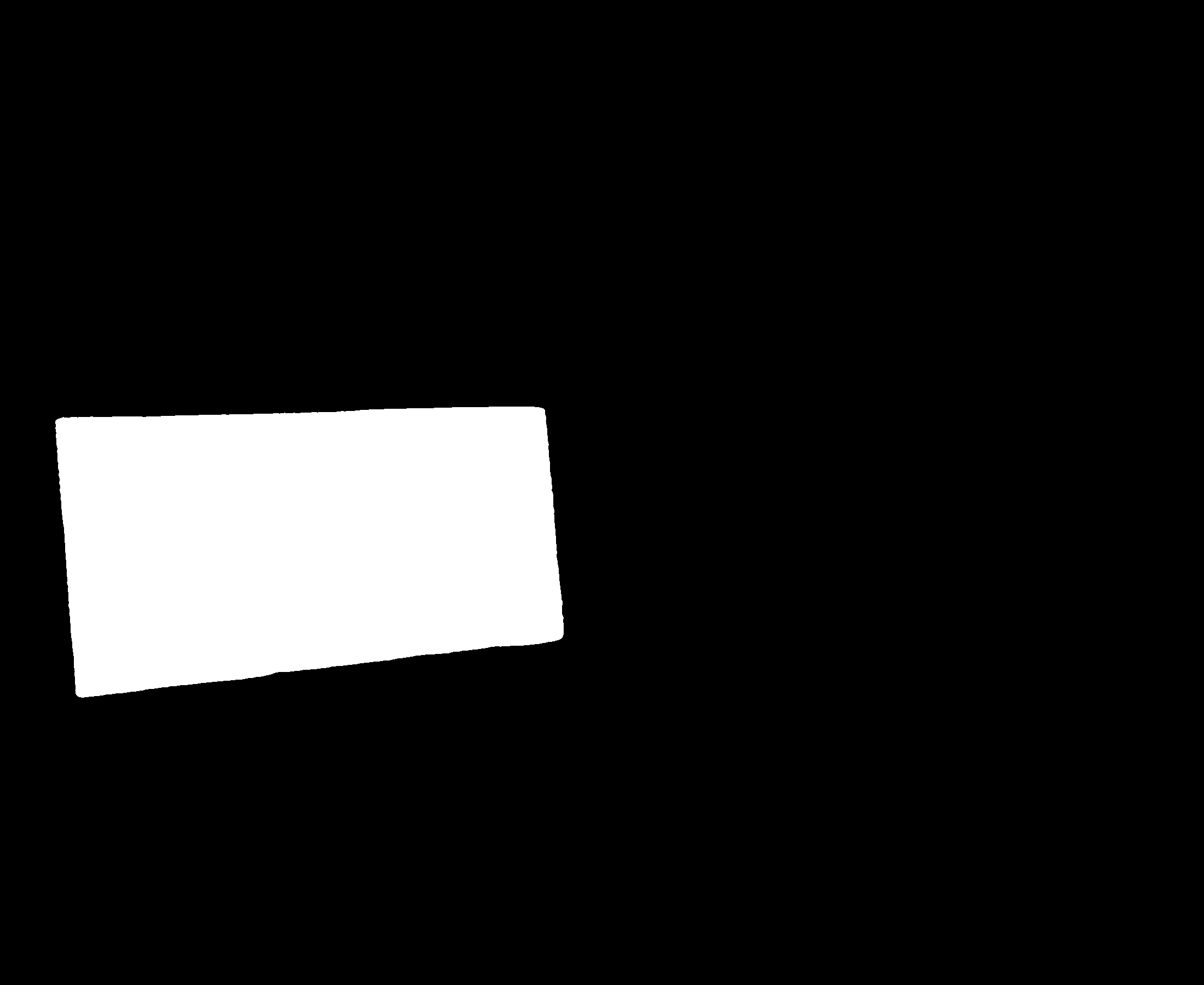}};
	\node(SRDE)[right of=Mask, xshift=3cm]{\includegraphics[width=0.15\textwidth]{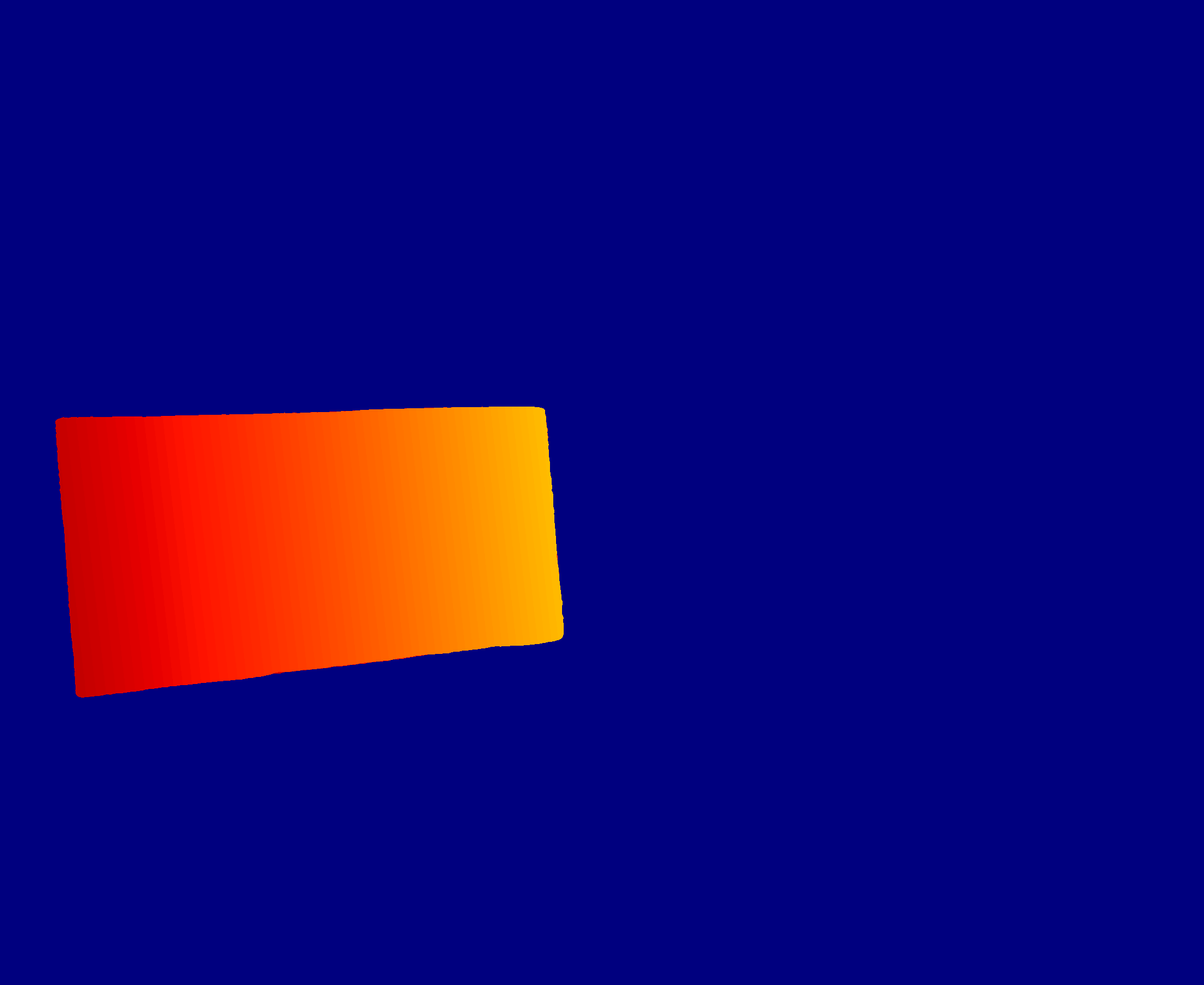}};
	\node(IGEV)[below of=Mask, yshift=-2cm]{\includegraphics[width=0.15\textwidth]{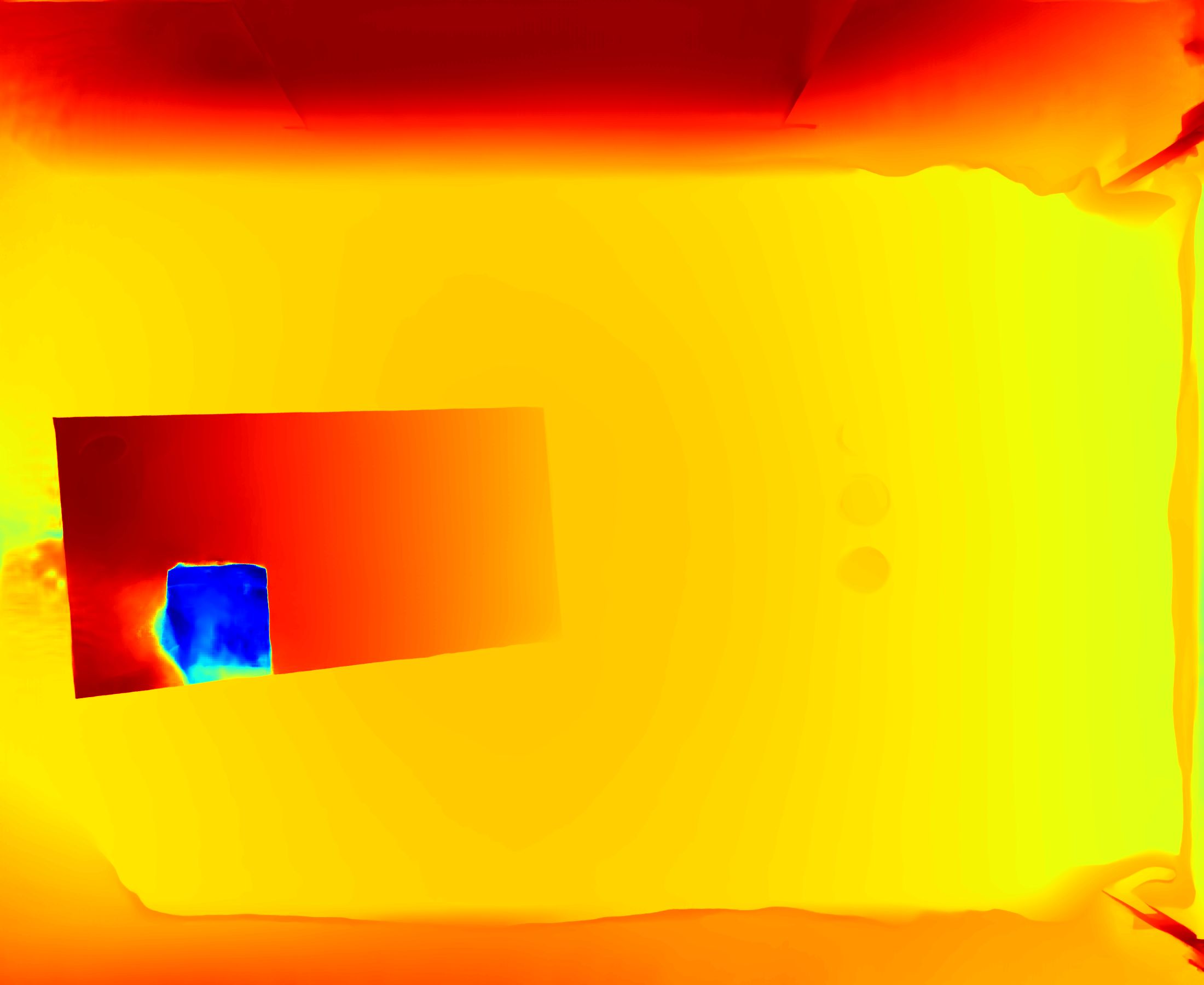}};
	\node(Integration)[right of=IGEV, xshift=3cm]{\includegraphics[width=0.15\textwidth]{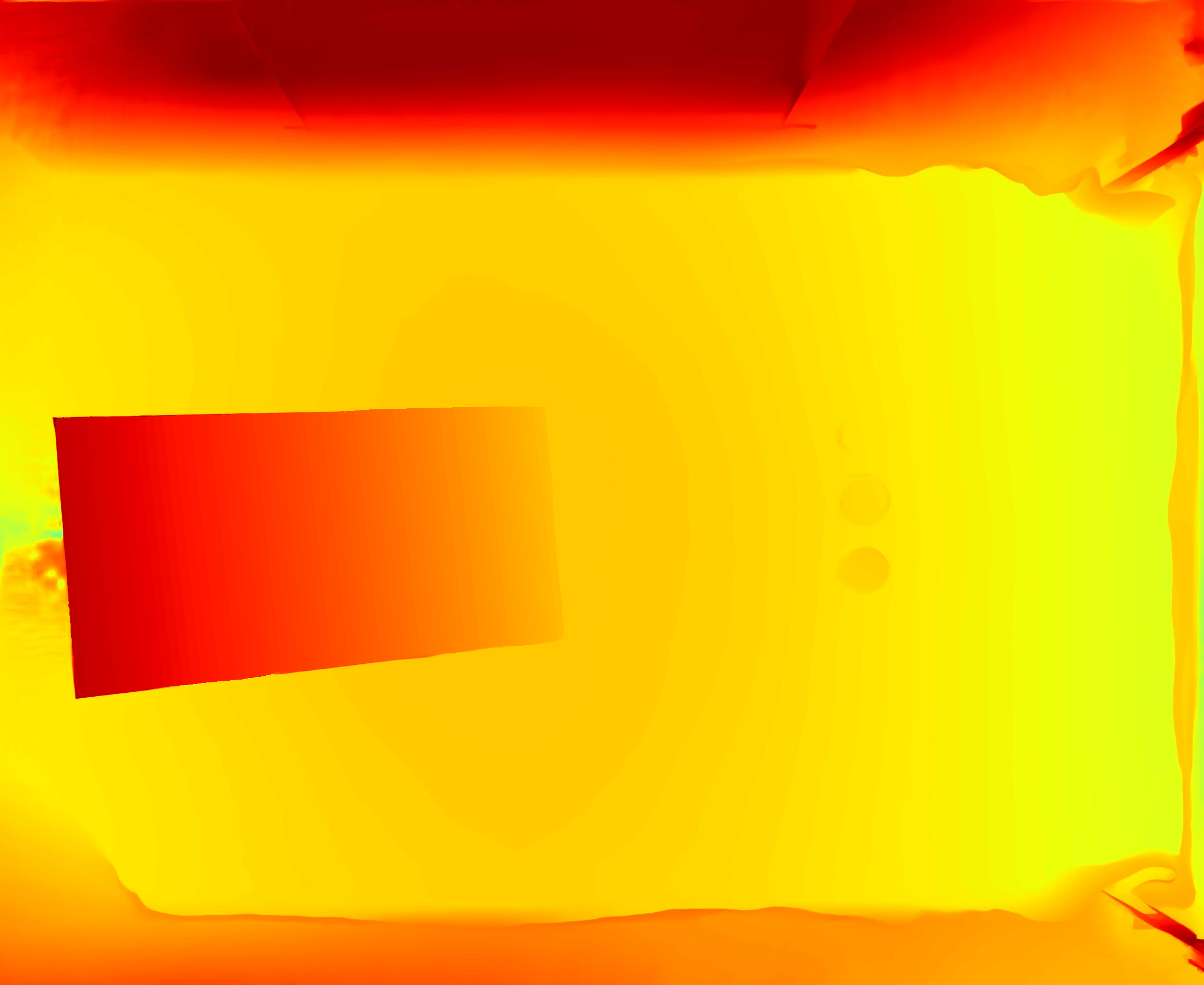}};
	\draw[->, line width=1.1pt](Mask)--(SRDE);
	\draw[->, line width=1.1pt](IGEV)--(Integration);
	\draw[->, line width=1.1pt](SRDE)--(Integration);
	\draw[->, line width=1.1pt](Mask)--(IGEV);
	\node(mask)[above of=Mask, yshift=0.3cm]{Mask};
	\node(srde)[above of=SRDE, yshift=0.3cm]{Output SRDE};
	\node(igev)[below of=IGEV, yshift=-0.4cm]{Output IGEV};
	\node(integrated)[below of=Integration, yshift=-0.4cm, ]{Output Integration};
	\end{tikzpicture}
	\vspace{-0.1cm}
	\caption{Integration of SRDE into IGEV. In case a specular reflection occurs, disparity values are replaced by the SRDE output, while IGEV provides the background estimation.}
	\label{fig:Integration}
\end{figure}
\begin{table}[t!]
	\centering
	\caption{Quantitative evaluation of complex synthetic scenes with specular reflections. The hybrid approach outperforms the baseline IGEV across all metrics. \textbf{Bold:} Best.}
	\vspace{-0.2cm}
	\renewcommand{\arraystretch}{1.2} 
	\begin{tabular}{lcc}
		\toprule[1.5pt]
		& \multicolumn{1}{c}{\parbox{1.5cm}{\centering \textbf{IGEV} \cite{IgevDE}}} & \multicolumn{1}{c}{\parbox{1.5cm}{\centering \textbf{Hybrid}\\}}\\
		\midrule[1.2pt]
		\textbf{EPE} & 0.0817 & \textbf{0.0776} \\
		\textbf{bmp0.1} & 0.1489 & \textbf{0.1419} \\
		\textbf{bmp0.01} & 0.9428 & \textbf{0.9407} \\
		\textbf{bmp0.001} & 0.9880 & \textbf{0.9878} \\
		\textbf{RMSE} & 0.1518 & \textbf{0.1429} \\ 
		\bottomrule[1.5pt]
	\end{tabular} 
	\label{tab:HybridBlender}
	\vspace{-0.2cm}
\end{table} 
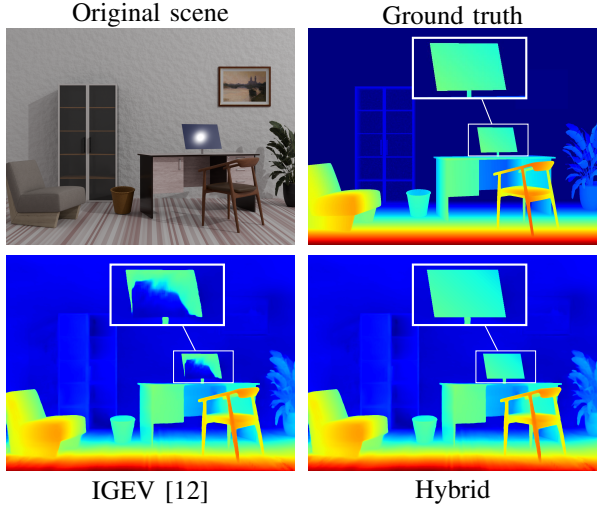
\begin{figure}[t]
	\centering
	\begin{tikzpicture}[spy using outlines={
			rectangle,                
			lens={scale=1.9},      
			width=1.5cm, height=0.8cm,           
			connect spies,         
			white
		}]
		
		\node(Scene)[]{\includegraphics[width=0.21\textwidth]{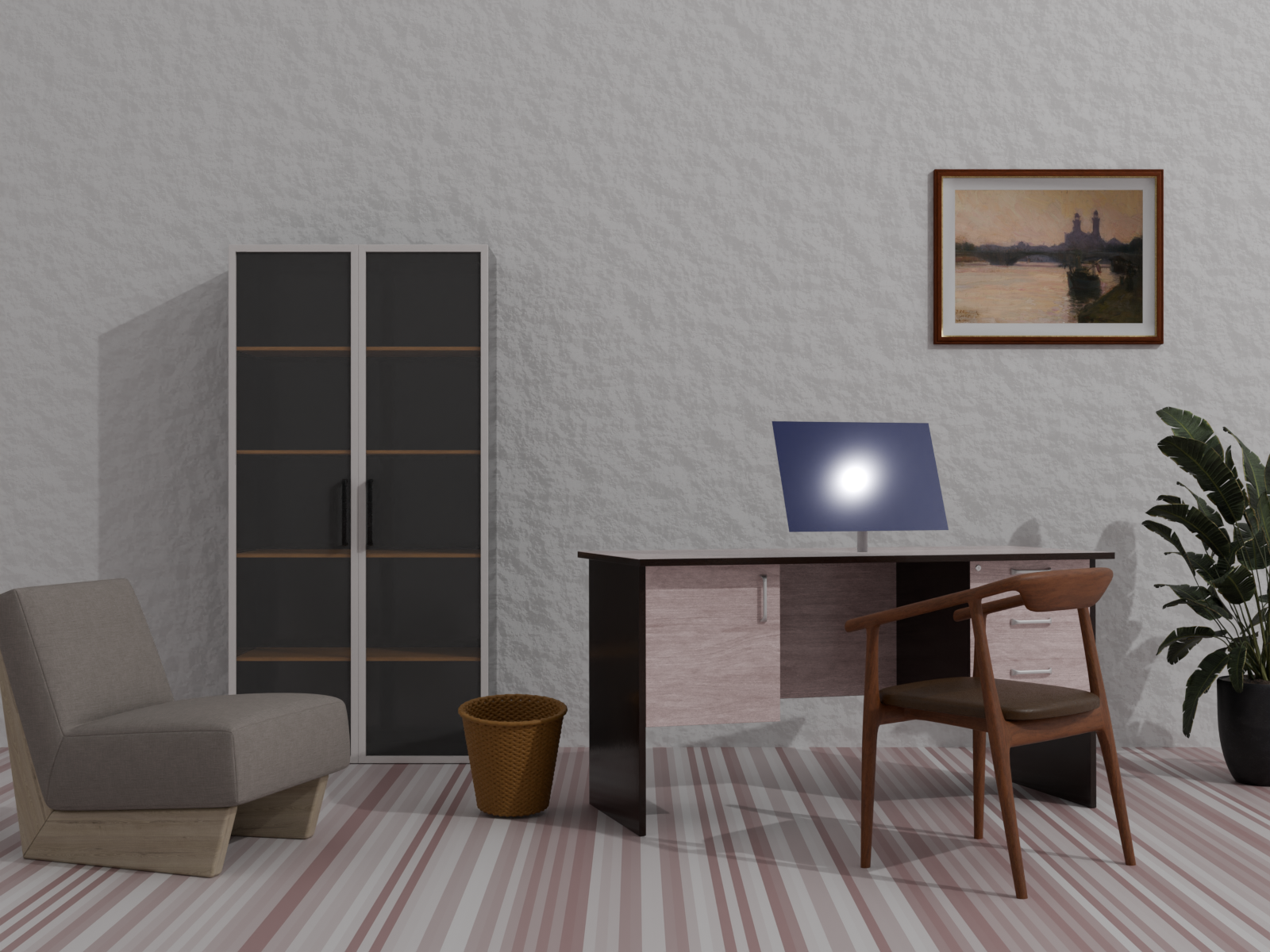}};
		\node(GT)[right of=Scene, xshift=3cm]{\includegraphics[width=0.21\textwidth]{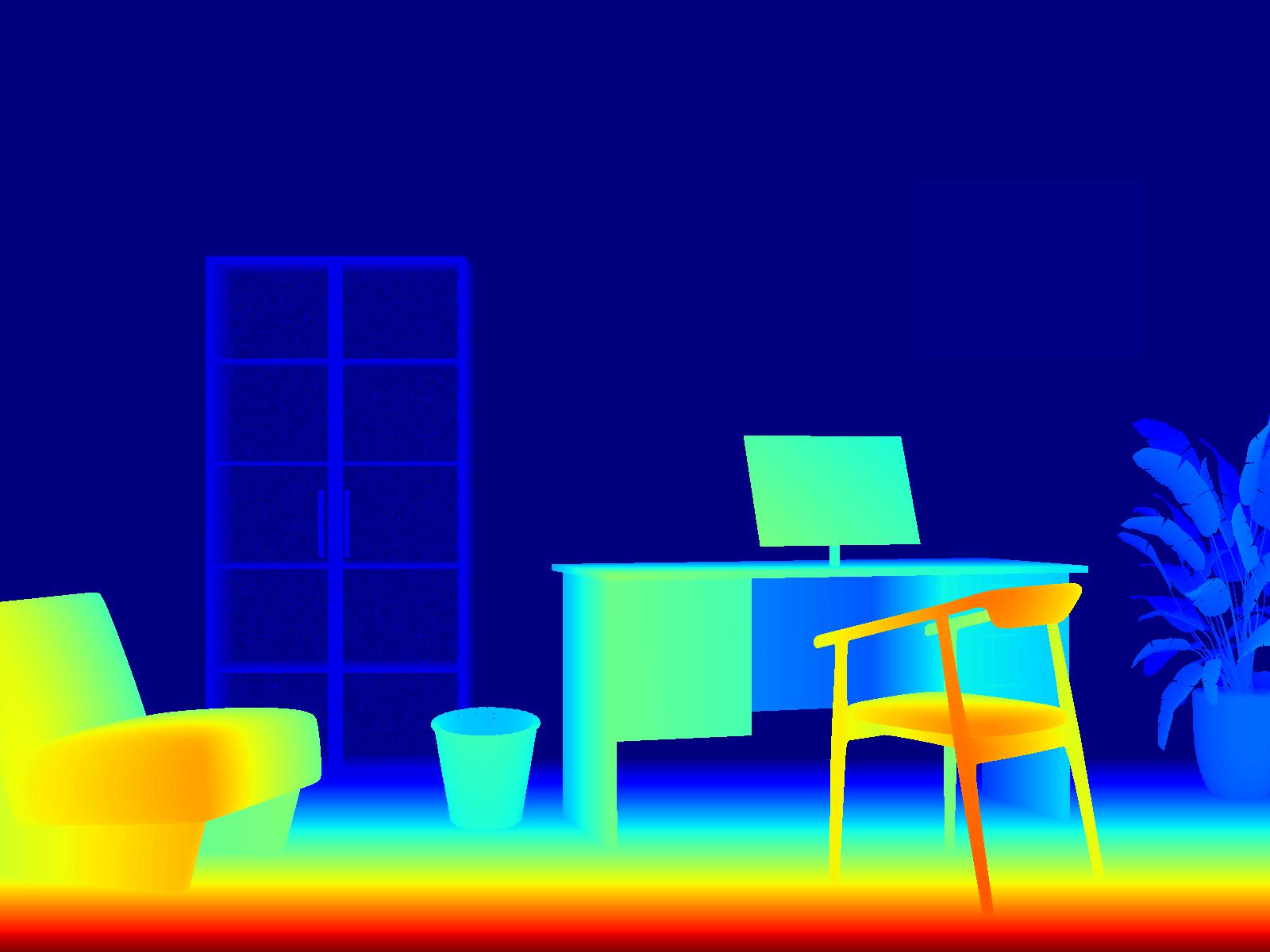}};
		\node(IGEV)[below of=Scene, yshift=-2cm]{\includegraphics[width=0.21\textwidth]{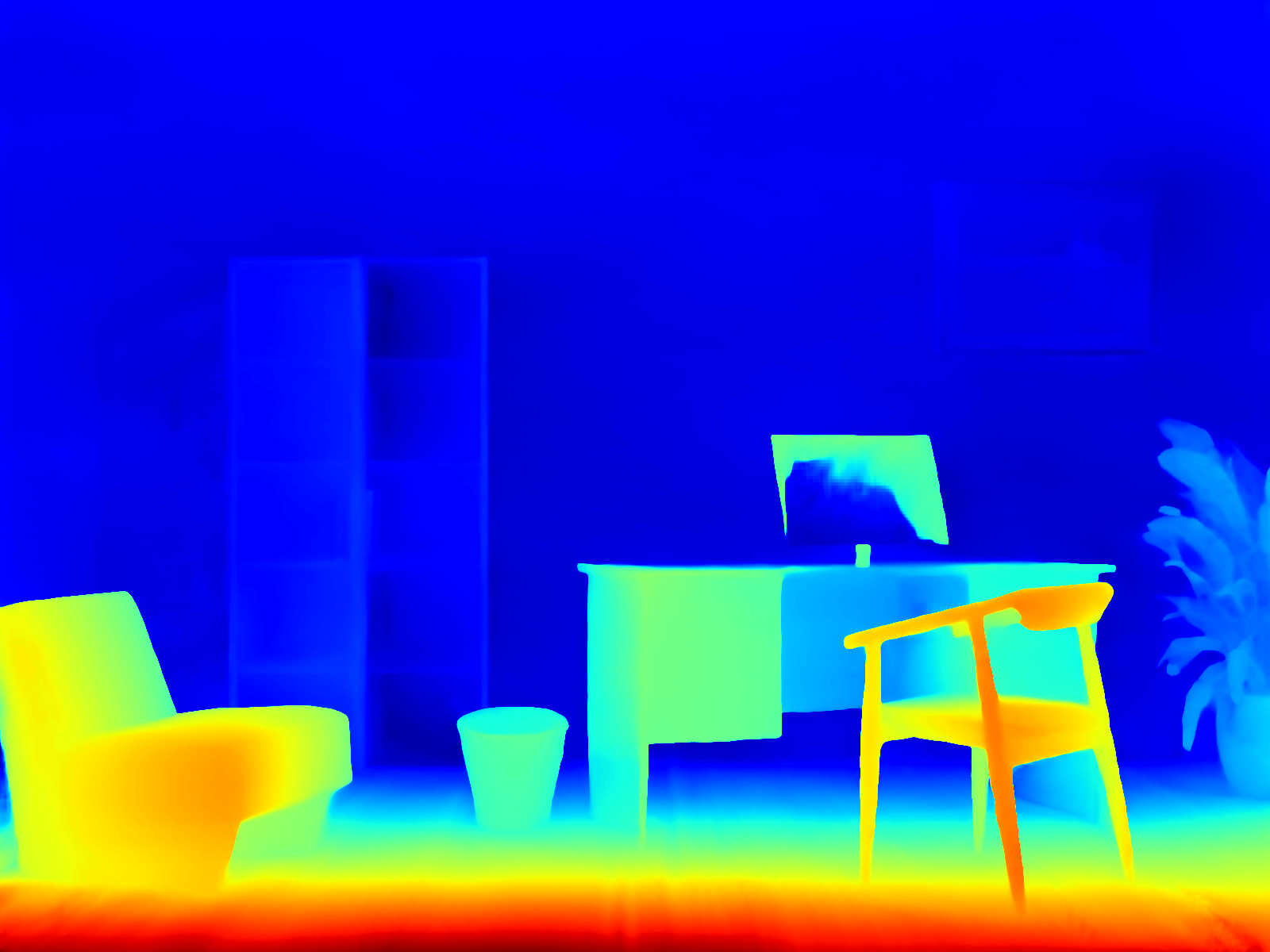}};
		\node(Hybrid)[right of=IGEV, xshift=3cm]{\includegraphics[width=0.21\textwidth]{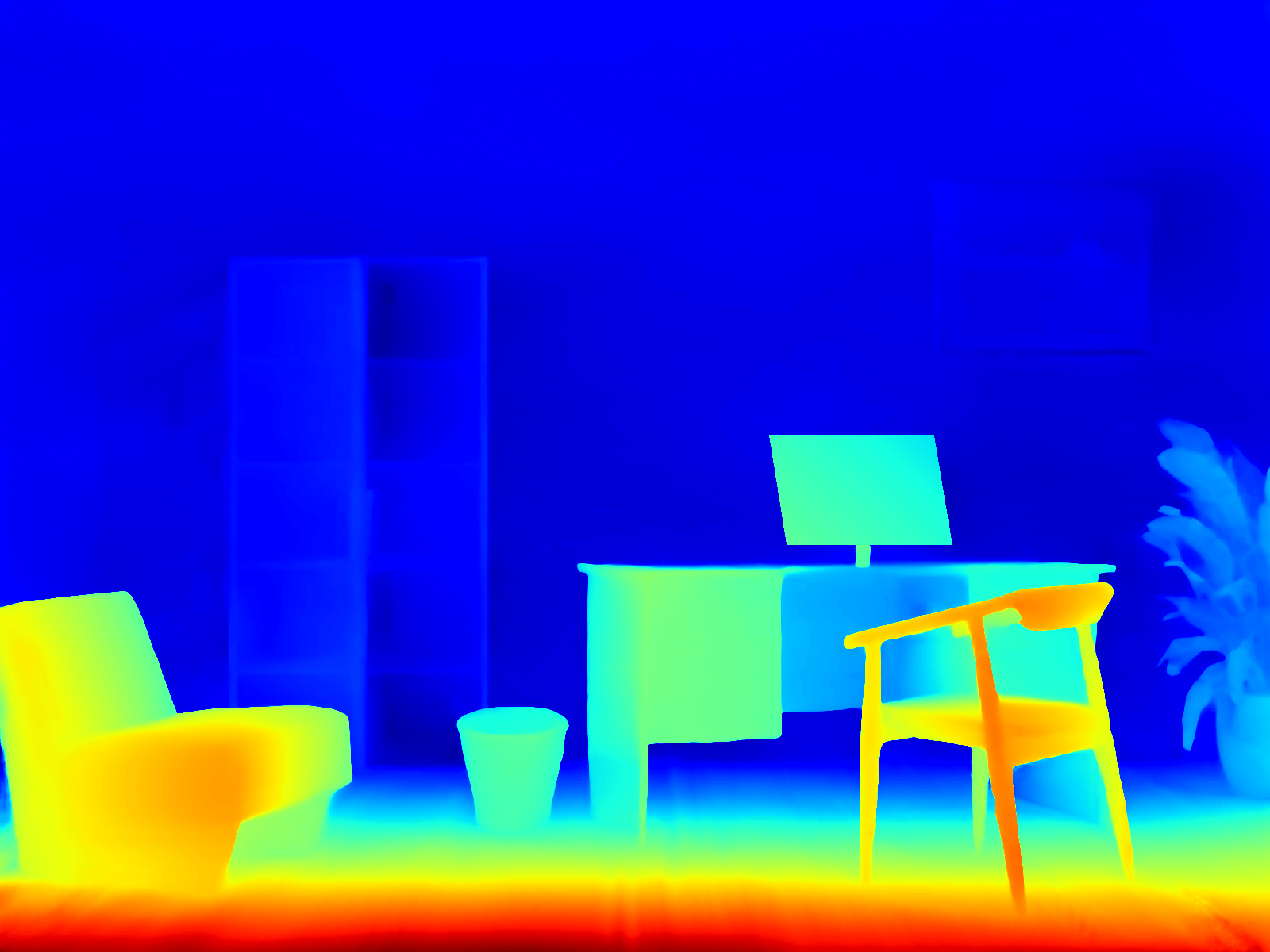}};
		\node(os)[above of=Scene, yshift=0.6cm]{Original scene};
		\node(gt)[above of=GT, yshift=0.6cm]{Ground truth};
		\node(igev)[below of=IGEV, yshift=-0.7cm]{IGEV \cite{IgevDE}};
		\node(hyb)[below of=Hybrid, yshift=-0.7cm]{Hybrid};
		
		\spy on ([xshift=0.6cm, yshift=-0.05cm]GT.center)    
		in node [left] at ([xshift=3cm, yshift=-0.65cm]GT.north west);
		\spy on ([xshift=0.7cm, yshift=-0.05cm]IGEV.center) 
		in node [left] at ([xshift=3cm, yshift=-0.65cm]IGEV.north west);
		\spy on ([xshift=0.7cm,yshift=-0.05cm]Hybrid.center)
		in node [left] at ([xshift=3cm, yshift=-0.65cm]Hybrid.north west);
	\end{tikzpicture}
	\vspace{-0.2cm}
	\caption{Qualitative disparity maps of a synthetic scene (top left). The ground truth (top right) is displayed along with the results for IGEV \cite{IgevDE} (bottom left) and the hybrid approach (bottom right). Circled insets magnify a detail area to show fine structural differences.}
	\label{fig:IntegrationExample}
	\vspace{-0.4cm}
\end{figure}
\noindent A key advantage of the proposed SRDE approach lies in its compatibility with existing disparity estimation models. Since SRDE operates on the geometric properties of specular reflections and requires only the reflective-region mask in addition to the camera array input images, it can be incorporated into neural-network-based pipelines with minimal architectural modifications.
			
To demonstrate this flexibility, we integrated SRDE as a complementary module within the state-of-the-art neural network IGEV \cite{IgevDE}. The pipeline is shown in Fig. \ref{fig:Integration}. In regions identified as the flat, reflective object using RePoSeg \cite{RePoSeg}, disparity values are computed using SRDE. For all remaining pixels, which lie outside the object mask, the disparity prediction of IGEV \cite{IgevDE} is used. This hybrid strategy leverages the strengths of both components, where standard neural networks handle textured, geometrically complex image regions and SRDE concentrates on the specular reflection objects. Importantly, the integration does not require retraining of the backbone network or modifications to its inference procedure, making SRDE a practical drop-in enhancement for existing stereo and multi-view systems.
	
To quantify how much overall gain this brings for whole images, we evaluated the hybrid pipeline on synthetic scenes generated with Blender \cite{Blender}. These scenes feature highly textured backgrounds coupled with a reflective foreground object. Our experiments show that this combined approach yields consistent improvements, as detailed in Table \ref{tab:HybridBlender}. The integration of SRDE reduces the overall scene End Point Error on average by 5.0\% compared to the standalone IGEV \cite{IgevDE} network. A visual representation of the improvement is provided in Fig. \ref{fig:IntegrationExample}. While the disparity of the reflective monitor is incorrectly estimated by IGEV \cite{IgevDE} alone, the hybrid approach corrects these errors by applying SRDE explicitly to reflective regions while leveraging IGEV \cite{IgevDE} elsewhere.

\vspace{-0.4cm}
\subsection{Model Limitations}
\label{sec:Model Assumption Violations}
\noindent As SRDE relies on specific geometric and physical premises, its operational boundaries can be systematically categorized into detection, geometric, and illumination constraints. If these conditions are violated, the underlying geometric model can no longer be applied for disparity estimation.

\subsubsection{Detection Boundaries}
\label{sec:Detection Boundaries}
For calculating the disparity maps using SRDE, the position of the light source has to be known and should (ideally) not deviate from the real position. Hence, the center of the specular reflection must be detected as accurately as possible. Weak specular reflections can be processed successfully as long as their gradient magnitudes exceed the adaptive gradient threshold $\vartheta$. However, if the reflection is too weak or diffuse to be distinguished from the background, accurate center detection fails. Consequently, the geometric model cannot be applied and disparity estimation fails.

\subsubsection{Geometric Boundaries}
\label{sec:Geometric Boundaries}
Another assumption of SRDE is a single dominant specular reflection on the object surface that can be approximated by a single Euclidean plane. Small surface irregularities are tolerable, as long as the surface can be reasonably approximated by a single plane. If the object has a slight continuous curvature but is still treated as flat, the assumption will lead to a systematic disparity error. This error increases towards the object's edges, as the modeled plane diverges from the true curved surface. For small objects, the method remains theoretically applicable, but only if the specular reflection is fully contained within the object's surface. If the entire object is illuminated to the point where no reflection boundary is visible on the surface, disparity estimation fails.
\begin{figure*}[t!]
	\centering
	\begin{tikzpicture}
		\node(rgb1)[]{\includegraphics[width=0.15\textwidth]{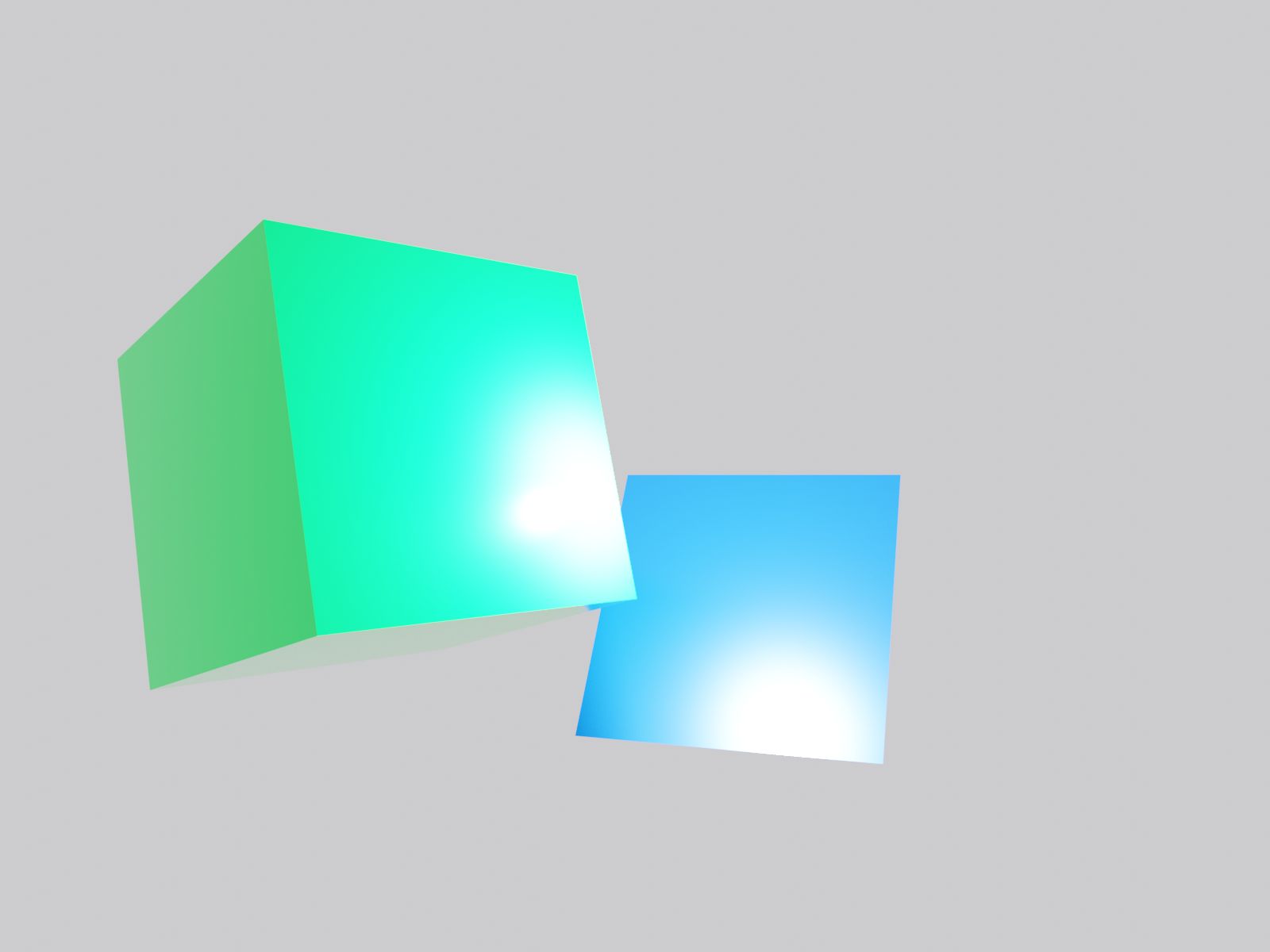}};
		\node(rgb2)[below of=rgb1, yshift=-1.1cm]{\includegraphics[width=0.15\textwidth]{Images/Sphere1.jpg}};
		\node(rgb3)[below of=rgb2,yshift=-1.12cm]{\includegraphics[width=0.15\textwidth]{Images/Cube3.jpg}};
		
		\node(gt1)[right of=rgb1, xshift=1.8cm]{\includegraphics[width=0.15\textwidth]{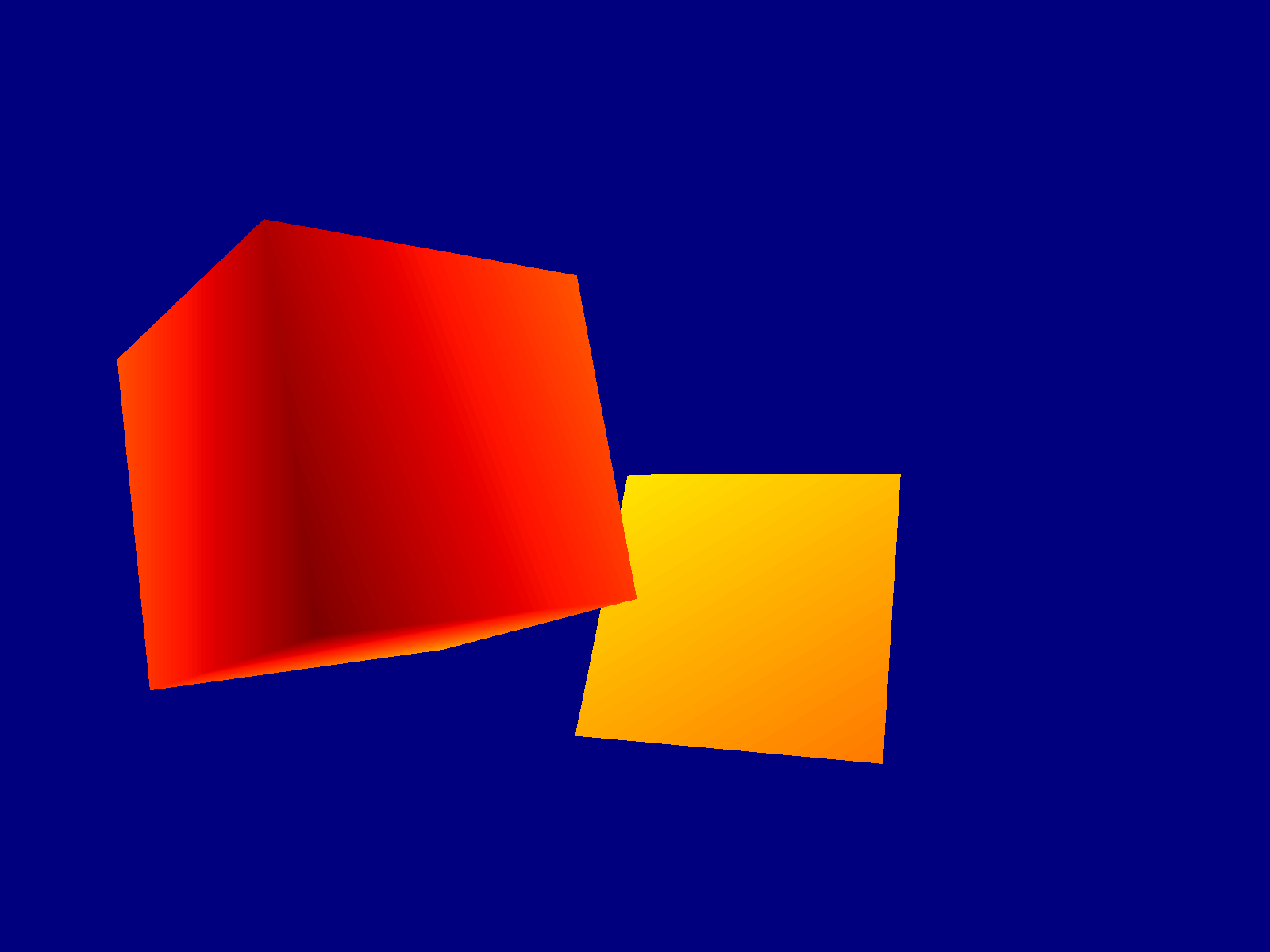}};
		\node(gt2)[right of=rgb2,xshift=1.8cm]{\includegraphics[width=0.15\textwidth]{Images/GT_Sphere1.jpg}};
		\node(gt3)[right of=rgb3, xshift=1.8cm]{\includegraphics[width=0.15\textwidth]{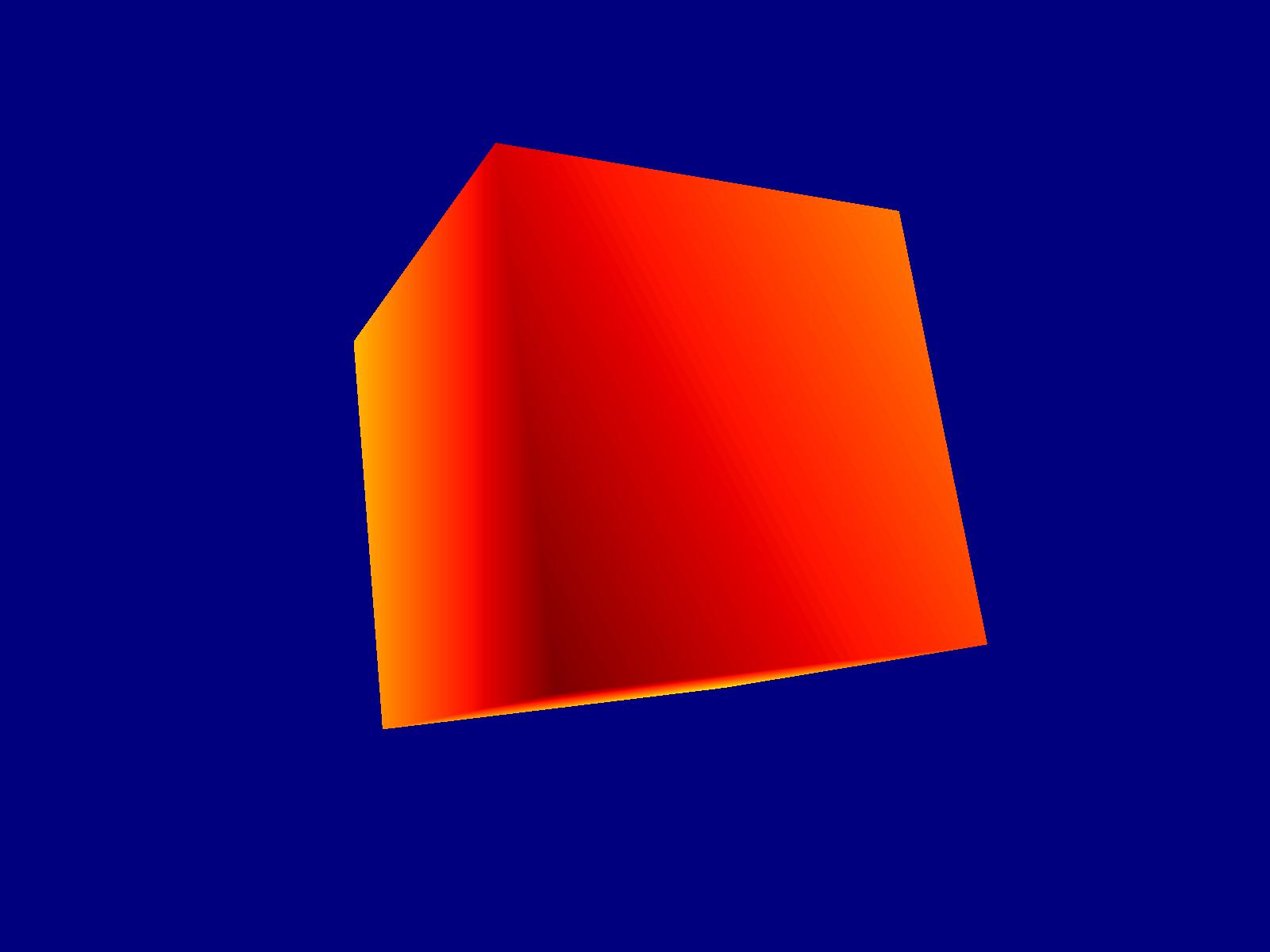}};
		
		\node(NMRF1)[right of=gt1,xshift=1.8cm]{\includegraphics[width=0.15\textwidth]{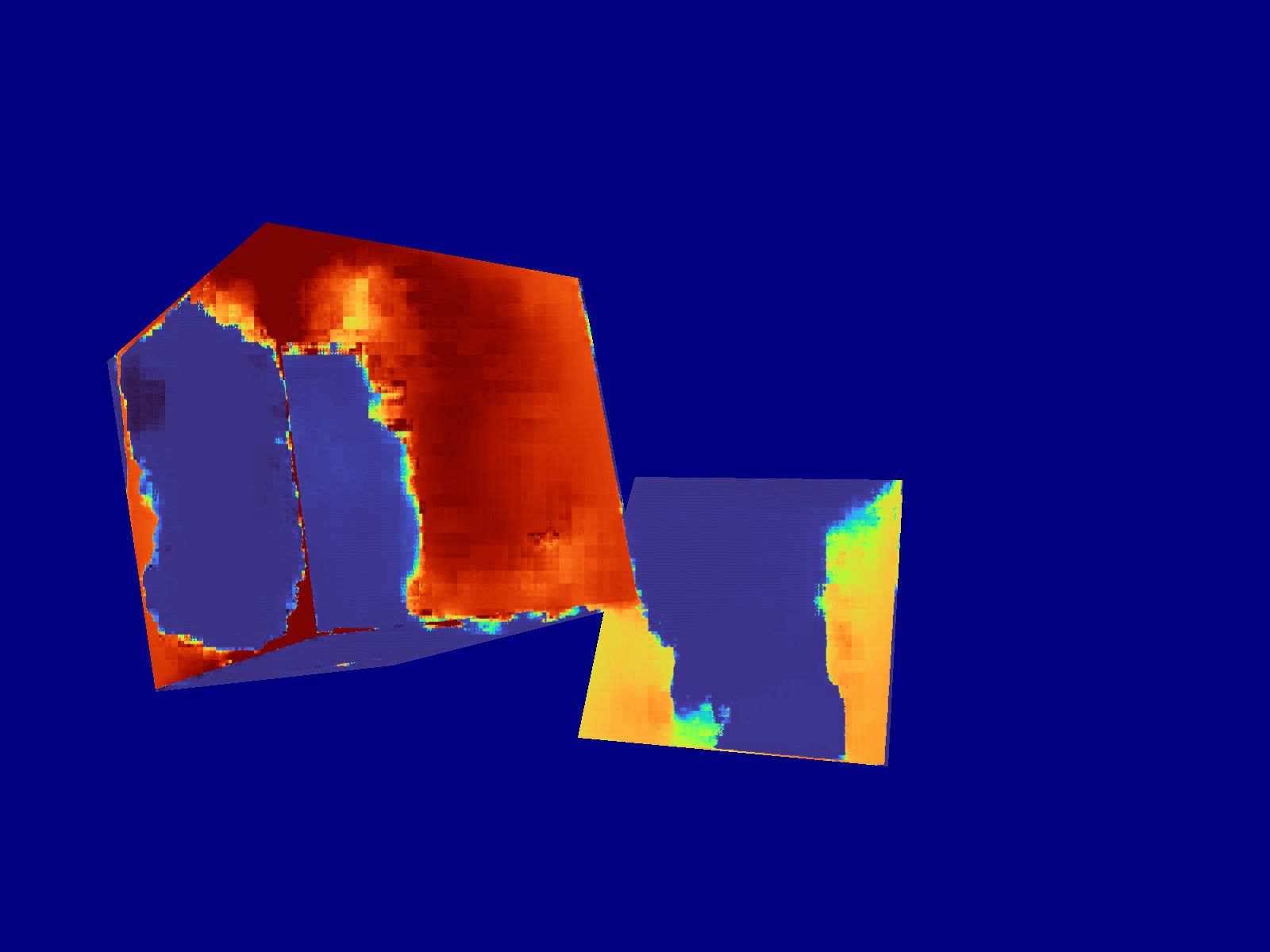}};
		\node(NMRF2)[right of=gt2,xshift=1.8cm]{\includegraphics[width=0.15\textwidth]{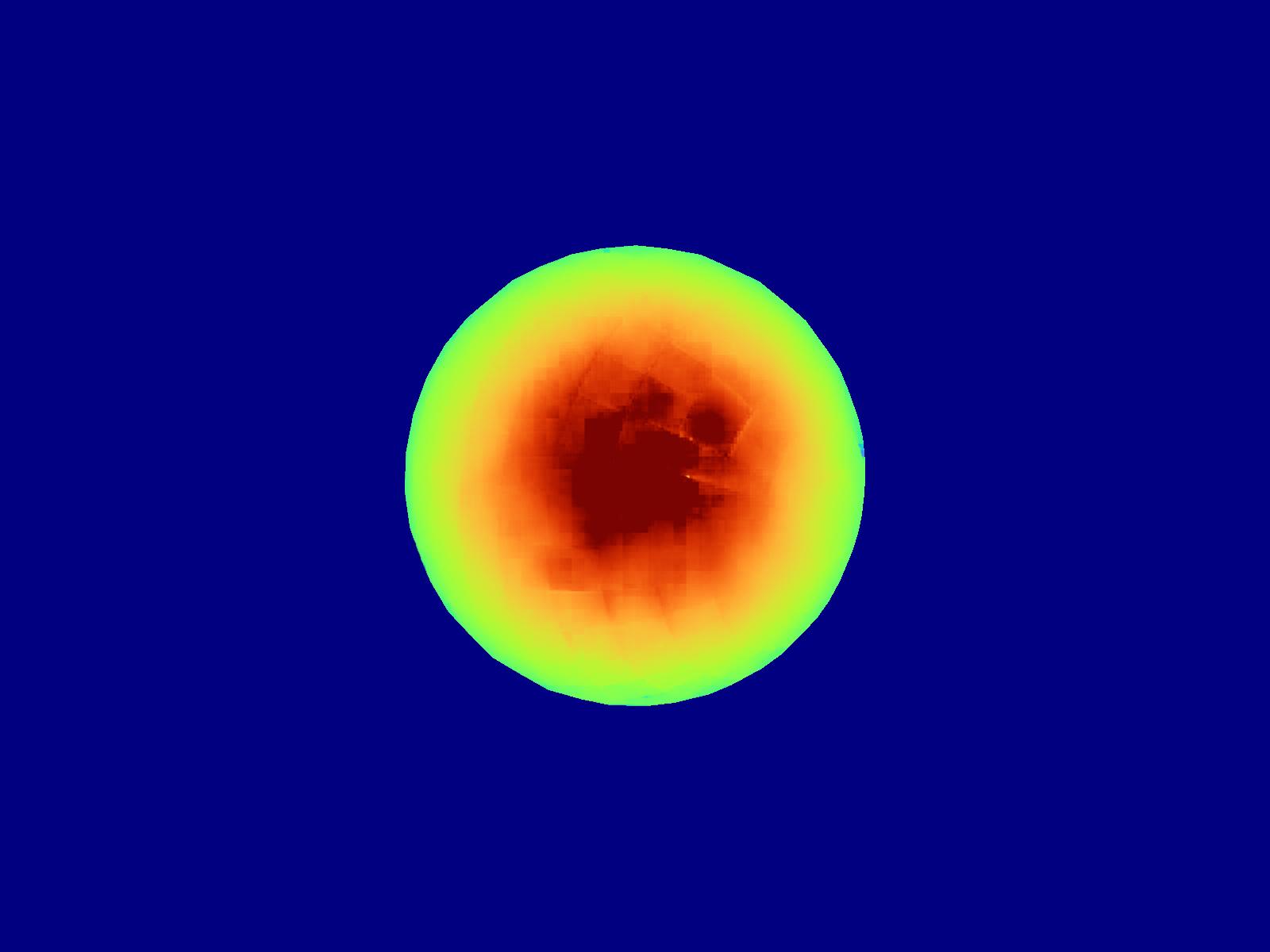}};
		\node(NMRF3)[right of=gt3,xshift=1.8cm]{\includegraphics[width=0.15\textwidth]{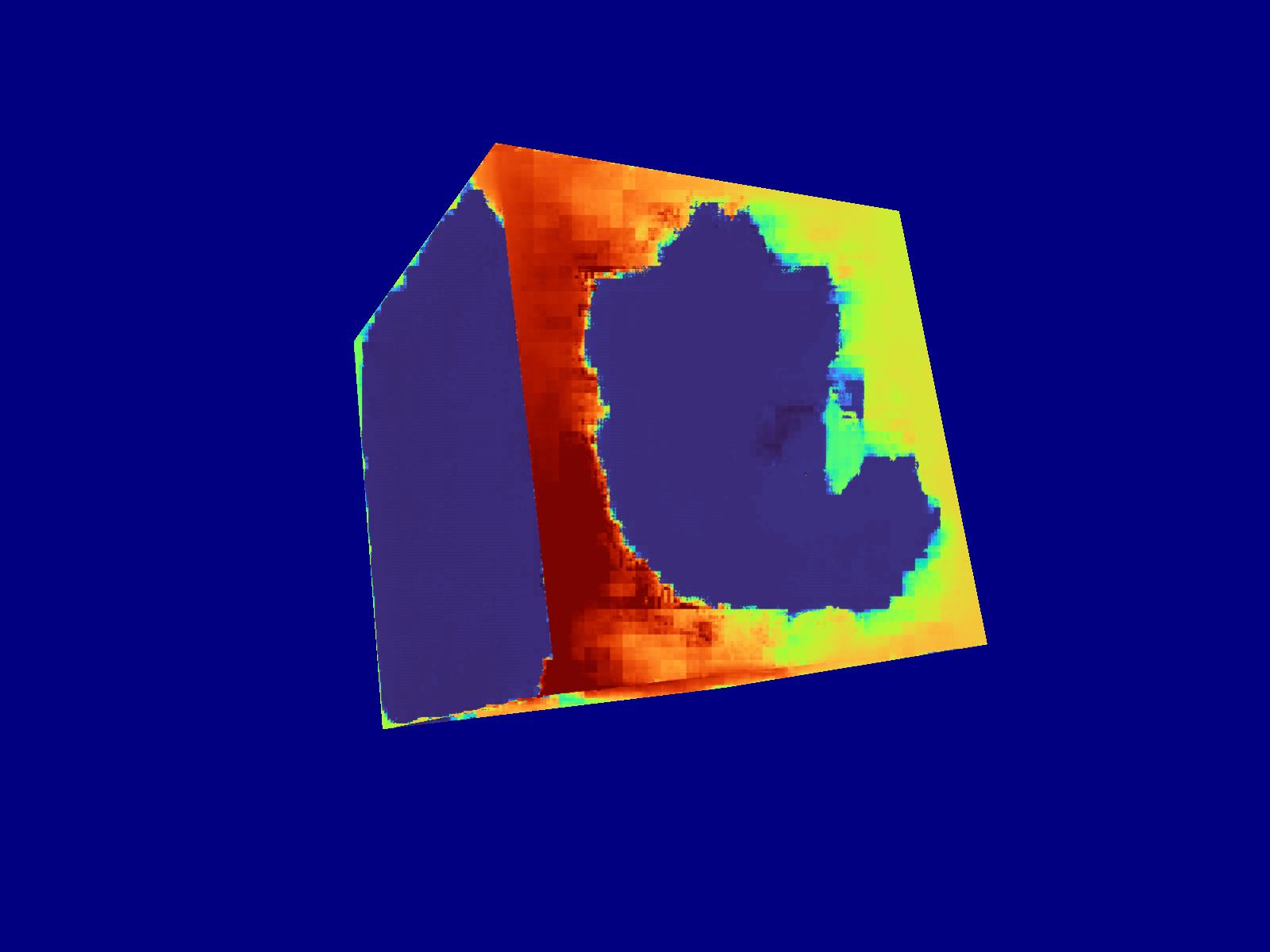}};
		
		\node(CRE1)[right of=NMRF1,xshift=1.8cm]{\includegraphics[width=0.15\textwidth]{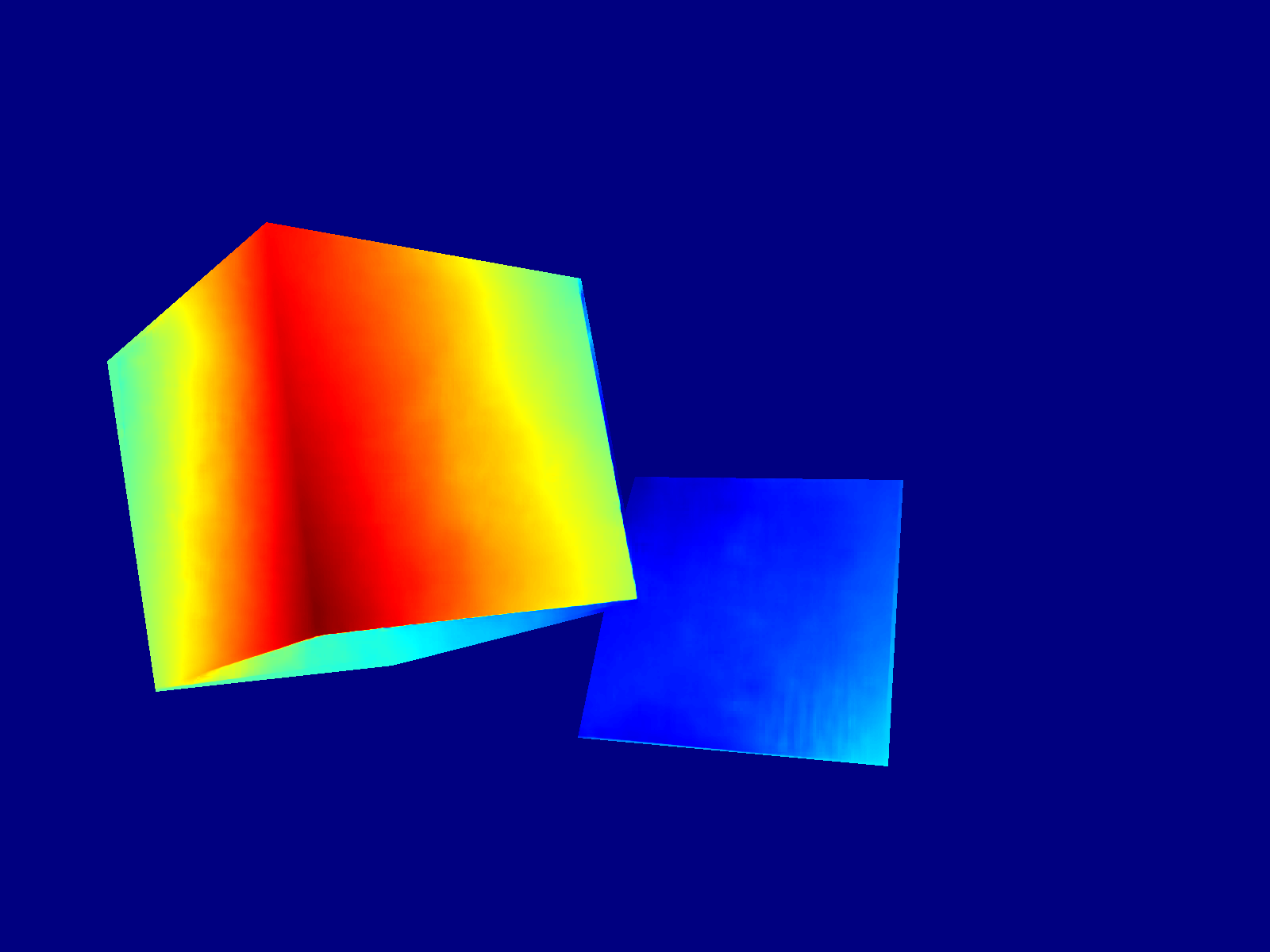}};
		\node(CRE2)[right of=NMRF2,xshift=1.8cm]{\includegraphics[width=0.15\textwidth]{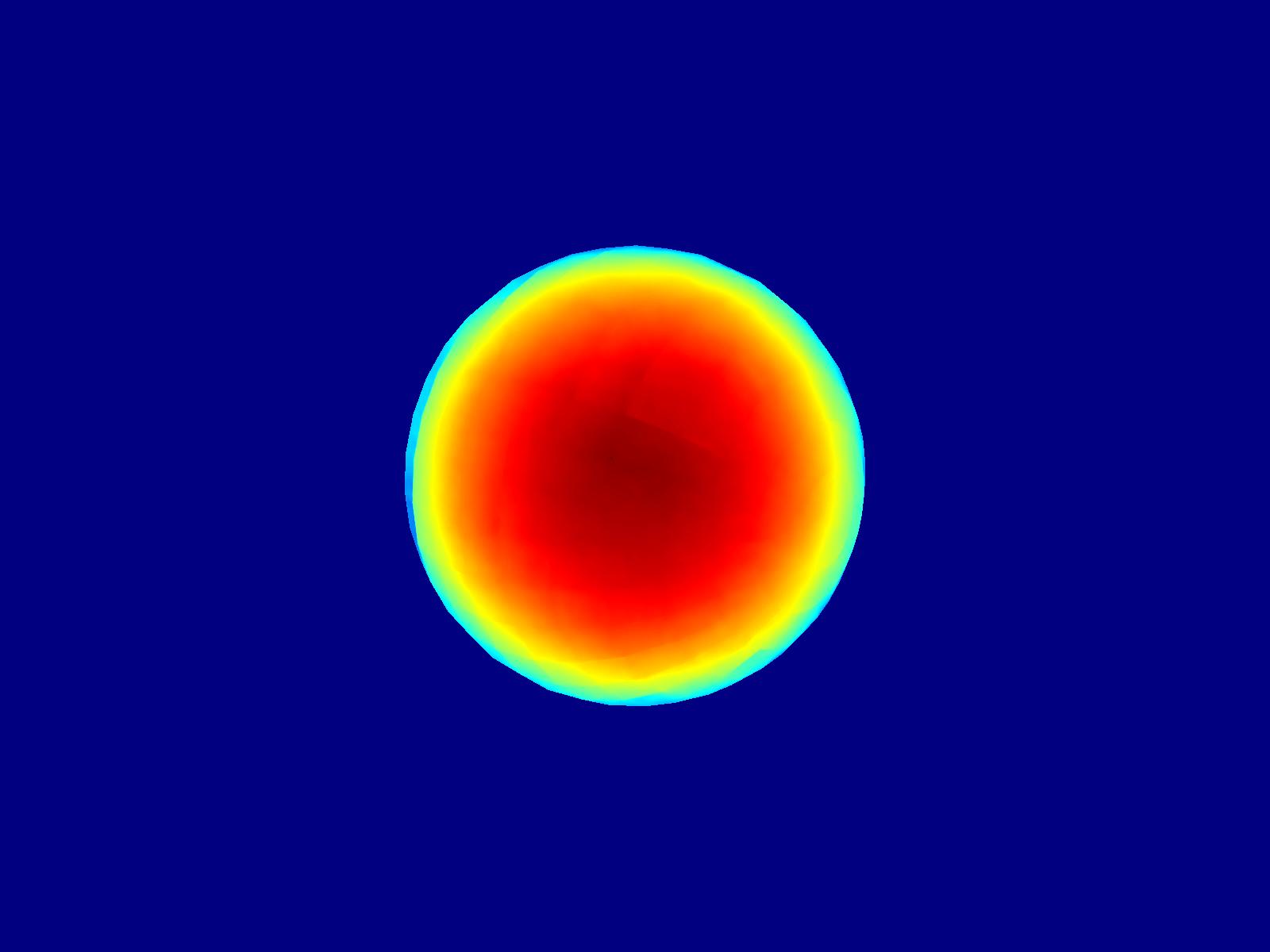}};
		\node(CRE3)[right of=NMRF3,xshift=1.8cm]{\includegraphics[width=0.15\textwidth]{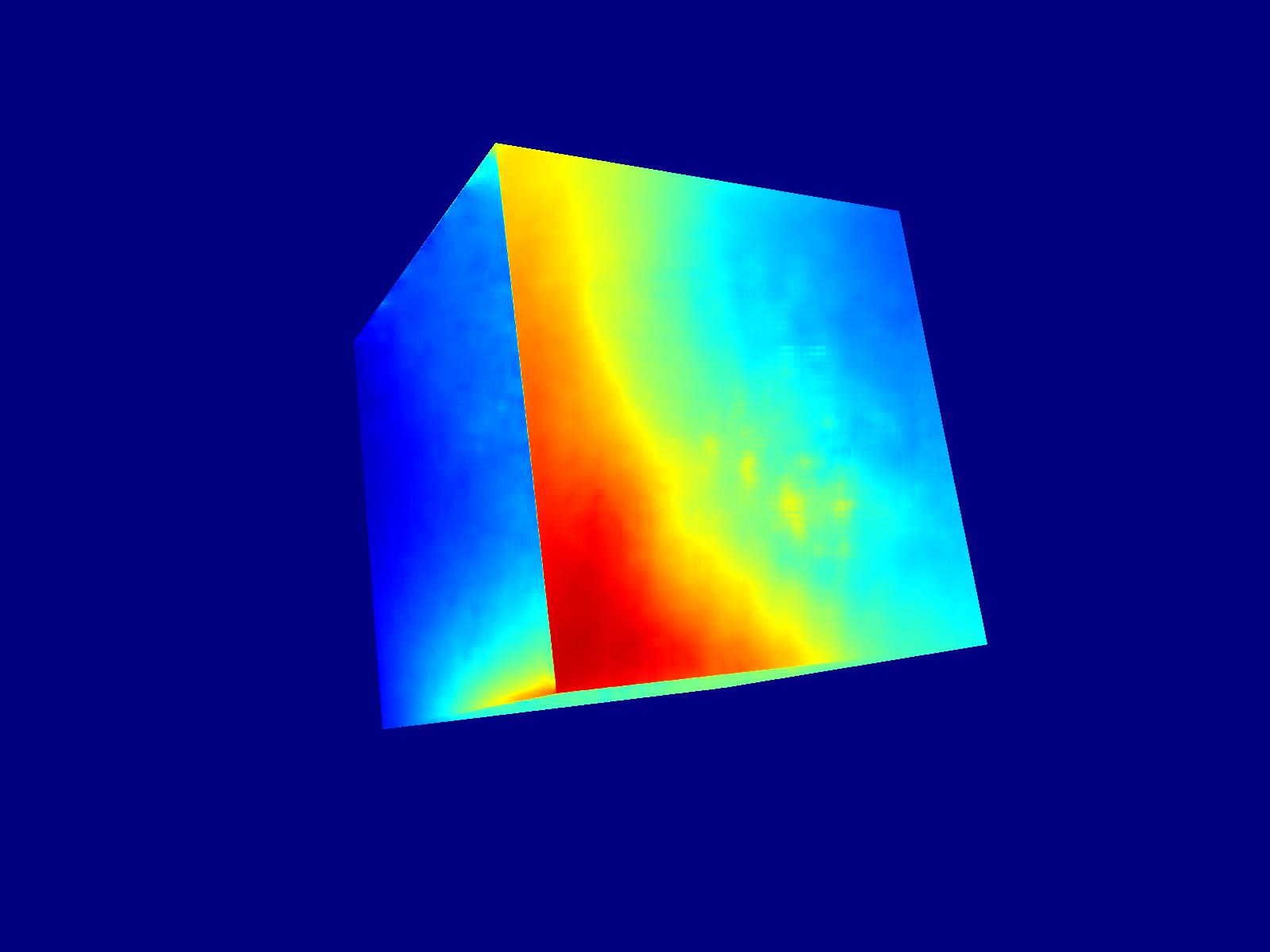}};
		
		\node(IGEV1)[right of=CRE1,xshift=1.8cm]{\includegraphics[width=0.15\textwidth]{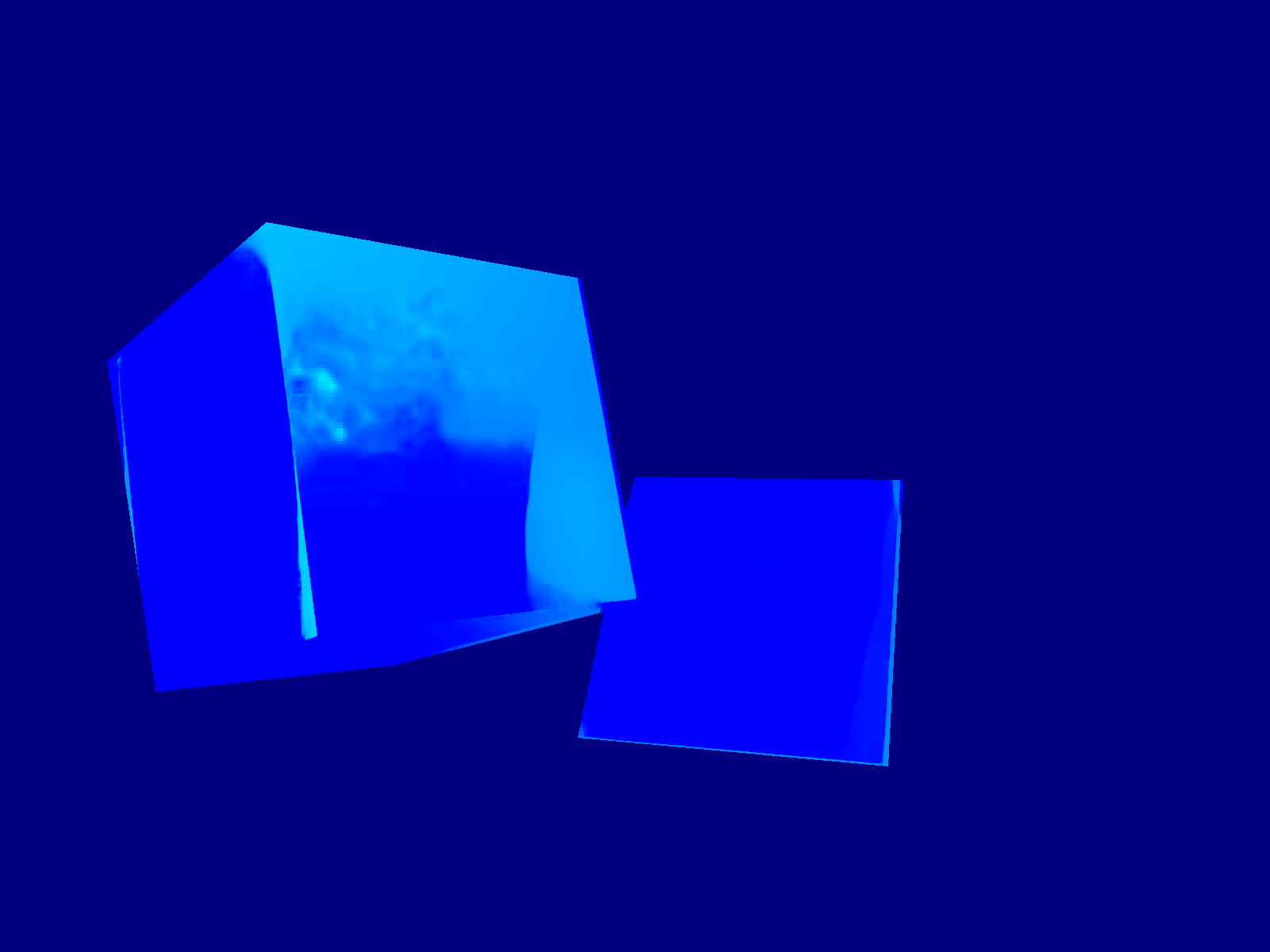}};
		\node(IGEV2)[right of=CRE2,xshift=1.8cm]{\includegraphics[width=0.15\textwidth]{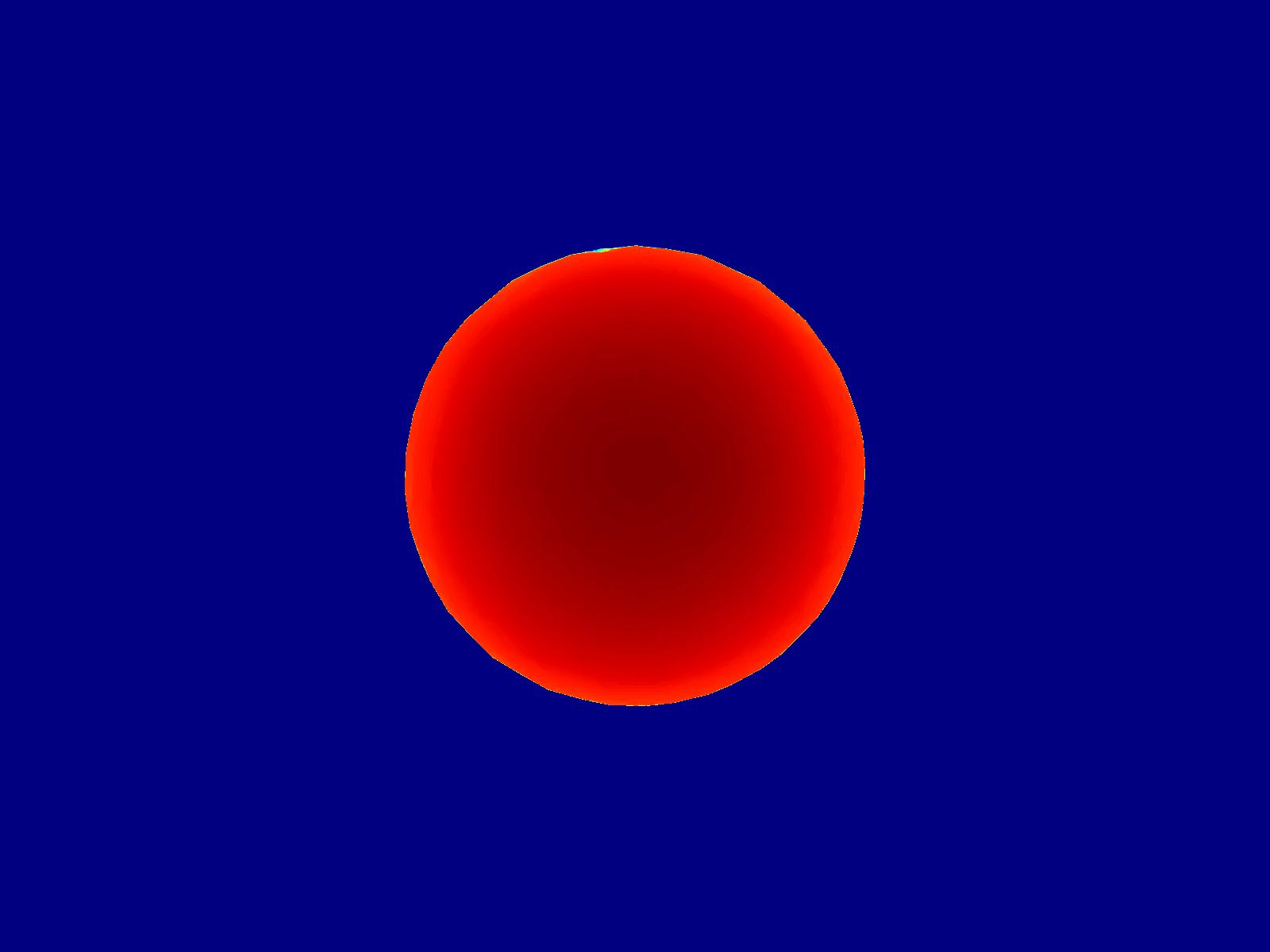}};
		\node(IGEV3)[right of=CRE3,xshift=1.8cm]{\includegraphics[width=0.15\textwidth]{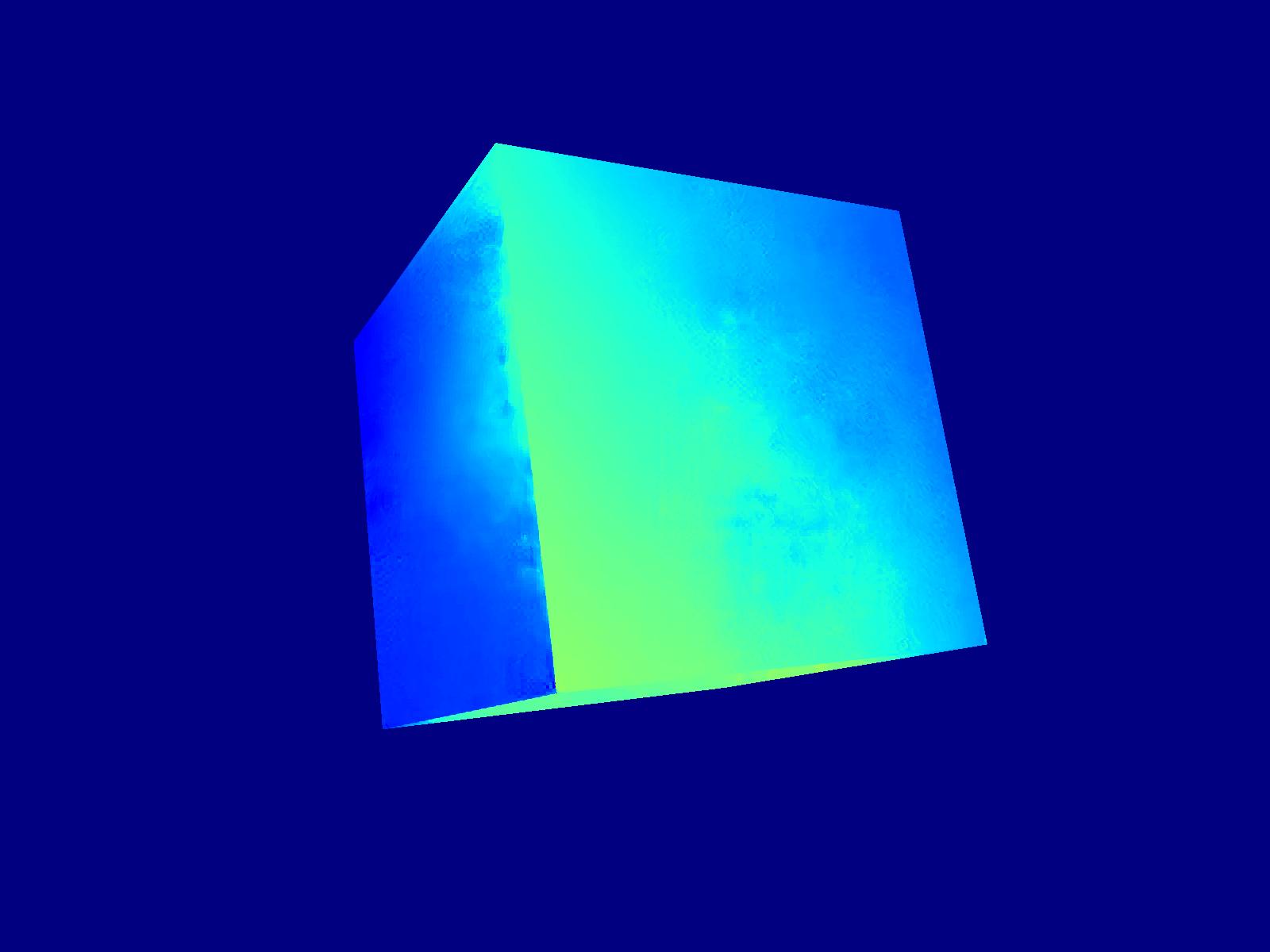}};
		
		\node(Ours1)[right of=IGEV1,xshift=1.8cm]{\includegraphics[width=0.15\textwidth]{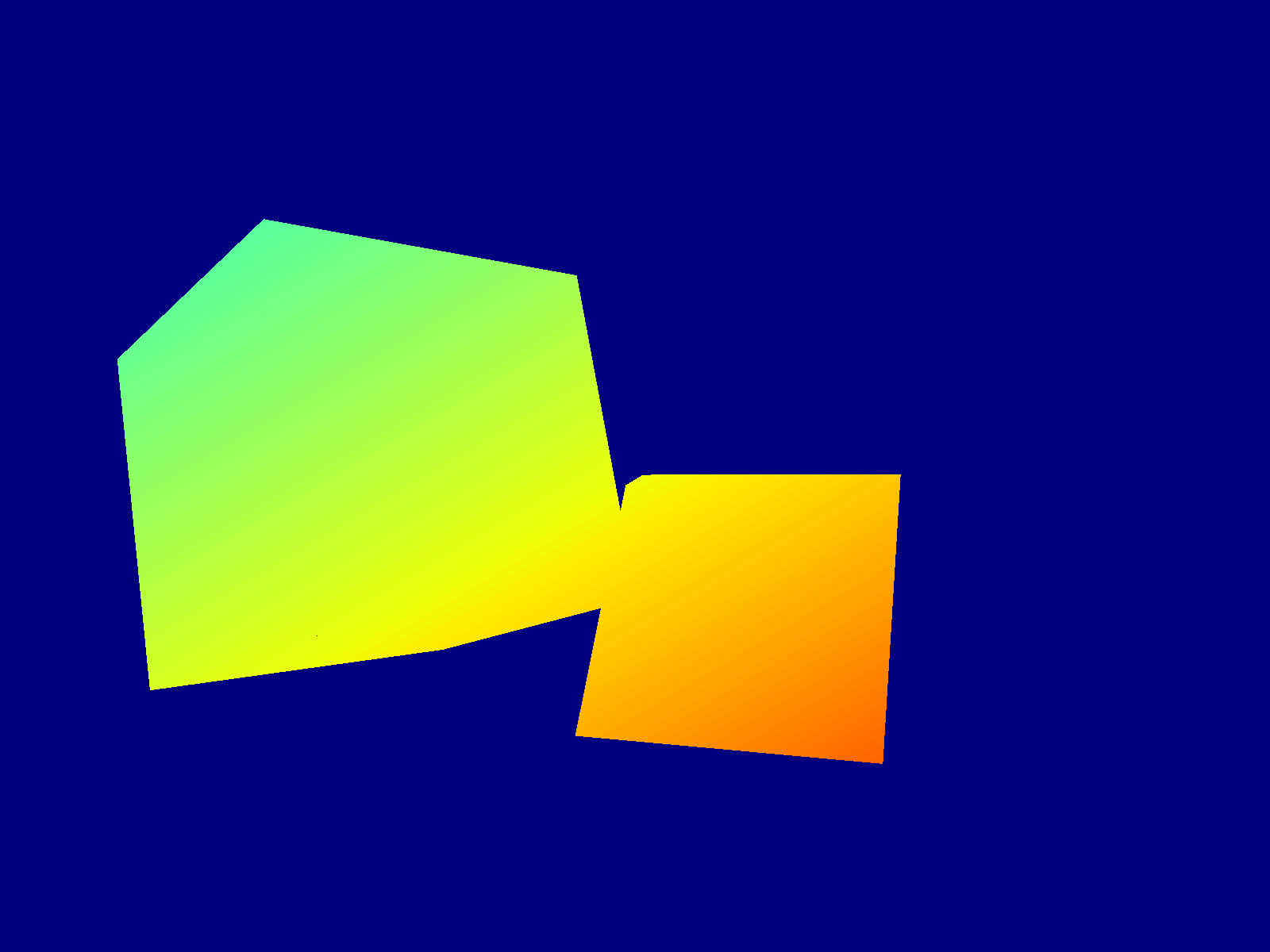}};
		\node(Ours2)[right of=IGEV2,xshift=1.8cm]{\includegraphics[width=0.15\textwidth]{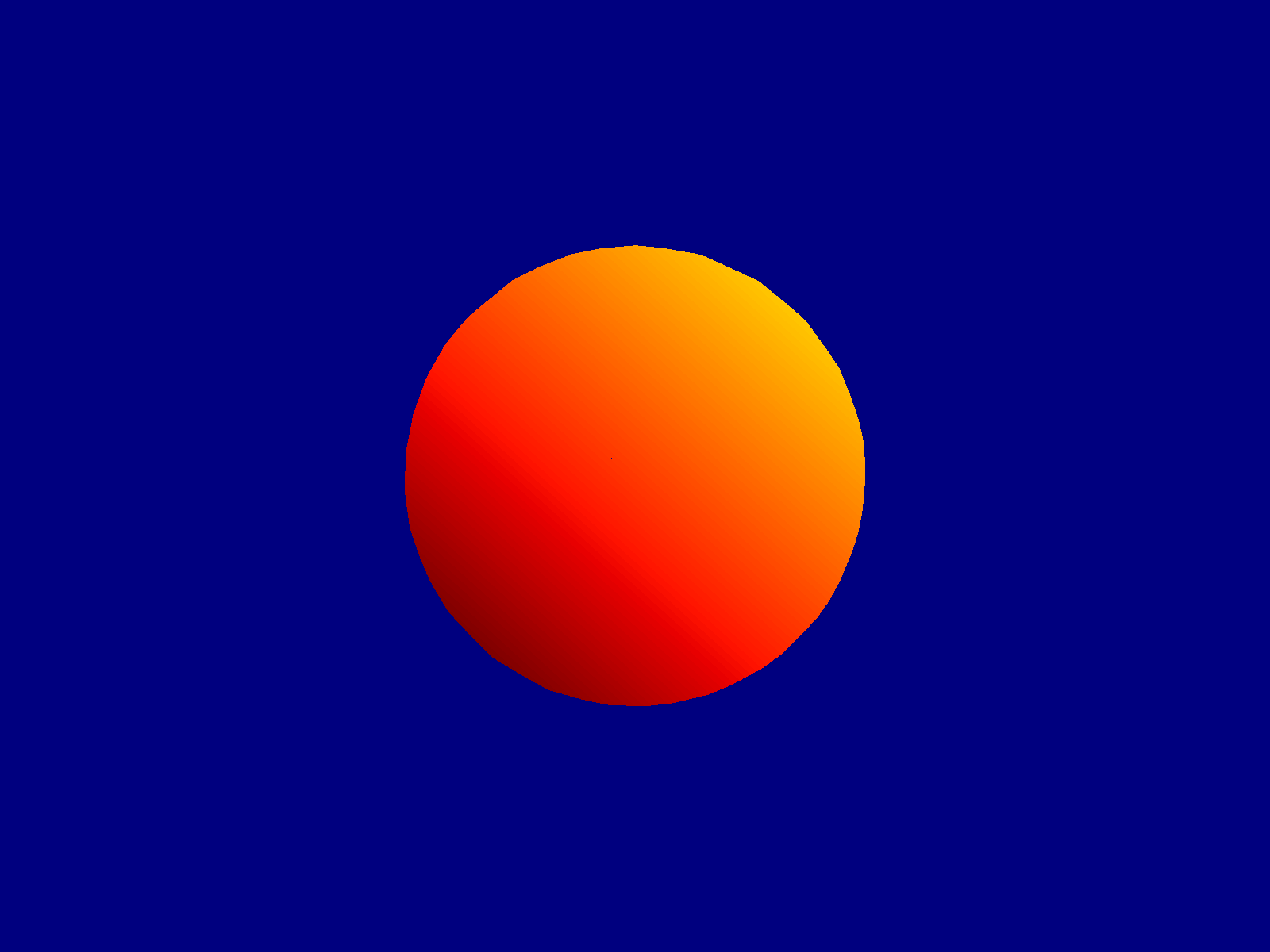}};
		\node(Ours3)[right of=IGEV3,xshift=1.8cm]{\includegraphics[width=0.15\textwidth]{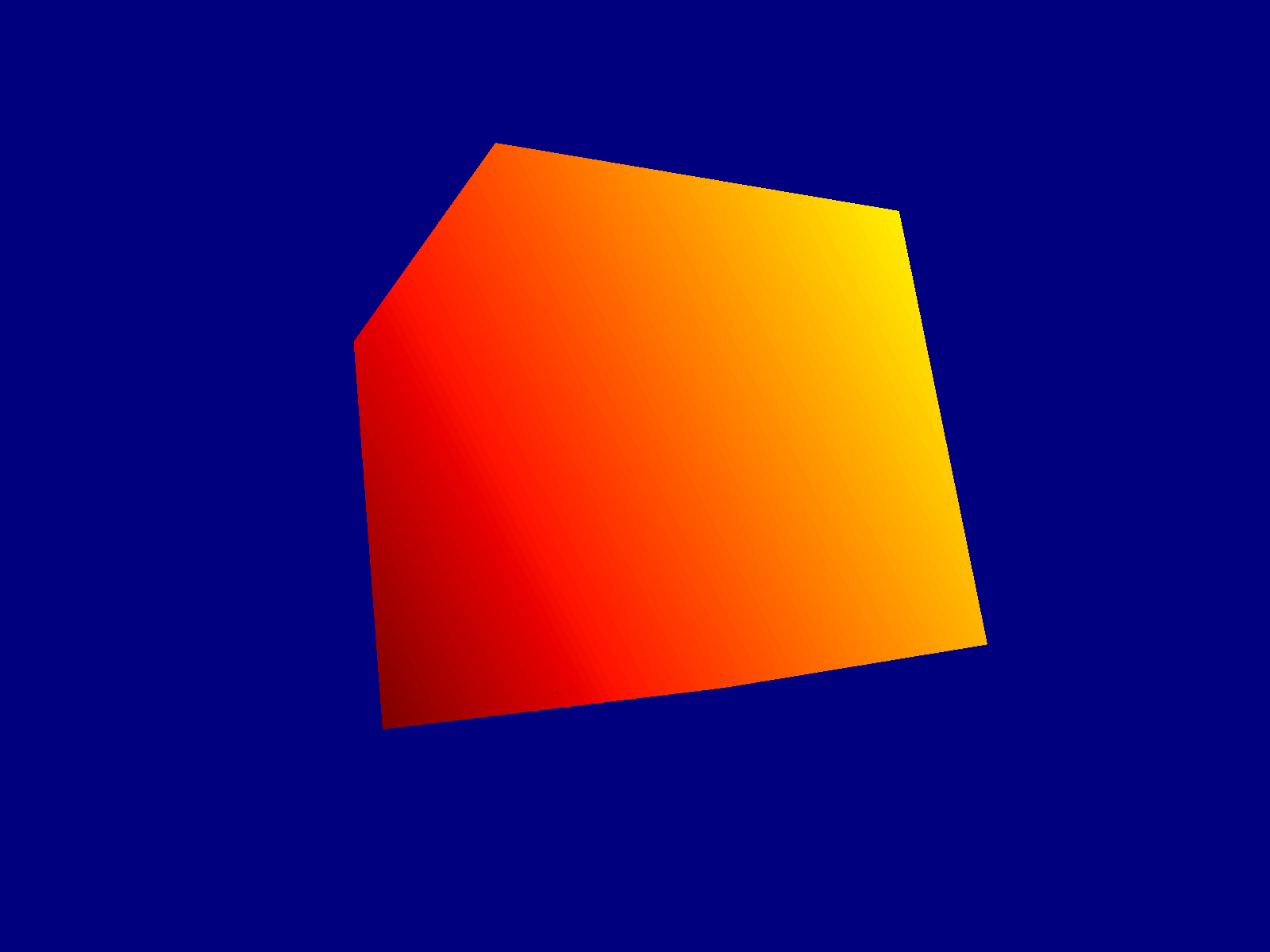}};
		
		\node(title1)[above of=rgb1,yshift=0.2cm]{Input image};
		\node(title2)[above of=gt1,yshift=0.2cm]{Ground truth};
		\node(title3)[above of=NMRF1,yshift=0.2cm]{NMRF \cite{NMRF}};
		\node(title4)[above of=CRE1,yshift=0.2cm]{CREStereo \cite{CREStereo}};
		\node(title5)[above of=IGEV1,yshift=0.2cm]{IGEV \cite{IgevDE}};
		\node(title6)[above of=Ours1,yshift=0.2cm]{SRDE};
	\end{tikzpicture}
	\vspace{-0.2cm}
	\caption{Qualitative disparity map results on multiple objects (top row), spherical (middle row), and cubic (bottom row) non-planar synthetic data. Shown are the input image, ground truth, and disparity maps produced by NMRF \cite{NMRF}, CREStereo \cite{CREStereo}, IGEV \cite{IgevDE} and SRDE (ours). Colors are normalized to depth.}
	\label{fig:ResultImagesDispAssumpVio}
	\vspace{-0.4cm}
\end{figure*}

Scenes containing multiple objects or non-planar surfaces, such as cubes or spheres, introduce ambiguities that violate the single global plane assumption. Although masking techniques like RePoSeg \cite{RePoSeg} can help isolate individual flat objects, accurately modeling complex, curved geometries requires multiple light sources and a more complex reconstruction strategy to capture multiple surface normals. An example is shown in Fig. \ref{fig:ResultImagesDispAssumpVio} in the top row, where the flat rectangle is correctly estimated but the cube with two visible faces is not. Similarly, the disparity map of the sphere is also not correctly reconstructed. 
These geometric boundaries are also explicitly reflected in the quantitative results in Table \ref{tab:Violation}. As expected, the performance of SRDE decreases compared to flat surfaces (Table \ref{tab:ResultsBenchmarkModels}).  However, the performance of the evaluated learning-based approaches degrades as well. Even though these networks can theoretically exploit geometric edges for pixel matching, they still struggle with the lack of texture and the complexities of non-planar specular reflections. Therefore, this is not only a challenging condition for SRDE, but remains an open research problem across the entire field.

\begin{table}[t!]
	\centering
	\caption{Quantitative evaluation of pretrained benchmark models on multiple objects (MO) and non-planar (NP) synthetic objects. \textbf{Bold:} Best.}
	\vspace{-0.2cm}
	\renewcommand{\arraystretch}{1.2}
	\begin{tabular}{llcccc}
		\toprule[1.5pt]
		& & \multicolumn{1}{c}{\parbox{1.1cm}{\centering \textbf{NMRF}\\\cite{NMRF}}} & \multicolumn{1}{c}{\parbox{1.1cm}{\centering \textbf{CREStereo}\\\cite{CREStereo}}} & \multicolumn{1}{c}{\parbox{1.1cm}{\centering \textbf{IGEV}\\\cite{IgevDE}}} & \multicolumn{1}{c}{\parbox{1.1cm}{\centering \textbf{SRDE}\\(Ours)}}\\
		\midrule[1.2pt]
		\multirow{5}{*}{MO}&\textbf{EPE} & 0.0853 & \textbf{0.0417} & 0.0654 & 0.0484 \\
		&\textbf{bmp0.1} & 0.2006 & \textbf{0.1064} & 0.1675 & 0.1605\\
		&\textbf{bmp0.01} & 0.2258 & 0.2160 & 0.2238 & \textbf{0.1912} \\
		&\textbf{bmp0.001} & 0.9998 & 0.9999 & 0.9978 & \textbf{0.1968} \\
		&\textbf{RMSE} & 0.1922 & 0.1143 & 0.1660 & \textbf{0.1036} \\ \midrule
		\multirow{5}{*}{NP}&\textbf{EPE} & 0.0533 & 0.0428 & \textbf{0.0416} & 0.0937 \\
		&\textbf{bmp0.1} & 0.1438 & 0.1318 & \textbf{0.1316} & 0.2171 \\
		&\textbf{bmp0.01} & 0.1901 & 0.1840 & \textbf{0.1823} & 0.2208\\
		&\textbf{bmp0.001} & 0.9960 & 0.9980 & 0.9957 & \textbf{0.2210} \\
		&\textbf{RMSE} & \textbf{0.1110} & 0.1284 & 0.1154 & 0.2093 \\
		\bottomrule[1.5pt]
	\end{tabular}
	\label{tab:Violation}
	\vspace{-0.5cm}
\end{table} 

\subsubsection{Illumination Boundaries}
\label{sec:Illumination Boundaries}
The current method is specifically designed for a single light source. Environments with complex illumination, such as ambient lighting or multiple light sources, violate this assumption and will lead to ambiguous or overlapping reflection shifts. Such scenes would require detecting multiple distinct specular reflections to compute several surface normals. This introduces a correspondence problem, as it must be determined which specular reflection originates from which light source. Thus, a more complex reconstruction strategy is required, and hence remains a subject for future work. 
				
\section{Conclusion and Future Work}
\label{sec:conclusion}
\noindent Despite the general success of neural networks, smooth textureless objects prone to specular reflections remain challenging for disparity estimation. In this work, we addressed this gap by introducing SRDE, which is specialized for flat, reflective surfaces. By detecting the positions of specular reflections, SRDE computes the normal vector and position of the object's plane, enabling the reconstruction of its 3D orientation. Thus, SRDE focuses on specific challenging cases rather than general applicability across all scene types. Furthermore, SRDE was integrated into the state-of-the-art disparity estimation network IGEV to form a hybrid approach that combines the strengths of model-based and learning-based methods. Experiments confirmed that SRDE improves disparity accuracy for flat objects with a single dominant specular reflection, while the hybrid framework extends its applicability to more complex scenes.
				
Future work will focus on extending SRDE to multi-faceted or curved surfaces. We will explore the integration of multiple distinct light sources. By capturing multiple specular reflections simultaneously, it would be possible to estimate several local surface normals and reconstruct curved geometries or multi-faceted objects more accurately. Furthermore, the robustness of the framework in scenarios involving low-light conditions or complex illumination will be investigated.
				
\vspace{-0.2cm}				
\subsection{Acknowledgments}
\noindent The authors gratefully acknowledge that this research was supported by the Bayerische Forschungsstiftung (BFS, Bavarian Research Foundation) under project number AZ-1547-22, and the Deutsche Forschungsgemeinschaft (DFG, German Research Foundation) under project number 491814627.
			
\bibliographystyle{IEEEtran}
\bibliography{SpecularDepth.bib}

\balance
			
\end{document}